\documentclass[11pt]{article}

\usepackage[T1]{fontenc}
\usepackage[utf8]{inputenc}
\usepackage{textcomp}
\usepackage{tgtermes}

\usepackage[version=3]{mhchem}
\usepackage{authblk}
\usepackage[colorlinks=false]{hyperref}
\usepackage[capitalize,nosort]{cleveref}
\usepackage{booktabs}
\usepackage[referable]{threeparttablex}
\usepackage{longtable}
\usepackage{amssymb,wasysym}
\usepackage{amsmath}

\usepackage{setspace}

\usepackage{multirow}
\usepackage{adjustbox}
\usepackage{tabularx}

\usepackage{geometry}
\usepackage{pdflscape}

\usepackage[title]{appendix}
\usepackage{chngcntr}

\usepackage{lineno}

\usepackage{natbib}
\bibpunct[]{(}{)}{;}{a}{,}{,}

\usepackage{bibunits}

\graphicspath{{Fig/}}

\title{\textbf{Earth and Mars -- distinct inner solar system products}}
\author[1]{Takashi Yoshizaki\thanks{Corresponding author. E-mail: \href{takashiy@tohoku.ac.jp}{takashiy@tohoku.ac.jp}}}
\author[1,2,3]{William F. McDonough}
\affil[1]{Department of Earth Science, Graduate School of Science, Tohoku University, Sendai, Miyagi 980-8578, Japan}
\affil[2]{Department of Geology, University of Maryland, College Park, MD 20742, USA}
\affil[3]{Research Center for Neutrino Science, Tohoku University, Sendai, Miyagi 980-8578, Japan}

\begin{document}

\maketitle


\begin{bibunit}[elsarticle-harv]

\section*{Abstract}

Composition of terrestrial planets records planetary accretion, core-mantle and crust-mantle differentiation, and surface processes. Here we compare the compositional models of Earth and Mars to reveal their characteristics and formation processes. Earth and Mars are equally enriched in refractory elements (1.9 $ \times $ CI), although Earth is more volatile-depleted and less oxidized than Mars. Their chemical compositions were established by nebular fractionation, with negligible contributions from post-accretionary losses of moderately volatile elements. The degree of planetary volatile element depletion might correlate with the abundances of chondrules in the accreted materials, planetary size, and their accretion timescale, which provides insights into composition and origin of Mercury, Venus, the Moon-forming giant impactor, and the proto-Earth. During its formation before and after the nebular disk's lifetime, the Earth likely accreted more chondrules and less matrix-like materials than Mars and chondritic asteroids, establishing its marked volatile depletion. A giant impact of an oxidized, differentiated Mars-like (i.e., composition and mass) body into a volatile-depleted, reduced proto-Earth produced a Moon-forming debris ring with mostly a proto-Earth's mantle composition. Chalcophile and some siderophile elements in the silicate Earth added by the Mars-like impactor were extracted into the core by a sulfide melt ($ \sim $0.5\% of the mass of the Earth's mantle). In contrast, the composition of Mars indicates its rapid accretion of lesser amounts of chondrules under nearly uniform oxidizing conditions. Mars' rapid cooling and early loss of its dynamo likely led to the absence of plate tectonics and surface water, and the present-day low surface heat flux. These similarities and differences between the Earth and Mars made the former habitable and the other inhospitable to uninhabitable. \bigskip


\section*{Keywords}

Earth; Mars; chondrites; solar system; cosmochemistry

\section{Introduction}
\label{sec:intro}

Earth and Mars share many chemical and physical attributes, but are distinct in size and inventory of volatile elements. Chemical data from surface rocks and meteorites combined with seismological and geodetic observations (e.g., mass, density, moment of inertia (MOI)) provide a multiply constrained compositional model of the terrestrial planets \citep[e.g.,][]{ringwood1966chemical,morgan1980chemical,wanke1981constitution,longhi1992bulk}. The physicochemical similarities and differences between Earth and Mars provide useful insights on formation and evolution processes of these bodies, and potentially provide insights into the present-day properties and origin of other rocky bodies (Venus, Mercury, and exoplanets). \bigskip

By definition, Earth and Mars are located within a habitable zone in which liquid water is available \citep{cockell2016habitability,ehlmann2016sustainability}. However, currently life exists only on Earth, and it remains unclear if Mars were inhabited or uninhabited in its history. The occurrence of life only on Earth indicates that there are compositional limits for a habitable planet formation, which can be revealed by an Earth-Mars comparison. Previous comparative studies of chemical compositions of the interiors of Earth and Mars described their differences including abundances of volatile or siderophile elements and redox states \citep{anders1977mars,dreibus1987volatiles,wanke1994chemistry}, but the recent advances in compositional modeling of both Earth and Mars will provide further insights into their characters and origins. \bigskip

Isotopic compositions of solar system materials provide strong constraints on the sources of the terrestrial planets. Mass-independent, nucleosynthetic isotope anomalies in meteorites reveal a heterogeneous distribution of distinct presolar materials and provide the basis for classifying the non-carbonaceous (NC) and carbonaceous meteorite (CC) groups \citep{,trinquier2007widespread,trinquier2009origin,warren2011stable,kruijer2017age}. The NC meteorites appear to be from the inner solar system, whereas the CC meteorites are considered to be samples from the outer solar system, including the Trojan asteroids and possibly beyond Jupiter's orbit \citep{walsh2011low,kruijer2017age}. These isotopic observations demonstrate a link between Earth and enstatite chondrites, whereas Mars is thought to be most closely related to ordinary chondrites \citep[e.g.,][]{trinquier2007widespread,trinquier2009origin,javoy2010chemical,warren2011stable,dauphas2017isotopic}. \bigskip

Here we compare the compositional models of Earth \citep[e.g.,][]{mcdonough1995composition,palme2014cosmochemical} and Mars \citep[e.g.,][]{wanke1994chemistry,taylor2013bulk,yoshizaki2019mars_long} to clarify their characteristics and formation processes. We explore the nature of the building blocks of the terrestrial planets, based on their compositional similarities and differences with chondritic meteorites. These comparisons provide a basis for insights into the conditions for habitable planet formation and evolution.

\section{Comparisons}
\label{sec:comparisons}

\subsection{Bulk planet}
\label{sec:comparisons_bulk}

The geochemical classification of elements consists of four element groups: lithophile (rock-loving), siderophile (metal-loving), chalcophile (sulfide-loving), and atmophile elements. Lithophile elements (e.g., Si, Mg, Ca, Al, Ti, Na, K, rare earth elements) are preferentially incorporated into oxide phases, and thus concentrated in a silicate shell in a differentiated planetary body. Therefore, their abundances in the mantle can be converted to the bulk composition if we also know a mass fraction of a metallic core in a planet. In addition, elements are also classified based on their volatilities in a gas of solar composition at 10 Pa \citep{lodders2003solar}. \bigskip

Historically, there have been three families of compositional models for the bulk silicate Earth, in terms of the proportions of the refractory elements relative to Mg and Si, which, together with Fe and O, make up $ \sim $90\% of rocky planets. These models have low \citep{javoy1995integral,javoy2010chemical,warren2008depleted,caro2010non}, medium \citep{ringwood1975composition,jagoutz1979abundances,wanke1981constitution,hart1986search,allegre1995chemical,mcdonough1995composition,palme2014cosmochemical}, and high \citep{wasserburg1963isotopic,turcotte2001thorium,turcotte2014geodynamics} proportion of refractory elements \citep[][ and references therein]{mcdonough2016composition}. Here we estimate a primitive (i.e., the least melt-depleted) mantle composition based on compositional trends from basalts and mantle rocks and propose a model with medium refractory element abundance \citep[e.g.,][; see also \cref{sec:updates_comp}]{mcdonough1995composition,palme2014cosmochemical}. As summarized in \citet{palme2014cosmochemical}, these geochemical models of Earth are basically similar to each other, and differences between the models do not affect comparison of Earth and Mars discussed in the current paper. \bigskip

A similar geochemical approach has been applied in constraining compositional models of Mars \citep{wanke1994chemistry,taylor2013bulk,yoshizaki2019mars_long}. Here we adopt the compositional model by \citet{yoshizaki2019mars_long}, which is compositionally and mineralogically similar to those of \citet{wanke1994chemistry} and \citet{taylor2013bulk} \citep[\cref{fig:planet_comp_mars_comparison}; see also][ for Martian mantle mineralogy models]{bertka1997mineralogy,khan2018geophysical,smrekar2019pre}, but differ in model development. For example, \citet{wanke1994chemistry} assumed that Mn and more refractory elements (including Fe, Mg and Si) are in chondritic relative abundances in bulk Mars, which is denied by \citet{yoshizaki2019mars_long}. The details of previous Mars models are summarized by \citet{taylor2013bulk} and \citet{yoshizaki2019mars_long}. \bigskip

\begin{figure}[p]
	\centering
	\includegraphics[width=0.8\linewidth]{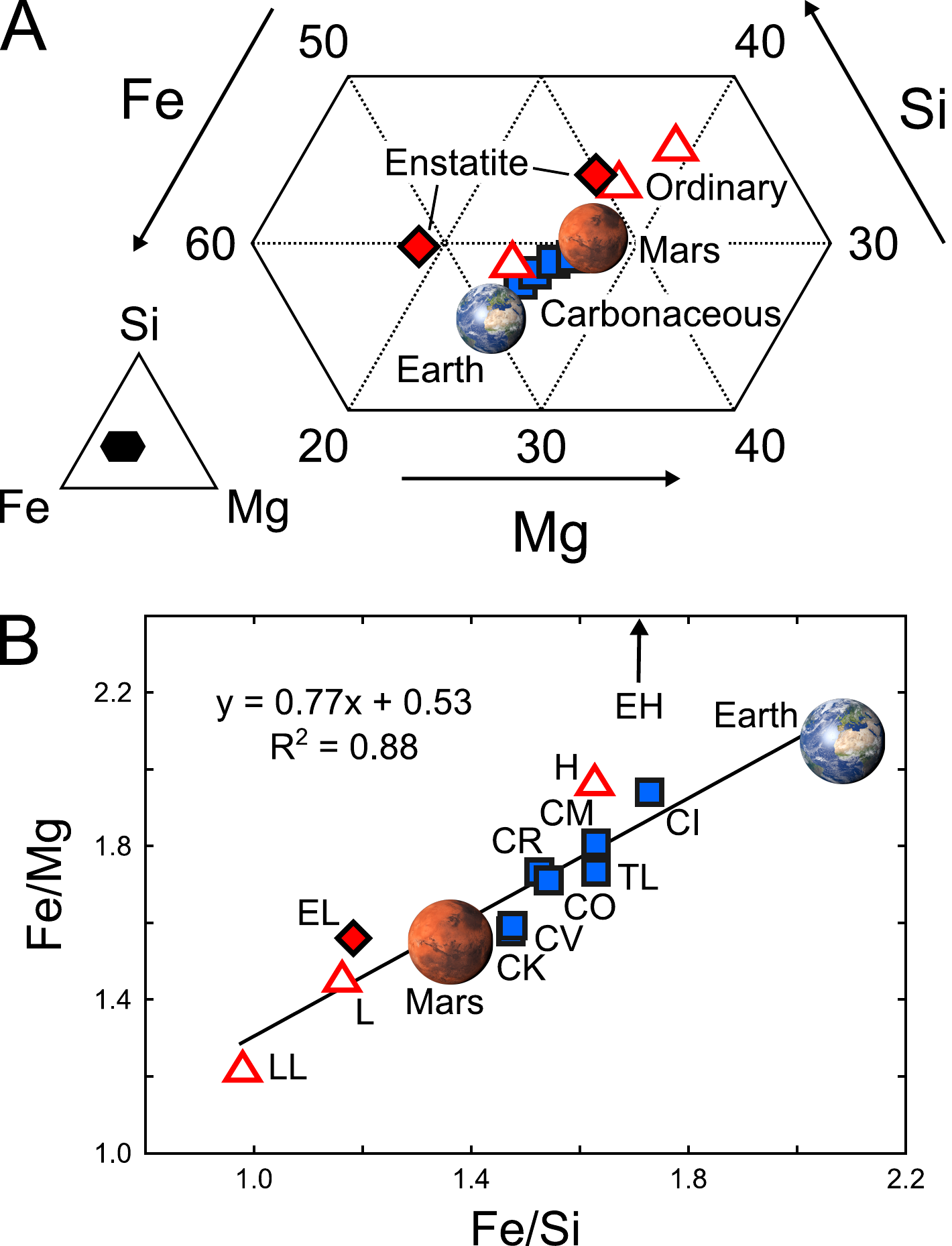}
	\caption{(A) Ternary plot and (B) scatter ratio plot of major elements Si, Fe and Mg in bulk Earth \citep{mcdonough2014compositional}, bulk Mars \citep{yoshizaki2019mars_long} and chondritic meteorites \citep{alexander2019quantitative_CC,alexander2019quantitative_NC}. The regression line for chondrites (except for sulfide-rich EH) is also shown in the lower panel. TL--Tagish Lake.}
	\label{fig:major_comparison}
\end{figure}

By establishing the planet's budget of the 36 refractory elements, recognizing that ratios of refractory elements (e.g., Ca/Al, Th/U) and Fe/Ni are constant ($ \pm $15\% or better) in the chondritic building blocks \citep[e.g.,][]{wasson1988compositions,alexander2019quantitative_CC,alexander2019quantitative_NC}, and either knowing the relative mass of the planet's core or the mantle's Mg\# (atomic ratio of Mg/(Mg + Fe)), limit the range of acceptable Mg/Si (i.e., olivine/pyroxene) values for planetary compositional models (\cref{fig:major_comparison,fig:MgSi_AlSi}A). The Fe/Mg vs Fe/Si correlation seen in both the NC and CC chondrites is suggestive of metal-silicate gradients in the solar system (\cref{fig:major_comparison,fig:MgSi_AlSi}B); interestingly, Earth and Mars also follow this correlation. \bigskip

The approach used here finds that the bulk Earth contains 1.85 $\times$ CI abundances for the refractory elements \citep[e.g.,][]{mcdonough1995composition,palme2014cosmochemical}, it has an olivine/pyroxene proportion equivalent to that of pyrolite \citep{ringwood1966chemical}, and is depleted in volatile elements \citep[i.e., sub-CI Rb/Sr, K/U, and S; \cref{fig:volatility_trend}A;][]{gast1960limitations,wasserburg1964relative}. Mars also contains $ \sim $ 1.9 $ \times$ CI abundances for the refractory elements and is depleted in volatile elements, but less so than Earth \citep[\cref{fig:volatility_trend}B;][]{yoshizaki2019mars_long}. The net atomic oxygen/(metallic Fe) values of Earth and Mars (3.7 and 8.7, respectively; \cref{tab:main_table1}) provide a measure of its average oxidation state. These compositional models for Earth and Mars provide a time-integrated perspective for materials available for accretion at 1 and 1.5 AU, respectively. This spatial sampling contrasts with the temporal comparison, as the mean timescales for Mars' formation \citep[$ \mathrm{\tau_{Mars}^{accretion}} \sim $ 2 Myr;][]{dauphas2011hf} differs from that for Earth \citep[$ \mathrm{\tau_{Earth}^{accretion}} \gtrsim $ 30 Myr;][]{kleine2009hf}. \bigskip

\subsection{Bulk silicate planet}
\label{sec:comparisons_mantle}

\subsubsection{Geochemistry and mineralogy}
\label{sec:comparisons_geochem_mineral}

\begin{figure}[p]
	\centering
	\includegraphics[width=0.9\linewidth]{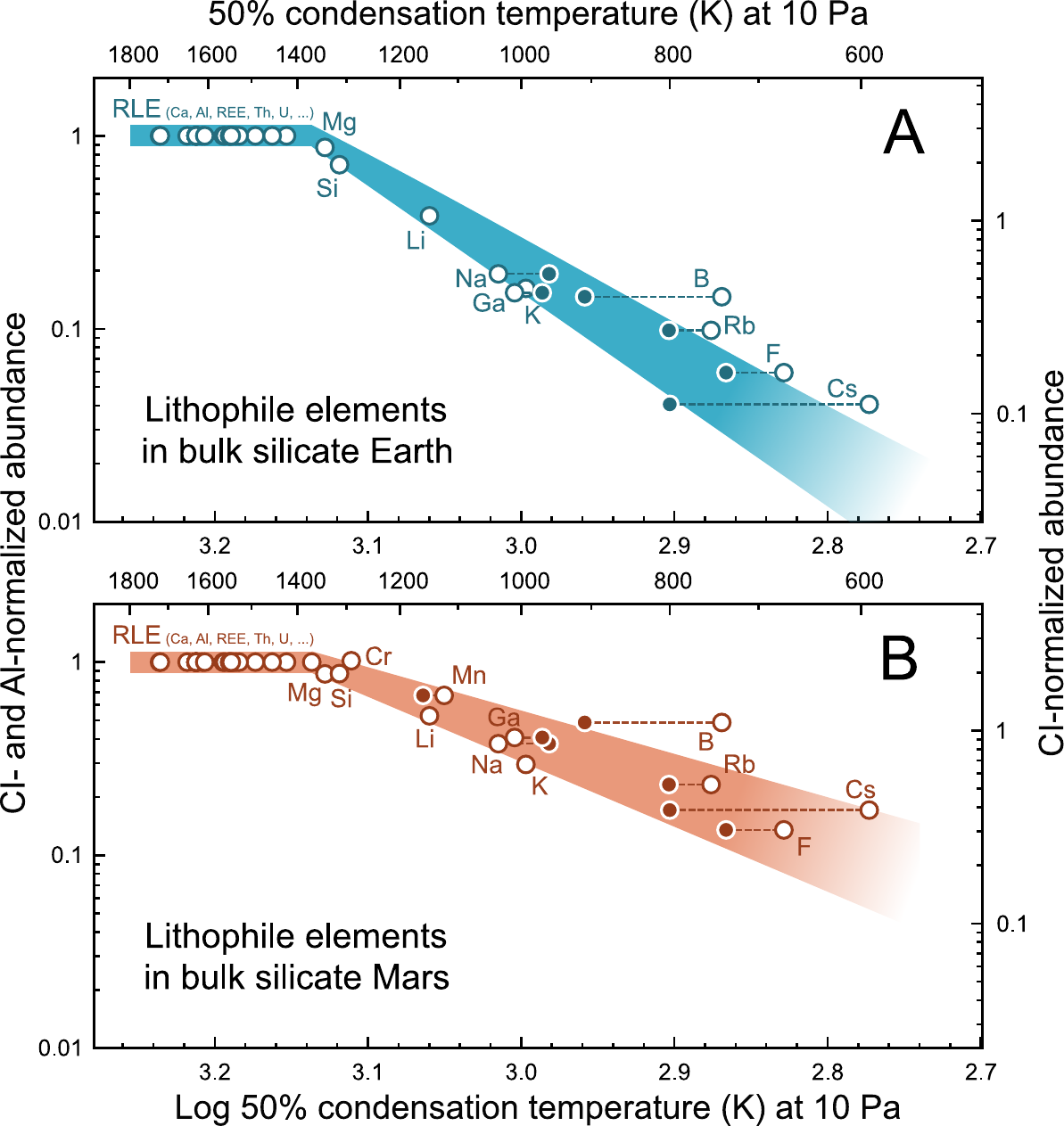}
	\caption{Lithophile element abundances in the bulk silicate Earth and Mars normalized to CI abundances and refractory lithophile element Al. The scale of right ordinate shows the CI-normalized abundances. The values are plotted against 50\% condensation temperature (K) of elements at 10 Pa \citep[open;][]{wood2019condensation} and \citep[filled;][]{lodders2003solar}. CI abundance is from \citet{lodders2020solar}. Other data sources are as in \cref{fig:major_comparison}.}
		\label{fig:volatility_trend}
	\end{figure}

Chondritic ratios of refractory lithophile elements in the bulk silicate Earth and Mars (BSE and BSM, respectively) are further confirmed and constrained by Lu-Hf, Sm-Nd and La-Ce isotopic systematics of these planetary materials \citep[e.g.,][]{bouvier2008lu,burkhardt2016nucleosynthetic,willig2020constraints}. The moderately volatile, lithophile elements (lithophile MVE; e.g., alkali metals) are depleted in Earth and Mars when compared to chondritic meteorites, both planets showing comparable correlations with their condensation temperatures (\cref{fig:volatility_trend}). Earth, with K/U of 14000 and Rb/Sr of 0.032 (i.e., MVE/refractory ratios; cf., 68000 and 0.30 in CI chondrites, respectively), shows such a strong depletion trend in these elements (\cref{fig:volatility_trend}A). Mars, with a K/U of 20000 and Rb/Sr of 0.068, also shows this trend (\cref{fig:volatility_trend}B), albeit less depleted than the Earth's. \bigskip

\begin{figure}[p]
	\centering
	\includegraphics[width=0.9\linewidth]{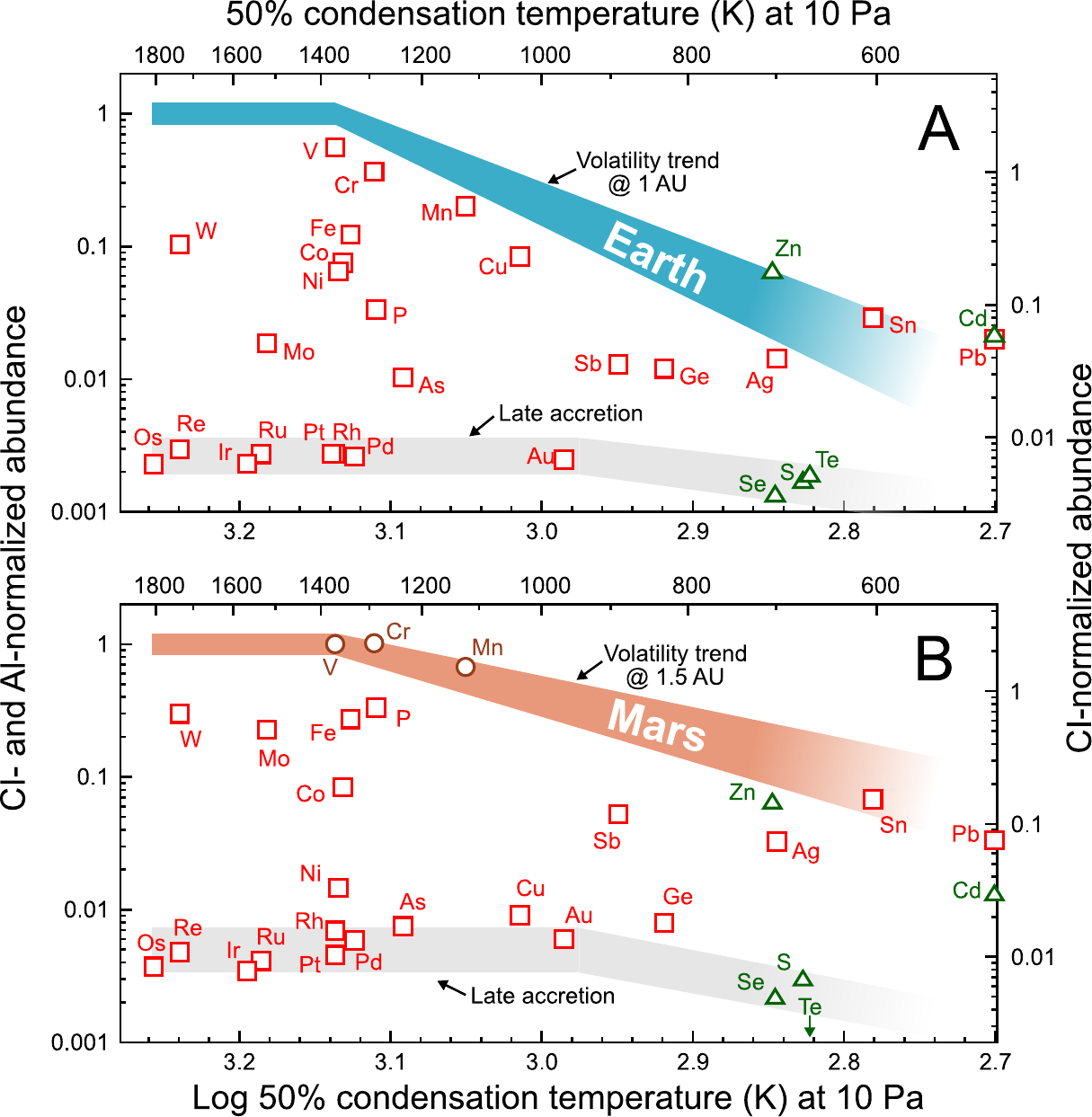}
	\caption{Siderophile (red square) and chalcophile (green triangle) element abundances in the bulk silicate Earth and Mars plotted against 50\% condensation temperature (K) of elements at 10 Pa \citep{wood2019condensation}. The values for y axes follow the same convention as \cref{fig:volatility_trend}. Also shown are time-integrated planetary volatility trends at 1 AU (Earth) and 1.5 AU (Mars) defined by lithophile elements (\cref{fig:volatility_trend}). Data sources are as in \cref{fig:major_comparison}.}
	\label{fig:volatility_trend_siderophile}
	\end{figure}

Siderophile and chalcophile elements are depleted in the BSE and BSM compared to lithophile elements with similar condensation temperatures, due to their incorporation into the metallic cores (\cref{fig:volatility_trend_siderophile}). The abundances of siderophile elements in the BSE and BSM do not show a sub-parallel trend with that defined by lithophiles (\cref{fig:volatility_trend_siderophile}). The degree of siderophile and chalcophile element depletion reflects a combination of planetary building block compositions and element fractionation processes during planetary accretion and differentiation \citep{mcdonough2014compositional}. \bigskip

The BSE and BSM both have high concentrations and chondritic relative proportions of highly siderophile elements (HSE: Re, Os, Ir, Pt, Ru, Rh, Pd and Au; \cref{fig:volatility_trend_siderophile}), which is at odds with any combination of high-pressure and temperature conditions for the partitioning of these elements into a core forming melt \citep{brandon2012evolution,walker2015search,day2016highly,tait2018chondritic}. For example, core-mantle equilibration model in Martian interior \citep{righter2015highly} predicts sub-CI Re/Os and super-CI Ir/Os and Ir/Ru values in silicates, which are inconsistent with the chondritic HSE pattern in the BSM \citep{brandon2012evolution,tait2018chondritic}. In addition, \citet{righter2015highly} assumes a high S concentration ($ > $10 wt\%) in the Martian core, which is much higher than a value predicted by the Martian volatility trend of lithophile elements \citep[$ \leq $7 wt\%;][]{yoshizaki2019mars_long}. These shared features of the HSE abundances in Earth and Mars are considered as evidence for late accretion of chondritic materials during the final stage (e.g., after $ \geq $98\% accretion) of planetary growth in the inner solar system \citep{kimura1974distribution,brandon2012evolution,day2016highly}. Delivery of these HSE was also accompanied by an addition of volatile gases and fluids \citep[H, C, N and O;][]{albarede2009volatile,alexander2012provenances,marty2012origins}, although the late-accreted material to Mars might have been volatile-depleted \citep{wang2017chalcophile,righter2019effect}. \bigskip

On the other hand, the BSE is characterized superchondritic Ru/Ir and possibly Pd/Ir, which cannot be easily accommodated by the late addition of volatile-rich materials \citep{becker2006highly}. Experimental studies showed that these high ratios in the BSE might reflect $ P $, $ T $, and composition-dependent changes in partitioning behaviors of Ru and Pd, and proposed no need for the late addition of these elements \citep{righter2008partitioning,wheeler2011pd,laurenz2016importance,righter2018effect}. Alternatively, these inconsistencies between the BSE and chondrites show an unrepresentative sampling of asteroidal materials in our meteorite collections \citep{walker2015search}. \bigskip

The average, time-integrated, planet-scale redox condition (i.e., metal-silicate equilibrium) is recorded in the Mg\# of its mantle, with the BSE and BSM having an Mg\# of $ \sim $0.89 and 0.75--0.8, respectively, and its core mass fraction (i.e., Mars $ \sim $20\% and Earth $ \sim $ 32\%; \cref{sec:comparisons_core}). These attributes document Mars being more oxidized than Earth. In addition, Mars lacks depletion in redox-sensitive, nominally lithophile elements (i.e., V, Cr, and Mn), and its mantle has a lower Hf/W value \citep[$ \sim $3.5;][]{dauphas2011hf,yoshizaki2019mars_long}, which contrast with those for Earth (\cref{fig:volatility_trend_siderophile}). The more oxidized conditions for the Martian mantle are also indicated by mineralogy, trace element compositions and valence states of Fe and Eu in Martian meteorites. The Martian mantle, however, as recorded in Martian meteorites, shows a heterogeneous redox state \citep[e.g.,][]{herd2001oxygen,herd2002oxygen,wadhwa2001redox,wadhwa2008redox,goodrich2003spinels,mccanta2009expanding,shearer2011ree,righter2016redox,herd2019reconciling}. Oxygen might be one of the light elements in the Earth's core \citep{ohtani1984composition}, but its limited solubility in iron liquids at high $ P $-$ T $ conditions \citep{oneill1998oxide} indicates small effects of the core formation in the BSE's Mg\#, and supports the more reduced state of the BSE as compared to the BSM.  \bigskip

\begin{figure}[p]
\centering
\includegraphics[width=0.65\linewidth]{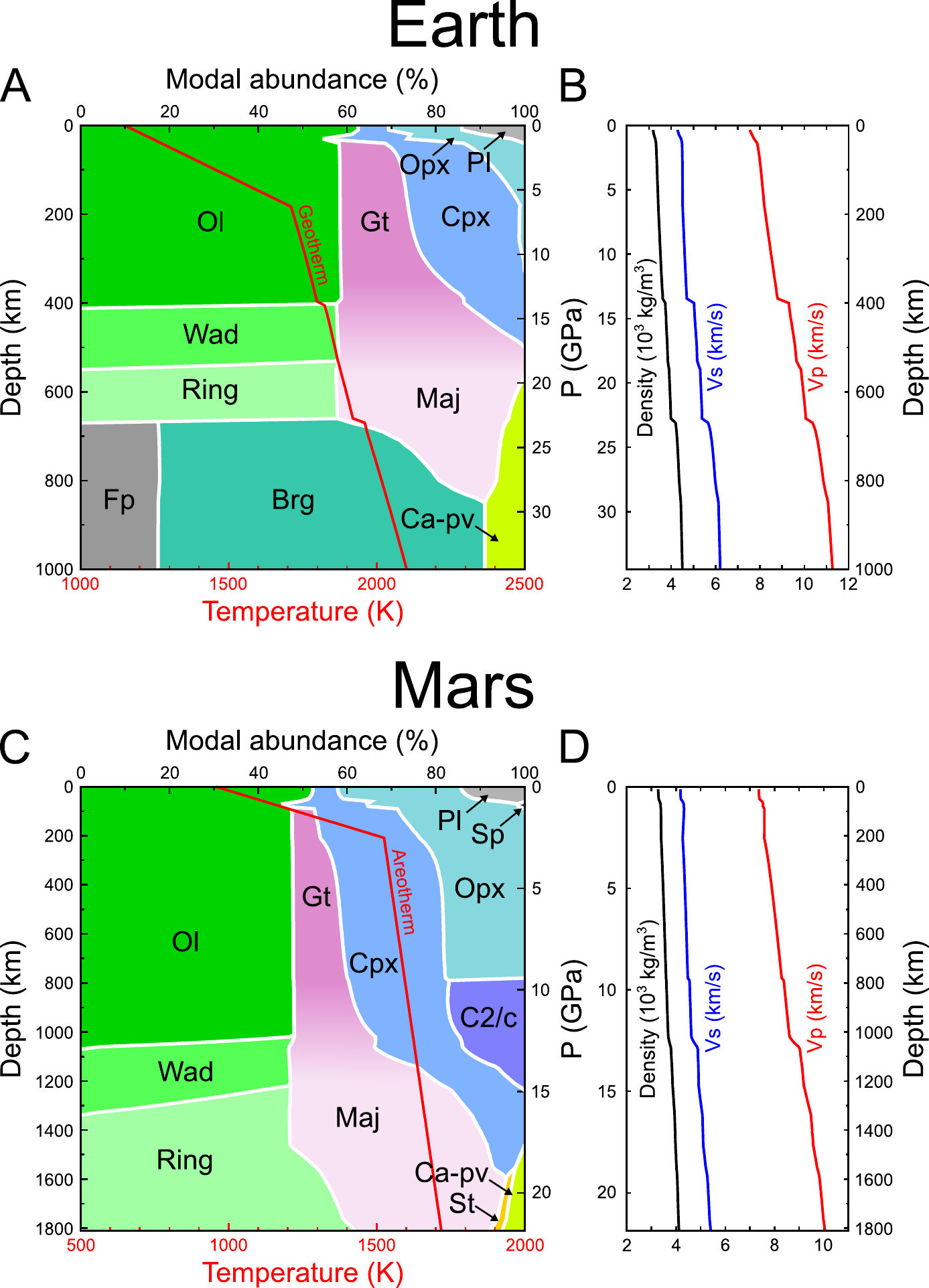}
    \caption{Mineralogy and physical properties of planetary mantles. (A, C) Phase transitions and (B, D) P- and S-wave velocities and density from surface to the core-mantle boundary in the Earth's and Martian mantle. Red lines show temperature profile in the mantle. Figures for Earth are based on the mantle compositional model of \citet{mcdonough2014compositional} and model geotherm from \citet{katsura2010adiabatic}. Figures for Mars are reprinted from T. Yoshizaki and W.F. McDonough (2020) The composition of Mars, Geochimica et Cosmochimica Acta 273, 137--162, Copyright (2020), with permission from Elsevier.
    Abbreviations: 
	Brg--bridgmanite;
	Ca-pv--Ca-perovskite;			
	Cpx--clinopyroxene;
	C2/c--high-pressure clinopyroxene;
	Fp--ferropericlase;
	Gt--garnet;
	Ol--olivine; 
	Opx--orthopyroxene;
	Pl--plagioclase;
	Ring--ringwoodite;
	Sp--spinel;
	St--stishovite;		
	Wad--wadsleyite.}
\label{fig:mantle_mineralogy}
\end{figure}

Earth and Mars have similar "upper mantle" mineralogies (\cref{fig:mantle_mineralogy}). Both contain olivine, ortho- and clino-pyroxenes, and garnet, with the Earth's mantle being richer in olivine than the Martian mantle ($ \sim $60\% and $ \sim $45\% in modal proportion, respectively). Differences in composition (\cref{tab:main_table1}) and interior thermal gradient \citep{breuer2003early,katsura2010adiabatic} of Earth and Mars results in different depths of the olivine-wadsleyite and wadsleyite-ringwoodite phase transitions. For Earth, these transitions are at 410 km and 520 km depth, respectively, and in Mars, they occur at 1000 km and 1300 km depth, respectively. These phase transitions are sharper in the Earth's mantle than the Martian mantle, because of the former's higher Mg\#, greater garnet abundance, and hotter temperature \citep{frost2003structure,filiberto2011fe2+,putirka2016rates}. The Martian mantle might not have a layer of bridgmanite, which is a dominant phase in the Earth's lower mantle (\cref{fig:pie_chart}).	\bigskip

\subsubsection{Heat producing elements (HPE)}	
\label{sec:comparisons_HPE}

The rocky planets are powered by both primordial accretion energy (and lesser amounts from core formation) and radiogenic heat by radioactive decays of heat producing elements (HPE: K, Th, U). All three of the HPE are incompatible, lithophile elements that have been excluded from their cores and concentrated into planetary crusts \citep[e.g.,][]{corgne2007how,blanchard2017solubility,wipperfurth2018earth}, which is beneficial for carrying out remote gamma-ray surveys of planetary surfaces \citep[e.g.,][]{peplowski2011radioactive,surkov1987uranium,prettyman2015concentrations}. The Martian crust is thicker \citep[estimated to be 30--60 km thick with an average value of $ \sim $50 km;][]{zuber2000internal,mcgovern2002localized,wieczorek2004thickness,humayun2013origin}, less enriched in incompatible elements, and contains $ \sim $50\% of the HPE budget of Mars, whereas Earth has a thinner, more chemically evolved crust, which contains only $ \sim $35\% of the planet's HPE budget \citep[\cref{tab:main_table1,sec:heat_production_Mars_calc};][]{rudnick2014composition}. Mars' internal heating Rayleigh number of $ > $10$ ^5 $ (\cref{sec:Ra_calc}) is consistent with a convecting Martian mantle and a dynamically stabilized Tharsis bulge \citep{mckenzie2002relationship,kiefer2003melting}. \bigskip	

With nearly a factor of 2 greater surface to volume ratio for Mars than Earth, the former cooled much faster \citep{filiberto2011fe2+,baratoux2011thermal,baratoux2013petrological,filiberto2017geochemistry,breuer2015dynamics,putirka2016rates}. Mars' surface heat flux is 19 $ \pm $ 1 mW/m$ ^2 $ \citep{parro2017present} \citep[cf., Earth's average is 90 mW/m$^2$;][]{jaupart2015temperatures}, equivalent to a global heat flux of 2.75 $ \pm $ 0.15 TW, of which 2.5 TW comes from radioactive decay (\cref{tab:main_table1,sec:heat_production_Mars_calc}). Mars' planetary Urey ratio ($ Ur $: total radioactive heat production relative to total surface heat loss) is 0.92 $ \pm $ 0.05, which is higher than that estimated for Earth (0.43) and its mantle $ Ur_{\mathrm{Mars}} $ (planetary $ Ur\mathrm{_{Mars}} $ minus the crustal fraction) is 0.8, more than double of that of the Earth's mantle (0.33). The high $ Ur_{\mathrm{Mars}} $ demonstrates a minor contribution from secular cooling to the total Martian heat flux. Importantly, the modeled Martian surface heat flow varies between 14 and 25 mW/m$ ^2 $ \citep{parro2017present}. A direct measurement the Martian surface heat flux by NASA's Interior Exploration using Seismic Investigations, Geodesy and Heat Transport (InSight) mission \citep{banerdt2020initial} will provide important constraints on its heat production and thermal history.
\bigskip

\subsection{Core}
\label{sec:comparisons_core}

The mantles of both Earth and Mars are depleted in siderophile and chalcophile elements due to core extraction (\cref{fig:volatility_trend_siderophile}). Their core compositions are modeled using the following constraints: the planet's chondritic Fe/Ni value, temperature, mass, density, MOI, and for Earth, seismic profile \citep[see][]{mcdonough2014compositional,yoshizaki2019mars_long}. Also, based on iron meteorite mineralogy \citep{scott2014chondrites} and cosmic abundance of elements \citep{lodders2020solar}, planetary cores might contain sulfides, carbides and phosphides, and possibly other phases (e.g., silicides if formed under highly reduced conditions). \bigskip 

\begin{figure}[p]
	\centering
	\includegraphics[width=1\linewidth]{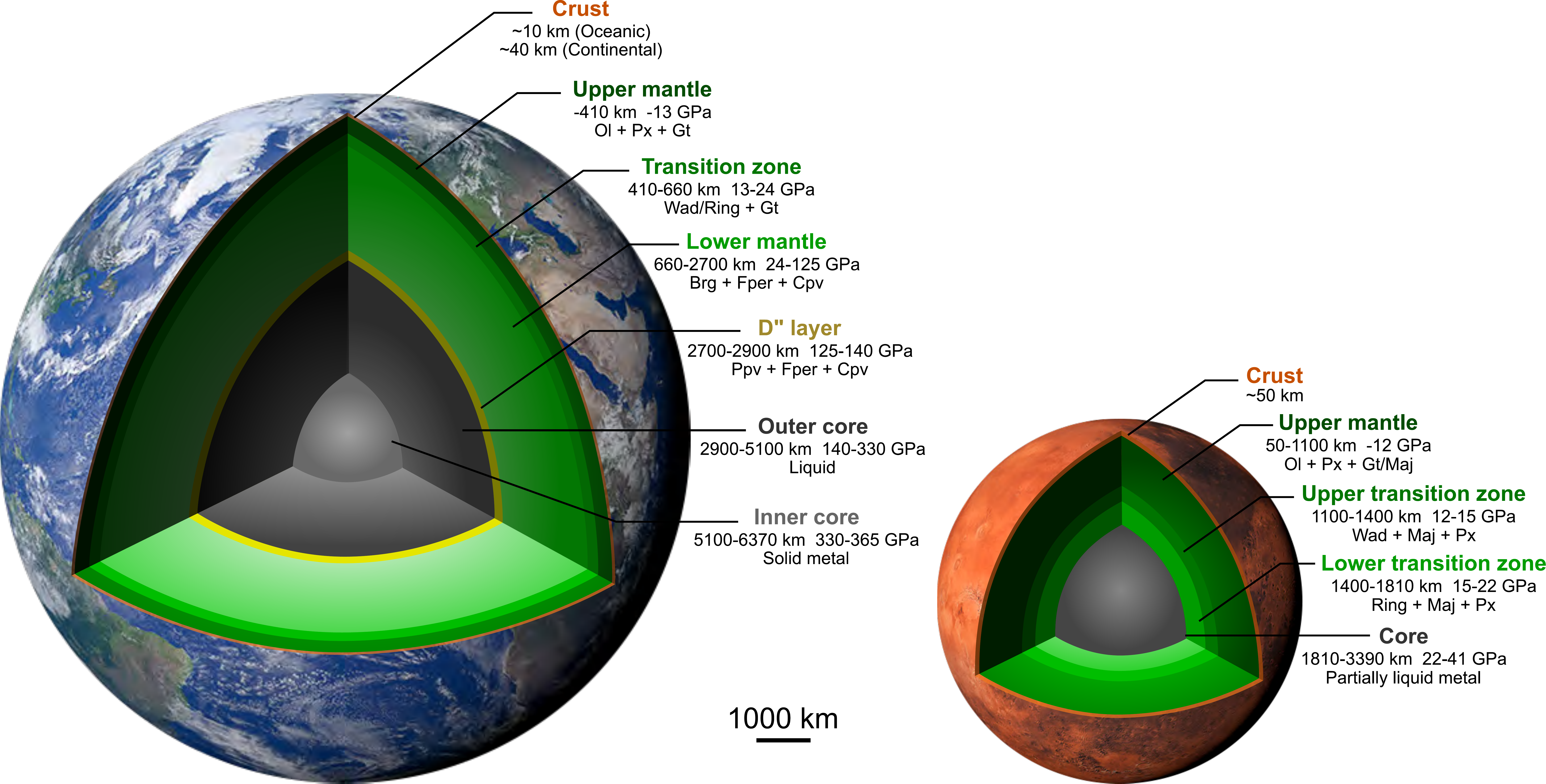}
	\caption{Interior structure of Earth and Mars. Abbreviations: Ol--olivine; Px--pyroxene; Gt--garnet; Wad--wadsleyite; Ring--ringwoodite; Brg--bridgmanite; Cpv--Ca-perovskite; Ppv--post-perovskite; Fper--ferropericlase. The Mars figure is reprinted from T. Yoshizaki and W.F. McDonough (2020) The composition of Mars, Geochimica et Cosmochimica Acta 273, 137--162, Copyright (2020), with permission from Elsevier.}
	\label{fig:pie_chart}
\end{figure}

The Earth's core (32\% by mass) has an outer liquid layer and a solid inner core ($ \sim $5 wt\%) (\cref{fig:pie_chart}). Estimates of the size of the Martian core ranges from 18--25 wt\% and 1500--1900 km, respectively (\cref{sec:S_in_Mars_core}). Data from multiply orbiting satellites provide a precise ($ \pm $2\%) measure of Mars' Love number solution ($ k_2 $) and document a possible existence of a partially molten core \citep{yoder2003fluid,genova2016seasonal,konopliv2016improved}. This observation indicates significant amounts of light elements in the core that lowers its solidus temperature. \bigskip

Sulfur is a prime candidate for light elements in a planetary core. The volatile element depletion trends for both Earth and Mars (\cref{fig:volatility_trend_siderophile}) constrain the core's S content of $ \sim $1.8\% in Earth and $ \sim $7\% in Mars \citep[][; see also \cref{sec:S_in_Earth,sec:S_in_Mars_core}]{yoshizaki2019mars_long}. Since the estimated S contents of the Earth's and Mars' core are not high enough to explain their respected core density deficits, additional light elements in these cores are needed (e.g., Si, O, H and C). The additional light elements are likely to be different in these cores because metal-silicate partitioning behavior of elements depends on conditions of core formation (e.g., timing, $ P $, $ T $, $ f_{\ce{O2}} $) and compositions of coexisting silicate and metallic phases. \bigskip

\begin{figure}[p]
	\centering
	\includegraphics[width=1\linewidth]{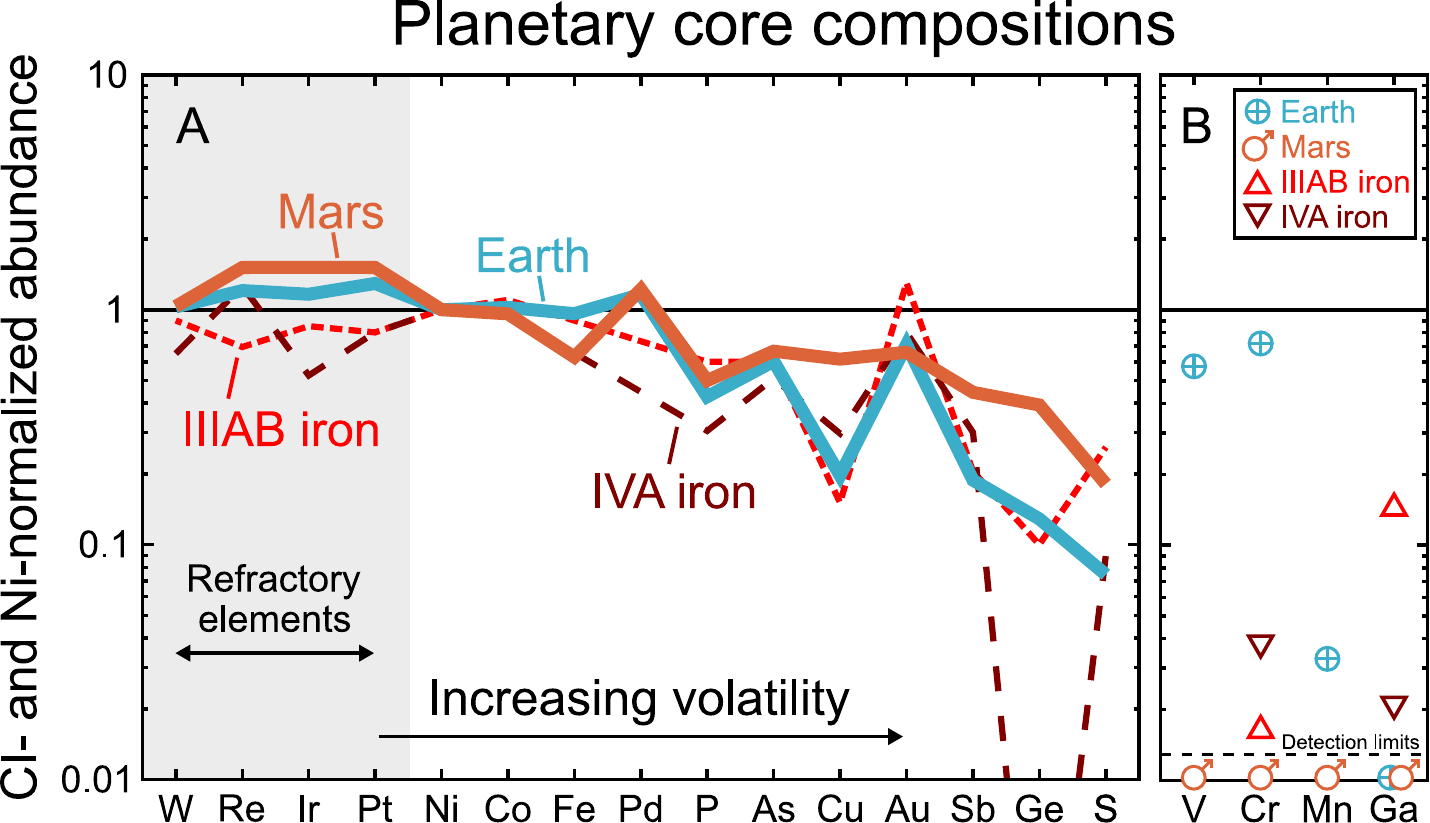}
	\caption{Composition of metallic cores of Earth \citep{mcdonough2014compositional}, Mars \citep{yoshizaki2019mars_long} and NC-group iron meteorite parent bodies \citep{wasson2001fractionation,chabot2018composition}. (A) Siderophile and chalcophile element and (B) nominally lithophile element abundances. Elemental abundances are normalized to CI chondrite composition and Ni. Elements are arranged by their 50\% nebular condensation temperatures \citep{wood2019condensation}, while W and Re are replaced to highlight chondritic highly siderophile element abundance in planetary cores. Type IIIAB and IVA iron meteorites have fractional crystallization (i.e., magmatic) origin and are isotopically classified as NC group meteorites \citep{burkhardt2011molybdenum,kruijer2017age,poole2017nucleosynthetic}.}
\label{fig:core_comparison}
\end{figure}

\begin{figure}[p]
\centering
\includegraphics[width=1\linewidth]{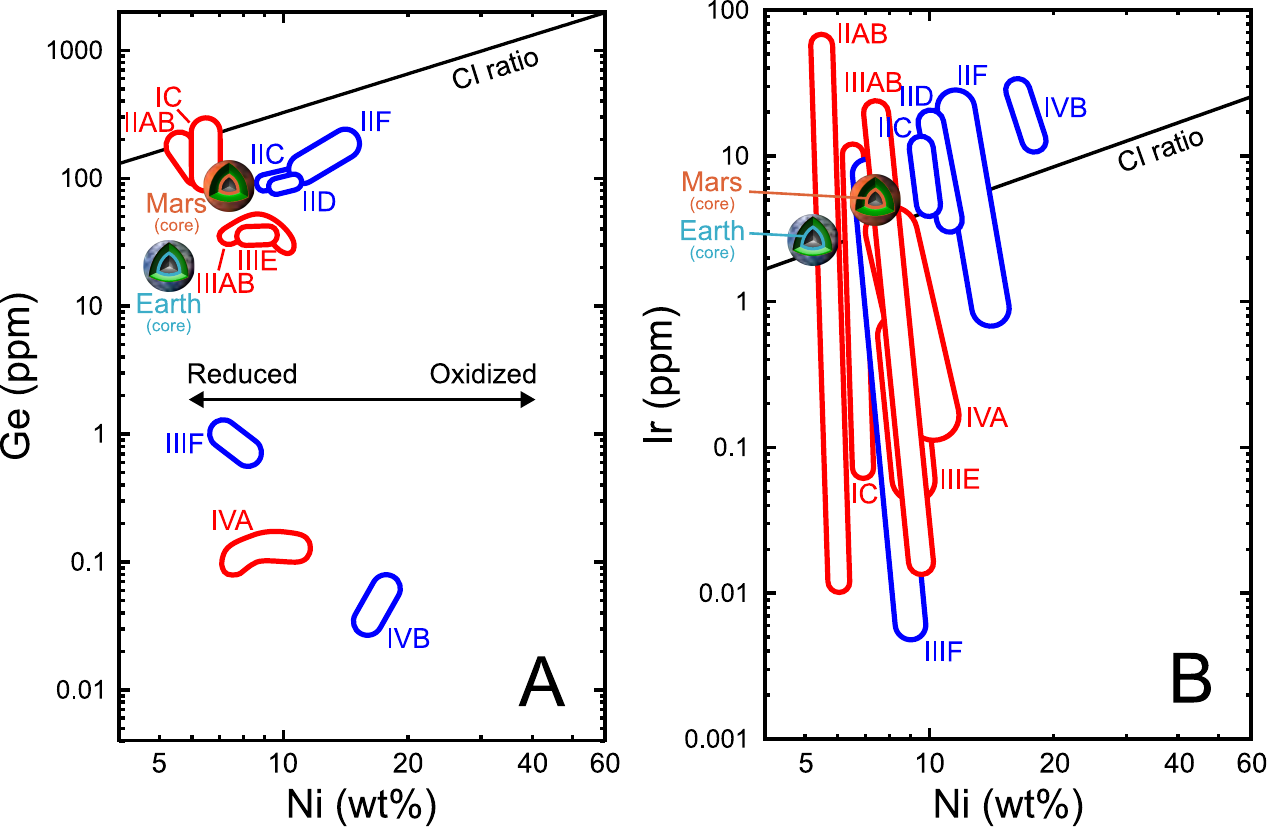}
    \caption{Composition of metallic cores of Earth \citep{mcdonough2014compositional}, Mars \citep{yoshizaki2019mars_long} and fractionally crystallized (i.e., magmatic)  iron meteorites \citep{goldstein2009iron}. (A) Ge versus Ni. (B) Ir versus Ni. Iron meteorites are classified into non-carbonaceous (NC: in red) and carbonaceous (CC: in blue) groups based on their Mo and W isotopic compositions \citep{burkhardt2011molybdenum,kruijer2017age,poole2017nucleosynthetic}. Silicate-bearing (non-magmatic) iron groups are not shown. The IIG iron meteorites are also excluded from the plot because isotopic data to classify them into NC or CC group are not available. CI chondritic ratios are from \citet{lodders2020solar}.}
\label{fig:iron_GeNi_IrNi}
\end{figure}

As described above, a planet's volatility trend sets expectations for the absolute amount of siderophile and chalcophile elements in the planet and defines the proportion of these elements that were partitioned into the core. The degree of depletion below the volatility trend defines an element's empirically established metal/silicate partition coefficient \citep{mcdonough2014compositional,yoshizaki2019mars_long}. The Martian core is enriched in siderophile and chalcophile MVE (e.g., P, Ge, S) as compared to the Earth's, and there appears to be little to no Ga in the cores of terrestrial planets, which contrasts with that seen in iron meteorite compositions (\cref{fig:core_comparison,fig:iron_GeNi_IrNi}). Compared to the Earth's core, the Martian core is smaller, it might contain larger amounts of H and O, and it has a lower Fe/Ni value, with its mantle being enriched in Fe (\cref{tab:main_table1}). Overall, these factors contribute to Mars' lower uncompressed density as compared to the Earth's (3750 vs 4060 kg/m$^3$; \cref{tab:main_table1}). There is no unique model for core compositions of Earth and Mars that satisfies the geodetic and geochemical constraints \citep{mcdonough2014compositional,yoshizaki2019mars_long}, and further experimental and theoretical efforts are needed to constrain the light element budget in the planetary cores. \bigskip

\section{Discussion}
\label{sec:discussion}

\subsection{Origin of volatile depletion in terrestrial planets}
\label{sec:discussion_origin_volatile_depletion}

From Mercury to Mars and beyond to Vesta, the second biggest asteroid, their surface K/Th values are significantly lower relative to chondrites \citep{surkov1987uranium,taylor2006bulk,peplowski2011radioactive,prettyman2015concentrations}. The origins of this depletion and that of other MVE remain elusive, with possible explanations including post-nebular volatile loss due to internal or impact-induced heating \citep[e.g.,][]{oneill2008collisional,norris2017earth} and incomplete condensation of nebular gas \citep[e.g.,][]{palme2014cosmochemical,siebert2018chondritic}. Since planetary K/Th ratios and relative amounts of more refractory olivine (\ce{Mg2SiO4}) and less refractory pyroxene (\ce{MgSiO3}) do not correlate with their heliocentric distances in both the solar and extra-solar systems \citep[e.g., \cref{fig:MgSi_AlSi}A;][]{van2004building,kessler2006c2d,bouwman2008formation,bouwman2010protoplanetary,sargent2008silica}, these volatile depletions do not solely reflect an outward temperature decrease in the protoplanetary disk. \bigskip

The absence of heavy isotope enrichment in rocks from Earth and Mars for multiple isotope systems \citep[e.g., K, Zn, Rb, Fe, Cd;][]{humayun1995potassium,nebel2011rubidium,paniello2012zinc,sossi2016on,sossi2018zinc,pringle2017rubidium,wombacher2008cadmium} indicates negligible post-accretionary evaporative loss of the MVE. In contrast, isotopically heavier siderophile or chalcophile compositions (e.g., Fe, Ga, Sn) of terrestrial and lunar mantles might reflect (1) core-mantle differentiation, (2) evaporation of the giant impactor's core and addition of the metal/sulfur-loving elements to the terrestrial mantle during the giant impact (see \cref{sec:formation_planets}), or (3) planetary surface processes \citep[e.g.,][]{poitrasson2004iron,kato2017gallium_Moon,creech2019tin}. The \ce{^{53}Mn}-\ce{^{53}Cr} and \ce{^{87}Rb}-\ce{^{87}Sr} isotope systematics of meteorites and terrestrial samples are consistent with a volatility-dependent, gas-solid fractionation during first few Myr of the solar system \citep[e.g.,][]{shukolyukov2006manganese,trinquier200853mn,hans2013rb,moynier2012planetary}, which might be prior to the dissipation of the nebular gas \citep[$ \sim t_0 $ + 5 Myr;][]{wang2017lifetime}. \bigskip

There is negligible evidence for post-accretionary MVE losses in the chemical compositions of these planets. The bulk Earth and Mars, together with chondrites, show negligible evidence of evaporative losses in ratios of Na, Mn and Ti \citep[\cref{fig:MnNa_NaTi,sec:Mn_in_BSE};][]{oneill2008collisional,siebert2018chondritic}, which have distinct relative volatilities during condensation and evaporation. Values of Mn/Na and Na/Ti in the bulk Earth, Mars and chondrites are consistent with an incomplete nebular condensation, in which earlier condensates are removed from a nebular gas before a completion of more volatile species \citep[e.g.,][]{larimer1967chemical,larimer1967chemical2} \citep[cf.][]{oneill2008collisional}. The plot of Mn/Na versus Na/Ti indicates that a formation of precursors of Earth and Mars at higher temperatures than their isotopically linked counterparts (enstatite and ordinary chondrites, respectively), which is also suggested by the former's higher Mg/Si (i.e., olivine/pyroxene), Al/Si, and Al/Mg ratios \citep{kerridge1979fractionation,larimer1979condensation,dauphas2015planetary,morbidelli2020subsolar}. \bigskip


Collectively, the compositions of rocky planets, as compared to their chondritic relatives, likely reflects volatility-dependent chemical fractionation in the protoplanetary disk, rather than the post-accretionary losses of MVE. In contrast, small differentiated asteroids (e.g., parent bodies of Eucrite and Angrite) show clear evidence for preferential loss of MVE \citep[\cref{fig:MnNa_NaTi};][]{oneill2008collisional} and heavy isotope enrichment \citep[e.g.,][]{paniello2012zinc_HED,pringle2017origin,tian2019potassium}, which are the hallmarks of evaporative losses during or after accretion. Volatile elements might have escaped from these small bodies during and/or after their accretion, due to their weak gravity field. \bigskip

\subsection{The building blocks of the terrestrial planets}
\label{sec:planetary_building_blocks}

\begin{figure}[p]
	\centering
	\includegraphics[width=1\linewidth]{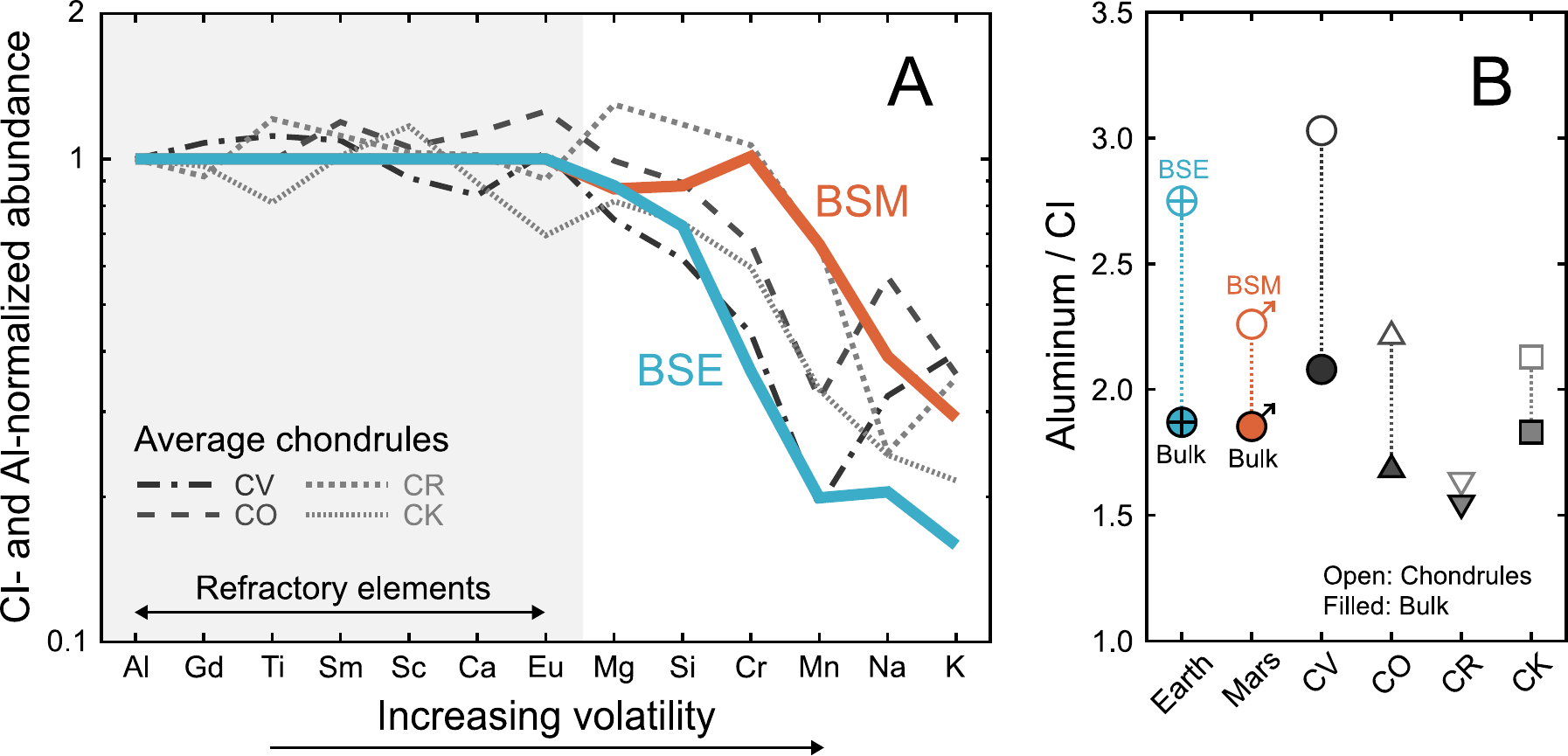}
	\caption{Lithophile element composition of Earth, Mars, and carbonaceous chondrites and their components. (A) Lithophile element abundance in the bulk silicate Mars (BSM), bulk silicate Earth (BSE) and bulk chondrules from carbonaceous chondrites. Elemental abundances are normalized to CI chondrite composition and Al. Elements are arranged by their 50\% nebular condensation temperatures. (B) CI-normalized Al abundances in the BSE, BSM, bulk planets, bulk chondrules and bulk chondrites. Chemical composition of chondrules is from \citet{hezel2018what} and \citet{metbase}. Other data sources are as in \cref{fig:major_comparison}.}
	\label{fig:planet_chondrules}
\end{figure}

The planetary building blocks appeared to be made up of high-temperature materials, dominantly chondrules, which are silicate droplets formed by transient heating events within first few Myr of the solar system evolution \citep{connelly2012absolute,bollard2017early}, and are an essential component of chondrites \citep[e.g.,][]{scott2014chondrites,russell2018chondrules}. MVE composition of chondrules provides a record of incomplete nebular condensation rather than evaporation processes \citep{humayun1995potassium,alexander2000lack,alexander2008formation,galy2000formation,pringle2017origin}. The refractory element enrichment and MVE depletion in the BSE and BSM are comparable to those observed for chondrules from carbonaceous chondrites \citep[\cref{fig:planet_chondrules};][]{hewins1996nebular,mahan2018volatile}. Significantly, mass-dependent Ca \citep{huang2010calcium,magna2015calcium,amsellem2017testing,simon2017calcium,bermingham2018origins} and Mg \citep{bizzarro2004mg,young2004isotope,wiechert2007non,bouvier2013magnesium,olsen2016magnesium,hin2017magnesium} isotopic compositions of chondrules, BSE and BSM also suggest that they inherited fractionated isotopic signatures from similar precursor materials, which have experienced high-temperature gas-solid fractionation processes. \bigskip

Recent theoretical models of planetary growth favor formation of terrestrial planets via accretion of chondrule-sized pebbles \citep[e.g.,][]{johansen2015new,johansen2015growth,levison2015growing}. These models predict a rapid accretion of Mars-sized bodies under the presence of a nebular disk. Thus, Mars, with a $ \mathrm{\tau_{Mars}^{accretion}} $ of $ \sim $2 Myr \citep{dauphas2011hf,kruijer2017early,bouvier2018evidence} \citep[cf.][]{marchi2020compositionally}, formed within the nebular disk lifetime \citep[$ \sim $5 Myr;][]{wang2017lifetime}, whereas Earth is suggested to have $ \mathrm{\tau_{Earth}^{accretion}} $ $ \geq $30 Myr \citep[e.g.,][]{kleine2009hf}, which documents its accretion stretched beyond the lifetime of the nebular disk. \bigskip

Differences in planetary MVE abundances (\cref{fig:planet_vs_chondrite}) likely reflect aspects of their accretion histories and sizes. Chondrites, the least MVE-depleted bodies' materials, contain fine-grained, volatile-rich matrix that accounts for the largest fraction of the inventory of volatiles \citep[e.g.,][]{alexander2005re,alexander2019quantitative_CC,alexander2019quantitative_NC,bland2005volatile,zanda2018chondritic}. Although a relationship between chondrules and matrix remains poorly understood, their coexistence in the protoplanetary disk before planetesimal accretion is accepted \citep[e.g.,][]{hezel2018composition,zanda2018chondritic}. Some carbonaceous chondritic asteroids accreted more chondrules and less matrix, resulting in more MVE-depleted compositions (Figs.~\ref{fig:K-Th_parameters}B and \ref{fig:Rb_Sr_parameters}). According to the pebble accretion model \citep{johansen2015new,johansen2015growth,levison2015growing}, planetesimal growth starts off by preferentially accreting the smallest particles, and as the body grows, it prefers to accrete larger and larger size particles. This mechanism would lead to a growing planetesimal having a larger chondrule/matrix ratio, and becoming more MVE-depleted as its mass increases. Thus, small chondritic asteroids likely co-accreted chondrules and matrix, whereas Mars, with its intermediate volatile depletion, size and accretion timescale, likely accreted a greater fraction of chondrules to matrix (\cref{fig:K-Th_parameters}). Finally, the Earth's prolonged accretion history was dominated by chondrule accretion, resulting in its significant MVE-depleted composition. The MVE depletion scales with size of the chondritic parent body and planet (\cref{fig:K-Th_parameters}), implying an accretion driven process from an undepleted nebula (i.e., CI (solar) composition). Exceptions are small differentiated bodies (e.g., the Moon and Vesta) that experienced post-nebular volatile loss due to internal or impact-induced heating (\cref{sec:discussion_origin_volatile_depletion}). Thus, the chondrule-driven planetary growth plays a critical role in establishing the planetary MVE-depleted characteristics. \bigskip

\begin{figure}[p]
    \centering
    \includegraphics[width=1\linewidth]{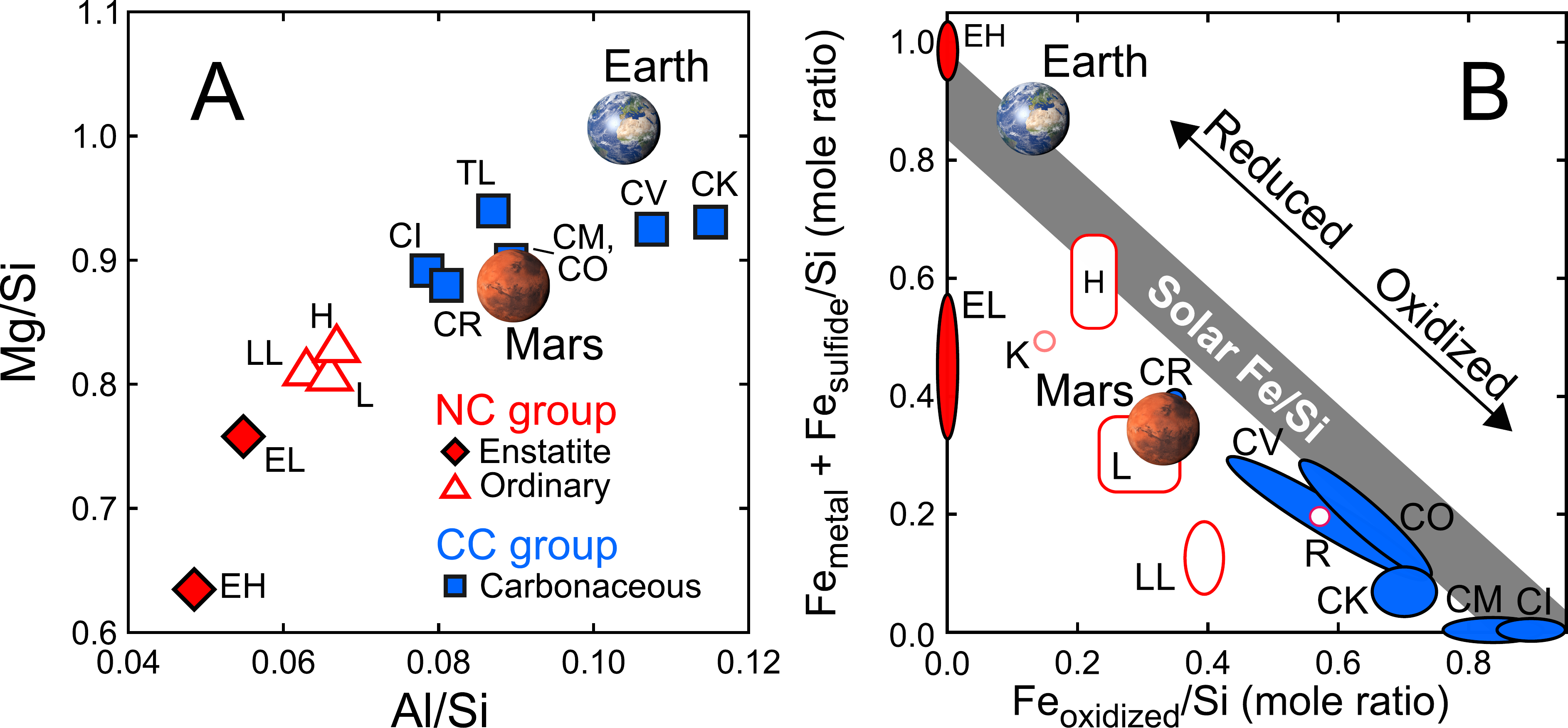}
    \caption{(A) Weight ratios of Al/Si vs Mg/Si values for planets and chondrites. Earth and Mars show higher RLE abundances (i.e., Al) and Mg/Si value (i.e., olivine/pyroxene ratio) compared to enstatite and ordinary chondrites, which are respectively their isotopically identified relatives. (B) The Urey-Craig diagram \citep[after][]{urey1953composition} illustrates relative redox condition for Earth, Mars, and chondrites. Chondrite classification (non-carbonaceous (NC; red) vs carbonaceous (CC; blue) groups) follows \citet{warren2011stable} and \citet{kruijer2017age}. TL--Tagish Lake. Data sources are as in \cref{fig:major_comparison}.}
	\label{fig:MgSi_AlSi}
\end{figure}

In contrast, enstatite and ordinary chondrites, whose isotopic composition and redox state are most related to Earth and Mars, respectively \citep[e.g.,][; \cref{fig:MgSi_AlSi}B]{trinquier2007widespread,trinquier2009origin,warren2011stable,dauphas2017isotopic}, do not show a clear refractory enrichment nor MVE depletion (Figs.~\ref{fig:MgSi_AlSi}A, \ref{fig:K-Th_parameters} and \ref{fig:Rb_Sr_parameters}). In addition, chondrules from these NC chondrites are less depleted in MVE than the BSE and BSM (\cref{fig:planet_chondrules_NC}). Thus, the chondrule-driven MVE-depletion scenario discussed above cannot be applied to the NC chondrites we have in our collections. These observations indicate that the NC chondrites represent refractory-poor, volatile-rich counterparts of the inner rocky planets. \citet{morbidelli2020subsolar} showed that the low Mg/Si and Al/Si solids, which are comparable to those of NC chondrites (\cref{fig:MgSi_AlSi}A), condense after removal of early-formed, high-temperature condensates from the system. \bigskip

Differences in the timing of planetary accretion might also be important in establishing their relative abundances of MVE. Chronology of meteorites combined with thermal modeling of asteroids \citep{sugiura2014correlated,kruijer2017age,zhu2020dating} indicates that the undifferentiated NC planetesimals accreted 1--2 Myr after formation of differentiated NC bodies (i.e., iron meteorite parent bodies and terrestrial planets). The variation in the absolute ages of chondrules \citep{connelly2012absolute,bollard2017early}, occurrence of relict grains \citep{jones1994relict,weisberg2011petrology,tenner2018oxygen} and igneous rims \citep{krot1995igneous} in chondrules, and the presence of compound chondrules \citep{wasson1995compound} indicate that
chondrule formation was a repeated event. Thus, the longer chondrules remained in the accretionary disk, the more opportunities it has being recycled by later events. Ordinary chondrite chondrules records an admixing of MVE-rich CC-like materials to MVE-poor chondrule precursors into the NC reservoir \citep{mahan2018volatile,schiller2018isotopic,bollard2019combined}. Additions of MVE-rich materials and repeated chondrule recycling produce younger chondrules with higher MVE abundance \citep{mahan2018volatile}. Thus, terrestrial planets, which are dominated by earlier materials that experienced fewer recycle events and MVE addition, are more depleted in MVE as compared to the NC chondrites, which accreted the younger MVE-enriched chondrules. \bigskip

In contrast, chondrule formation scenarios predict less frequent chondrule formation/recycle events in the outer solar system \citep[e.g.,][]{morris2012chondrule,johnson2015impact,sanders2018making_splash,pilipp1998chondrule}. This prediction is consistent with chemical and isotopic signatures of CC chondrules \citep[e.g.,][]{hewins2012chondrules,tenner2018oxygen,mahan2018volatile}. Consequently, the CC chondrules could have preserved their MVE depletion until the accretion of CC chondritic asteroids, and thus are chemically comparable to the early NC materials which formed the inner rocky planets. \bigskip

This chondrule-rich accretion model for the terrestrial planets reveals the limited ability to reach greater levels of enrichment in refractory elements (Figs.~\ref{fig:MgSi_AlSi}A and \ref{fig:planet_chondrules}B), which is a concern for planetary compositional models promoting high refractory element abundance (\cref{sec:comparisons_bulk}). To reach higher levels of refractory element enrichment beyond that seen in CV chondrites and their chondrules, larger amounts of refractory inclusions are needed, which is inconsistent with chondritic REE ratios of the BSE \citep{stracke2012refractory,dauphas2015thulium}. \bigskip

\subsection{Formation models of the terrestrial planets}
\label{sec:formation_planets}

\subsubsection{Previous models of the Earth's accretion and Moon formation}
\label{sec:previos_fm_model}

Depletion of moderately to highly siderophile and chalcophile elements in the BSE is consistent with Earth's initial accretion from highly reduced, volatile-depleted materials, that was later oxidized by volatile-rich additions \citep[e.g.,][]{wanke1981constitution,wanke1984mantle,wanke1988chemical,oneil1991origin_Earth,rubie2011heterogeneous,rubie2015accretion}. Support for this temporal evolution in volatile accretion is found in multiple radiogenic isotope systems \citep[e.g., Rb-Sr, U-Pb, Ag-Pd, I-Pu-Xe;][]{halliday2001in,albarede2009volatile,schonbachler2010heterogeneous,mukhopadhyay2012early,ballhaus2013u,maltese2020pb}, metal-silicate partitioning behaviors of elements such as W, Mo, S and C \citep[e.g.,][]{wade2012metal,li2016carbon,suer2017sulfur,tsuno2018core,ballhaus2017great}, and \textit{N}-body simulations of planetary formation \citep[e.g.,][]{raymond2006high,morbidelli2012building}. However, the degree of volatile element depletion in the pre-impact proto-Earth remains unconstrained. \bigskip

The two-component accretion models for the growth of Earth envisage mixing of highly reduced, volatile-depleted materials with oxidized, volatile-rich, "CI-chondritic" materials in some proportion, not well defined, but generally conceived their mass ratios to be in the range of 60:40 to 90:10 \citep{wanke1984mantle,wanke1988chemical,oneil1991origin_Earth}. Some mass estimates of the late oxidized addition are comparable to a Mars-sized, Moon-forming impactor \citep[e.g.,][]{canup2001origin,canup2004simulations,canup2008accretion}. The impactor's composition has often been assumed to be CI-chondritic \citep[e.g.,][]{oneil1991origin,oneil1991origin_Earth,maltese2020pb}. \bigskip

However, the CI-chondritic impactor model has been multiply challenged. The common MVE depletion in rocky differentiated bodies in the solar system (e.g., sub-solar K/Th ratios; \cref{sec:discussion_origin_volatile_depletion}) might indicate a similar MVE depletion in the Mars-sized impactor. In addition, the CI-like impactor model requires a significant volatile depletion in the proto-Earth, perhaps at levels seen in angrites and calcium-aluminum-rich inclusions (CAIs), which show heavy Mg- and Si-isotope enrichment and significant depletion of moderately volatile elements \citep[e.g.,][]{grossman2000major,grossman2008primordial,pringle2014silicon} due to significant evaporative losses of the major and more volatile elements by impact-induced or transient nebular heating events \citep[e.g.,][]{stolper1986crystallization,richter2002elemental,pringle2014silicon,yoshizaki2019nebular,young2019near}. However, such isotopic signatures are not recognized for Earth (\cref{sec:discussion_origin_volatile_depletion}). Furthermore, models predicting compositional zoning in the protoplanetary disk have a region of CI-like material accreting at $ \geq $15 AU \citep{desch2017effect}, where it is predicted that the disk mass is too low to form a Mars-sized body. \bigskip

Compositional similarities of Earth and Moon in multiple isotope systems set strict constraints on the nature of both the impactor and proto-Earth, that is, they are derived from a similar NC isotopic reservoir that appears to be restricted to inner solar system sources \citep[e.g.,][]{wiechert2001oxygen,trinquier2009origin,warren2011stable,zhang2012proto,greenwood2018oxygen,dauphas2017isotopic,kruijer2017age}. This observation is consistent with dynamical simulations which predict low probabilities of a Moon-forming impactor originating from $ > $10 AU \citep[e.g.,][]{jackson2018constraints}. Thus, these constraints exclude the isotopically distinct CI chondrite as an impactor candidate. In contrast, recent Mo and O isotopic data from a broad range of NC and CC materials challenge this exclusion and find support for a CC-like impactor and/or vigorous mixing of proto-Earth and impactor, requesting further constraints to reveal the origin of the Earth-Moon system \citep{young2016oxygen,budde2019molybdenum,cano2020distinct}. Furthermore, the recently-proposed synestia model \citep{lock2018origin} has the Moon forming in a vapor cloud surrounding Earth. This vapor cloud was produced by a large impact, resulting in a well-mixed, chemically equilibrated proto-earth and the vapor cloud.
\bigskip

\subsubsection{Mars-like Moon-forming giant impactor model}
\label{sec:Mars-like_impactor}

\begin{figure}[h]
	\centering
	\includegraphics[width=.9\linewidth]{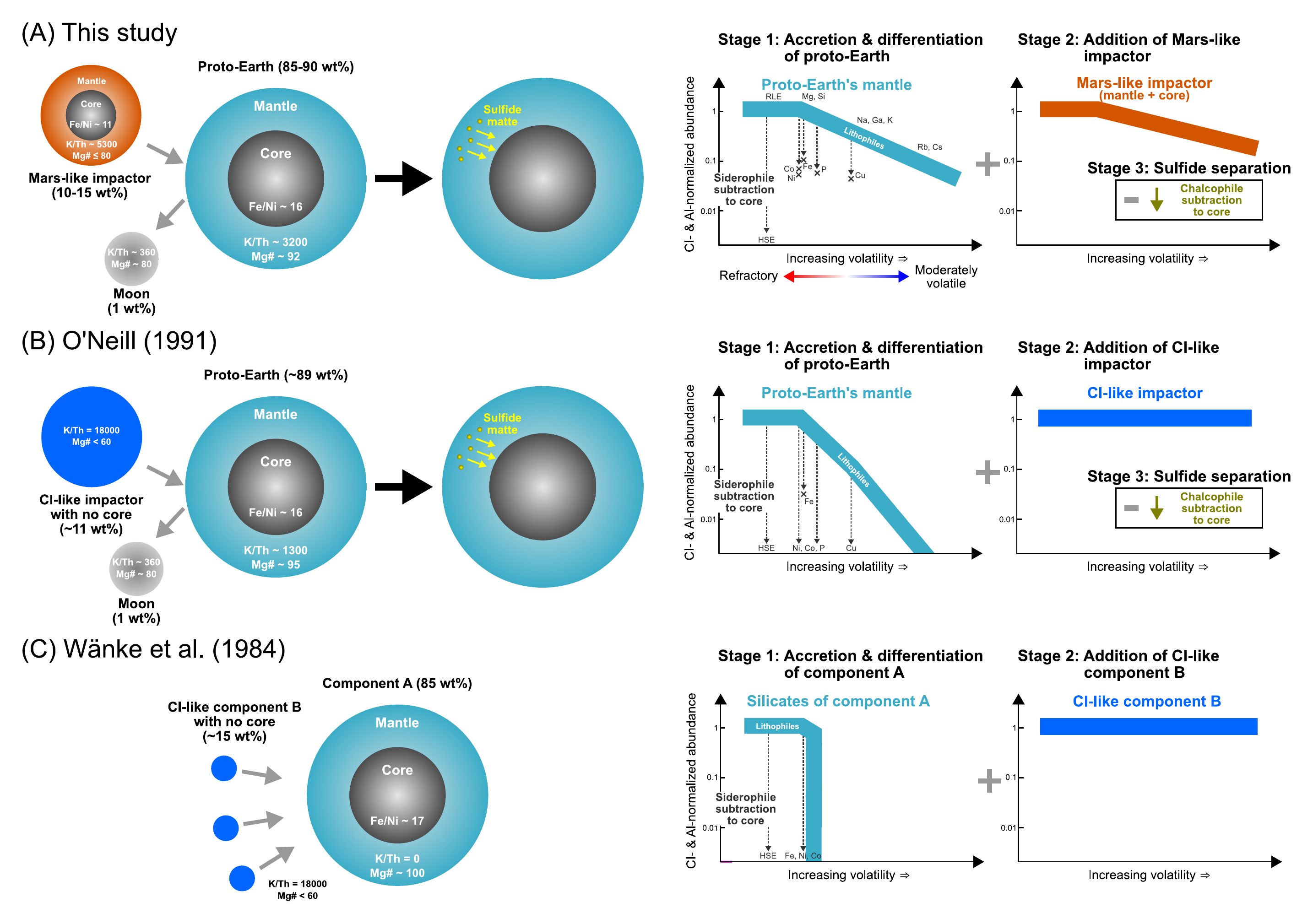}
	\caption{A cartoon summarizing three models of the Earth's accretion: (A) this study; (B) \citet{oneil1991origin_Earth}; and (C) \citet{wanke1984mantle}. The later models envisage Earth’s formation as composed of four stages: (1) accretion and differentiation of a proto-Earth, (2) addition of a giant impactor to the proto-Earth's mantle, (3) loss of a sulfide matte from this mantle, and (4) late accretion of volatile-rich chondritic materials (not shown).}	\label{fig:Earth_accretion_models}
\end{figure}

Here we propose a model for the origin of the Moon. Our model is a modification of the original \citet{wanke1984mantle}'s model and later modified by \citet{oneil1991origin,oneil1991origin_Earth}. In our version of this model, we envision the Earth's formational history in 4 steps and use a differentiated Mars-like composition \citep{yoshizaki2019mars_long}, instead of CI composition, for the giant impactor. \cref{fig:Earth_accretion_models,fig:volatility_trend_impactor} show the details of our model: \cref{fig:Earth_accretion_models} specifically highlights the differences between our model and those previously presented. Our model starts with
\begin{enumerate}
    \item the accretion of the proto-Earth (reduced and volatile-depleted) accompanied by continuous core-mantle differentiation ($ \sim $90\% of Earth's mass),
    \item followed by a late-stage Moon-forming giant impact event \citep[30--100 Myr after $ t_0 $;][]{kleine2009hf} that adds the final $ \sim $10\% mass (oxidized and volatile-enriched) to Earth and forms a protolunar accretion disk,
    \item subsequently, the mantle loses a Fe-Ni ($ \pm $O) sulfide liquid \citep[sulfide matte;][]{,oneil1991origin_Earth} to the core ($ \sim $0.5\% BSE mass),
    \item and finally, the BSE receives the addition of ($ \sim $0.5\% of the BSE mass) a late accretion component that brings the highly siderophile and chalcophile elements in chondritic proportions and highly volatile gases and fluids (see \cref{sec:model_Earth_accretion_SM} for details of the modeling).
\end{enumerate} \bigskip

In this scenario, the bulk proto-Earth already contains $ \geq $80\% of the present-day Earth's budgets of most of the MVE (e.g., K/Th $ \sim $ 3200; Rb/Sr $ \sim $ 0.026; \cref{tab:proto-Earth}), since a Mars-like impactor contributes only a limited amount of additional MVE (\cref{fig:volatility_trend_impactor}). The addition of MVE by the impactor leads to sulfur saturation in the magma ocean,
generating an Fe-Ni ($ \pm$ O) sulfide liquid \citep[post-impact sulfide matte;][]{oneil1991origin_Earth,rubie2016highly}. The sulfide matte precipitates through a crystallizing mantle into the core due to its high immiscibility, low wetting angle, and high density \citep{gaetani1999wetting,rose2001wetting}. Assuming that the sulfide matte, with a present-day mantle sulfide-like composition \citep[][]{lorand1983contribution}, extracted all S from the post-impact Earth’s mantle; its mass is estimated to be $\sim$0.5 wt\% of the present-day Earth's silicate mantle (\cref{sec:model_Earth_accretion_SM}). \bigskip

The siderophile and chalcophile element abundances of the BSE are reproduced when 10--15\% of a planetary mass is added by the impactor and most of the impactor's core equilibrates with the proto-Earth's mantle (\cref{fig:volatility_trend_impactor,sec:model_Earth_accretion_SM}). In this scenario, the bulk proto-Earth contains $ \geq $80\% of the present-day Earth's budget of most of the MVE (e.g., K/Th $ \sim $ 3200; Rb/Sr $ \sim $ 0.026; \cref{tab:proto-Earth}), with the Mars-like impactor contributing a limited amount of MVE (\cref{fig:volatility_trend_impactor}). We envision the proto-Earth as having a MVE abundance comparable to that of chondrules (\cref{fig:planet_chondrules,fig:K-Th_parameters}B). Thus, the proto-Earth's composition is consistent with the chondrule-rich accretion scenario (\cref{sec:planetary_building_blocks}) and requires no need for a post-accretionary loss of MVE from the proto-Earth before the Moon-forming event (\cref{sec:discussion_origin_volatile_depletion}). \bigskip

\begin{figure}[p]
	\centering
	\includegraphics[width=1\linewidth]{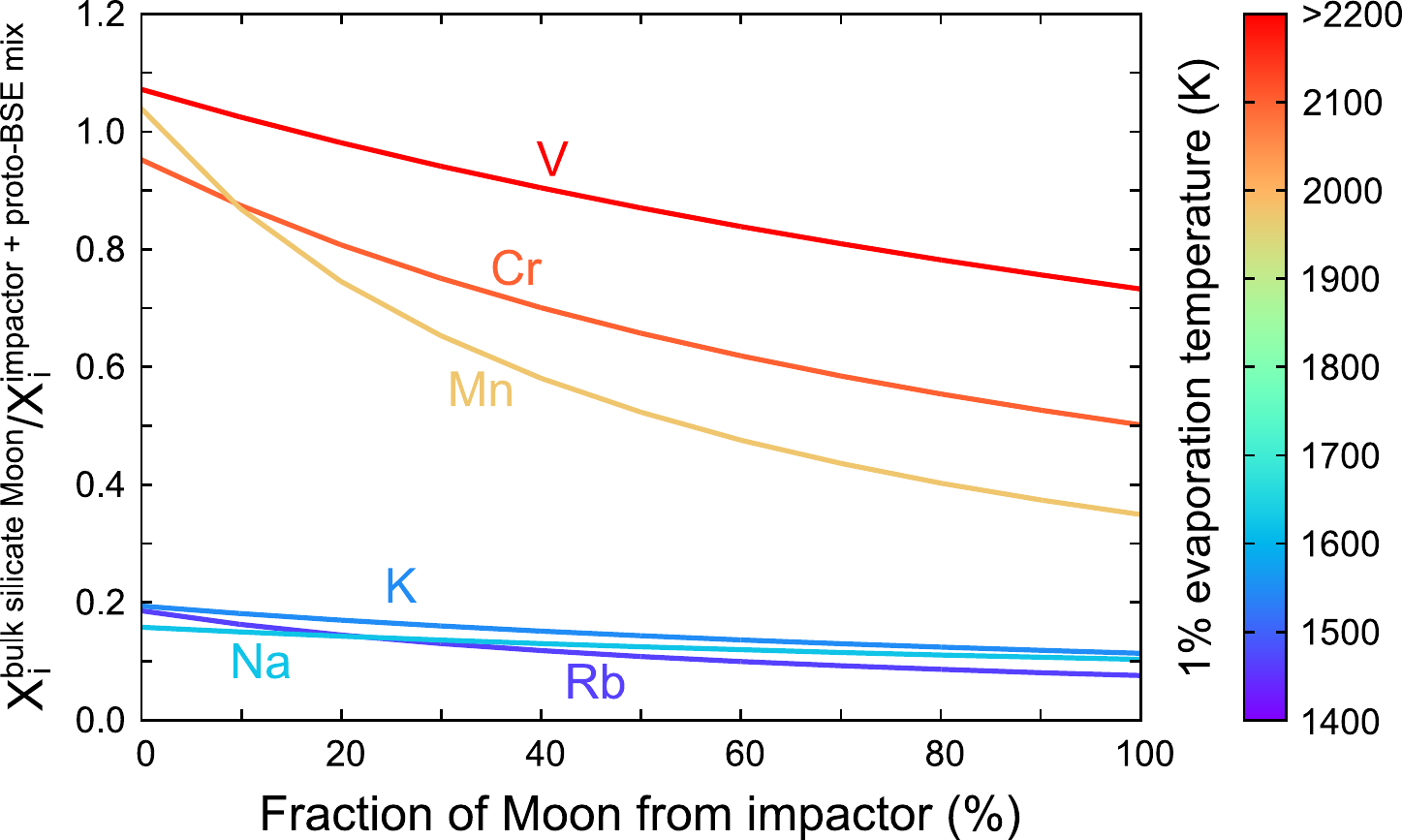}
	\caption{Abundances of moderately volatile elements ($ X\mathrm{_i} $) in bulk silicate Moon and mixture of the Mars-like Moon-forming impactor's mantle and proto-BSE, plotted against mass fraction of the impactor in Moon. $ X\mathrm{^{bulk~silicate~Moon}_i} $ are based on \citet{warren2005new}, \citet{mcdonough1992potassium}, and \citet{oneil1991origin}. Colors of curves correspond 1\% evaporation temperatures of elements from a silicate melt at $ \log f_{\mathrm{O_2}} = -10$ at 1 bar \citep{sossi2019evaporation}.}
	\label{fig:mix_imp_protoE}
\end{figure}

The mass fraction of the impactor contributing to the lunar composition provides a critical constraint on the lunar formation models. The lunar mantle is depleted in nominally lithophile V, Cr and Mn \citep{dreibus1979chemical}. Such lunar mantle depletion can be achieved by high-$T$ or S-rich conditions during lunar core formation, but these conditions seem unlikely \citep{steenstra2016new}. If a Mars-sized impactor with no V, Cr or Mn-depletion in its mantle contributed $ > $70\% of Moon, as predicted by the canonical giant impact models \citep{canup2001origin,canup2004simulations,canup2008accretion}, $ \geq $40\% of evaporative losses of Mn and Cr are needed to produce their depletion in the lunar mantle (\cref{fig:mix_imp_protoE}). Such significant evaporation of these elements is inconsistent with their least volatile nature among MVE \citep{gellissen2019heating,sossi2019evaporation}. Thus, the proto-Earth's V, Cr and Mn-depleted mantle should be a primary source of the Moon-forming materials, as supported by their isotopic similarities \citep[e.g.,][]{,wiechert2001oxygen,warren2011stable,zhang2012proto,greenwood2018oxygen}. A recent geochemical model of Earth's Moon formation and differentiation prefers a present-day BSE-like MVE composition for the bulk proto-Moon \citep[i.e., lunar source materials before gas-melt fractionation;][]{righter2019volatile}, which is consistent with the proto-Earth's MVE abundance and its large contribution to the lunar source materials (\cref{fig:mix_imp_protoE,tab:proto-Earth}). The proto-Earth origin of the Moon is also consistent with recent particle hydrodynamic collision simulations \citep{hosono2019terrestrial} which showed that a terrestrial magma ocean might be a source of lunar building blocks. \bigskip

\subsubsection{Mars' accretion and core formation}
\label{sec:formation_Mars}

Our compositional estimate of Mars requires a much simpler formation history as compared to Earth's multi-stage formation scenario. The lack of depletion in nominally lithophile elements in the BSM (V, Cr, and Mn; \cref{fig:volatility_trend_siderophile}) is consistent with its accretion under uniformly oxidized and low-$ T $ conditions \citep[e.g.,][]{wood2008core}. Mars might have experienced a large impact(s) that produced its hemispheric crustal dichotomy and possibly Martian moons \citep[e.g.,][]{marinova2008mega,canup2018origin}, but estimates of a putative small impactor \citep[$ \sim $0.1\% of the Mars' mass;][]{canup2018origin} would result in negligible changes in its bulk composition. The 0.5--1\% mass addition of chondritic late accretion material provides highly siderophile and chalcophile elements to the Martian mantle \citep{tait2018chondritic,yoshizaki2019mars_long}. \bigskip

\citet{righter2011moderately} and \citet{yang2015siderophile} estimated Martian core formation happening at 14 $ \pm $ 3 GPa and 2100 $ \pm $ 200 K based on its previous compositional models. Following the approaches of \citet{righter2011moderately} and \citet{yang2015siderophile}, the Martian siderophile element distribution of \citet{yoshizaki2019mars_long} can be modeled by single stage $ P $-$ T $ and redox conditions (\cref{fig:Mars_D_model}), whereas there is no unique solution for Martian core formation (\cref{fig:calc_core-mantle}). \bigskip

The simple formation history of Mars, combined with its rapid and early accretion \citep{dauphas2011hf,kruijer2017early,bouvier2018evidence,marchi2020compositionally}, is consistent with its status as a planetary embryo. Given compositional and redox state gradients in the protoplanetary disk, Mars might record the chemical characteristics of nebular materials in the Mars' orbit in the first few years of the solar system, whereas Earth might have incorporated oxidized materials from greater heliocentric distance in its later accretion stage \citep[\cref{sec:Mars-like_impactor};][]{rubie2015accretion}. The Mars' status as a planetary embryo suggest that a Mars-sized body commonly has a Mars-like composition, supporting the Mars-like Moon-forming impactor scenario (\cref{sec:Mars-like_impactor}). \bigskip

\subsection{Conditions for a habitable planet formation}
\label{sec:habitable_planet_condition}

The early $ \mathrm{\tau_{Mars}^{accretion}} $ age implies that Mars underwent global-scale silicate melting and rapid core formation due to heating from short-lived \ce{^{26}Al} and \ce{^{60}}Fe \citep[\cref{fig:Mars_growth_model};][]{dauphas2011hf,mcdonough2020radiogenic}. The peak radiogenic heating occurs at about 1 to 5 Myr after $ t_0 $, well within the time scale for Mars accretion. With $ \mathrm{\tau_{Mars}^{accretion}} = $ 2 Myr, the radiogenic energy supplied by $^{26}$Al is comparable to Mars' gravitational binding energy ($ \sim $ 7 $ \times $ 10$ ^{29} $ J and $ \sim $10$ ^{30} $ J, respectively). During the first 10 Myr, radiogenic heating (\ce{^{26}Al}, \ce{^{60}}Fe and \ce{^{40}}K, in order of significance) is comparable to the planet's primordial energy and is a major control on its thermal evolution. \bigskip

Mars has the attributes needed for a rocky planet to be biologically available in its early history \citep[e.g.,][]{ehlmann2016sustainability}, and it has a higher bulk heat production than Earth's (3.9 vs 3.3 pW/kg; \cref{tab:main_table1}). Nonetheless, it has rapidly lost much of its primordial energy (i.e., accretion and core differentiation) due to its larger surface to volume ratio (a factor of 2) and smaller core size (i.e., reduced bottom heating), and it is in waning stages of limited fuel resources \citep{parro2017present,yoshizaki2019mars_long}. Basal heating of the Martian mantle by its core enhances its thermal evolution, while the transfer of hydrogen from the adjacent ringwoodite to the core \citep{shibazaki2009hydrogen,yoshizaki2019mars_long} reduces the lifetime of the dynamo \citep{orourke2019hydrogenation}. Collectively, these processes likely contributed to dynamo termination at $ \sim $4 Ga and loss of the protective magnetosphere \citep{acuna1999global,arkani2004timing,lillis2008rapid}. This magnetosphere shields the planet from atmospheric losses, enhances its surface UV radiation, and leads to dramatic climate changes \citep{ehlmann2016sustainability}. \bigskip

Volatile elements may play a critical role in establishing the amount of light elements in and solidus of the core. The amount of water and other volatile species in the planet's interior and surface may potentially create the appropriate conditions for the initiation of plate tectonics \citep[e.g.,][]{albarede2009volatile,ehlmann2016sustainability}. Likewise, the heat-producing elements and a reduced core solidus keep the metallic core convecting and lead to the creation of a magnetic field, which shields a planet's surface from cosmic rays. Together, heat-producing and volatile elements regulate a planet's cooling history, drive crustal differentiation, and make it habitable \citep{ehlmann2016sustainability}. The simple formation history of Mars \citep[\cref{sec:formation_Mars};][]{rubie2015accretion} emphasizes the uniqueness of Earth, the sole habitable planet in our solar system. In turn, the Earth-Mars comparison indicates that high-temperature nebular chemical processes and timescales of planetary accretion are essential in making habitable planets. \bigskip

Depletion in volatile elements appears to be a common feature of the terrestrial planets \citep[\cref{fig:volatility_trend};][]{surkov1987uranium,peplowski2011radioactive}, and may likely be so for rocky exoplanets \citep{harrison2018polluted,harrison2021evidence}. The relationship between the planetary volatile depletion, size, accretion timescale and abundance of chondrules (\cref{fig:K-Th_parameters}) predict accretion timescales of Venus and Mercury of 30--100 Myr and 2--10 Myr, respectively, based on their Earth- and Mars-like K/Th ratios, respectively \citep{surkov1987uranium,peplowski2011radioactive}. The predicted ages of Mercury and Venus provide a foundation for future investigation of their thermal history and habitability. \bigskip

Venus and Earth are quite similar in their physical and chemical properties \citep[size, bulk K/Th ratio, Mg\# in basalt and possibly MOI; e.g.,][]{surkov1987uranium,dumoulin2017tidal}, but their present-day surface conditions are distinct. Venus does not show a clear evidence for a giant impact, and its size and K/Th ratio \citep[$ \sim $3000;][]{surkov1987uranium} are comparable to those of the proto-Earth (K/Th $ \sim $ 3200; \cref{tab:proto-Earth}). Therefore, the present-day Venus might be comparable to the pre-impact proto-Earth. Further observational, cosmochemical and theoretical investigations of Venus may provide useful insights into the pre- and post-formational history of Earth, the only habitable planet in the today's solar system. \bigskip

\section{Conclusions}

The refractory element enhancement and volatile depletion of Earth and Mars were established by a nebular chemical fractionation. Post-accretionary losses of moderately volatile elements are negligible. The degree of volatile element depletion correlates with the abundance of accreted chondrules, planetary size, and their accretion timescale. The present-day bulk silicate Earth composition is consistent with a Moon-forming impactor having a Mars-like size and composition. Planetary chemistry, which is related to many factors including the building block composition, the timing, duration and sequence of accretion and its differentiation history, play an essential role in making a planet habitable. \bigskip

\subsection*{Acknowledgments}
We thank many our colleagues who have listened to various versions of this project and given feedback, especially Eiji Ohtani, Nick Schmerr, Beda Roskovec, Ond\v{r}ej \v{S}r\'{a}mek, Sarah Stewart-Mukhopadhyay, Kevin Righter, and Henri Samuel. We also thank Attilio Rivoldini for helpful comments. We greatly appreciate Hugh O'Neill, Bernard Wood, and an anonymous referee for their constructive reviews, which helped improve the manuscript. We thank the journal editor Astrid Holzheid for editorial efforts. TY acknowledges supports from the Japanese Society for the Promotion of Science (JP18J20708) and GP-EES and DIARE research grants. WFM gratefully acknowledges NSF support (EAR1650365).

\subsection*{Author contributions}

TY and WFM proposed and conceived various portions of this study and together calculated the compositional models of planets. The manuscript was jointly written by TY and WFM and they read and approved the final manuscript.

\subsection*{Competing interests}

The authors declare no competing interests.

\subsection*{Additional information}

Correspondence and requests for materials should be addressed to TY.

\subsection*{Data and materials availability}

Materials used in this study are available within the paper or supplementary materials.

\clearpage


\begin{table}[p]
\centering
\footnotesize
\begin{threeparttable}
	\caption{Physical and chemical properties of Earth and Mars. Modeled values are in normal and reference values in italic fonts.}
	\label{tab:main_table1}%
	\begin{tabularx}{480pt}{ccccccc}
	\toprule
	Observation & Unit & Crust & Mantle & Core  & Bulk planet  & Reference value \\
	\midrule
	\textbf{Earth} &&&&&&\\
	Mass\tnotex{lab_chambat}\tnote{,}\tnotex{lab_wipp}  & kg & 3.12 $ \times  $ $10^{22}  $  & 4.00 $\times $ $10^{24} $  & 1.94 $  \times $ $ 10^{24} $  &  5.972 $\times $ $ 10^{24} $ &  \textit{5.97218(60)} $ \times  $ \textit{10}$ ^{24} $ \\
	Mean density\tnotex{lab_chambat}\tnote{,}\tnotex{lab_wipp}\tnote{,}\tnotex{lab_PREM} & kg/m$ ^3 $  & 2800  & 4400  & 11870  & 5514 & \textit{5514(2)}  \\
	Moment of inertia\tnotex{lab_chambat} & -- & 1\%    & 88\%    & 11\%     & 0.3308 & \textit{0.330690(9)} \\
	\multirow{2}[0]{*}{Heat production (K, Th, U)}\tnotex{lab_wipp}\tnote{,}\tnotex{lab_mcd14}\tnote{,}\tnotex{lab_jaup}\tnote{,}\tnotex{lab_rud}	& TW & \textit{7.3} & \textit{12.6} & 0 & \textit{19.9} & \textit{46(3)}\tnotex{lab_hf}  \\
	& pW/kg & 232  & 3.1  & 0 &  3.3  & --  \\	
	\midrule
	\textbf{Mars} &&&&&&\\
	Mass\tnotex{lab_kono} & kg &  2.56 $ \times $ $10^{22} $ &5.01 $\times $ $ 10^{23} $ & 1.17 $\times $ $ 10^{23} $ &  6.419 $\times$ $10^{23} $ &  \textit{6.417(3)} $\times $ \textit{10}$^{23} $  \\
	Mean density\tnotex{lab_riv} & kg/m$ ^3 $ & 3010  & 3640  & 6910  & 3936  & \textit{3935(1)}  \\
	Moment of inertia\tnotex{lab_kono} & -- & 7\%    & 89\%    & 4\%     & 0.3638 & \textit{0.3639(1)}  \\
	\multirow{2}[0]{*}{Heat production (K, Th, U)}\tnotex{lab_taylor}\tnote{,}\tnotex{lab_yoshi}\tnote{,}\tnotex{lab_parro}  & TW & 1.3 & 1.3 & 0 & 2.5 & \textit{2.7(2)}\tnotex{lab_hf} \\
	& pW/kg & 49 & 2.5   & 0 &  3.9 & -- \\
	\midrule
	&&&&&&\\
	\end{tabularx}
	\begin{tabular*}{480pt}{ @{\extracolsep{\fill}} cccc @{\extracolsep{\fill}} cccc  @{\extracolsep{\fill}} cccccc @{\extracolsep{\fill}} }
	\midrule
	& \multicolumn{2}{c}{Mantle + crust\tnotex{lab_mcd14}\tnote{,}\tnotex{lab_yoshi}} &  &       & \multicolumn{2}{c}{Core\tnotex{lab_mcd14}\tnote{,}\tnotex{lab_yoshi}} &  &       & \multicolumn{2}{c}{Bulk planet\tnotex{lab_mcd14}\tnote{,}\tnotex{lab_yoshi}\tnote{,}\tnotex{lab_current}} & \multicolumn{3}{c}{Chondrite\tnotex{lab_norm}} \\
	\cline{2-3}\cline{6-7}\cline{10-11}\cline{12-14}
	wt\%       & Earth &            Mars            &  & wt\%  & Earth &       Mars     &  & wt\%   &          Earth     & Mars &  EH  &  L  &                 CI                \\
	\midrule
	\ce{SiO2}  & 44.9  &            45.5            &  & Fe    & 85.5  &       79.5       &  & O      &           29.7       & 36.3    & 30.0     & 35.8    &          29.9             \\
	\ce{TiO2}  & 0.20  &            0.17            &  & Ni    &  5.1  &       7.4        &  & Fe    &           32.0      & 23.7       & 31.0  & 22.7     &       27.7               \\
	\ce{Al2O3} & 4.44  &            3.59            &  & O     &   2   &       5.2        &  & Mg     &           15.4    & 15.3       & 11.3 & 15.7       &         14.3              \\
	\ce{MnO}   & 0.14  &            0.37            &  & S     &  1.8  &       6.6        &  & Si    &           16.1    & 17.4        & 17.8  & 19.5       &        16.0               \\
	\ce{FeO}   & 8.06  &            14.7            &  & H     & 0.06  &       0.9        &  & Ni     &           1.82    & 1.4         & 1.87& 1.27 &               1.58               \\
	\ce{MgO}   & 37.8  &            31.0            &  & Co    & 0.25  &       0.33       &  & Ca    &           1.71     & 1.69       & 0.91& 1.38  &               1.36               \\
	\ce{CaO}   & 3.54  &            2.88            &  & P     & 0.20  &       0.33       &  & Al     &           1.59     & 1.56       & 0.86& 1.29 &               1.26               \\
	\ce{Na2O}  & 0.36  &            0.59            &  & Si    &   4   &        0         &  & S     &           0.59    & 1.2         & 6.2 & 2.3   &               8.0                \\
	\ce{K2O}   & 0.034 &           0.043            &  & Cr    & 0.75  &        0         &  &       &      &                          &      &      &                                  \\
	\ce{P2O5}  & 0.021 &            0.17            &  &       &       &                  &  & Total  &           99.0 & 98.5          & 100  & 100  &               100                \\
	\ce{NiO}   & 0.25  &           0.046            &  & Total & 99.7  &      100.2       &  &       &      &                          &      &      &                                  \\
	\ce{Cr2O3} & 0.15  &            0.88            &  &       &       &                  &  & Mg/Si &           0.96     & 0.88       & 0.63& 0.81  &               0.89               \\
	   &       &                            &  &       &       &                  &  & Al/Si &           0.10    & 0.09         & 0.05 & 0.07&               0.08               \\
	K (ppm)    &  280  &            360             &  &       &       &                  &  & Fe/Si  &           2.0   & 1.4        & 1.7 & 1.2 &               1.7               \\
	Th (ppb)   &  76   &             68             &  &       &       &                  &  &       &      &                          &      &      &                                  \\
	U (ppb)    &  20   &             18             &  &       &       &                  &  &       &      &                          &      &      &                                  \\
	           &       &                            &  &       &       &                  &  &       &      &                          &      &      &                                  \\
	Total      & 100.2 &            99.9            &  &       &       &                  &  &       &      &                          &      &      &                                  \\
	\bottomrule
	\end{tabular*}
	\begin{tablenotes}[para]
\item[a] \citet{chambat2010flattening}. \label{lab_chambat}
\item[b] \citet{wipperfurth2019reference}. \label{lab_wipp}
\item[c] \citet{dziewonski1981preliminary}. \label{lab_PREM}
\item[d] \citet{mcdonough2014compositional}. \label{lab_mcd14}
\item[e] \citet{jaupart2015temperatures}. \label{lab_jaup}
\item[f] \citet{rudnick2014composition}. \label{lab_rud}
\item[g] Global surface heat loss. \label{lab_hf}
\item[h] \citet{konopliv2016improved}. \label{lab_kono}
\item[i] \citet{rivoldini2011geodesy}. \label{lab_riv}
\item[j] \citet{taylor2009planetary}. \label{lab_taylor}
\item[k] \citet{yoshizaki2019mars_long}. \label{lab_yoshi}
\item[l] \citet{parro2017present}. \label{lab_parro}
\item[m] This study. \label{lab_current}
\item[n] \citet{alexander2019quantitative_CC,alexander2019quantitative_NC} (Volatile-free, normalized to total = 100 wt\%). \label{lab_norm}
	\end{tablenotes}
	\end{threeparttable}
\end{table}
\clearpage

\clearpage

\clearpage

	\clearpage
			
\clearpage

\clearpage

\clearpage

\clearpage

\clearpage

\begin{figure}[p]
	\centering
	\includegraphics[width=1\linewidth]{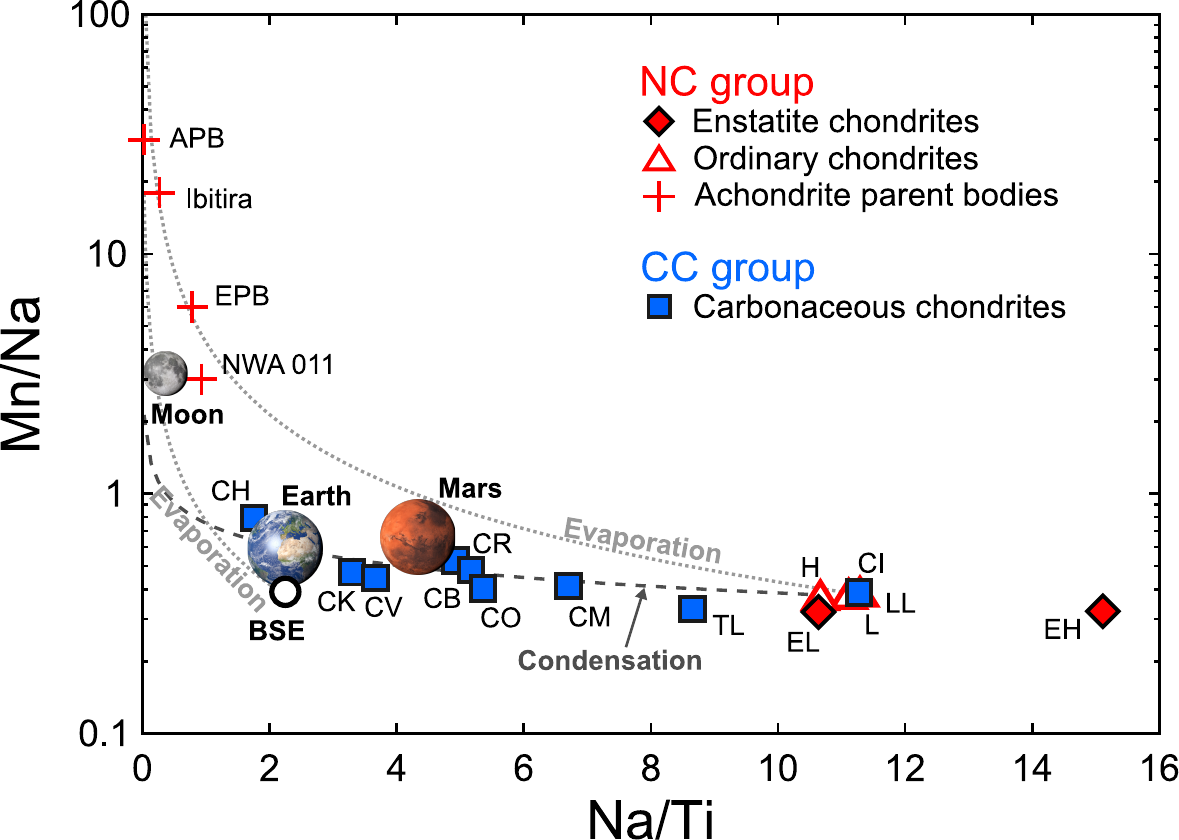}
	\caption{Na/Ti vs Mn/Na in chondrites \citep{alexander2019quantitative_CC,alexander2019quantitative_NC}, differentiated asteroids \citep{oneill2008collisional}, Moon \citep{dauphas2014geochemical}, Earth \citep[this study;][]{mcdonough2014compositional,siebert2018chondritic} and Mars \citep{yoshizaki2019mars_long}. Dark gray line shows incomplete condensation from a gas of CI composition. Light gray lines correspond to evaporative loss of Na and Mn from CI and the bulk silicate Earth (BSE) compositions, respectively. See \cref{sec:Mn_in_BSE} for details of model calculations.} \label{fig:MnNa_NaTi}
\end{figure}
\clearpage

\clearpage

\begin{figure}[p]
	\centering
	\includegraphics[width=1\linewidth]{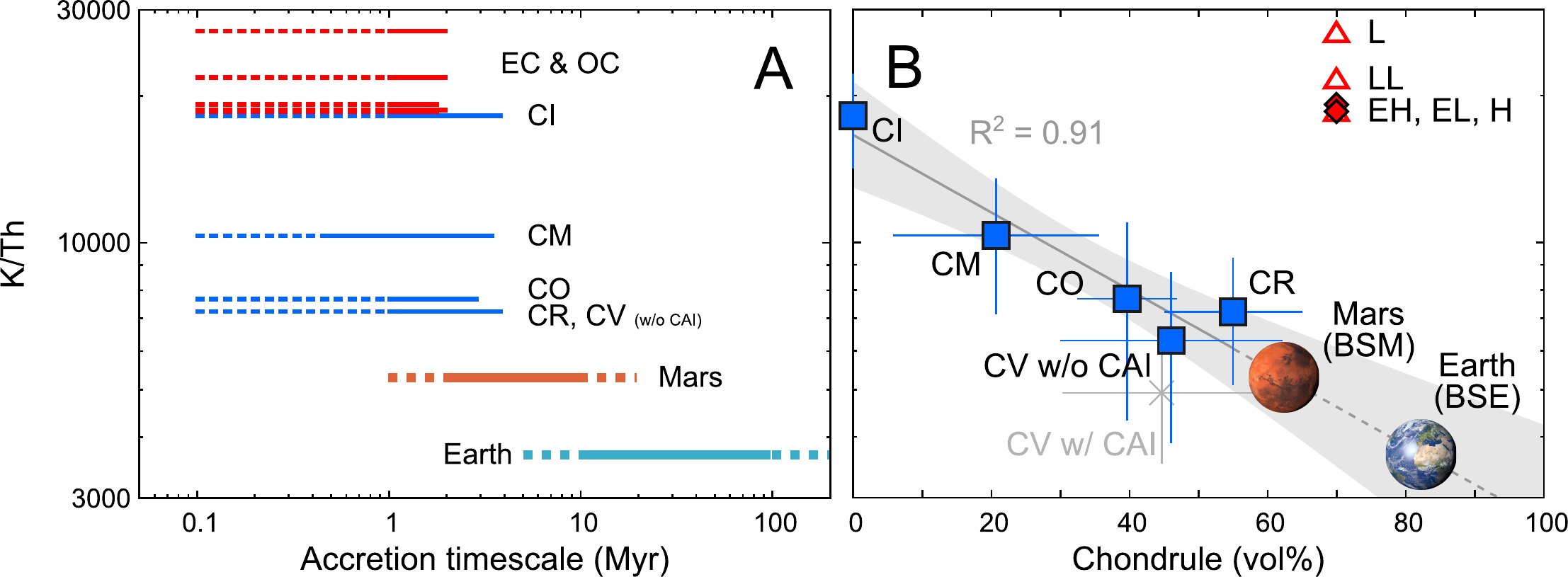}
	\caption[Potassium/Th ratios, accretion timescales, and chondrule abundances of chondrites,and planets.]{(A) Accretion timescale (Myr) vs K/Th ratio of chondrites and terrestrial planets. Duration of accretion of chondritic asteroids is based on the Pb-Pb, Hf-W and Al-Mg ages of chondrules \citep{amelin2002lead,amelin2007pb,connelly2008chronology,connelly2012absolute,connelly2009pb,bollard2017early,bollard2019combined,villeneuve2009homogeneous,nagashima201726al,schrader2017distribution,kita201326al,pape2019time,budde2016tungsten,budde2018hf}, Mn-Cr ages of asteroidal secondary alteration products \citep{fujiya2012evidence,doyle2015early}, and thermal modeling of asteroids \citep{sugiura2014correlated}. Accretion timescales of Earth and Mars are defined based on \citet{kruijer2017early,kruijer2019great}, \citet{kleine2009hf}, \citet{dauphas2011hf}, \citet{kruijer2017early}, and \citet{connelly2019pb}. (B) Abundance of chondrules (vol\%) vs K/Th value of chondrites. Error bars (shown only for carbonaceous chondrites) represent 2 standard deviations, and gray area corresponds to the 95\% confidence interval of the linear regression. The abundances of chondrules in chondrites are based on \citet{mcsween1977carbonaceous}, \citet{mcsween1977petrographic}, \citet{mcsween1979alteration} and  \citet{scott2014chondrites}. The amounts of chondrules in Earth and Mars are estimated based on their K/Th values and extrapolation of the carbonaceous chondritic trend. See \cref{sec:data_fig_9} for a method to calculate the CAI-free bulk CV composition.} 
	\label{fig:K-Th_parameters}
\end{figure}
\clearpage

\clearpage

\begin{figure}[p]
	\centering
	\includegraphics[width=0.9\linewidth]{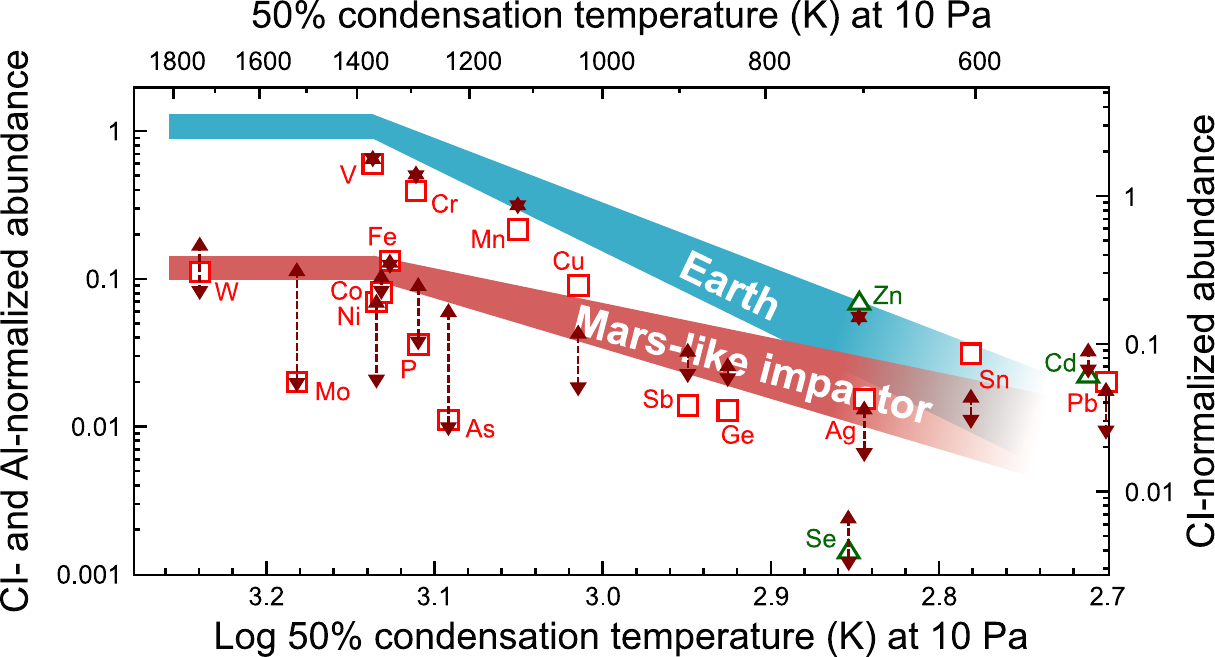}
	\caption{Siderophile (red square) and chalcophile (green triangle) element abundances in the BSE, and a model BSE composition after the sulfide matte subtraction (dark red arrow) predicted by a Mars-like (i.e., size and composition) Moon-forming impactor scenario. The dark red arrows represent the model compositional range based on sulfide-silicate partition coefficients of elements (\cref{tab:D-values}). Highly siderophile and chalcophile elements are not shown, as their abundances in the present-day BSE can be explained by the late addition of volatile-rich materials (Step 4). Also shown are the Mars-like impactor's contribution (red band) in the present-day Earth composition (blue band). The values for x and y axes follow the same conventions as \cref{fig:volatility_trend_siderophile}.}	\label{fig:volatility_trend_impactor}
\end{figure}
\clearpage

\clearpage

\putbib[myrefs]
\end{bibunit}

\clearpage

\setcounter{page}{1}

\begin{appendices}

\counterwithin{figure}{section}
\counterwithin{table}{section}
\counterwithin{equation}{section}

\begin{bibunit}[elsarticle-harv]

\begin{center}
{\Large \textbf{Supplementary materials for \\
\vspace{1ex}
Earth and Mars--distinct inner solar system products}} \\
\vspace{2ex}
Takashi Yoshizaki*$ ^1 $ and William F. McDonough$ ^{1,2,3} $
\vspace{1ex}

$ ^1 $Department of Earth Science, Graduate School of Science, Tohoku University, Sendai, Miyagi 980-8578, Japan \\
$ ^2 $Department of Geology, University of Maryland, College Park, MD 20742, USA \\
$ ^3 $Research Center for Neutrino Science, Tohoku University, Sendai, Miyagi 980-8578, Japan \\
(*Corresponding author. E-mail: takashiy@tohoku.ac.jp) \\
\vspace{1ex}

\today

\end{center}

\section{Supplementary materials}
		
\subsection{Updates in planetary compositions}
\label{sec:updates_comp}

Here we revise compositional models of Earth \citep{mcdonough1995composition,palme2014cosmochemical,mcdonough2014compositional} and Mars \citep{yoshizaki2019mars_long} for some elements based on recent literature \citep{braukmuller2018chemical,alexander2019quantitative_CC,alexander2019quantitative_NC,wood2019condensation,lodders2020solar}.

\subsubsection{Manganese in the bulk Earth and its core}
\label{sec:Mn_in_BSE} 

The BSE has a lower Mn/Na value than that predicted by a condensation model based on the BSE's Na/Ti value (\cref{fig:MnNa_NaTi}). This Mn depletion in the BSE reflects its partial incorporation into the metallic core \citep{mcdonough2014compositional}, which is consistent with moderately siderophile behavior of Mn under highly reduced condition in the proto-Earth \citep{siebert2018chondritic}. Thus, the Mn abundance in the bulk Earth includes contributions from the core and the mantle. \bigskip

The Mn abundance in the BSE is $ \sim $1,050 ppm \citep{mcdonough1995composition,palme2014cosmochemical}. \citet{mcdonough2014compositional} estimated Mn abundance of the bulk Earth and core of 800 ppm and 300 ppm, respectively, based on a linear trend in a plot of Na/Ti vs Mn/Na among chondrites. This estimate gives core-mantle partition coefficient of Mn ($ D\mathrm{^{core-mantle}_{Mn}} $) of $ \sim $0.3, lower than that predicted by high P-T experimental studies \citep[e.g.,][]{siebert2018chondritic}. Recently, \citet{siebert2011systematics} re-evaluated the bulk Earth's bulk Mn content of 900--1300 ppm based on moderately siderophile behavior of Mn under Earth's core formation conditions. \bigskip

Here we test if the ratios of Na, Mn and Ti of Earth \citep{mcdonough1995composition,palme2014cosmochemical,mcdonough2014compositional,siebert2018chondritic}, Mars \citep{yoshizaki2019mars_long} and other small solar system bodies \citep{oneill2008collisional,wasson1988compositions,braukmuller2018chemical} reflect incomplete condensation of a solar nebular gas or post-accretionary evaporative loss of moderately volatile elements (MVE). We modeled the incomplete condensation in the solar nebula, in which volatile depletion in condensates is strictly depending on the 50\% condensation temperature of elements, following \citet{cassen1996models} and \citet{cassen2001nebular}. We assume that a log of CI-normalized abundance of non-refractory elements in condensates shows a linear trend when plotted against the 50\% condensation temperature of elements during a partial condensation. Thus, the composition of condensates is formulated as:
\begin{equation}
    \log \frac{C\mathrm{^{cond}_i}}{C\mathrm{^{CI}_i}}   =  
    \begin{cases} - (\log{T_c\mathrm{(ref)}} - \log{T_c\mathrm{(i)}}) \times f & (T_c\mathrm{(i)} < T_c\mathrm{(ref)}) \\
    1 & (T_c\mathrm{(i)} \geq T_c\mathrm{(ref)})
    \end{cases}
    \label{eq:CI-norm_abn_model}
\end{equation}
where 
$ C\mathrm{^{cond}_i} $ and $ C\mathrm{^{CI}_i} $ are abundances of an element i in a condensate and CI chondrite, respectively,
$ T_c\mathrm{(i)} $ and $ T_c\mathrm{(ref)} $ are 50\% condensation temperatures of the element i and vanadium, which has the lowest condensation temperature among refractory elements \citep[1370 K at 10 Pa;][]{wood2019condensation}, respectively, and $ f $ is a slope of the linear volatility trend in a $ \log (C\mathrm{^{cond}}/C\mathrm{^{CI}}) $ vs $ \log{T_c} $ space (i.e., degree of volatile depletion in the condensate). Condensation temperature of elements from \citet{wood2019condensation} and CI abundance from \citet{lodders2020solar} are adopted in the calculation. The evaporative losses of elements from a high-temperature liquid were modeled using equations and thermochemical data from \citet{sossi2019evaporation} and \citet{chase1998nist}. \bigskip

Our simplified calculation of incomplete condensation well reproduces a chemical composition of chondrites and Mars (\cref{fig:MnNa_NaTi}). The bulk Earth composition \citep{mcdonough2014compositional} with the updated Mn abundance \citep{siebert2018chondritic} is consistent with the incomplete condensation model. In contrast, compositions of differentiated asteroids (e.g., parent bodies of angrites and eucrites) and Earth's moon are consistent with the evaporation from a CI and the BSE compositions, respectively. These observations, combined with a lack of heavy MVE isotope enrichment composition of terrestrial samples \citep[e.g., K, Fe, Si, Mg, Zn, Cd;][]{humayun1995potassium,poitrasson2004iron,norman2006magnesium,wombacher2008cadmium,herzog2009isotopic,paniello2012zinc,zambardi2013silicon,sossi2016on,sossi2018zinc}, document that the bulk Earth's MVE abundance is established by a nebular condensation, rather than post-accretionary evaporative losses. With the bulk Earth's Na/Ti $ \sim $ 2.2 \citep{mcdonough2014compositional}, the incomplete condensation model predicts Mn abundance of 1070 ppm and 1120 ppm in the bulk Earth and its core, respectively, with the $ D\mathrm{^{core-mantle}_{Mn}} $ of $ \sim $1.1, which are in agreement with the values predicted by \citet{siebert2018chondritic}. Thus, we update the Mn abundances in the bulk Earth and its core with these values. \bigskip

\subsubsection{Sulfur in the bulk Earth and its core}
\label{sec:S_in_Earth}

The S content in the bulk Earth is of particular interest when estimating a mass fraction of the post-impact sulfide matte. Sulfur is a moderately volatile, chalcophile element so that its abundances in the bulk Earth and core can be constrained by an abundance of lithophile element with similar volatility. Zinc abundance in the BSE has been used to constrain Earth's S content, since Zn does not show deviation from Earth's lithophile volatility trend and its condensation temperature is close to that of S \citep{mcdonough1995composition,dreibus1996cosmochemical}. \bigskip

\citet{mcdonough2014compositional} estimated sulfur in the BSE at 250 $ \pm $ 50 ppm and a bulk Earth with 6350 ppm, based on Earth's volatility trend and S abundance of CI chondrite of 5.40 wt\%. These estimates lead to the prediction of 1.9 wt\% S in Earth's core. The authors used 50\% condensation temperatures of elements from \citet{wasson1985meteorites}, in which $ T_c $ of S and Zn (treated as lithophile) are 648 K and 660 K, respectively. \bigskip

\citet{dreibus1996cosmochemical} assumed a CI-like Zn/S value for Earth given a nearly similar $ T_c $ of S and Zn \citep{wasson1985meteorites}, and estimated 5600 ppm S in the bulk Earth. They used a lower and higher S abundance in the BSE (130 ppm) and CI chondrite (5.9 wt\%) than the values taken by \citet{mcdonough2014compositional}, which gives an estimate of $ \sim $1.7 wt\% S in the Earth's core. \bigskip

Here we revise the S content of the bulk Earth and its core based on recently-revised $ T_c $ \citep{wood2019condensation} and CI abundance \citep{lodders2020solar}. The updated $ T_c\mathrm{(S)} $ and $ T_c\mathrm{(Zn)} $ are 672 K and 704 K, respectively, with a difference of 32 K, which is larger than that of the \citet{wasson1985meteorites}'s value. Thus, we consider that the Zn/S value of the bulk Earth might not be chondritic if these elements are depleted in Earth based strictly on their relative volatilities. We use the S abundance of 250 ppm in the BSE \citep{mcdonough1995composition,palme2014cosmochemical} and 5.36 wt\% in CI chondrite \citep{lodders2020solar}, and modeled a incomplete condensation of moderately volatile elements as we applied for Mn (\cref{sec:Mn_in_BSE}). With these values, we estimate S concentrations of 5900 ppm in the bulk Earth, and 1.8 wt\% in the core. The revised S abundance in Earth is used to estimate the mass of post-impact sulfide matte (\cref{sec:Mars-like_impactor}).
	
\subsubsection{Sulfur in the Martian core}	
\label{sec:S_in_Mars_core}

The S content in the Martian core estimated by \citet{yoshizaki2019mars_long} is much lower than that estimated by \citet{wanke1994chemistry} and \citet{taylor2013bulk}, who constrained the S content in the Martian core by assuming that moderately volatile elements are in chondritic abundances in the bulk Mars. However, the trend of depletion of moderately volatile elements in Mars is inconsistent with assumptions in \citep{yoshizaki2019mars_long}. \bigskip

Physical models of Martian interior \citep[e.g.,][]{khan2008constraining,rivoldini2011geodesy,khan2018geophysical} based on available observations (mass, density, MOI and tidal Love number $ k_2 $) typically use the compositional model of the Martian mantle established by \citet{wanke1994chemistry}. These studies generally predict a larger core (1700--1800 km in radius) with a high S concentration (15--20 wt\%), which are differ from the estimates by \citet{yoshizaki2019mars_long} ($r \sim $ 1600 km; $ \leq $7 wt\% S). These geophysical models use the Mars' $ k_2 $ Love number to estimate its core S content. However, a deconvolution of the $ k_2 $ Love number requires an accurate knowledge of the degree of partial melting and hydration in the mantle, phase- and composition-dependent changes in elastic properties of mantle minerals, and grain size variation \citep[e.g.,][]{nimmo2013dissipation,khan2018geophysical}. Thus, we use the planetary volatility trend to estimate the Mars' bulk and core S abundances \citep{yoshizaki2019mars_long}, following the practice used to constrain the S abundance in the Earth's core \citep{dreibus1996cosmochemical,mcdonough2014compositional}. A direct seismic observation of Martian interior by NASA's ongoing InSight mission will place strong constraints on the size and light element inventory of the Martian core. \bigskip

\begin{figure}[p]
	\centering
	\includegraphics[width=0.7\linewidth]{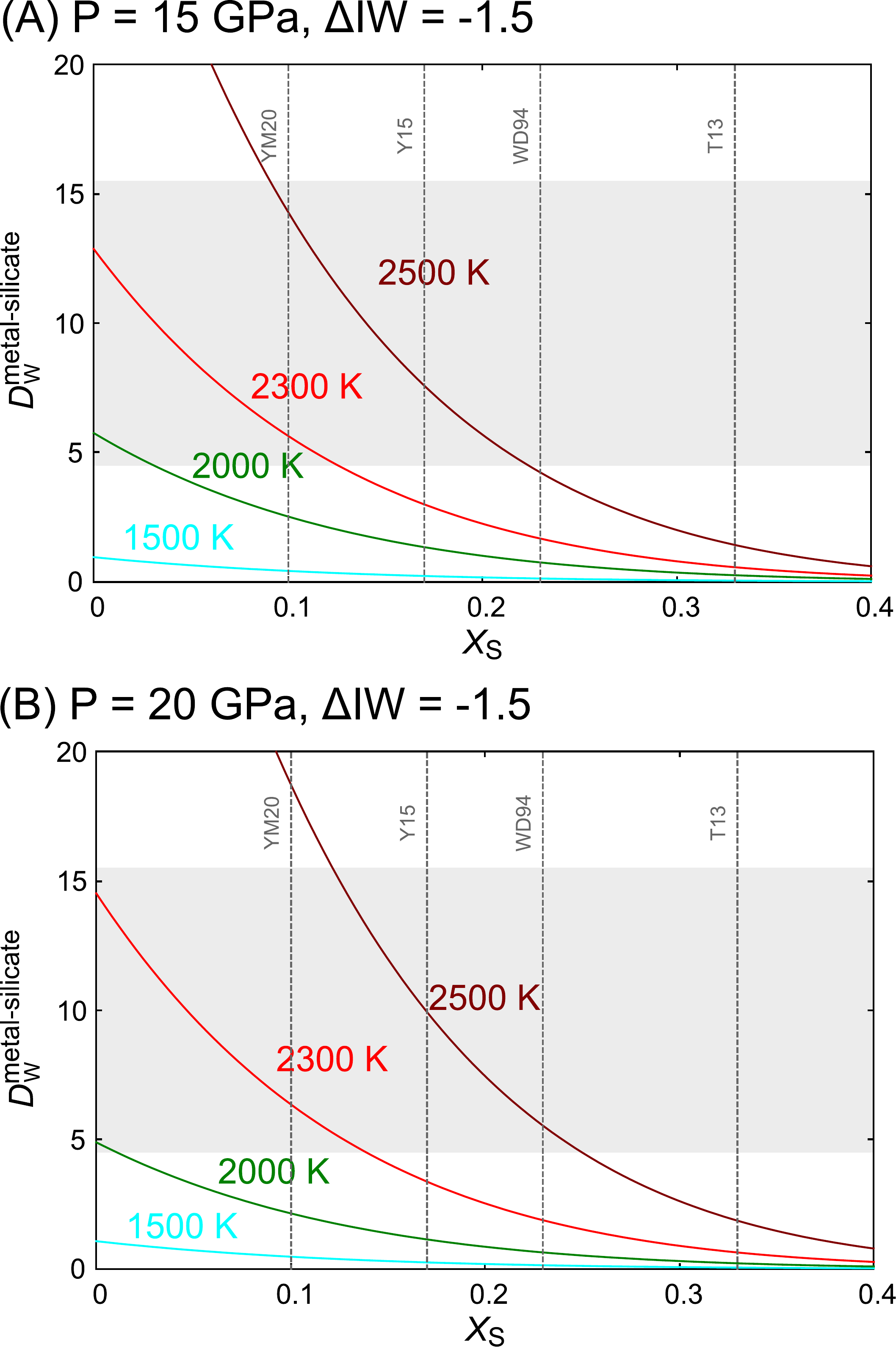}
	\caption{Metal-silicate partitioning coefficients of W at (A) 15 GPa and (B) 20 GPa as a function of mole fraction of S in the metal	($ X_{\mathrm{S}} $), calculated using parameters from \citet{righter2011prediction}. The redox condition is fixed at $ \Delta $ IW $ = -1.5 $. The gray band shows the core-mantle partitioning coefficients ($ \pm $1 standard deviation) of W based on the Mars model of \citet{yoshizaki2019mars_long}. The vertical dashed gray lines show the core models of \citet{yoshizaki2019mars_long} (YM20), \citet{yang2015siderophile} (Y15), \citet{wanke1994chemistry} (WD94), and \citet{taylor2013bulk} (T13).}
	\label{fig:D-W_comparison}
\end{figure}

\begin{figure}[p]
	\centering
	\includegraphics[width=1\linewidth]{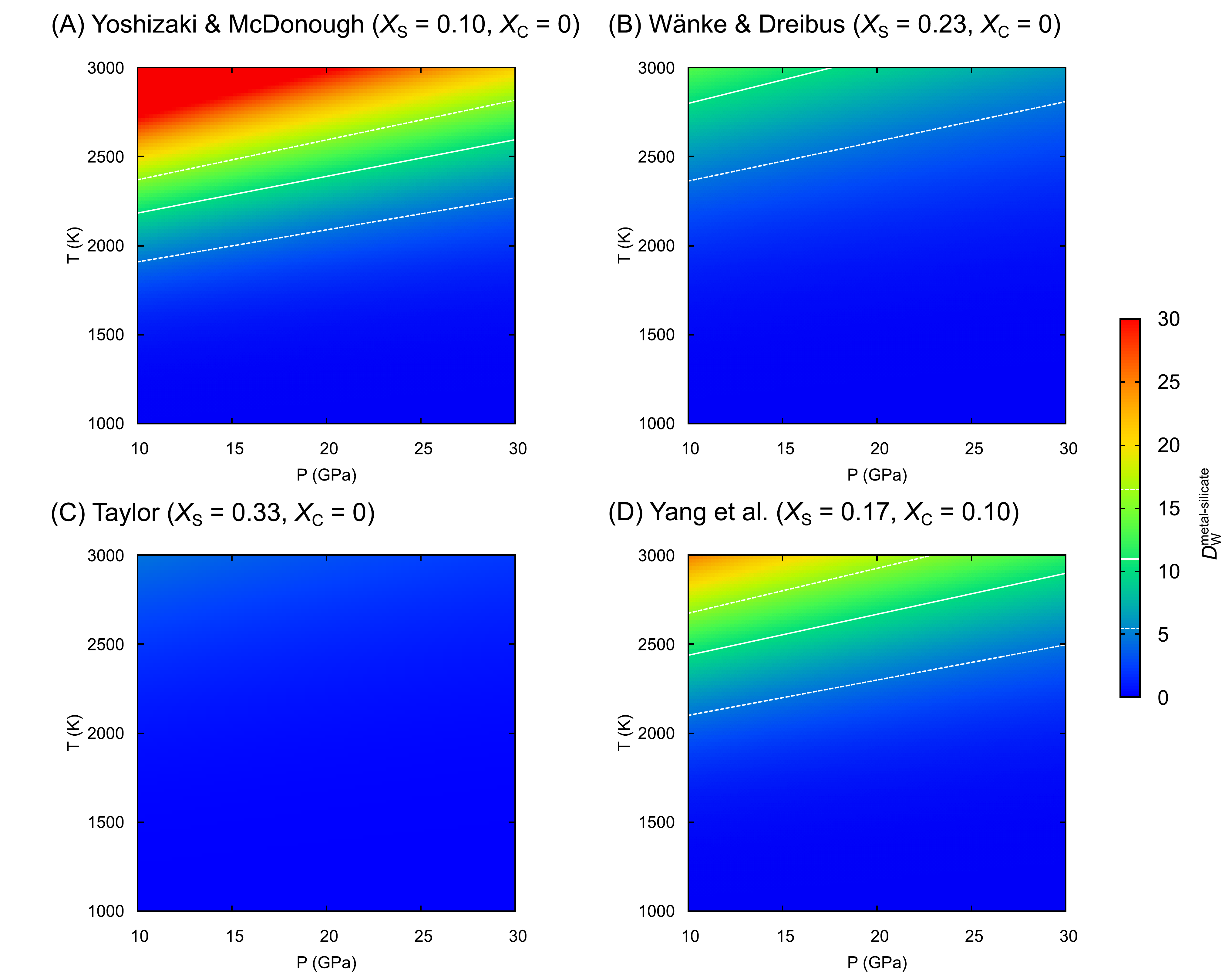}
	\caption{Metal-silicate partitioning coefficients of W calculated for the Martian compositional models of (A) \citet{yoshizaki2019mars_long},  (B) \citet{wanke1994chemistry}, (C) \citet{taylor2013bulk}, and (D) \citet{yang2015siderophile}. White solid and broken lines show $ P $-$ T $ conditions that are consistent with the core-mantle distribution of W in Mars within 1 standard deviations \citep{yoshizaki2019mars_long}.}
	\label{fig:D-W_comparison_PT}
\end{figure}

Core-mantle partitioning behaviors of refractory elements, which are sensitive to the S content of the metal \citep{righter2011prediction,wade2012metal}, provide further constraints on the S concentration in the Martian core. The compositional model of Mars \citep{yoshizaki2019mars_long} predict core-mantle $ D $-value of $ 10 \pm 5.5 $, based on the depletion of W in the BSM. \cref{fig:D-W_comparison,fig:D-W_comparison_PT} show the metal-silicate partitioning behavior of W at $ \Delta $IW $ = -1.5 $ for the Martian core models \citep{wanke1994chemistry,taylor2013bulk,yang2015siderophile,yoshizaki2019mars_long}, estimated based on a compilation of previous experimental constraints \citep{righter2011prediction}. \cref{fig:D-W_comparison} shows that S become less siderophile as the S concentration in the metal increases. For a S-rich core model (>20 wt\%; Taylor 2013), the $ D $-value of W is much lower than the Martian compositional model (\cref{fig:D-W_comparison_PT}C). Models with 10--15 wt\% S in the core \citep{wanke1994chemistry,yang2015siderophile} requires $ \sim $2500 K to explain the BSM's W depletion. The S-poor core composition (S $ \sim $ 7 wt\%; Yoshizaki \& McDonough, 2020) is consistent with a BSM model forming at $ > $2300 K. Thus, the predicted W abundance in the BSM is consistent with lower S models for the Martian core. We note that the temperature conditions required to explain the distribution of W in Mars decreases by $ \sim $200 K if the log $ f_{\ce{O2}} $ value decreased for 0.5 units in the $ \Delta $IW space. \bigskip

Similarly, the behavior of Mo is also sensitive to S contents in a metal \citep{wade2012metal}. Therefore, Mo is another useful element for constraining the S abundance in the Martian core. Unfortunately, the Mo abundance in the BSM is currently poorly constrained. \citet{yoshizaki2019mars_long} found no clear correlation of Mo with other elements in Martian meteorites. \citet{yang2015siderophile} proposed the Mo abundance in the BSM based on Mo-Ce correlation in two Martian meteorites, which were not identified by \citet{yoshizaki2019mars_long}. Future constraints on the Mo abundance in the Martian mantle will provide important constraints on the light element composition of the Martian core.

\subsubsection{Halogens in Earth and Mars}
\label{sec:halogens_Earth_Mars}

Abundances of heavy halogens (Cl, Br and I) in the BSE is enigmatic because of their large depletion in the BSE \citep{mcdonough1995composition,palme2014cosmochemical}, perhaps indicative of unique behaviors of these elements during and/or after the Earth's accretion \citep[e.g.,][]{kramers2003volatile,armytage2013metal,mcdonough2014compositional,sharp2013chlorine,zolotov2007hydrogen,clay2017halogens,jackson2018early,steenstra2020experimental}. The present bulk Earth model \citep{mcdonough2014compositional} attributed this heavy halogen depletion to their incorporation into the core, and estimated core/mantle enrichment factors of 10--15 for these elements. Limited numbers of experimental studies suggest that iodine becomes siderophile during core formation \citep{armytage2013metal,jackson2018early}, whereas Cl does not show such a metal-loving behavior \citep{sharp2013chlorine}. \bigskip

Recently, \citet{clay2017halogens} proposed a lower halogen abundance for chondritic meteorites than previous estimates \citep[e.g.,][]{dreibus1979halogens,lodders2003solar}. By normalizing the BSE abundance of halogens to their updated CI composition, \citet{clay2017halogens} showed that Br and I are plotted on the lithophile volatility trend and no need for any special processes of halogen fractionation. However, even if normalized to \citet{clay2017halogens}'s CI chondrite composition, the BSE's Cl abundance still shows a depletion from the volatility trend. \bigskip

The CI abundances of heavy halogens proposed by \citet{clay2017halogens} has been challenged by \citet{fegley2020volatile}, who showed that these updated CI abundances fall below a curve of nuclide abundance vs mass number. \citet{lodders2020solar} critically evaluated and updated the CI composition proposed by \citet{clay2017halogens} and brought it in line with previous estimates of the CI abundance \citep{dreibus1979halogens,lodders2003solar}. \bigskip

\begin{figure}[p]
	\centering
	\includegraphics[width=0.8\linewidth]{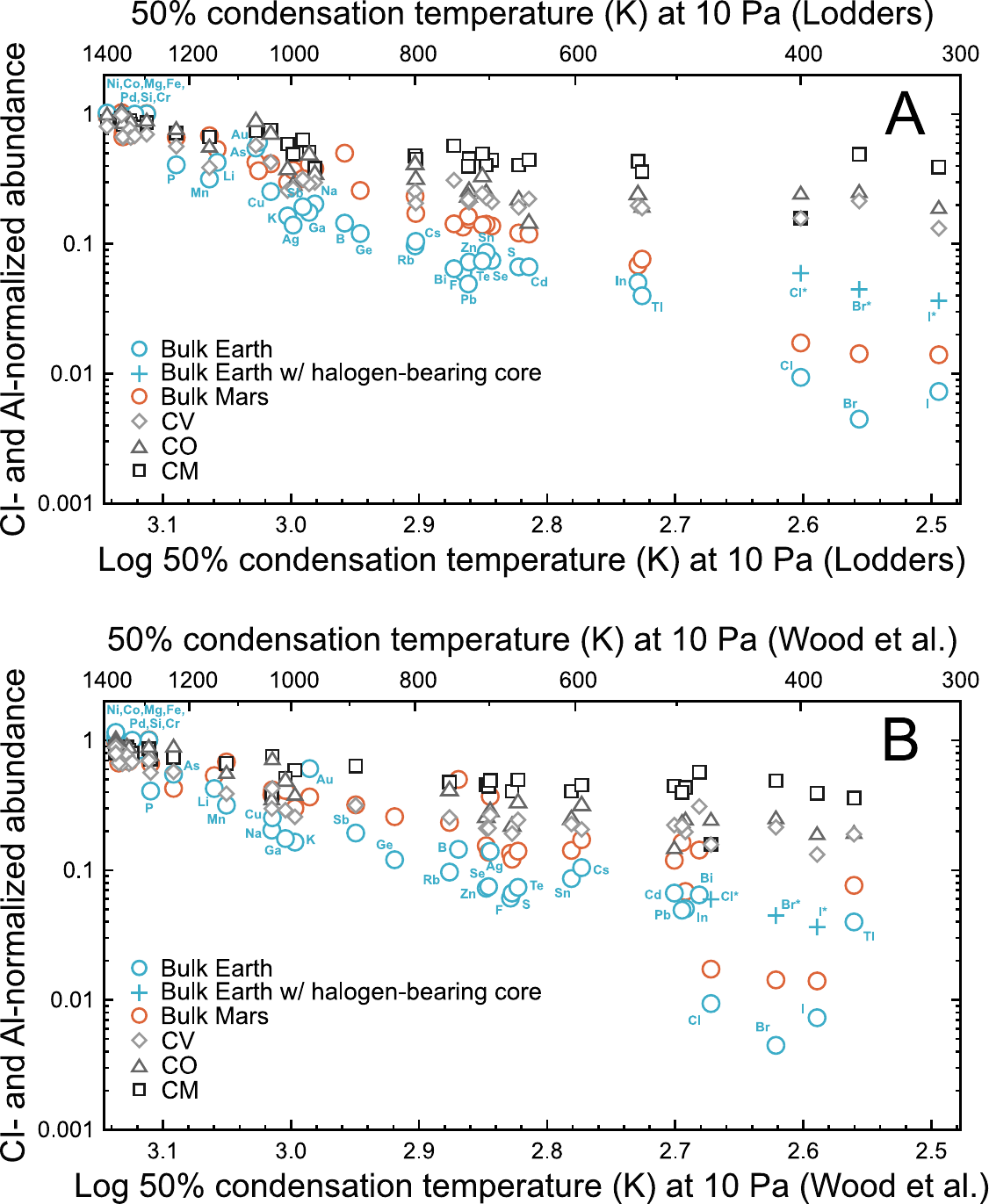}
	\caption{Abundances of moderately volatile and volatile elements in Earth \citep[this study;][]{mcdonough2014compositional}, Mars \citep[this study;][]{yoshizaki2019mars_long} and chondritic meteorites \citep{alexander2019quantitative_CC}. The bulk Earth model with a halogen-bearing core \citep{mcdonough2014compositional} is also shown. The x axis corresponds to the 50\% condensation temperature of elements at 10 Pa from \citet{lodders2003solar} and \citet{fegley2018volatile} (for halogens) in the top panel, and from \citet{wood2019condensation} in the bottom.}
\label{fig:planet_vs_chondrite}
\end{figure}

The revised 50\% condensation temperatures of the halogens \citep{fegley2018volatile,wood2019condensation} are lower than the previous values \citep{lodders2003solar}. Consequently, using condensation temperatures from \citet{fegley2018volatile} and CI abundance from \citet{lodders2020solar}, \citet{fegley2020volatile} showed that all the halogens plot on the Earth's volatility trend for the lithophile elements, thereby eliminating any need for any extra halogen fractionation (\cref{fig:planet_vs_chondrite}A). We note that a hockey-stick depletion pattern of MVE in the BSE \citep[e.g.,][]{braukmuller2019earth} is not observed when CI abundance from \citet{lodders2020solar} is used instead of that of \citet{clay2017halogens}. In this case, the Earth's halogen budgets (i.e., 12 ppm Cl, 0.03 ppm Br, and 0.01 ppm I in the bulk Earth) are hosted in its silicate fraction, and there is no need for their incorporation into the core. \bigskip

In contrast, a separate effort by \citet{wood2019condensation} derived higher 50\% condensation temperatures for these elements compared to \citet{fegley2018volatile}, which seem to be consistent with the halogen-bearing core model (\cref{fig:planet_vs_chondrite}B). Thus, further analytical, experimental and theoretical efforts are needed to better constrain the abundance and distribution of heavy halogens in Earth. \bigskip

\citet{yoshizaki2019mars_long} estimated halogen abundances in the BSM and bulk Mars based on the CI abundance of these elements from \citet{clay2017halogens}. Using CI chondritic abundance from \citet{lodders2020solar}, and assuming that the heavy halogens are hosted in the Mars' silicate fraction, the BSM and bulk Mars heavy halogen abundances are updated as follows: 28 ppm Cl, 0.13 ppm Br, and 0.03 ppm I in the BSM; and 23 ppm Cl, 0.10 ppm Br, and 0.02 ppm I in the bulk Mars. If the Martian core contained heavy halogens, the bulk Mars' halogen abundance might be 0.1--0.2 $ \times $ CI. \bigskip

\subsection{Heat production in terrestrial planets}
\label{sec:heat_production_Mars_calc}

Internal heat production of Mars is calculated using the BSM abundances of heat-producing elements (HPE) \ce{^{40}K}, \ce{^{232}Th}, \ce{^{235}U} and \ce{^{238}U} \citep{yoshizaki2019mars_long}, and their standard decay constants \citep[][ and references therein]{mcdonough2020radiogenic}. The Martian core contributes negligibly to the radiogenic heating of the planet as it is predicted to contain insignificant quantities of HPE. Heat production in the Martian crust is calculated for a crust with 3740 ppm K, 700 ppb Th and 180 ppb U \citep{taylor2009planetary}. Abundances of HPE in the Martian mantle are based on mass-balance considerations
\begin{equation}
	X\mathrm{_{Mm}^i} = \frac{X\mathrm{_{Mc}^i} \times M\mathrm{_{Mc}} - X\mathrm{_{BSM}^i} \times M\mathrm{_{BSM}}}{M\mathrm{_{Mm}}}	\label{eq:HPE_mantle}
\end{equation}
where $ X\mathrm{^i_j} $ is concentration of HPE i (\ce{^{40}K}, \ce{^{232}Th}, \ce{^{235}U} and \ce{^{238}U}) in the reservoir j (Mm, Mc, and BSM are Martian mantle, Martian crust and the bulk silicate Mars, respectively) and $ M\mathrm{_j} $ is mass of reservoir j (\cref{tab:main_table1}). For $ X\mathrm{_{BSM}} $ of 360 ppm K, 68 ppb Th and 18 ppb U \citep{yoshizaki2019mars_long} and $ X\mathrm{_{Mc}} $ from \citet{taylor2009planetary}, \cref{eq:HPE_mantle} yields 190 ppm K, 36 ppb Th and 10 ppb U in the Martian mantle. Martian planetary Urey ratio (\textit{Ur}) is given by 
\begin{equation}
Ur\mathrm{_{Mars}} = \frac{H\mathrm{_{BSM}}}{4 \pi R^2 F} \label{eq:Ur}
\end{equation}
where $ H\mathrm{_{BSM}}$ is heat production in the bulk silicate Mars, $ R $ is surface area of Mars, $ F\mathrm{_{Mars}} $ is the average surface heat flow \citep[19 $ \pm $ 1 mW/m$ ^2 $;][]{parro2017present}. Similarly, the internal heat production in Earth is calculated based on compositions of the BSE and Earth's crust from \citet{mcdonough2014compositional} and \citet{rudnick2014composition}, respectively, $ F\mathrm{_{Earth}} $ from \citet{jaupart2015temperatures}, and mass and density of the crust from \citet{wipperfurth2019reference}. \bigskip

\subsection{Rayleigh number of the Martian mantle}
\label{sec:Ra_calc}
		
The Rayleigh number (\textit{Ra}) for the internally-heated present-day Martian mantle is given by		
\begin{equation}
    Ra = \frac{\alpha \rho^2 g H d^5}{k \mu \kappa} \label{eq:Ra}
\end{equation}
where $ \alpha $ is thermal expansion coefficient, $ \rho $ is density of the Martian mantle, $ g $ is gravitational acceleration, $ H $ is heat production in the present-day Martian mantle, $ d $ is thickness of the Martian mantle, $ k $ is thermal conductivity, $ \mu $ is viscosity and $ \kappa $ is thermal diffusivity. We take $ \alpha $ = 3$ \times $10$ ^5 $ K$ ^{-1} $, $ g $ = 3.7 m/s$^{2}$, $ k $ = 4 W/m/K, $ \mu $ = 10$ ^{21}$--10$ ^{23} $ Pa s, and $ \kappa $ = 10$ ^{-6} $ m$ ^2 $/s \citep{turcotte2014geodynamics,samuel2019rheology} and obtain \textit{Ra} = 3$ \times $10$ ^5 $ -- 3$ \times $10$ ^7 $.

\subsection{CAI-free bulk composition of CV chondrite}
\label{sec:data_fig_9}

\begin{figure}[p]
	\centering
	\includegraphics[width=.7\linewidth]{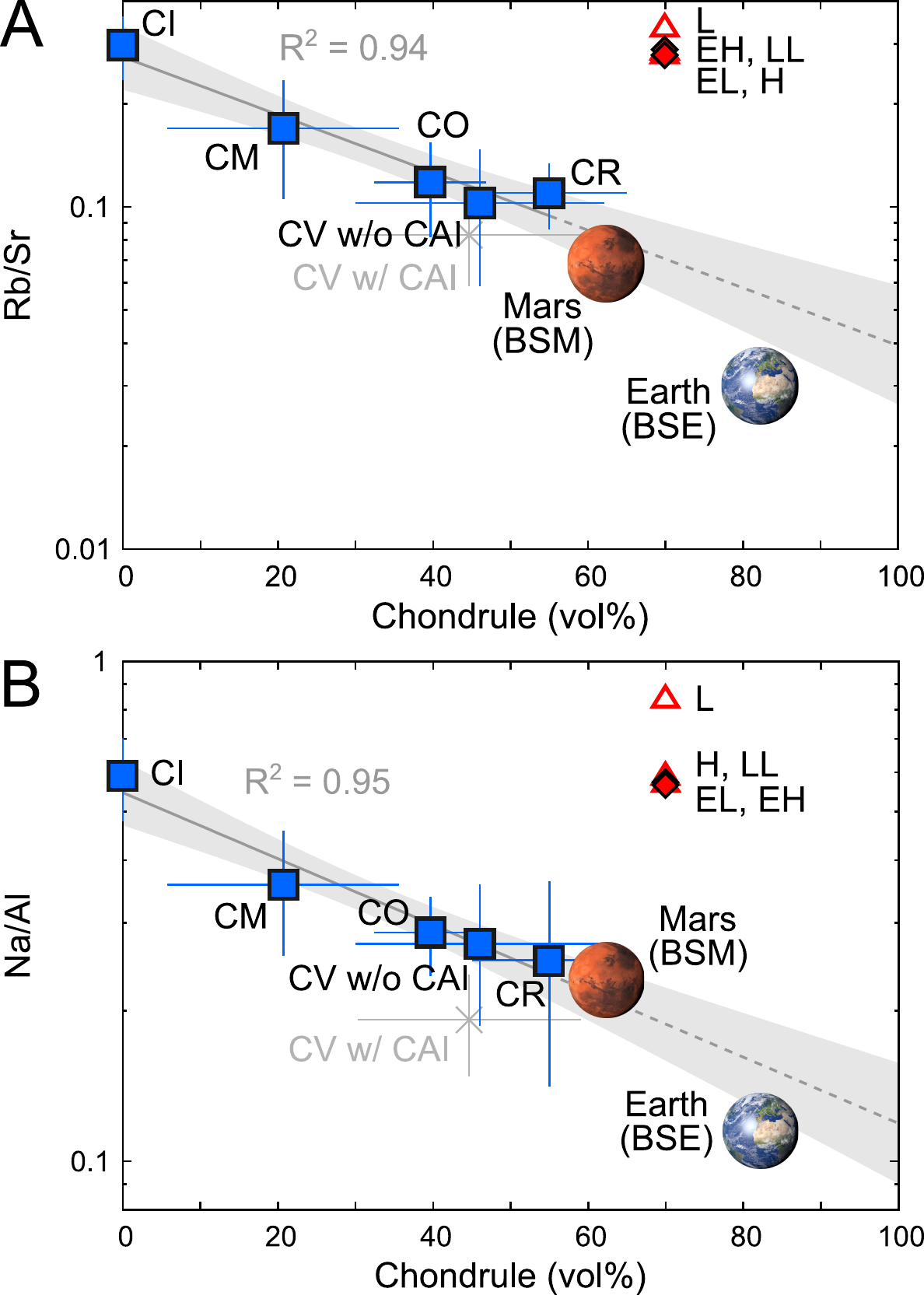}
	\caption{Abundance of chondrules (vol\%) vs Rb/Sr and Na/Al of chondrites and terrestrial planets. Compositions of chondrites, Earth, and Mars are from \citet{alexander2019quantitative_CC,alexander2019quantitative_NC}, \citet{mcdonough2014compositional}, and \citet{yoshizaki2019mars_long}, respectively. The abundances of chondrules in chondrites are taken from \citet{scott2014chondrites}. The amounts of chondrules in Earth and Mars are estimated based on their K/Th values (\cref{fig:K-Th_parameters}).}
	\label{fig:Rb_Sr_parameters}
\end{figure}

Abundances of the MVE are apparently lower in CV chondrites than other carbonaceous chondrites, because of their higher abundance of refractory inclusions. Since relative abundance of refractory lithophile elements in Earth and Mars are distinct from that of CAIs and bulk CV, it is unlikely that the planetary MVE fractionation is inherited from CV CAI-like materials \citep{stracke2012refractory,dauphas2015thulium,barrat2016evidence}. Therefore, to evaluate the contribution of CV chondrules in the bulk CV composition, we calculated a CAI-free chemical composition of bulk CV chondrites using mean composition of unaltered Allende CAIs \citep{mason1977geochemical}, modal abundance of CAIs in CV \citep{hezel2008modal,rubin2011origin,scott2014chondrites,desch2017effect}, and the bulk CV composition \citep{alexander2019quantitative_CC}. The calculated CAI-free bulk CV composition is shown in \cref{fig:K-Th_parameters,fig:Rb_Sr_parameters}. \bigskip

\subsection{Modeling the Earth's accretion}
\label{sec:model_Earth_accretion_SM}

Here we describe the details of formation model of the Earth (\cref{sec:Mars-like_impactor}). The basic sequence of the Earth's accretion is taken from \citet{wanke1984mantle} and \citet{oneil1991origin,oneil1991origin_Earth}, with modifications:
\begin{enumerate}
    \item the accretion of the proto-Earth (reduced and volatile-depleted) accompanied by continuous core-mantle differentiation ($ \sim $90\% of Earth's mass),
    \item followed by a late-stage \citep[e.g., 30--100 Myr after $ t_0 $;][]{kleine2009hf} Moon-forming giant impact event that adds the final $ \sim $10\% mass (oxidized and volatile-enriched) to Earth and forms a protolunar accretion disk,
    \item subsequently, the mantle loses a Fe-Ni ($ \pm $O) sulfide liquid \citep[sulfide matte;][]{,oneil1991origin_Earth} to the core ($ \sim $0.5\% BSE mass),
    \item and finally, the BSE receives the addition of ($ \sim $0.5\% of the BSE mass) a late accretion component that brings the highly siderophile and chalcophile elements in chondritic proportions and highly volatile gases and fluids.
\end{enumerate}

The bulk composition of the proto-Earth (\cref{tab:proto-Earth}) is calculated by subtracting contributions of the Mars-like impactor and late-added materials from the bulk composition of the present-day Earth \citep{,mcdonough2014compositional}. In this scenario, the bulk proto-Earth contains $ \geq $80\% of the present-day Earth's budgets of most of the MVE (e.g., K/Th $ \sim $ 3200; Rb/Sr $ \sim $ 0.026; \cref{tab:proto-Earth}), since a Mars-like impactor contributes only a limited amount of additional MVE (\cref{fig:volatility_trend_impactor}).  \bigskip

The core-mantle differentiation of proto-Earth is modeled using the calculated bulk proto-Earth composition and metal-silicate distribution coefficients of elements at high P-T conditions ($ \sim $30 GPa and $ \sim $3000 K) and $ \log f_{\ce{O2}} \sim \mathrm{IW} - 2 $ (\cref{tab:D-values}). \bigskip

\begin{figure}[p]
	\centering
	\includegraphics[width=0.9\linewidth]{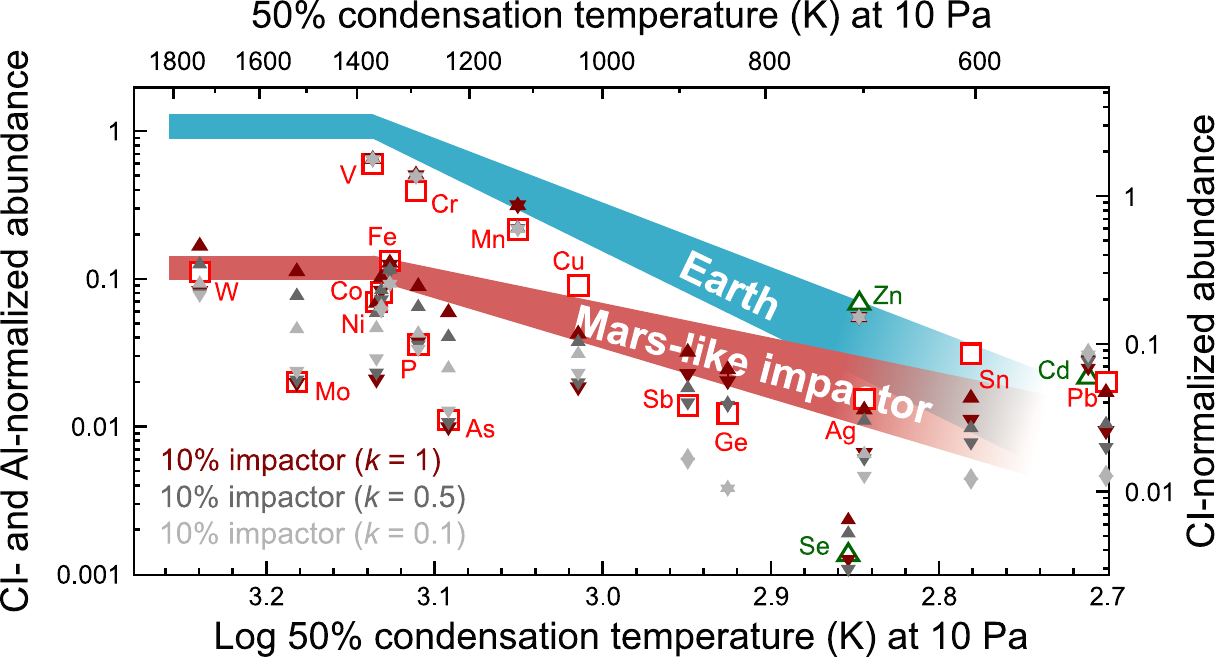}
	\caption{Same as \cref{fig:volatility_trend_impactor}, but different equilibrium factors ($ k $) between the impactor core and the proto-Earth's mantle.}	\label{fig:volatility_trend_impactor_k}
\end{figure}

The mass fraction of the impactor's core that equilibrated with the Earth's mantle ($ k $) is poorly constrained. While a small impactor might be efficiently emulsified and completely equilibrated with the proto-Earth's magma ocean, the behavior of Mars-sized impactor's core remains controversial \citep[e.g.,][]{rubie2003mechanisms,rubie2015accretion,dahl2010turbulent,samuel2012reevaluation,deguen2014turbulent}. The Hf-W and U-Pb isotopic systematics and abundances of moderately siderophile element in the BSE is consistent with $ 0.4 \leq k \leq 1  $ \citep{jacobsen2005hf,nimmo2006isotopic,rubie2015accretion,rubie2016highly}. Recently, \citet{budde2019molybdenum} showed that nucleosynthetic Mo isotopic composition of the BSE is well reproduced if 20--100\% of equilibrium was achieved between the impactor's core and the proto-Earth's mantle \citep[see also][]{kleine2020non}. We found that $  0.5 \leq k \leq 1 $ is consistent with the BSE abundances of most siderophile and chalcophile elements (\cref{fig:volatility_trend_impactor_k}). \bigskip

As a consequence of the Moon-forming event, the impactor's mantle and core are emulsified and equilibrated within the Earth's mantle \citep[e.g.,][]{oneil1991origin_Earth,rubie2016highly}, providing not only lithophile, but also siderophile and chalcophile elements to the post-impact silicate Earth (\cref{fig:volatility_trend_impactor}). The Earth's mantle might reach S saturation, given that the addition of the impactor increases the mantle S abundance, and the S concentration at sulfide saturation decreases dramatically as the mantle cools \citep{oneil1991origin_Earth,rubie2016highly}. Thus, an immiscible Fe-Ni ($ \pm $O) sulfide liquid (post-impact sulfide matte) precipitates through a crystallizing mantle into the core due to its high immiscibility, low wetting angle, and high density \citep{gaetani1999wetting,rose2001wetting}. \bigskip

The mass of post-impact sulfide matte is calculated by assuming that all S in the post-impact BSE was stripped by a sulfide phase, and all S in the present-day BSE is derived from the late-accreted materials  \citep{yi2000cadmium,rose2009effect,wang2013ratios}. Using a metal-silicate partition coefficient of S ($ D^{\mathrm{met-sil}}_{\mathrm{S}} $) of 100 \citep{rose2009effect,boujibar2014metal}, the proto-Earth's mantle is estimated to contain $ \sim $170 ppm S. The bulk Mars-like, Moon-forming impactor has 1.2 wt\% S \citep{yoshizaki2019mars_long}, so its addition increases S concentration in the post-impact Earth's mantle to $ \sim $0.2 wt\%. If all S in the post-impact BSE is segregated by a sulfide liquid with a composition of sulfides in spinel lherzolite xenoliths \citep[37.7 wt\% Fe, 21.1 wt\% Ni, 0.3 wt\% Co, 1.6 wt\% Cu, and 38.3 wt\% S;][]{lorand1983contribution}, the amount of the sulfide liquid is $\sim$0.5 wt\% of the present-day Earth's silicate mantle. Fractionation of other siderophile or chalcophile elements due to the sulfide matte precipitation is modeled using the $ D^{\mathrm{sul-sil}} $ values listed in \cref{tab:D-values}. Since there are limited numbers of experimental dataset on $ D^{\mathrm{sul-sil}} $ values, we do not specify P-T conditions of the sulfide matte formation. The errors accompanied with the wide variation in the expected $ D^{\mathrm{sul-sil}} $ values are much larger than those from other parameters (e.g., planetary compositional models, P, T-dependent $ D^{\mathrm{met-sil}} $ values), requiring additional experimental efforts in future. \bigskip

The elevated abundances of highly siderophile and chalcophile elements in the Earth's mantle indicate an addition of $ \sim $0.5\% of volatile-rich materials at the final stage of Earth's accretion \citep[e.g.,][]{kimura1974distribution,chou1978fractionation,walker2015search,rubie2016highly,wang2013ratios,righter2018effect}. The absence of meteorite class that matches relative abundances and isotopic composition of highly siderophile or chalcophile elements in the BSE keeps the nature of this late-accreted material controversial, but there is a general consensus that the late-added material had a chondritic elemental composition \citep{,walker2002comparative,albarede2009volatile,alexander2012provenances,wang2013ratios,varas2019selenium,fischer2020ruthenium}. In this modeling, 
we do not specify the type of chondritic materials added during the final stage. Note that all classes of chondrites are similarly enriched in volatiles compared to the rocky planets, and mass fraction of  the late-accreted material in the Earth's mantle is only $ \sim $0.5 \%. Therefore, the choice of other type of chondrites (e.g., enstatite or ordinary) as the late-accreted material do not affect the main results of this modeling. \bigskip

\begin{figure}[p]
	\centering
	\includegraphics[width=0.9\linewidth]{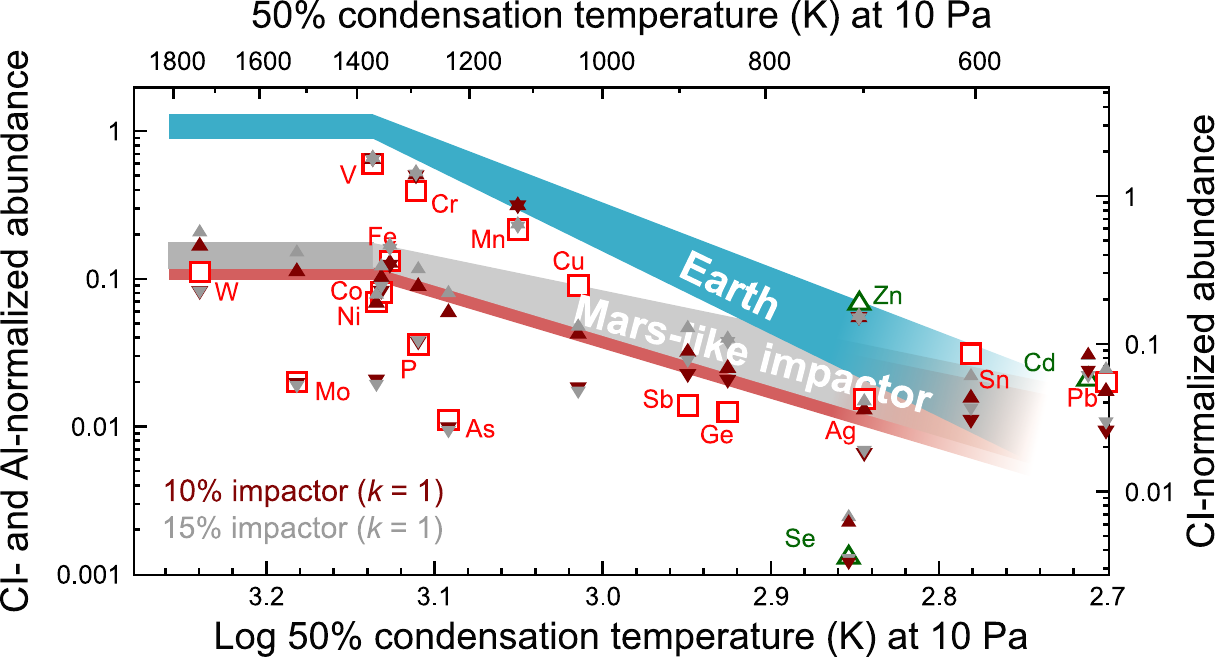}
	\caption{Same as \cref{fig:volatility_trend_impactor}, but different mass fractions of the impactor (10\% and 15\%).}	\label{fig:volatility_trend_impactor_10-15}
\end{figure}

The sulfide matte segregation decreases the abundances of the siderophile and chalcophile elements to the present-day BSE levels (\cref{fig:volatility_trend_impactor}). It also removes HSE like Au, Pd, Pt from the Earth's mantle, decreasing their mantle abundance $ > $10 times smaller than the present-day values \citep{rubie2016highly,righter2018effect}. As a final step, the late addition of $ \sim $0.5 wt\% of chondritic materials to the Earth's mantle after the sulfide matte segregation increases mantle abundances of HSE and chalcogens to the present-day BSE level \citep{wang2013ratios,walker2015search} (not shown in \cref{fig:volatility_trend_impactor,fig:volatility_trend_impactor_10-15,fig:volatility_trend_impactor_k}). Thus, this accretion scenario successfully reproduces the abundances of the chalcophile and highly siderophile elements in the BSE. \bigskip

A challenge to our model is the unexplained higher levels of Cu and Sn and strong depletion Sb and Ge in the BSE. These misfits can be due to our limited understanding of chalcophile behavior of these elements at the high P and T conditions. The problem with the Cu abundance is also found even in the CI-like Impactor model \citep{oneil1991origin_Earth}, and it would be relaxed if the lower BSE abundance of Cu \citep[20 ppm instead of 30 ppm;][]{oneil1991origin_Earth,mcdonough2014compositional} and/or lower $ D\mathrm{^{sul-sil}_{Cu}} $ under oxidizing or high-T conditions \citep[e.g.,][]{li2015effects} are considered. 

\begin{table}[p]
  \centering
  \begin{threeparttable}
  \caption{Metal-silicate and sulfide-silicate partition coefficients of elements (\textit{D}) adopted in the modeling.}
    \begin{tabular}{ccccc}
    \toprule
    Element & $ D\mathrm{^{met-sil}} $ & Reference\tnotex{lab:ref_D-value} & $ D\mathrm{^{sul-sil}} $ & Reference\tnotex{lab:ref_D-value} \\
    \midrule
    P     & 20    & R10      & 10--300 & JD86, R97 \\
    S     & 100   & R09, B14, S17  & $ \gg $1000 & See note\tnotex{lab:S_assumption} \\
    V     & 1.3   & R11     & 0.1--2 & GG97 \\
    Cr    & 2     & R11      & 2--4  & KW13 \\
    Mn    & 0.8   & R11, S18      & 1--3  & KW13 \\
    Fe    & 20    & Mass balance\tnotex{lab:mass_balance_met} & 9  & Mass balance\tnotex{lab:mass_balance_sul} \\
    Co    & 30    & R11      & 40--100 & KW13, P13 \\
    Ni    & 40   & R11      & 100--800 & KW13, P13, WW17 \\
    Cu    & 15    & R11      & 100--500 & KW13, B15 \\
    Zn    & 0.5   & Y15      & 2--5  & KW13, P13 \\
    Ge & 200 &  Y15 &  0--5& KW15 \\
    As    & 50    & R17      & 1--1000 & KP89, LA15 \\
    Se & 1000 & R09 & 1000--2000 & B15 \\
    Mo    & 50    & R11      & 10--1000 & KW13, P13, LA15 \\
    Ag    & 50    & W14      & 400--1000 & KW13, P13, B17 \\
    Cd & 2 & R18 & 40--100 & KW13 \\
    Sn    & 50    & R18      & 10--100 & P13, LA15, B17 \\
    Sb    & 100   & R17      & 10--100 & KW13, B15 \\
    W     & 20    & R11      & 1--200 & JD86, KP89, R97  \\
    Pb    & 25    & WH10      & 5--200 & KW13, P13, B15, B17 \\
    \bottomrule
    \end{tabular}%
  \label{tab:D-values}%
\begin{tablenotes}[b]
	\item[a] B14--\citet{boujibar2014metal};
	B15--\citet{brenan2015se}; 
	B17--\citet{ballhaus2017great};
	GG97--\citet{gaetani1997partitioning};
	JD86--\citet{jones1986geochemical};
	KP89--\citet{klock1988partitioning};
	KW13--\citet{kiseeva2013simple};
	LA15--\citet{li2015effects}; 
	LP91--\citet{lodders1991chalcophile}; 
	P13--\citet{patten2013partition}; 
	R09--\citet{rose2009effect};
	R97--\citet{righter1997metal};
	R10--\citet{righter2010partitioning}; 
	R11--\citet{righter2011prediction}; 
	R17--\citet{righter2017distribution}; 
	R18--\citet{righter2018volatile}; 
	S17--\citet{suer2017sulfur};
	S18--\citet{siebert2018chondritic}; 
	Y15--\citet{yang2015siderophile}; 
	W14--\citet{wood2014accretion};
	WH10--\citet{wood2010lead};
	WW17--\citet{wohlers2017uranium}. \label{lab:ref_D-value}
	\item[b] Assumed that all S in the post-impact Earth's mantle were subtracted by the sulfide matte with a present-day mantle sulfide composition \citep{lorand1983contribution}. \label{lab:S_assumption}
	\item[c] Based on mass balance between proto-Earth, Mars-like impactor, post-impact sulfide matte and present-day silicate Earth. \label{lab:mass_balance_met}
	\item[d] Based on mass balance between post-impact Earth's mantle and sulfide matte with a present-day mantle sulfide composition \citep{lorand1983contribution}.
	\label{lab:mass_balance_sul}
\end{tablenotes}
\end{threeparttable}
\end{table}%
\clearpage

\begin{table}[p]
  \centering
  \caption{Composition of the bulk proto-Earth predicted by the Mars-like Moon-forming impactor model. Concentrations are in ppm ($ \mu $g/g), otherwise noted.}
  \label{tab:proto-Earth}%
\begin{tabular}{ccccc}
\toprule
Element & Bulk proto-Earth &       &   Element  & Bulk proto-Earth \\
\midrule
Li    & 1.1   &       & Nb & 0.4 \\
Be    & 0.05  &       & Mo & 2 \\
B     & 0.1   &       & Ag & 0.04 \\
O (\%)    & 29.0 &       & Sn  & 0.2 \\
F     & 9     &       & Sb  & 0.04 \\
Na    & 1600  &       & Cs  & 0.03 \\
Mg (\%)    & 15.4 &       & Ba  & 4.5 \\
Al (\%)    & 1.59 &       & La  & 0.44 \\
Si (\%)    & 16.0 &       & Ce  & 1.1 \\
P     & 660   &       & Pr & 0.17 \\
S     & 5200  &       & Nd & 0.84 \\
K     & 180   &       & Sm & 0.27 \\
Ca (\%)    & 1.90 &       & Eu  & 0.10 \\
Sc    & 11    &       & Gd & 0.37 \\
Ti    & 810   &       & Tb  & 0.07 \\
V     & 110   &       & Dy & 0.46 \\
Cr    & 4680  &       & Ho & 0.10 \\
Mn    & 920   &       & Er & 0.30 \\
Fe (\%)    & 32.9 &       & Tm & 0.046 \\
Co    & 900   &       & Yb & 0.30 \\
Ni (\%)    & 1.87 &       & Lu  & 0.046 \\
Cu    & 60    &       & Hf  & 0.19 \\
Zn    & 30    &       & Ta  & 0.025 \\
As    & 1.7   &       & W & 0.17 \\
Rb    & 0.3   &       & Pb  & 0.17 \\
Sr    & 13    &       & Th  & 0.055 \\
Y     & 2.9   &       & U & 0.015 \\
Zr    & 7.1   &       &       &  \\
\bottomrule
\end{tabular}%
\end{table}%
\clearpage

\begin{figure}[p]
	\centering
	\includegraphics[width=.8\linewidth]{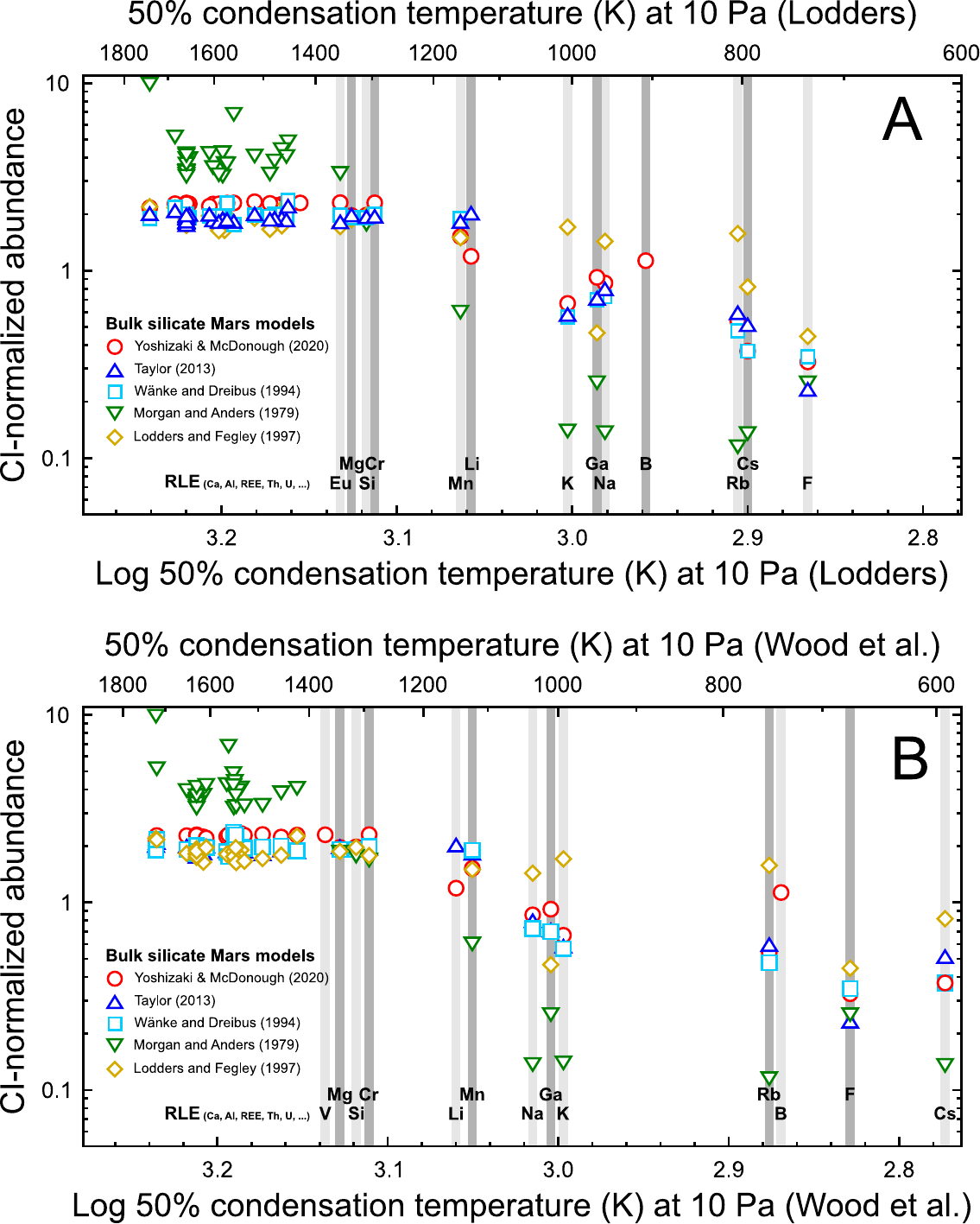}
	\caption{Comparison of lithophile element abundances in models of the bulk silicate Mars composition \citep{yoshizaki2019mars_long,taylor2013bulk,wanke1994chemistry,morgan1979chemical,lodders1997oxygen}. Condensation temperatures of elements are from \citet{lodders2003solar} (A) and \citet{wood2019condensation} (B), and CI composition is from \citet{lodders2020solar}. RLE--refractory lithophile elements.}
	\label{fig:planet_comp_mars_comparison}
\end{figure}
\clearpage

\begin{figure}[p]
	\centering
	\includegraphics[width=0.9\linewidth]{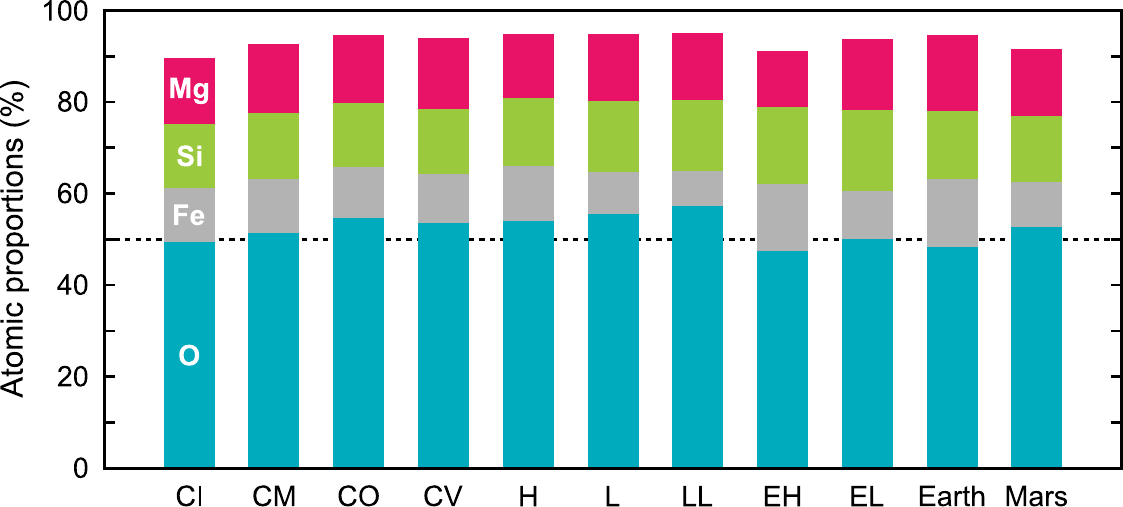}
	\caption{Bulk major element compositions by atomic proportions for Earth \citep{mcdonough2014compositional}, Mars \citep{yoshizaki2019mars_long} and chondritic meteorites \citep{alexander2019quantitative_CC,alexander2019quantitative_NC}.}
	\label{fig:planet_comp_stack}
\end{figure}
\clearpage

\clearpage

\clearpage

\begin{figure}[p]
	\centering
	\includegraphics[width=1\linewidth]{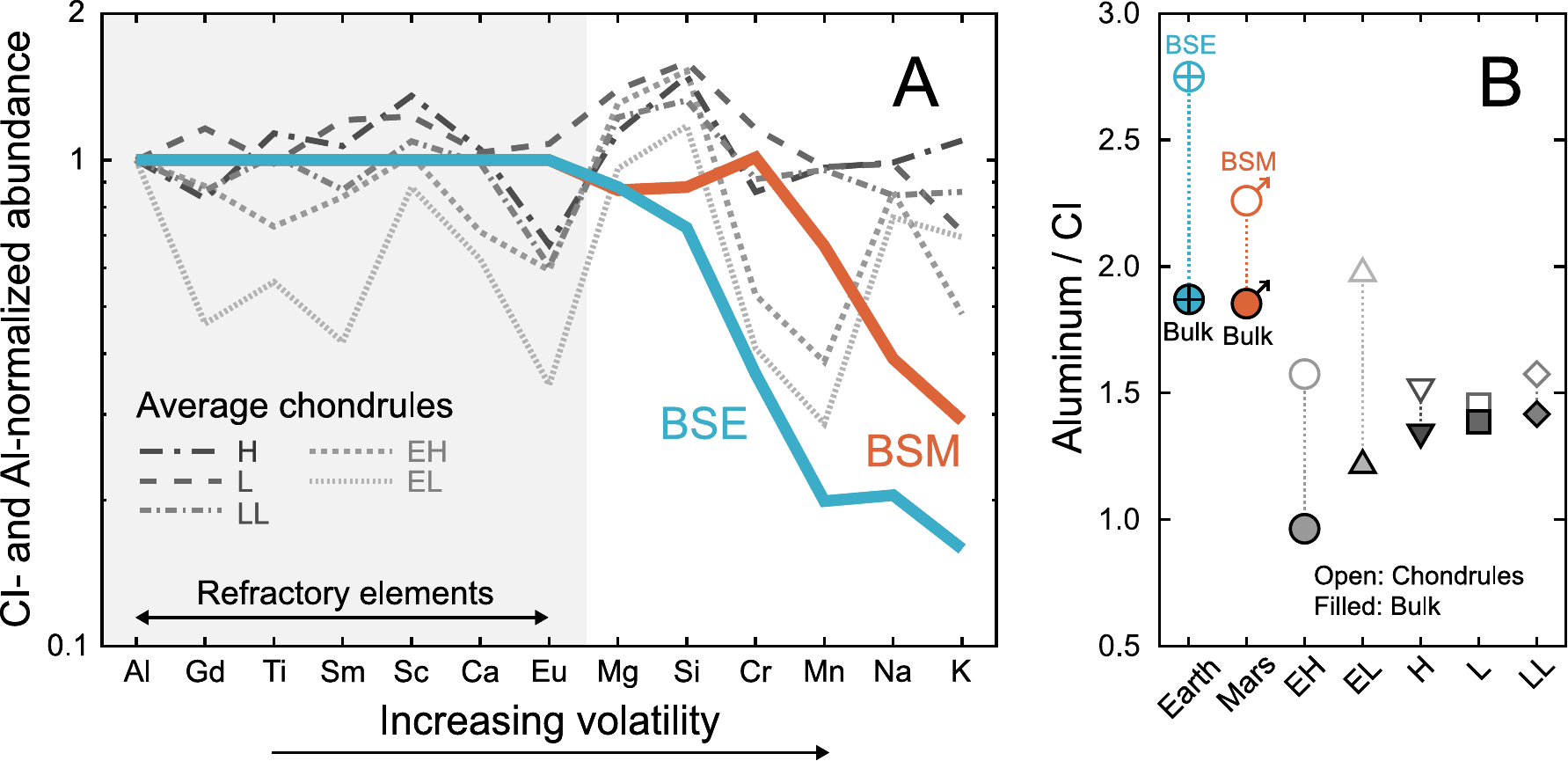}
	\caption{Lithophile element composition of Earth, Mars and non-carbonaceous chondrites and their components. (A) Lithophile element abundance in the bulk silicate Mars (BSM), bulk silicate Earth (BSE) and bulk chondrules from non-carbonaceous chondrites. Elemental abundances are normalized to CI chondrite composition and Al. Elements are arranged by their 50\% nebular condensation temperatures. Note that Cr and Mn depletion in the BSE and enstatite chondrite chondrules reflects less lithophile behavior of these elements under reduced conditions. (B) CI-normalized Al abundances in the BSE, BSM, bulk planets, bulk chondrules and bulk chondrites. Chemical composition of chondrules is from \citet{hezel2018what}, \citet{yoshizaki2018chemically} and \citet{metbase}. Other data sources are as in \cref{fig:major_comparison}.}
	\label{fig:planet_chondrules_NC}
\end{figure}
\clearpage

\clearpage

\clearpage

\begin{figure}[p]
	\centering
	\includegraphics[width=0.8\linewidth]{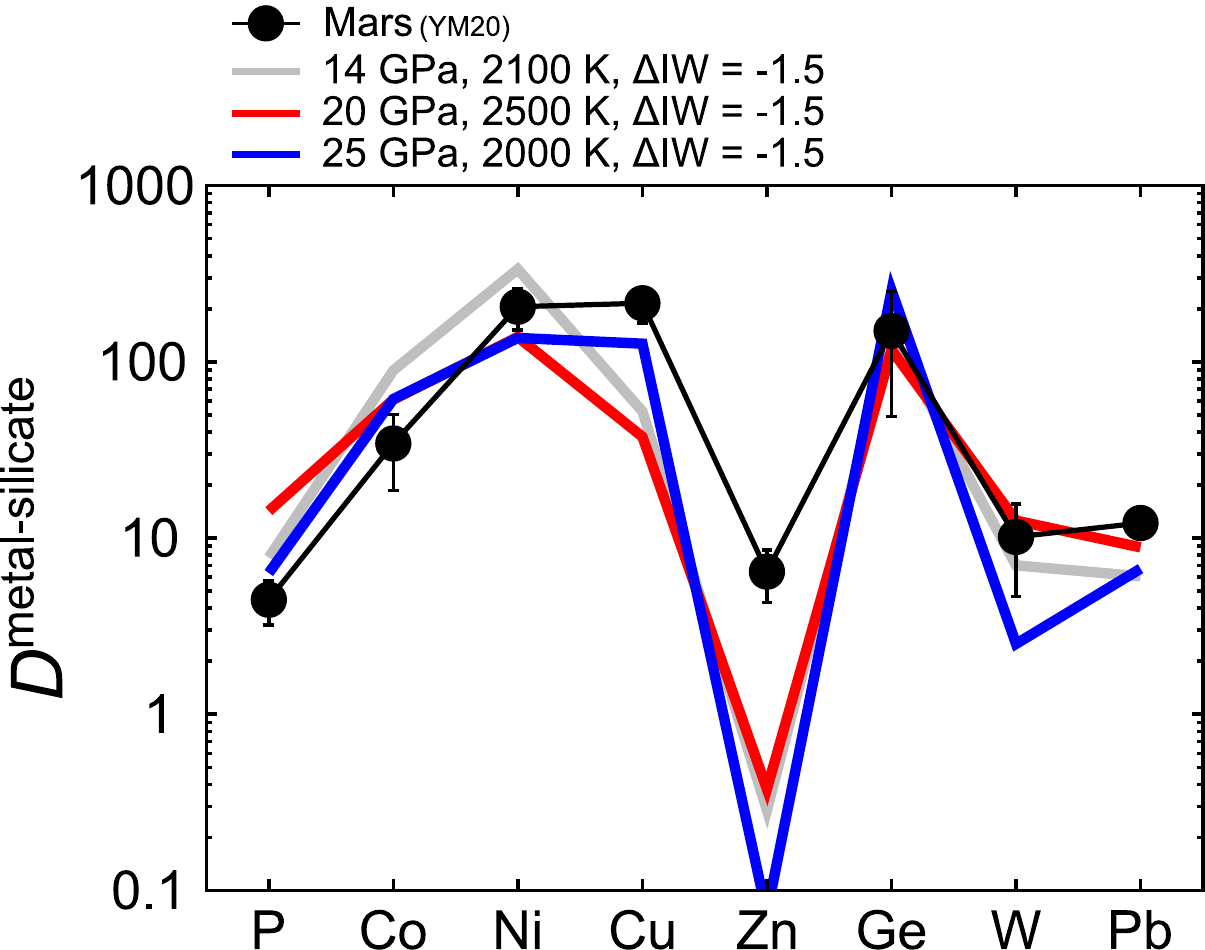}
	\caption{Core/mantle enrichment factors of siderophile elements in Mars \citep{yoshizaki2019mars_long} and metal-silicate distribution coefficients of these elements modeled at 14 GPa and 2100 K (gray), 20 GPa and 2500 K (red), and 25 GPa, 2000 K (blue) following \citet{righter2011moderately} and \citet{yang2015siderophile}. Redox conditions are fixed at $ \log f_{\ce{O2}} $ = IW $-$ 1.5. Error bars on the Mars' values reflect 1 standard deviations in the Martian bulk silicate composition model \citep{yoshizaki2019mars_long}.}
	\label{fig:Mars_D_model}
\end{figure}
\clearpage

\begin{figure}[p]
	\centering
	\includegraphics[width=1\linewidth]{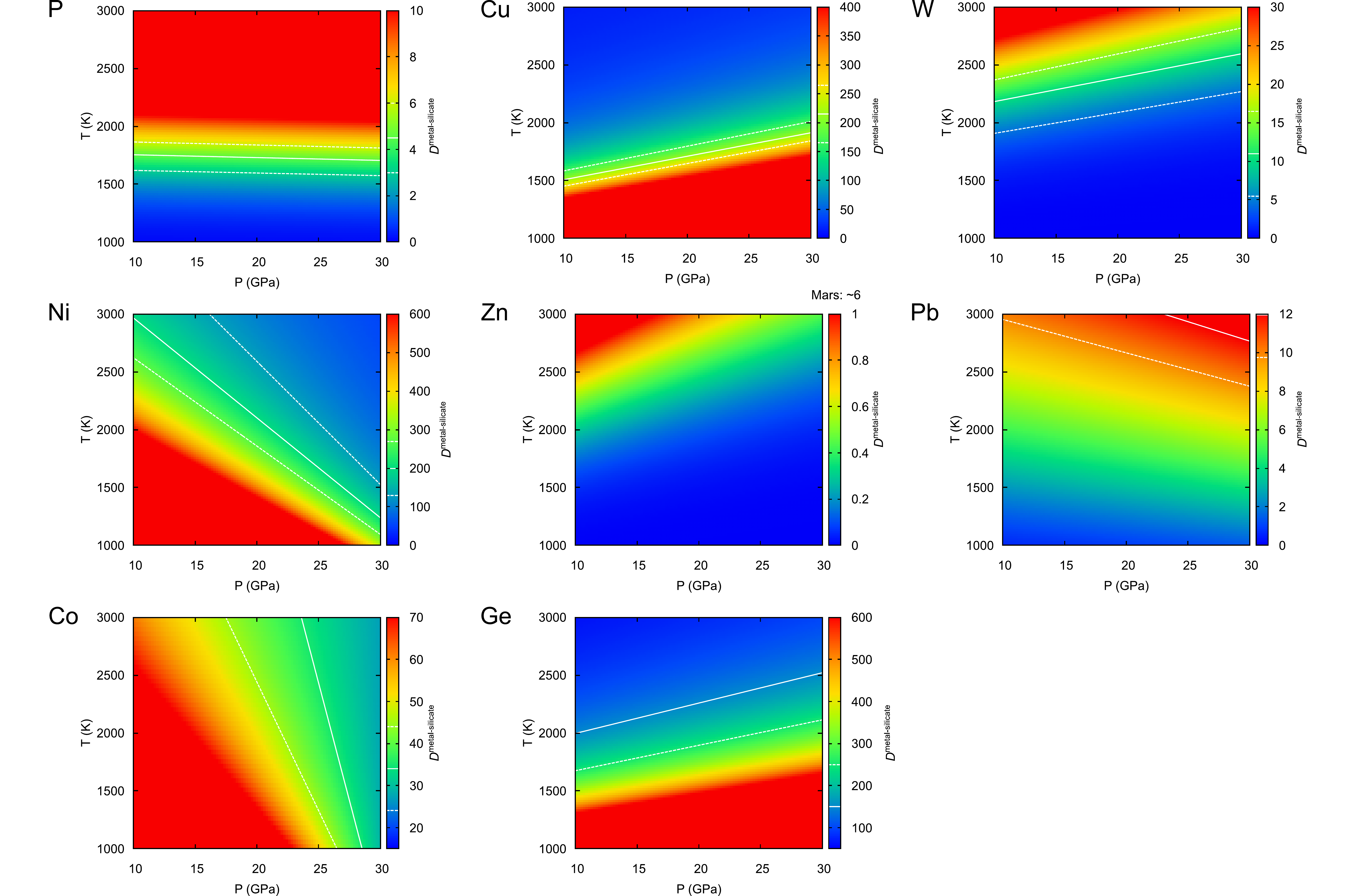}
	\caption{Metal-silicate partitioning coefficients of siderophile elements calculated for the Martian core model of \citet{yoshizaki2019mars_long}. Redox conditions are fixed at $ \log f_{\ce{O2}} $ = IW $-$ 1.5. White solid and broken lines show $ P $-$ T $ conditions that are consistent with the core-mantle distribution of these elements in Mars within 1 standard deviations \citep{yoshizaki2019mars_long}.}
	\label{fig:calc_core-mantle}
\end{figure}
\clearpage

\clearpage

\clearpage

\begin{figure}[p]
	\centering
	\includegraphics[width=1\linewidth]{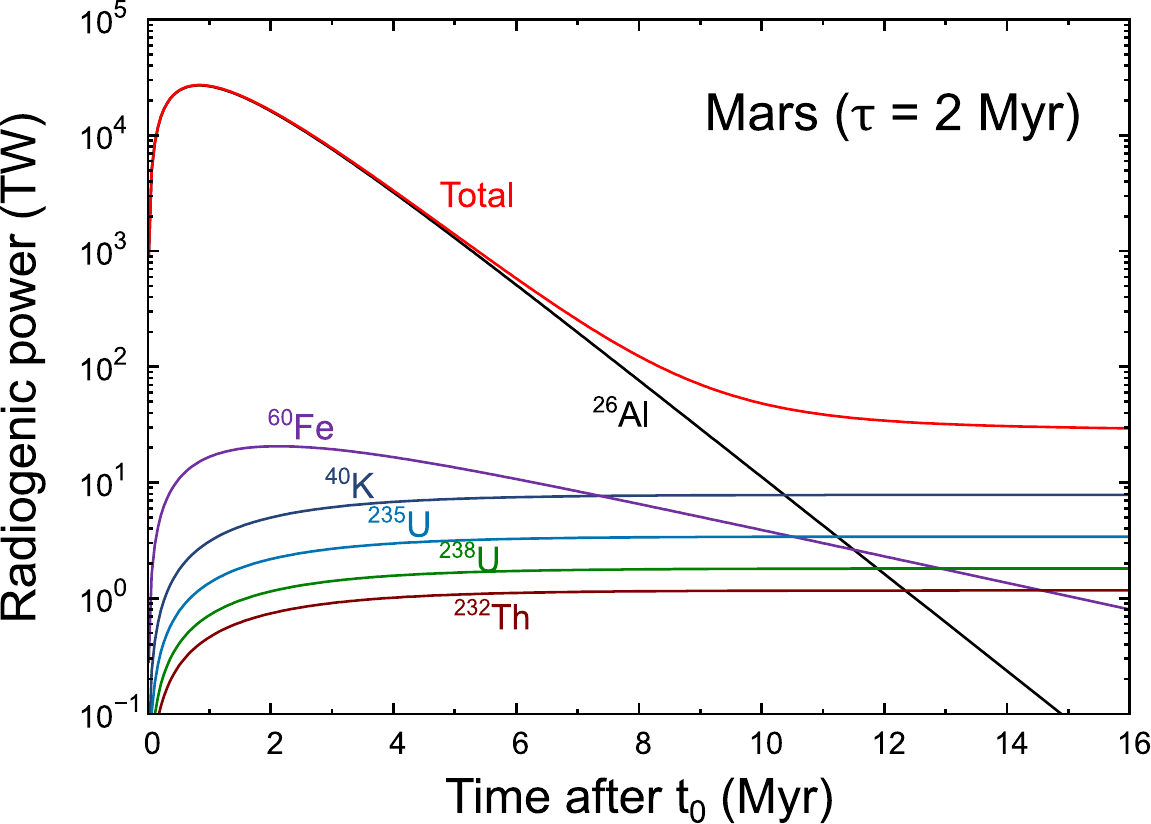}
	\caption{Relative contributions of radiogenic heat to Mars during accretion over the first 16 Myr of the solar system history. The compositional model for the bulk Mars is from \citet{yoshizaki2019mars_long}. The $ \tau\mathrm{^{accretion}_{Mars}} $ age of 2 Myr is adopted in the modeling \citep{dauphas2011hf}.}
	\label{fig:Mars_growth_model}
\end{figure}
\clearpage


\begin{thebibliography}{277}
\expandafter\ifx\csname natexlab\endcsname\relax\def\natexlab#1{#1}\fi
\providecommand{\url}[1]{\texttt{#1}}
\providecommand{\href}[2]{#2}
\providecommand{\path}[1]{#1}
\providecommand{\DOIprefix}{doi:}
\providecommand{\ArXivprefix}{arXiv:}
\providecommand{\URLprefix}{URL: }
\providecommand{\Pubmedprefix}{pmid:}
\providecommand{\doi}[1]{\href{http://dx.doi.org/#1}{\path{#1}}}
\providecommand{\Pubmed}[1]{\href{pmid:#1}{\path{#1}}}
\providecommand{\bibinfo}[2]{#2}
\ifx\xfnm\relax \def\xfnm[#1]{\unskip,\space#1}\fi
\bibitem[{Acu{\~{n}}a et~al.(1999)Acu{\~{n}}a, Connerney, Ness, Lin, Mitchell,
  Carlson, McFadden, Anderson, R{\`{e}}me, Mazelle, Vignes, Wasilewski and
  Cloutier}]{acuna1999global}
\bibinfo{author}{Acu{\~{n}}a, M.H.}, \bibinfo{author}{Connerney, J.E.},
  \bibinfo{author}{Ness, N.F.}, \bibinfo{author}{Lin, R.P.},
  \bibinfo{author}{Mitchell, D.}, \bibinfo{author}{Carlson, C.W.},
  \bibinfo{author}{McFadden, J.}, \bibinfo{author}{Anderson, K.A.},
  \bibinfo{author}{R{\`{e}}me, H.}, \bibinfo{author}{Mazelle, C.},
  \bibinfo{author}{Vignes, D.}, \bibinfo{author}{Wasilewski, P.},
  \bibinfo{author}{Cloutier, P.}, \bibinfo{year}{1999}.
\newblock \bibinfo{title}{{Global distribution of crustal magnetization
  discovered by the Mars Global Surveyor MAG/ER experiment}}.
\newblock \bibinfo{journal}{Science} \bibinfo{volume}{284},
  \bibinfo{pages}{790--793}.
\newblock \DOIprefix\doi{10.1126/science.284.5415.790}.
\bibitem[{Albarede(2009)}]{albarede2009volatile}
\bibinfo{author}{Albarede, F.}, \bibinfo{year}{2009}.
\newblock \bibinfo{title}{{Volatile accretion history of the terrestrial
  planets and dynamic implications}}.
\newblock \bibinfo{journal}{Nature} \bibinfo{volume}{461},
  \bibinfo{pages}{1227--1233}.
\newblock \DOIprefix\doi{10.1038/nature08477}.
\bibitem[{Alexander(2005)}]{alexander2005re}
\bibinfo{author}{Alexander, C.M.O'D.}, \bibinfo{year}{2005}.
\newblock \bibinfo{title}{{Re-examining the role of chondrules in producing the
  elemental fractionations in chondrites}}.
\newblock \bibinfo{journal}{Meteoritics \& Planetary Science Archives}
  \bibinfo{volume}{40}, \bibinfo{pages}{943--965}.
\newblock \DOIprefix\doi{10.1111/j.1945-5100.2005.tb00166.x}.
\bibitem[{Alexander(2019a)}]{alexander2019quantitative_CC}
\bibinfo{author}{Alexander, C.M.O'D.}, \bibinfo{year}{2019}a.
\newblock \bibinfo{title}{{Quantitative models for the elemental and isotopic
  fractionations in chondrites: The carbonaceous chondrites}}.
\newblock \bibinfo{journal}{Geochimica et Cosmochimica Acta}
  \bibinfo{volume}{254}, \bibinfo{pages}{277--309}.
\newblock \DOIprefix\doi{10.1016/j.gca.2019.02.008}.
\bibitem[{Alexander(2019b)}]{alexander2019quantitative_NC}
\bibinfo{author}{Alexander, C.M.O'D.}, \bibinfo{year}{2019}b.
\newblock \bibinfo{title}{{Quantitative models for the elemental and isotopic
  fractionations in the chondrites: The non-carbonaceous chondrites}}.
\newblock \bibinfo{journal}{Geochimica et Cosmochimica Acta}
  \bibinfo{volume}{254}, \bibinfo{pages}{246--276}.
\newblock \DOIprefix\doi{10.1016/j.gca.2019.01.026}.
\bibitem[{Alexander et~al.(2012)Alexander, Bowden, Fogel, Howard, Herd and
  Nittler}]{alexander2012provenances}
\bibinfo{author}{Alexander, C.M.O'D.}, \bibinfo{author}{Bowden, R.},
  \bibinfo{author}{Fogel, M.L.}, \bibinfo{author}{Howard, K.T.},
  \bibinfo{author}{Herd, C.D.K.}, \bibinfo{author}{Nittler, L.R.},
  \bibinfo{year}{2012}.
\newblock \bibinfo{title}{{The provenances of asteroids, and their
  contributions to the volatile inventories of the terrestrial planets}}.
\newblock \bibinfo{journal}{Science} \bibinfo{volume}{337},
  \bibinfo{pages}{721--723}.
\newblock \DOIprefix\doi{10.1126/science.1223474}.
\bibitem[{Alexander et~al.(2008)Alexander, Grossman, Ebel and
  Ciesla}]{alexander2008formation}
\bibinfo{author}{Alexander, C.M.O'D.}, \bibinfo{author}{Grossman, J.N.},
  \bibinfo{author}{Ebel, D.S.}, \bibinfo{author}{Ciesla, F.J.},
  \bibinfo{year}{2008}.
\newblock \bibinfo{title}{{The formation conditions of chondrules and
  chondrites}}.
\newblock \bibinfo{journal}{Science} \bibinfo{volume}{320},
  \bibinfo{pages}{1617--1619}.
\newblock \DOIprefix\doi{10.1126/science.1156561}.
\bibitem[{Alexander et~al.(2000)Alexander, Grossman, Wang, Zanda, Bourot-Denise
  and Hewins}]{alexander2000lack}
\bibinfo{author}{Alexander, C.M.O'D.}, \bibinfo{author}{Grossman, J.N.},
  \bibinfo{author}{Wang, J.}, \bibinfo{author}{Zanda, B.},
  \bibinfo{author}{Bourot-Denise, M.}, \bibinfo{author}{Hewins, R.H.},
  \bibinfo{year}{2000}.
\newblock \bibinfo{title}{{The lack of potassium-isotopic fractionation in
  Bishunpur chondrules}}.
\newblock \bibinfo{journal}{Meteoritics \& Planetary Science}
  \bibinfo{volume}{35}, \bibinfo{pages}{859--868}.
\newblock \DOIprefix\doi{10.1111/j.1945-5100.2000.tb01469.x}.
\bibitem[{All{\`e}gre et~al.(1995)All{\`e}gre, Poirier, Humler and
  Hofmann}]{allegre1995chemical}
\bibinfo{author}{All{\`e}gre, C.J.}, \bibinfo{author}{Poirier, J.P.},
  \bibinfo{author}{Humler, E.}, \bibinfo{author}{Hofmann, A.W.},
  \bibinfo{year}{1995}.
\newblock \bibinfo{title}{{The chemical composition of the Earth}}.
\newblock \bibinfo{journal}{Earth and Planetary Science Letters}
  \bibinfo{volume}{134}, \bibinfo{pages}{515--526}.
\newblock \DOIprefix\doi{10.1016/0012-821X(95)00123-T}.
\bibitem[{Amelin and Krot(2007)}]{amelin2007pb}
\bibinfo{author}{Amelin, Y.}, \bibinfo{author}{Krot, A.}, \bibinfo{year}{2007}.
\newblock \bibinfo{title}{{Pb isotopic age of the Allende chondrules}}.
\newblock \bibinfo{journal}{Meteoritics \& Planetary Science}
  \bibinfo{volume}{42}, \bibinfo{pages}{1321--1335}.
\newblock \DOIprefix\doi{10.1111/j.1945-5100.2007.tb00577.x}.
\bibitem[{Amelin et~al.(2002)Amelin, Krot, Hutcheon and
  Ulyanov}]{amelin2002lead}
\bibinfo{author}{Amelin, Y.}, \bibinfo{author}{Krot, A.N.},
  \bibinfo{author}{Hutcheon, I.D.}, \bibinfo{author}{Ulyanov, A.A.},
  \bibinfo{year}{2002}.
\newblock \bibinfo{title}{{Lead isotopic ages of chondrules and
  calcium-aluminum-rich inclusions}}.
\newblock \bibinfo{journal}{Science} \bibinfo{volume}{297},
  \bibinfo{pages}{1678--1683}.
\newblock \DOIprefix\doi{10.1126/science.1073950}.
\bibitem[{Amsellem et~al.(2017)Amsellem, Moynier, Pringle, Bouvier, Chen and
  Day}]{amsellem2017testing}
\bibinfo{author}{Amsellem, E.}, \bibinfo{author}{Moynier, F.},
  \bibinfo{author}{Pringle, E.A.}, \bibinfo{author}{Bouvier, A.},
  \bibinfo{author}{Chen, H.}, \bibinfo{author}{Day, J.M.D.},
  \bibinfo{year}{2017}.
\newblock \bibinfo{title}{{Testing the chondrule-rich accretion model for
  planetary embryos using calcium isotopes}}.
\newblock \bibinfo{journal}{Earth and Planetary Science Letters}
  \bibinfo{volume}{469}, \bibinfo{pages}{75--83}.
\newblock \DOIprefix\doi{10.1016/j.epsl.2017.04.022}.
\bibitem[{Anders and Owen(1977)}]{anders1977mars}
\bibinfo{author}{Anders, E.}, \bibinfo{author}{Owen, T.}, \bibinfo{year}{1977}.
\newblock \bibinfo{title}{{Mars and Earth: Origin and abundance of volatiles}}.
\newblock \bibinfo{journal}{Science} \bibinfo{volume}{198},
  \bibinfo{pages}{453--465}.
\newblock \DOIprefix\doi{10.1126/science.198.4316.453}.
\bibitem[{Arkani-Hamed(2004)}]{arkani2004timing}
\bibinfo{author}{Arkani-Hamed, J.}, \bibinfo{year}{2004}.
\newblock \bibinfo{title}{{Timing of the Martian core dynamo}}.
\newblock \bibinfo{journal}{Journal of Geophysical Research: Planets}
  \bibinfo{volume}{109}, \bibinfo{pages}{E03006}.
\newblock \DOIprefix\doi{10.1029/2003JE002195.}
\bibitem[{Ballhaus et~al.(2017)Ballhaus, Fonseca, M{\"u}nker, Rohrbach, Nagel,
  Speelmanns, Helmy, Zirner, Vogel and Heuser}]{ballhaus2017great}
\bibinfo{author}{Ballhaus, C.}, \bibinfo{author}{Fonseca, R.O.C.},
  \bibinfo{author}{M{\"u}nker, C.}, \bibinfo{author}{Rohrbach, A.},
  \bibinfo{author}{Nagel, T.}, \bibinfo{author}{Speelmanns, I.M.},
  \bibinfo{author}{Helmy, H.M.}, \bibinfo{author}{Zirner, A.},
  \bibinfo{author}{Vogel, A.K.}, \bibinfo{author}{Heuser, A.},
  \bibinfo{year}{2017}.
\newblock \bibinfo{title}{{The great sulfur depletion of Earth's mantle is not
  a signature of mantle--core equilibration}}.
\newblock \bibinfo{journal}{Contributions to Mineralogy and Petrology}
  \bibinfo{volume}{172}, \bibinfo{pages}{68}.
\newblock \DOIprefix\doi{10.1007/s00410-017-1388-3}.
\bibitem[{Ballhaus et~al.(2013)Ballhaus, Laurenz, M{\"u}nker, Fonseca,
  Albar{\`e}de, Rohrbach, Lagos, Schmidt, Jochum, Stoll, Weis and
  Helmy}]{ballhaus2013u}
\bibinfo{author}{Ballhaus, C.}, \bibinfo{author}{Laurenz, V.},
  \bibinfo{author}{M{\"u}nker, C.}, \bibinfo{author}{Fonseca, R.O.C.},
  \bibinfo{author}{Albar{\`e}de, F.}, \bibinfo{author}{Rohrbach, A.},
  \bibinfo{author}{Lagos, M.}, \bibinfo{author}{Schmidt, M.W.},
  \bibinfo{author}{Jochum, K.P.}, \bibinfo{author}{Stoll, B.},
  \bibinfo{author}{Weis, U.}, \bibinfo{author}{Helmy, H.M.},
  \bibinfo{year}{2013}.
\newblock \bibinfo{title}{{The U/Pb ratio of the Earth's mantle--A signature of
  late volatile addition}}.
\newblock \bibinfo{journal}{Earth and Planetary Science Letters}
  \bibinfo{volume}{362}, \bibinfo{pages}{237--245}.
\newblock \DOIprefix\doi{10.1016/j.epsl.2012.11.049}.
\bibitem[{{Banerdt} et~al.(2020){Banerdt}, {Smrekar}, {Banfield}, {Giardini},
  {Golombek}, {Johnson}, {Lognonn{\'e}}, {Spiga}, {Spohn}, {Perrin},
  {St{\"a}hler}, {Antonangeli}, {Asmar}, {Beghein}, {Bowles}, {Bozdag}, {Chi},
  {Christensen}, {Clinton}, {Collins}, {Daubar}, {Dehant}, {Drilleau},
  {Fillingim}, {Folkner}, {Garcia}, {Garvin}, {Grant}, {Grott}, {Grygorczuk},
  {Hudson}, {Irving}, {Kargl}, {Kawamura}, {Kedar}, {King}, {Knapmeyer-Endrun},
  {Knapmeyer}, {Lemmon}, {Lorenz}, {Maki}, {Margerin}, {McLennan}, {Michaut},
  {Mimoun}, {Mittelholz}, {Mocquet}, {Morgan}, {Mueller}, {Murdoch},
  {Nagihara}, {Newman}, {Nimmo}, {Panning}, {Pike}, {Plesa}, {Rodriguez},
  {Rodriguez-Manfredi}, {Russell}, {Schmerr}, {Siegler}, {Stanley},
  {Stutzmann}, {Teanby}, {Tromp}, {van Driel}, {Warner}, {Weber} and
  {Wieczorek}}]{banerdt2020initial}
\bibinfo{author}{{Banerdt}, W.B.}, \bibinfo{author}{{Smrekar}, S.E.},
  \bibinfo{author}{{Banfield}, D.}, \bibinfo{author}{{Giardini}, D.},
  \bibinfo{author}{{Golombek}, M.}, \bibinfo{author}{{Johnson}, C.L.},
  \bibinfo{author}{{Lognonn{\'e}}, P.}, \bibinfo{author}{{Spiga}, A.},
  \bibinfo{author}{{Spohn}, T.}, \bibinfo{author}{{Perrin}, C.},
  \bibinfo{author}{{St{\"a}hler}, S.C.}, \bibinfo{author}{{Antonangeli}, D.},
  \bibinfo{author}{{Asmar}, S.}, \bibinfo{author}{{Beghein}, C.},
  \bibinfo{author}{{Bowles}, N.}, \bibinfo{author}{{Bozdag}, E.},
  \bibinfo{author}{{Chi}, P.}, \bibinfo{author}{{Christensen}, U.},
  \bibinfo{author}{{Clinton}, J.}, \bibinfo{author}{{Collins}, G.S.},
  \bibinfo{author}{{Daubar}, I.}, \bibinfo{author}{{Dehant}, V.},
  \bibinfo{author}{{Drilleau}, M.}, \bibinfo{author}{{Fillingim}, M.},
  \bibinfo{author}{{Folkner}, W.}, \bibinfo{author}{{Garcia}, R.F.},
  \bibinfo{author}{{Garvin}, J.}, \bibinfo{author}{{Grant}, J.},
  \bibinfo{author}{{Grott}, M.}, \bibinfo{author}{{Grygorczuk}, J.},
  \bibinfo{author}{{Hudson}, T.}, \bibinfo{author}{{Irving}, J.C.E.},
  \bibinfo{author}{{Kargl}, G.}, \bibinfo{author}{{Kawamura}, T.},
  \bibinfo{author}{{Kedar}, S.}, \bibinfo{author}{{King}, S.},
  \bibinfo{author}{{Knapmeyer-Endrun}, B.}, \bibinfo{author}{{Knapmeyer}, M.},
  \bibinfo{author}{{Lemmon}, M.}, \bibinfo{author}{{Lorenz}, R.},
  \bibinfo{author}{{Maki}, J.N.}, \bibinfo{author}{{Margerin}, L.},
  \bibinfo{author}{{McLennan}, S.M.}, \bibinfo{author}{{Michaut}, C.},
  \bibinfo{author}{{Mimoun}, D.}, \bibinfo{author}{{Mittelholz}, A.},
  \bibinfo{author}{{Mocquet}, A.}, \bibinfo{author}{{Morgan}, P.},
  \bibinfo{author}{{Mueller}, N.T.}, \bibinfo{author}{{Murdoch}, N.},
  \bibinfo{author}{{Nagihara}, S.}, \bibinfo{author}{{Newman}, C.},
  \bibinfo{author}{{Nimmo}, F.}, \bibinfo{author}{{Panning}, M.},
  \bibinfo{author}{{Pike}, W.T.}, \bibinfo{author}{{Plesa}, A.C.},
  \bibinfo{author}{{Rodriguez}, S.}, \bibinfo{author}{{Rodriguez-Manfredi},
  J.A.}, \bibinfo{author}{{Russell}, C.T.}, \bibinfo{author}{{Schmerr}, N.},
  \bibinfo{author}{{Siegler}, M.}, \bibinfo{author}{{Stanley}, S.},
  \bibinfo{author}{{Stutzmann}, E.}, \bibinfo{author}{{Teanby}, N.},
  \bibinfo{author}{{Tromp}, J.}, \bibinfo{author}{{van Driel}, M.},
  \bibinfo{author}{{Warner}, N.}, \bibinfo{author}{{Weber}, R.},
  \bibinfo{author}{{Wieczorek}, M.}, \bibinfo{year}{2020}.
\newblock \bibinfo{title}{{Initial results from the InSight mission on Mars}}.
\newblock \bibinfo{journal}{Nature Geoscience} \bibinfo{volume}{13},
  \bibinfo{pages}{1--7}.
\newblock \DOIprefix\doi{10.1038/s41561-020-0544-y}.
\bibitem[{Baratoux et~al.(2011)Baratoux, Toplis, Monnereau and
  Gasnault}]{baratoux2011thermal}
\bibinfo{author}{Baratoux, D.}, \bibinfo{author}{Toplis, M.J.},
  \bibinfo{author}{Monnereau, M.}, \bibinfo{author}{Gasnault, O.},
  \bibinfo{year}{2011}.
\newblock \bibinfo{title}{{Thermal history of Mars inferred from orbital
  geochemistry of volcanic provinces}}.
\newblock \bibinfo{journal}{Nature} \bibinfo{volume}{472},
  \bibinfo{pages}{338--341}.
\newblock \DOIprefix\doi{10.1038/nature09903}.
\bibitem[{Baratoux et~al.(2013)Baratoux, Toplis, Monnereau and
  Sautter}]{baratoux2013petrological}
\bibinfo{author}{Baratoux, D.}, \bibinfo{author}{Toplis, M.J.},
  \bibinfo{author}{Monnereau, M.}, \bibinfo{author}{Sautter, V.},
  \bibinfo{year}{2013}.
\newblock \bibinfo{title}{{The petrological expression of early Mars
  volcanism}}.
\newblock \bibinfo{journal}{Journal of Geophysical Research: Planets}
  \bibinfo{volume}{118}, \bibinfo{pages}{59--64}.
\newblock \DOIprefix\doi{10.1029/2012JE004234}.
\bibitem[{Becker et~al.(2006)Becker, Horan, Walker, Gao, Lorand and
  Rudnick}]{becker2006highly}
\bibinfo{author}{Becker, H.}, \bibinfo{author}{Horan, M.F.},
  \bibinfo{author}{Walker, R.J.}, \bibinfo{author}{Gao, S.},
  \bibinfo{author}{Lorand, J.P.}, \bibinfo{author}{Rudnick, R.L.},
  \bibinfo{year}{2006}.
\newblock \bibinfo{title}{{Highly siderophile element composition of the
  Earth's primitive upper mantle: constraints from new data on peridotite
  massifs and xenoliths}}.
\newblock \bibinfo{journal}{Geochimica et Cosmochimica Acta}
  \bibinfo{volume}{70}, \bibinfo{pages}{4528--4550}.
\newblock \DOIprefix\doi{10.1016/j.gca.2006.06.004}.
\bibitem[{Bermingham et~al.(2018)Bermingham, Gussone, Mezger and
  Krause}]{bermingham2018origins}
\bibinfo{author}{Bermingham, K.R.}, \bibinfo{author}{Gussone, N.},
  \bibinfo{author}{Mezger, K.}, \bibinfo{author}{Krause, J.},
  \bibinfo{year}{2018}.
\newblock \bibinfo{title}{{Origins of mass-dependent and mass-independent Ca
  isotope variations in meteoritic components and meteorites}}.
\newblock \bibinfo{journal}{Geochimica et Cosmochimica Acta}
  \bibinfo{volume}{226}, \bibinfo{pages}{206--223}.
\newblock \DOIprefix\doi{10.1016/j.gca.2018.01.034}.
\bibitem[{Bertka and Fei(1997)}]{bertka1997mineralogy}
\bibinfo{author}{Bertka, C.M.}, \bibinfo{author}{Fei, Y.},
  \bibinfo{year}{1997}.
\newblock \bibinfo{title}{{Mineralogy of the Martian interior up to core-mantle
  boundary pressures}}.
\newblock \bibinfo{journal}{Journal of Geophysical Research: Solid Earth}
  \bibinfo{volume}{102}, \bibinfo{pages}{5251--5264}.
\newblock \DOIprefix\doi{10.1029/96JB03270}.
\bibitem[{Bizzarro et~al.(2004)Bizzarro, Baker and Haack}]{bizzarro2004mg}
\bibinfo{author}{Bizzarro, M.}, \bibinfo{author}{Baker, J.A.},
  \bibinfo{author}{Haack, H.}, \bibinfo{year}{2004}.
\newblock \bibinfo{title}{{Mg isotope evidence for contemporaneous formation of
  chondrules and refractory inclusions}}.
\newblock \bibinfo{journal}{Nature} \bibinfo{volume}{431},
  \bibinfo{pages}{275--278}.
\newblock \DOIprefix\doi{10.1038/nature02882}.
\bibitem[{Blanchard et~al.(2017)Blanchard, Siebert, Borensztajn and
  Badro}]{blanchard2017solubility}
\bibinfo{author}{Blanchard, I.}, \bibinfo{author}{Siebert, J.},
  \bibinfo{author}{Borensztajn, S.}, \bibinfo{author}{Badro, J.},
  \bibinfo{year}{2017}.
\newblock \bibinfo{title}{{The solubility of heat-producing elements in Earth's
  core}}.
\newblock \bibinfo{journal}{Geochemical Perspectives Letters}
  \bibinfo{volume}{5}, \bibinfo{pages}{1--5}.
\newblock \DOIprefix\doi{10.7185/geochemlet.1737}.
\bibitem[{Bland et~al.(2005)Bland, Alard, Benedix, Kearsley, Menzies, Watt and
  Rogers}]{bland2005volatile}
\bibinfo{author}{Bland, P.A.}, \bibinfo{author}{Alard, O.},
  \bibinfo{author}{Benedix, G.K.}, \bibinfo{author}{Kearsley, A.T.},
  \bibinfo{author}{Menzies, O.N.}, \bibinfo{author}{Watt, L.E.},
  \bibinfo{author}{Rogers, N.W.}, \bibinfo{year}{2005}.
\newblock \bibinfo{title}{{Volatile fractionation in the early solar system and
  chondrule/matrix complementarity}}.
\newblock \bibinfo{journal}{Proceedings of the National Academy of Sciences of
  the United States of America} \bibinfo{volume}{102},
  \bibinfo{pages}{13755--13760}.
\newblock \DOIprefix\doi{10.1073/pnas.0501885102}.
\bibitem[{van Boekel et~al.(2004)van Boekel, Min, Leinert, Waters, Richichi,
  Chesneau, Dominik, Jaffe, Dutrey, Graser, , {Henning}, {de Jong},
  {K{\"o}hler}, {de Koter}, {Lopez}, {Malbet}, {Morel}, {Paresce}, {Perrin},
  {Preibisch}, {Przygodda}, {Sch{\"o}ller} and {Wittkowski}}]{van2004building}
\bibinfo{author}{van Boekel, R.J.H.M.}, \bibinfo{author}{Min, M.},
  \bibinfo{author}{Leinert, C.}, \bibinfo{author}{Waters, L.B.F.M.},
  \bibinfo{author}{Richichi, A.}, \bibinfo{author}{Chesneau, O.},
  \bibinfo{author}{Dominik, C.}, \bibinfo{author}{Jaffe, W.},
  \bibinfo{author}{Dutrey, A.}, \bibinfo{author}{Graser, U.}, ,
  \bibinfo{author}{{Henning}, T.}, \bibinfo{author}{{de Jong}, J.},
  \bibinfo{author}{{K{\"o}hler}, R.}, \bibinfo{author}{{de Koter}, A.},
  \bibinfo{author}{{Lopez}, B.}, \bibinfo{author}{{Malbet}, F.},
  \bibinfo{author}{{Morel}, S.}, \bibinfo{author}{{Paresce}, F.},
  \bibinfo{author}{{Perrin}, G.}, \bibinfo{author}{{Preibisch}, T.},
  \bibinfo{author}{{Przygodda}, F.}, \bibinfo{author}{{Sch{\"o}ller}, M.},
  \bibinfo{author}{{Wittkowski}, M.}, \bibinfo{year}{2004}.
\newblock \bibinfo{title}{{The building blocks of planets within the
  ‘terrestrial’ region of protoplanetary disks}}.
\newblock \bibinfo{journal}{Nature} \bibinfo{volume}{432},
  \bibinfo{pages}{479--482}.
\newblock \DOIprefix\doi{10.1038/nature03088}.
\bibitem[{Bollard et~al.(2017)Bollard, Connelly, Whitehouse, Pringle, Bonal,
  J{\o}rgensen, Nordlund, Moynier and Bizzarro}]{bollard2017early}
\bibinfo{author}{Bollard, J.}, \bibinfo{author}{Connelly, J.N.},
  \bibinfo{author}{Whitehouse, M.J.}, \bibinfo{author}{Pringle, E.A.},
  \bibinfo{author}{Bonal, L.}, \bibinfo{author}{J{\o}rgensen, J.K.},
  \bibinfo{author}{Nordlund, {\AA}.}, \bibinfo{author}{Moynier, F.},
  \bibinfo{author}{Bizzarro, M.}, \bibinfo{year}{2017}.
\newblock \bibinfo{title}{{Early formation of planetary building blocks
  inferred from Pb isotopic ages of chondrules}}.
\newblock \bibinfo{journal}{Science Advances} \bibinfo{volume}{3},
  \bibinfo{pages}{e1700407}.
\newblock \DOIprefix\doi{10.1126/sciadv.1700407}.
\bibitem[{Bollard et~al.(2019)Bollard, Kawasaki, Sakamoto, Olsen, Itoh, Larsen,
  Wielandt, Schiller, Connelly, Yurimoto and Bizzarroa}]{bollard2019combined}
\bibinfo{author}{Bollard, J.}, \bibinfo{author}{Kawasaki, N.},
  \bibinfo{author}{Sakamoto, N.}, \bibinfo{author}{Olsen, M.},
  \bibinfo{author}{Itoh, S.}, \bibinfo{author}{Larsen, K.},
  \bibinfo{author}{Wielandt, D.}, \bibinfo{author}{Schiller, M.},
  \bibinfo{author}{Connelly, J.N.}, \bibinfo{author}{Yurimoto, H.},
  \bibinfo{author}{Bizzarroa, M.}, \bibinfo{year}{2019}.
\newblock \bibinfo{title}{{Combined U-corrected Pb-Pb dating and
  \ce{^{26}Al}-\ce{^{26}Mg} systematics of individual chondrules--Evidence for
  a reduced initial abundance of \ce{^{26}Al} amongst inner Solar System
  chondrules}}.
\newblock \bibinfo{journal}{Geochimica et Cosmochimica Acta}
  \bibinfo{volume}{260}, \bibinfo{pages}{62--83}.
\newblock \DOIprefix\doi{10.1016/j.gca.2019.06.025}.
\bibitem[{Bouvier et~al.(2008)Bouvier, Vervoort and Patchett}]{bouvier2008lu}
\bibinfo{author}{Bouvier, A.}, \bibinfo{author}{Vervoort, J.D.},
  \bibinfo{author}{Patchett, P.J.}, \bibinfo{year}{2008}.
\newblock \bibinfo{title}{{The Lu--Hf and Sm--Nd isotopic composition of CHUR:
  Constraints from unequilibrated chondrites and implications for the bulk
  composition of terrestrial planets}}.
\newblock \bibinfo{journal}{Earth and Planetary Science Letters}
  \bibinfo{volume}{273}, \bibinfo{pages}{48--57}.
\newblock \DOIprefix\doi{10.1016/j.epsl.2008.06.010}.
\bibitem[{Bouvier et~al.(2013)Bouvier, Wadhwa, Simon and
  Grossman}]{bouvier2013magnesium}
\bibinfo{author}{Bouvier, A.}, \bibinfo{author}{Wadhwa, M.},
  \bibinfo{author}{Simon, S.B.}, \bibinfo{author}{Grossman, L.},
  \bibinfo{year}{2013}.
\newblock \bibinfo{title}{{Magnesium isotopic fractionation in chondrules from
  the Murchison and Murray CM2 carbonaceous chondrites}}.
\newblock \bibinfo{journal}{Meteoritics \& Planetary Science}
  \bibinfo{volume}{48}, \bibinfo{pages}{339--353}.
\newblock \DOIprefix\doi{10.1111/maps.12059}.
\bibitem[{Bouvier et~al.(2018)Bouvier, Costa, Connelly, Jensen, Wielandt,
  Storey, Nemchin, Whitehouse, Snape, Bellucci, {Moynier}, {Agranier},
  {Gueguen}, {Sch{\"o}nb{\"a}chler} and {Bizzarro}}]{bouvier2018evidence}
\bibinfo{author}{Bouvier, L.C.}, \bibinfo{author}{Costa, M.M.},
  \bibinfo{author}{Connelly, J.N.}, \bibinfo{author}{Jensen, N.K.},
  \bibinfo{author}{Wielandt, D.}, \bibinfo{author}{Storey, M.},
  \bibinfo{author}{Nemchin, A.A.}, \bibinfo{author}{Whitehouse, M.J.},
  \bibinfo{author}{Snape, J.F.}, \bibinfo{author}{Bellucci, J.J.},
  \bibinfo{author}{{Moynier}, F.}, \bibinfo{author}{{Agranier}, A.},
  \bibinfo{author}{{Gueguen}, B.}, \bibinfo{author}{{Sch{\"o}nb{\"a}chler},
  M.}, \bibinfo{author}{{Bizzarro}, M.}, \bibinfo{year}{2018}.
\newblock \bibinfo{title}{{Evidence for extremely rapid magma ocean
  crystallization and crust formation on Mars}}.
\newblock \bibinfo{journal}{Nature} \bibinfo{volume}{558},
  \bibinfo{pages}{586--589}.
\newblock \DOIprefix\doi{10.1038/s41586-018-0222-z}.
\bibitem[{Bouwman et~al.(2008)Bouwman, Henning, Hillenbrand, Meyer, Pascucci,
  Carpenter, Hines, Kim, Silverstone, Hollenbach and
  Wolf}]{bouwman2008formation}
\bibinfo{author}{Bouwman, J.}, \bibinfo{author}{Henning, T.},
  \bibinfo{author}{Hillenbrand, L.A.}, \bibinfo{author}{Meyer, M.R.},
  \bibinfo{author}{Pascucci, I.}, \bibinfo{author}{Carpenter, J.},
  \bibinfo{author}{Hines, D.}, \bibinfo{author}{Kim, J.S.},
  \bibinfo{author}{Silverstone, M.D.}, \bibinfo{author}{Hollenbach, D.},
  \bibinfo{author}{Wolf, S.}, \bibinfo{year}{2008}.
\newblock \bibinfo{title}{{The formation and evolution of planetary systems:
  Grain growth and chemical processing of dust in T Tauri systems}}.
\newblock \bibinfo{journal}{The Astrophysical Journal} \bibinfo{volume}{683},
  \bibinfo{pages}{479--498}.
\newblock \DOIprefix\doi{10.1086/587793}.
\bibitem[{Bouwman et~al.(2010)Bouwman, Lawson, Juh{\'a}sz, Dominik, Feigelson,
  Henning, Tielens and Waters}]{bouwman2010protoplanetary}
\bibinfo{author}{Bouwman, J.}, \bibinfo{author}{Lawson, W.A.},
  \bibinfo{author}{Juh{\'a}sz, A.}, \bibinfo{author}{Dominik, C.},
  \bibinfo{author}{Feigelson, E.D.}, \bibinfo{author}{Henning, T.},
  \bibinfo{author}{Tielens, A.G.G.M.}, \bibinfo{author}{Waters, L.B.F.M.},
  \bibinfo{year}{2010}.
\newblock \bibinfo{title}{{The protoplanetary disk around the M4 star RECX 5:
  witnessing the influence of planet formation?}}
\newblock \bibinfo{journal}{The Astrophysical Journal Letters}
  \bibinfo{volume}{723}, \bibinfo{pages}{L243--L247}.
\newblock \DOIprefix\doi{10.1088/2041-8205/723/2/L243}.
\bibitem[{Brandon et~al.(2012)Brandon, Puchtel, Walker, Day, Irving and
  Taylor}]{brandon2012evolution}
\bibinfo{author}{Brandon, A.D.}, \bibinfo{author}{Puchtel, I.S.},
  \bibinfo{author}{Walker, R.J.}, \bibinfo{author}{Day, J.M.D.},
  \bibinfo{author}{Irving, A.J.}, \bibinfo{author}{Taylor, L.A.},
  \bibinfo{year}{2012}.
\newblock \bibinfo{title}{{Evolution of the martian mantle inferred from the
  \ce{^{187}Re}--\ce{^{187}Os} isotope and highly siderophile element abundance
  systematics of shergottite meteorites}}.
\newblock \bibinfo{journal}{Geochimica et Cosmochimica Acta}
  \bibinfo{volume}{76}, \bibinfo{pages}{206--235}.
\newblock \DOIprefix\doi{10.1016/j.gca.2011.09.047}.
\bibitem[{Breuer and Moore(2015)}]{breuer2015dynamics}
\bibinfo{author}{Breuer, D.}, \bibinfo{author}{Moore, W.B.},
  \bibinfo{year}{2015}.
\newblock \bibinfo{title}{{Dynamics and Thermal History of the Terrestrial
  Planets, the Moon, and Io}}, in: \bibinfo{editor}{Schubert, G.} (Ed.),
  \bibinfo{booktitle}{Treatise on Geophysics (Second Edition)}.
  \bibinfo{publisher}{Elsevier}, \bibinfo{address}{Oxford}, pp.
  \bibinfo{pages}{255--305}.
\newblock \DOIprefix\doi{10.1016/B978-0-444-53802-4.00173-1}.
\bibitem[{Breuer and Spohn(2003)}]{breuer2003early}
\bibinfo{author}{Breuer, D.}, \bibinfo{author}{Spohn, T.},
  \bibinfo{year}{2003}.
\newblock \bibinfo{title}{{Early plate tectonics versus single-plate tectonics
  on Mars: Evidence from magnetic field history and crust evolution}}.
\newblock \bibinfo{journal}{Journal of Geophysical Research: Planets}
  \bibinfo{volume}{108}, \bibinfo{pages}{5072}.
\newblock \DOIprefix\doi{10.1029/2002JE001999}.
\bibitem[{Budde et~al.(2019)Budde, Burkhardt and Kleine}]{budde2019molybdenum}
\bibinfo{author}{Budde, G.}, \bibinfo{author}{Burkhardt, C.},
  \bibinfo{author}{Kleine, T.}, \bibinfo{year}{2019}.
\newblock \bibinfo{title}{{Molybdenum isotopic evidence for the late accretion
  of outer Solar System material to Earth}}.
\newblock \bibinfo{journal}{Nature Astronomy} \bibinfo{volume}{3},
  \bibinfo{pages}{736--741}.
\newblock \DOIprefix\doi{10.1038/s41550-019-0779-y}.
\bibitem[{Budde et~al.(2016)Budde, Kleine, Kruijer, Burkhardt and
  Metzler}]{budde2016tungsten}
\bibinfo{author}{Budde, G.}, \bibinfo{author}{Kleine, T.},
  \bibinfo{author}{Kruijer, T.S.}, \bibinfo{author}{Burkhardt, C.},
  \bibinfo{author}{Metzler, K.}, \bibinfo{year}{2016}.
\newblock \bibinfo{title}{{Tungsten isotopic constraints on the age and origin
  of chondrules}}.
\newblock \bibinfo{journal}{Proceedings of the National Academy of Sciences}
  \bibinfo{volume}{113}, \bibinfo{pages}{2886--2891}.
\newblock \DOIprefix\doi{10.1073/pnas.1524980113}.
\bibitem[{Budde et~al.(2018)Budde, Kruijer and Kleine}]{budde2018hf}
\bibinfo{author}{Budde, G.}, \bibinfo{author}{Kruijer, T.S.},
  \bibinfo{author}{Kleine, T.}, \bibinfo{year}{2018}.
\newblock \bibinfo{title}{{Hf-W chronology of CR chondrites: Implications for
  the timescales of chondrule formation and the distribution of \ce{^{26}Al} in
  the solar nebula}}.
\newblock \bibinfo{journal}{Geochimica et Cosmochimica Acta}
  \bibinfo{volume}{222}, \bibinfo{pages}{284--304}.
\newblock \DOIprefix\doi{10.1016/j.gca.2017.10.014}.
\bibitem[{Burkhardt et~al.(2016)Burkhardt, Borg, Brennecka, Shollenberger,
  Dauphas and Kleine}]{burkhardt2016nucleosynthetic}
\bibinfo{author}{Burkhardt, C.}, \bibinfo{author}{Borg, L.E.},
  \bibinfo{author}{Brennecka, G.A.}, \bibinfo{author}{Shollenberger, Q.R.},
  \bibinfo{author}{Dauphas, N.}, \bibinfo{author}{Kleine, T.},
  \bibinfo{year}{2016}.
\newblock \bibinfo{title}{{A nucleosynthetic origin for the Earth's anomalous
  \ce{^{142}Nd} composition}}.
\newblock \bibinfo{journal}{Nature} \bibinfo{volume}{537},
  \bibinfo{pages}{394--398}.
\newblock \DOIprefix\doi{10.1038/nature18956}.
\bibitem[{Burkhardt et~al.(2011)Burkhardt, Kleine, Oberli, Pack, Bourdon and
  Wieler}]{burkhardt2011molybdenum}
\bibinfo{author}{Burkhardt, C.}, \bibinfo{author}{Kleine, T.},
  \bibinfo{author}{Oberli, F.}, \bibinfo{author}{Pack, A.},
  \bibinfo{author}{Bourdon, B.}, \bibinfo{author}{Wieler, R.},
  \bibinfo{year}{2011}.
\newblock \bibinfo{title}{{Molybdenum isotope anomalies in meteorites:
  Constraints on solar nebula evolution and origin of the Earth}}.
\newblock \bibinfo{journal}{Earth and Planetary Science Letters}
  \bibinfo{volume}{312}, \bibinfo{pages}{390--400}.
\newblock \DOIprefix\doi{10.1016/j.epsl.2011.10.010}.
\bibitem[{Cano et~al.(2020)Cano, Sharp and Shearer}]{cano2020distinct}
\bibinfo{author}{Cano, E.J.}, \bibinfo{author}{Sharp, Z.D.},
  \bibinfo{author}{Shearer, C.K.}, \bibinfo{year}{2020}.
\newblock \bibinfo{title}{{Distinct oxygen isotope compositions of the Earth
  and Moon}}.
\newblock \bibinfo{journal}{Nature Geoscience} \bibinfo{volume}{13},
  \bibinfo{pages}{270--274}.
\newblock \DOIprefix\doi{10.1038/s41561-020-0550-0}.
\bibitem[{Canup and Salmon(2018)}]{canup2018origin}
\bibinfo{author}{Canup, R.}, \bibinfo{author}{Salmon, J.},
  \bibinfo{year}{2018}.
\newblock \bibinfo{title}{{Origin of Phobos and Deimos by the impact of a
  Vesta-to-Ceres sized body with Mars}}.
\newblock \bibinfo{journal}{Science Advances} \bibinfo{volume}{4},
  \bibinfo{pages}{eaar6887}.
\newblock \DOIprefix\doi{10.1126/sciadv.aar6887}.
\bibitem[{Canup(2004)}]{canup2004simulations}
\bibinfo{author}{Canup, R.M.}, \bibinfo{year}{2004}.
\newblock \bibinfo{title}{{Simulations of a late lunar-forming impact}}.
\newblock \bibinfo{journal}{Icarus} \bibinfo{volume}{168},
  \bibinfo{pages}{433--456}.
\newblock \DOIprefix\doi{10.1016/j.icarus.2003.09.028}.
\bibitem[{Canup(2008)}]{canup2008accretion}
\bibinfo{author}{Canup, R.M.}, \bibinfo{year}{2008}.
\newblock \bibinfo{title}{{Accretion of the Earth}}.
\newblock \bibinfo{journal}{Philosophical Transactions of the Royal Society A:
  Mathematical, Physical and Engineering Sciences} \bibinfo{volume}{366},
  \bibinfo{pages}{4061--4075}.
\newblock \DOIprefix\doi{10.1098/rsta.2008.0101}.
\bibitem[{Canup and Asphaug(2001)}]{canup2001origin}
\bibinfo{author}{Canup, R.M.}, \bibinfo{author}{Asphaug, E.},
  \bibinfo{year}{2001}.
\newblock \bibinfo{title}{{Origin of the Moon in a giant impact near the end of
  the Earth's formation}}.
\newblock \bibinfo{journal}{Nature} \bibinfo{volume}{412},
  \bibinfo{pages}{708--712}.
\newblock \DOIprefix\doi{10.1038/35089010}.
\bibitem[{Caro and Bourdon(2010)}]{caro2010non}
\bibinfo{author}{Caro, G.}, \bibinfo{author}{Bourdon, B.},
  \bibinfo{year}{2010}.
\newblock \bibinfo{title}{{Non-chondritic Sm/Nd ratio in the terrestrial
  planets: Consequences for the geochemical evolution of the mantle--crust
  system}}.
\newblock \bibinfo{journal}{Geochimica et Cosmochimica Acta}
  \bibinfo{volume}{74}, \bibinfo{pages}{3333--3349}.
\newblock \DOIprefix\doi{10.1016/j.gca.2010.02.025}.
\bibitem[{Chabot(2018)}]{chabot2018composition}
\bibinfo{author}{Chabot, N.L.}, \bibinfo{year}{2018}.
\newblock \bibinfo{title}{{Composition of metallic cores in the early Solar
  System}}, in: \bibinfo{booktitle}{Lunar and Planetary Science Conference}, p.
  \bibinfo{pages}{1532}.
\bibitem[{Chambat et~al.(2010)Chambat, Ricard and
  Valette}]{chambat2010flattening}
\bibinfo{author}{Chambat, F.}, \bibinfo{author}{Ricard, Y.},
  \bibinfo{author}{Valette, B.}, \bibinfo{year}{2010}.
\newblock \bibinfo{title}{{Flattening of the Earth: further from hydrostaticity
  than previously estimated}}.
\newblock \bibinfo{journal}{Geophysical Journal International}
  \bibinfo{volume}{183}, \bibinfo{pages}{727--732}.
\newblock \DOIprefix\doi{10.1111/j.1365-246X.2010.04771.x}.
\bibitem[{Cockell et~al.(2016)Cockell, Bush, Bryce, Direito, Fox-Powell,
  Harrison, Lammer, Landenmark, Martin-Torres, Nicholson, Noack,
  O'Malley-James, Payler, Rushby, Samuels, Schwendner, Wadsworth and
  Zorzano}]{cockell2016habitability}
\bibinfo{author}{Cockell, C.}, \bibinfo{author}{Bush, T.},
  \bibinfo{author}{Bryce, C.}, \bibinfo{author}{Direito, S.},
  \bibinfo{author}{Fox-Powell, M.}, \bibinfo{author}{Harrison, J.},
  \bibinfo{author}{Lammer, H.}, \bibinfo{author}{Landenmark, H.},
  \bibinfo{author}{Martin-Torres, J.}, \bibinfo{author}{Nicholson, N.},
  \bibinfo{author}{Noack, L.}, \bibinfo{author}{O'Malley-James, J.},
  \bibinfo{author}{Payler, S.}, \bibinfo{author}{Rushby, A.},
  \bibinfo{author}{Samuels, T.}, \bibinfo{author}{Schwendner, P.},
  \bibinfo{author}{Wadsworth, J.}, \bibinfo{author}{Zorzano, M.},
  \bibinfo{year}{2016}.
\newblock \bibinfo{title}{{Habitability: a review}}.
\newblock \bibinfo{journal}{Astrobiology} \bibinfo{volume}{16},
  \bibinfo{pages}{89--117}.
\newblock \DOIprefix\doi{10.1089/ast.2015.1295}.
\bibitem[{Connelly et~al.(2008)Connelly, Amelin, Krot and
  Bizzarro}]{connelly2008chronology}
\bibinfo{author}{Connelly, J.N.}, \bibinfo{author}{Amelin, Y.},
  \bibinfo{author}{Krot, A.N.}, \bibinfo{author}{Bizzarro, M.},
  \bibinfo{year}{2008}.
\newblock \bibinfo{title}{{Chronology of the solar system's oldest solids}}.
\newblock \bibinfo{journal}{The Astrophysical Journal Letters}
  \bibinfo{volume}{675}, \bibinfo{pages}{L121--L124}.
\newblock \DOIprefix\doi{10.1086/533586}.
\bibitem[{Connelly and Bizzarro(2009)}]{connelly2009pb}
\bibinfo{author}{Connelly, J.N.}, \bibinfo{author}{Bizzarro, M.},
  \bibinfo{year}{2009}.
\newblock \bibinfo{title}{{Pb--Pb dating of chondrules from CV chondrites by
  progressive dissolution}}.
\newblock \bibinfo{journal}{Chemical Geology} \bibinfo{volume}{259},
  \bibinfo{pages}{143--151}.
\newblock \DOIprefix\doi{10.1016/j.chemgeo.2008.11.003}.
\bibitem[{Connelly et~al.(2012)Connelly, Bizzarro, Krot, Nordlund, Wielandt and
  Ivanova}]{connelly2012absolute}
\bibinfo{author}{Connelly, J.N.}, \bibinfo{author}{Bizzarro, M.},
  \bibinfo{author}{Krot, A.N.}, \bibinfo{author}{Nordlund, {\AA}.},
  \bibinfo{author}{Wielandt, D.}, \bibinfo{author}{Ivanova, M.A.},
  \bibinfo{year}{2012}.
\newblock \bibinfo{title}{{The absolute chronology and thermal processing of
  solids in the solar protoplanetary disk}}.
\newblock \bibinfo{journal}{Science} \bibinfo{volume}{338},
  \bibinfo{pages}{651--655}.
\newblock \DOIprefix\doi{10.1126/science.1226919}.
\bibitem[{Connelly et~al.(2019)Connelly, Schiller and
  Bizzarro}]{connelly2019pb}
\bibinfo{author}{Connelly, J.N.}, \bibinfo{author}{Schiller, M.},
  \bibinfo{author}{Bizzarro, M.}, \bibinfo{year}{2019}.
\newblock \bibinfo{title}{{Pb isotope evidence for rapid accretion and
  differentiation of planetary embryos}}.
\newblock \bibinfo{journal}{Earth and Planetary Science Letters}
  \bibinfo{volume}{525}, \bibinfo{pages}{115722}.
\newblock \DOIprefix\doi{10.1016/j.epsl.2019.115722}.
\bibitem[{Corgne et~al.(2007)Corgne, Keshav, Fei and McDonough}]{corgne2007how}
\bibinfo{author}{Corgne, A.}, \bibinfo{author}{Keshav, S.},
  \bibinfo{author}{Fei, Y.}, \bibinfo{author}{McDonough, W.F.},
  \bibinfo{year}{2007}.
\newblock \bibinfo{title}{{How much potassium is in the Earth's core? New
  insights from partitioning experiments}}.
\newblock \bibinfo{journal}{Earth and Planetary Science Letters}
  \bibinfo{volume}{256}, \bibinfo{pages}{567--576}.
\newblock \DOIprefix\doi{10.1016/j.epsl.2007.02.012}.
\bibitem[{Creech and Moynier(2019)}]{creech2019tin}
\bibinfo{author}{Creech, J.B.}, \bibinfo{author}{Moynier, F.},
  \bibinfo{year}{2019}.
\newblock \bibinfo{title}{{Tin and zinc stable isotope characterisation of
  chondrites and implications for early Solar System evolution}}.
\newblock \bibinfo{journal}{Chemical Geology} \bibinfo{volume}{511},
  \bibinfo{pages}{81--90}.
\newblock \DOIprefix\doi{10.1016/j.chemgeo.2019.02.028}.
\bibitem[{Dauphas(2017)}]{dauphas2017isotopic}
\bibinfo{author}{Dauphas, N.}, \bibinfo{year}{2017}.
\newblock \bibinfo{title}{{The isotopic nature of the Earth's accreting
  material through time}}.
\newblock \bibinfo{journal}{Nature} \bibinfo{volume}{541},
  \bibinfo{pages}{521--524}.
\newblock \DOIprefix\doi{10.1038/nature20830}.
\bibitem[{Dauphas et~al.(2014)Dauphas, Burkhardt, Warren and
  Fang-Zhen}]{dauphas2014geochemical}
\bibinfo{author}{Dauphas, N.}, \bibinfo{author}{Burkhardt, C.},
  \bibinfo{author}{Warren, P.H.}, \bibinfo{author}{Fang-Zhen, T.},
  \bibinfo{year}{2014}.
\newblock \bibinfo{title}{{Geochemical arguments for an Earth-like Moon-forming
  impactor}}.
\newblock \bibinfo{journal}{Philosophical Transactions of the Royal Society A:
  Mathematical, Physical and Engineering Sciences} \bibinfo{volume}{372},
  \bibinfo{pages}{20130244}.
\newblock \DOIprefix\doi{10.1098/rsta.2013.0244}.
\bibitem[{Dauphas et~al.(2015)Dauphas, Poitrasson, Burkhardt, Kobayashi and
  Kurosawa}]{dauphas2015planetary}
\bibinfo{author}{Dauphas, N.}, \bibinfo{author}{Poitrasson, F.},
  \bibinfo{author}{Burkhardt, C.}, \bibinfo{author}{Kobayashi, H.},
  \bibinfo{author}{Kurosawa, K.}, \bibinfo{year}{2015}.
\newblock \bibinfo{title}{{Planetary and meteoritic Mg/Si and
  $\delta$\ce{^{30}}Si variations inherited from solar nebula chemistry}}.
\newblock \bibinfo{journal}{Earth and Planetary Science Letters}
  \bibinfo{volume}{427}, \bibinfo{pages}{236--248}.
\newblock \DOIprefix\doi{10.1016/j.epsl.2015.07.008}.
\bibitem[{Dauphas and Pourmand(2011)}]{dauphas2011hf}
\bibinfo{author}{Dauphas, N.}, \bibinfo{author}{Pourmand, A.},
  \bibinfo{year}{2011}.
\newblock \bibinfo{title}{{Hf--W--Th evidence for rapid growth of Mars and its
  status as a planetary embryo}}.
\newblock \bibinfo{journal}{Nature} \bibinfo{volume}{473},
  \bibinfo{pages}{489--492}.
\newblock \DOIprefix\doi{10.1038/nature10077}.
\bibitem[{Dauphas and Pourmand(2015)}]{dauphas2015thulium}
\bibinfo{author}{Dauphas, N.}, \bibinfo{author}{Pourmand, A.},
  \bibinfo{year}{2015}.
\newblock \bibinfo{title}{{Thulium anomalies and rare earth element patterns in
  meteorites and Earth: Nebular fractionation and the nugget effect}}.
\newblock \bibinfo{journal}{Geochimica et Cosmochimica Acta}
  \bibinfo{volume}{163}, \bibinfo{pages}{234--261}.
\newblock \DOIprefix\doi{10.1016/j.gca.2015.03.037}.
\bibitem[{Day et~al.(2016)Day, Brandon and Walker}]{day2016highly}
\bibinfo{author}{Day, J.M.D.}, \bibinfo{author}{Brandon, A.D.},
  \bibinfo{author}{Walker, R.J.}, \bibinfo{year}{2016}.
\newblock \bibinfo{title}{{Highly siderophile elements in Earth, Mars, the
  Moon, and asteroids}}.
\newblock \bibinfo{journal}{Reviews in Mineralogy and Geochemistry}
  \bibinfo{volume}{81}, \bibinfo{pages}{161--238}.
\newblock \DOIprefix\doi{10.2138/rmg.2016.81.04}.
\bibitem[{Desch et~al.(2018)Desch, Kalyaan and Alexander}]{desch2017effect}
\bibinfo{author}{Desch, S.J.}, \bibinfo{author}{Kalyaan, A.},
  \bibinfo{author}{Alexander, C.M.O'D.}, \bibinfo{year}{2018}.
\newblock \bibinfo{title}{{The effect of Jupiter's formation on the
  distribution of refractory elements and inclusions in meteorites}}.
\newblock \bibinfo{journal}{The Astrophysical Journal Supplement Series}
  \bibinfo{volume}{238}, \bibinfo{pages}{11}.
\newblock \DOIprefix\doi{10.3847/1538-4365/aad95f}.
\bibitem[{Doyle et~al.(2015)Doyle, Jogo, Nagashima, Krot, Wakita, Ciesla and
  Hutcheon}]{doyle2015early}
\bibinfo{author}{Doyle, P.M.}, \bibinfo{author}{Jogo, K.},
  \bibinfo{author}{Nagashima, K.}, \bibinfo{author}{Krot, A.N.},
  \bibinfo{author}{Wakita, S.}, \bibinfo{author}{Ciesla, F.J.},
  \bibinfo{author}{Hutcheon, I.D.}, \bibinfo{year}{2015}.
\newblock \bibinfo{title}{{Early aqueous activity on the ordinary and
  carbonaceous chondrite parent bodies recorded by fayalite}}.
\newblock \bibinfo{journal}{Nature Communications} \bibinfo{volume}{6},
  \bibinfo{pages}{7444}.
\newblock \DOIprefix\doi{10.1038/ncomms8444}.
\bibitem[{Dreibus and W{\"a}nke(1979)}]{dreibus1979chemical}
\bibinfo{author}{Dreibus, G.}, \bibinfo{author}{W{\"a}nke, H.},
  \bibinfo{year}{1979}.
\newblock \bibinfo{title}{{On the chemical composition of the Moon and the
  Eucrite parent body and a comparison with the composition of the Earth, the
  case of Mn, Cr, and V}}, in: \bibinfo{booktitle}{Lunar and Planetary Science
  Conference}, pp. \bibinfo{pages}{315--317}.
\bibitem[{Dreibus and W{\"a}nke(1987)}]{dreibus1987volatiles}
\bibinfo{author}{Dreibus, G.}, \bibinfo{author}{W{\"a}nke, H.},
  \bibinfo{year}{1987}.
\newblock \bibinfo{title}{{Volatiles on Earth and Mars: A comparison}}.
\newblock \bibinfo{journal}{Icarus} \bibinfo{volume}{71},
  \bibinfo{pages}{225--240}.
\newblock \DOIprefix\doi{10.1016/0019-1035(87)90148-5}.
\bibitem[{Dumoulin et~al.(2017)Dumoulin, Tobie, Verhoeven, Rosenblatt and
  Rambaux}]{dumoulin2017tidal}
\bibinfo{author}{Dumoulin, C.}, \bibinfo{author}{Tobie, G.},
  \bibinfo{author}{Verhoeven, O.}, \bibinfo{author}{Rosenblatt, P.},
  \bibinfo{author}{Rambaux, N.}, \bibinfo{year}{2017}.
\newblock \bibinfo{title}{{Tidal constraints on the interior of Venus}}.
\newblock \bibinfo{journal}{Journal of Geophysical Research: Planets}
  \bibinfo{volume}{122}, \bibinfo{pages}{1338--1352}.
\newblock \DOIprefix\doi{10.1002/2016JE005249}.
\bibitem[{Dziewonski and Anderson(1981)}]{dziewonski1981preliminary}
\bibinfo{author}{Dziewonski, A.M.}, \bibinfo{author}{Anderson, D.L.},
  \bibinfo{year}{1981}.
\newblock \bibinfo{title}{{Preliminary reference Earth model}}.
\newblock \bibinfo{journal}{Physics of the Earth and Planetary Interiors}
  \bibinfo{volume}{25}, \bibinfo{pages}{297--356}.
\newblock \DOIprefix\doi{10.1016/0031-9201(81)90046-7}.
\bibitem[{Ehlmann et~al.(2016)Ehlmann, Anderson, Andrews-Hanna, Catling,
  Christensen, Cohen, Dressing, Edwards, Elkins-Tanton, Farley, Fassett,
  Fischer, Fraeman, Golombek, Hamilton, Hayes, Herd, Horgan, Hu, Jakosky,
  Johnson, Kasting, Kerber, Kinch, Kite, Knutson, Lunine, Mahaffy, Mangold,
  McCubbin, Mustard, Niles, Quantin-Nataf, Rice, Stack, Stevenson, Stewart,
  Toplis, Usui, Weiss, Werner, Wordsworth, Wray, Yingst, Yung and
  Zahnle}]{ehlmann2016sustainability}
\bibinfo{author}{Ehlmann, B.L.}, \bibinfo{author}{Anderson, F.S.},
  \bibinfo{author}{Andrews-Hanna, J.}, \bibinfo{author}{Catling, D.C.},
  \bibinfo{author}{Christensen, P.R.}, \bibinfo{author}{Cohen, B.A.},
  \bibinfo{author}{Dressing, C.D.}, \bibinfo{author}{Edwards, C.S.},
  \bibinfo{author}{Elkins-Tanton, L.T.}, \bibinfo{author}{Farley, K.A.},
  \bibinfo{author}{Fassett, C.I.}, \bibinfo{author}{Fischer, W.W.},
  \bibinfo{author}{Fraeman, A.A.}, \bibinfo{author}{Golombek, M.P.},
  \bibinfo{author}{Hamilton, V.E.}, \bibinfo{author}{Hayes, A.G.},
  \bibinfo{author}{Herd, C.D.}, \bibinfo{author}{Horgan, B.},
  \bibinfo{author}{Hu, R.}, \bibinfo{author}{Jakosky, B.M.},
  \bibinfo{author}{Johnson, J.R.}, \bibinfo{author}{Kasting, J.F.},
  \bibinfo{author}{Kerber, L.}, \bibinfo{author}{Kinch, K.M.},
  \bibinfo{author}{Kite, E.S.}, \bibinfo{author}{Knutson, H.A.},
  \bibinfo{author}{Lunine, J.I.}, \bibinfo{author}{Mahaffy, P.R.},
  \bibinfo{author}{Mangold, N.}, \bibinfo{author}{McCubbin, F.M.},
  \bibinfo{author}{Mustard, J.F.}, \bibinfo{author}{Niles, P.B.},
  \bibinfo{author}{Quantin-Nataf, C.}, \bibinfo{author}{Rice, M.S.},
  \bibinfo{author}{Stack, K.M.}, \bibinfo{author}{Stevenson, D.J.},
  \bibinfo{author}{Stewart, S.T.}, \bibinfo{author}{Toplis, M.J.},
  \bibinfo{author}{Usui, T.}, \bibinfo{author}{Weiss, B.P.},
  \bibinfo{author}{Werner, S.C.}, \bibinfo{author}{Wordsworth, R.D.},
  \bibinfo{author}{Wray, J.J.}, \bibinfo{author}{Yingst, R.A.},
  \bibinfo{author}{Yung, Y.L.}, \bibinfo{author}{Zahnle, K.J.},
  \bibinfo{year}{2016}.
\newblock \bibinfo{title}{{The sustainability of habitability on terrestrial
  planets: Insights, questions, and needed measurements from Mars for
  understanding the evolution of Earth-like worlds}}.
\newblock \bibinfo{journal}{Journal of Geophysical Research: Planets}
  \bibinfo{volume}{121}, \bibinfo{pages}{1927--1961}.
\newblock \DOIprefix\doi{10.1002/2016JE005134}.
\bibitem[{Filiberto(2017)}]{filiberto2017geochemistry}
\bibinfo{author}{Filiberto, J.}, \bibinfo{year}{2017}.
\newblock \bibinfo{title}{{Geochemistry of Martian basalts with constraints on
  magma genesis}}.
\newblock \bibinfo{journal}{Chemical Geology} \bibinfo{volume}{466},
  \bibinfo{pages}{1--14}.
\newblock \DOIprefix\doi{10.1016/j.chemgeo.2017.06.009}.
\bibitem[{Filiberto and Dasgupta(2011)}]{filiberto2011fe2+}
\bibinfo{author}{Filiberto, J.}, \bibinfo{author}{Dasgupta, R.},
  \bibinfo{year}{2011}.
\newblock \bibinfo{title}{{\ce{Fe^{2+}}--Mg partitioning between olivine and
  basaltic melts: Applications to genesis of olivine-phyric shergottites and
  conditions of melting in the Martian interior}}.
\newblock \bibinfo{journal}{Earth and Planetary Science Letters}
  \bibinfo{volume}{304}, \bibinfo{pages}{527--537}.
\newblock \DOIprefix\doi{10.1016/j.epsl.2011.02.029}.
\bibitem[{Frost(2003)}]{frost2003structure}
\bibinfo{author}{Frost, D.J.}, \bibinfo{year}{2003}.
\newblock \bibinfo{title}{{The structure and sharpness of (Mg,Fe)\ce{_2SiO4}
  phase transformations in the transition zone}}.
\newblock \bibinfo{journal}{Earth and Planetary Science Letters}
  \bibinfo{volume}{216}, \bibinfo{pages}{313--328}.
\newblock \DOIprefix\doi{10.1016/S0012-821X(03)00533-8}.
\bibitem[{Fujiya et~al.(2012)Fujiya, Sugiura, Hotta, Ichimura and
  Sano}]{fujiya2012evidence}
\bibinfo{author}{Fujiya, W.}, \bibinfo{author}{Sugiura, N.},
  \bibinfo{author}{Hotta, H.}, \bibinfo{author}{Ichimura, K.},
  \bibinfo{author}{Sano, Y.}, \bibinfo{year}{2012}.
\newblock \bibinfo{title}{{Evidence for the late formation of hydrous asteroids
  from young meteoritic carbonates}}.
\newblock \bibinfo{journal}{Nature Communications} \bibinfo{volume}{3},
  \bibinfo{pages}{1--6}.
\newblock \DOIprefix\doi{10.1038/ncomms1635}.
\bibitem[{Gaetani and Grove(1999)}]{gaetani1999wetting}
\bibinfo{author}{Gaetani, G.A.}, \bibinfo{author}{Grove, T.L.},
  \bibinfo{year}{1999}.
\newblock \bibinfo{title}{{Wetting of mantle olivine by sulfide melt:
  implications for Re/Os ratios in mantle peridotite and late-stage core
  formation}}.
\newblock \bibinfo{journal}{Earth and Planetary Science Letters}
  \bibinfo{volume}{169}, \bibinfo{pages}{147--163}.
\newblock \DOIprefix\doi{10.1016/S0012-821X(99)00062-X}.
\bibitem[{Galy et~al.(2000)Galy, Young, Ash and O'Nions}]{galy2000formation}
\bibinfo{author}{Galy, A.}, \bibinfo{author}{Young, E.D.},
  \bibinfo{author}{Ash, R.D.}, \bibinfo{author}{O'Nions, R.K.},
  \bibinfo{year}{2000}.
\newblock \bibinfo{title}{{The formation of chondrules at high gas pressures in
  the solar nebula}}.
\newblock \bibinfo{journal}{Science} \bibinfo{volume}{290},
  \bibinfo{pages}{1751--1753}.
\newblock \DOIprefix\doi{10.1126/science.290.5497.1751}.
\bibitem[{Gast(1960)}]{gast1960limitations}
\bibinfo{author}{Gast, P.W.}, \bibinfo{year}{1960}.
\newblock \bibinfo{title}{{Limitations on the composition of the upper
  mantle}}.
\newblock \bibinfo{journal}{Journal of Geophysical Research}
  \bibinfo{volume}{65}, \bibinfo{pages}{1287--1297}.
\newblock \DOIprefix\doi{10.1029/JZ065i004p01287}.
\bibitem[{Gellissen et~al.(2019)Gellissen, Holzheid, Kegler and
  Palme}]{gellissen2019heating}
\bibinfo{author}{Gellissen, M.}, \bibinfo{author}{Holzheid, A.},
  \bibinfo{author}{Kegler, P.}, \bibinfo{author}{Palme, H.},
  \bibinfo{year}{2019}.
\newblock \bibinfo{title}{{Heating experiments relevant to the depletion of Na,
  K and Mn in the Earth and other planetary bodies}}.
\newblock \bibinfo{journal}{Geochemistry} \bibinfo{volume}{79},
  \bibinfo{pages}{125540}.
\newblock \DOIprefix\doi{10.1016/j.chemer.2019.125540}.
\bibitem[{Genova et~al.(2016)Genova, Goossens, Lemoine, Mazarico, Neumann,
  Smith and Zuber}]{genova2016seasonal}
\bibinfo{author}{Genova, A.}, \bibinfo{author}{Goossens, S.},
  \bibinfo{author}{Lemoine, F.G.}, \bibinfo{author}{Mazarico, E.},
  \bibinfo{author}{Neumann, G.A.}, \bibinfo{author}{Smith, D.E.},
  \bibinfo{author}{Zuber, M.T.}, \bibinfo{year}{2016}.
\newblock \bibinfo{title}{{Seasonal and static gravity field of Mars from MGS,
  Mars Odyssey and MRO radio science}}.
\newblock \bibinfo{journal}{Icarus} \bibinfo{volume}{272},
  \bibinfo{pages}{228--245}.
\newblock \DOIprefix\doi{10.1016/j.icarus.2016.02.050}.
\bibitem[{Goldstein et~al.(2009)Goldstein, Scott and
  Chabot}]{goldstein2009iron}
\bibinfo{author}{Goldstein, J.I.}, \bibinfo{author}{Scott, E.R.D.},
  \bibinfo{author}{Chabot, N.L.}, \bibinfo{year}{2009}.
\newblock \bibinfo{title}{{Iron meteorites: Crystallization, thermal history,
  parent bodies, and origin}}.
\newblock \bibinfo{journal}{Geochemistry} \bibinfo{volume}{69},
  \bibinfo{pages}{293--325}.
\newblock \DOIprefix\doi{10.1016/j.chemer.2009.01.002}.
\bibitem[{Goodrich et~al.(2003)Goodrich, Herd and Taylor}]{goodrich2003spinels}
\bibinfo{author}{Goodrich, C.A.}, \bibinfo{author}{Herd, C.D.K.},
  \bibinfo{author}{Taylor, L.A.}, \bibinfo{year}{2003}.
\newblock \bibinfo{title}{{Spinels and oxygen fugacity in olivine-phyric and
  lherzolitic shergottites}}.
\newblock \bibinfo{journal}{Meteoritics \& Planetary Science}
  \bibinfo{volume}{38}, \bibinfo{pages}{1773--1792}.
\newblock \DOIprefix\doi{10.1111/j.1945-5100.2003.tb00014.x}.
\bibitem[{Greenwood et~al.(2018)Greenwood, Barrat, Miller, Anand, Dauphas,
  Franchi, Sillard and Starkey}]{greenwood2018oxygen}
\bibinfo{author}{Greenwood, R.C.}, \bibinfo{author}{Barrat, J.A.},
  \bibinfo{author}{Miller, M.F.}, \bibinfo{author}{Anand, M.},
  \bibinfo{author}{Dauphas, N.}, \bibinfo{author}{Franchi, I.A.},
  \bibinfo{author}{Sillard, P.}, \bibinfo{author}{Starkey, N.A.},
  \bibinfo{year}{2018}.
\newblock \bibinfo{title}{{Oxygen isotopic evidence for accretion of Earth's
  water before a high-energy Moon-forming giant impact}}.
\newblock \bibinfo{journal}{Science Advances} \bibinfo{volume}{4},
  \bibinfo{pages}{eaao5928}.
\newblock \DOIprefix\doi{10.1126/sciadv.aao5928}.
\bibitem[{Grossman et~al.(2000)Grossman, Ebel, Simon, Davis, Richter and
  Parsad}]{grossman2000major}
\bibinfo{author}{Grossman, L.}, \bibinfo{author}{Ebel, D.S.},
  \bibinfo{author}{Simon, S.B.}, \bibinfo{author}{Davis, A.M.},
  \bibinfo{author}{Richter, F.M.}, \bibinfo{author}{Parsad, N.M.},
  \bibinfo{year}{2000}.
\newblock \bibinfo{title}{{Major element chemical and isotopic compositions of
  refractory inclusions in C3 chondrites: The separate roles of condensation
  and evaporation}}.
\newblock \bibinfo{journal}{Geochimica et Cosmochimica Acta}
  \bibinfo{volume}{64}, \bibinfo{pages}{2879--2894}.
\newblock \DOIprefix\doi{10.1016/S0016-7037(00)00396-3}.
\bibitem[{Grossman et~al.(2008)Grossman, Simon, Rai, Thiemens, Hutcheon,
  Williams, Galy, Ding, Fedkin, Clayton and Mayeda}]{grossman2008primordial}
\bibinfo{author}{Grossman, L.}, \bibinfo{author}{Simon, S.B.},
  \bibinfo{author}{Rai, V.K.}, \bibinfo{author}{Thiemens, M.H.},
  \bibinfo{author}{Hutcheon, I.D.}, \bibinfo{author}{Williams, R.W.},
  \bibinfo{author}{Galy, A.}, \bibinfo{author}{Ding, T.},
  \bibinfo{author}{Fedkin, A.V.}, \bibinfo{author}{Clayton, R.N.},
  \bibinfo{author}{Mayeda, T.K.}, \bibinfo{year}{2008}.
\newblock \bibinfo{title}{{Primordial compositions of refractory inclusions}}.
\newblock \bibinfo{journal}{Geochimica et Cosmochimica Acta}
  \bibinfo{volume}{72}, \bibinfo{pages}{3001--3021}.
\newblock \DOIprefix\doi{10.1016/j.gca.2008.04.002}.
\bibitem[{Halliday and Porcelli(2001)}]{halliday2001in}
\bibinfo{author}{Halliday, A.N.}, \bibinfo{author}{Porcelli, D.},
  \bibinfo{year}{2001}.
\newblock \bibinfo{title}{{In search of lost planets--The paleocosmochemistry
  of the inner solar system}}.
\newblock \bibinfo{journal}{Earth and Planetary Science Letters}
  \bibinfo{volume}{192}, \bibinfo{pages}{545--559}.
\newblock \DOIprefix\doi{10.1016/S0012-821X(01)00479-4}.
\bibitem[{Hans et~al.(2013)Hans, Kleine and Bourdon}]{hans2013rb}
\bibinfo{author}{Hans, U.}, \bibinfo{author}{Kleine, T.},
  \bibinfo{author}{Bourdon, B.}, \bibinfo{year}{2013}.
\newblock \bibinfo{title}{{Rb--Sr chronology of volatile depletion in
  differentiated protoplanets: BABI, ADOR and ALL revisited}}.
\newblock \bibinfo{journal}{Earth and Planetary Science Letters}
  \bibinfo{volume}{374}, \bibinfo{pages}{204--214}.
\newblock \DOIprefix\doi{10.1016/j.epsl.2013.05.029}.
\bibitem[{Harrison et~al.(2021)Harrison, Shorttle and
  Bonsor}]{harrison2021evidence}
\bibinfo{author}{Harrison, J.H.}, \bibinfo{author}{Shorttle, O.},
  \bibinfo{author}{Bonsor, A.}, \bibinfo{year}{2021}.
\newblock \bibinfo{title}{{Evidence for post-nebula volatilisation in an
  exo-planetary body}}.
\newblock \bibinfo{journal}{Earth and Planetary Science Letters}
  \bibinfo{volume}{554}, \bibinfo{pages}{116694}.
\newblock \DOIprefix\doi{10.1016/j.epsl.2020.116694}.
\bibitem[{Harrison et~al.(2018)Harrison, Bonsor and
  Madhusudhan}]{harrison2018polluted}
\bibinfo{author}{Harrison, J.H.D.}, \bibinfo{author}{Bonsor, A.},
  \bibinfo{author}{Madhusudhan, N.}, \bibinfo{year}{2018}.
\newblock \bibinfo{title}{{Polluted white dwarfs: constraints on the origin and
  geology of exoplanetary material}}.
\newblock \bibinfo{journal}{Monthly Notices of the Royal Astronomical Society}
  \bibinfo{volume}{479}, \bibinfo{pages}{3814--3841}.
\newblock \DOIprefix\doi{10.1093/mnras/sty1700}.
\bibitem[{Hart and Zindler(1986)}]{hart1986search}
\bibinfo{author}{Hart, S.R.}, \bibinfo{author}{Zindler, A.},
  \bibinfo{year}{1986}.
\newblock \bibinfo{title}{{In search of a bulk-Earth composition}}.
\newblock \bibinfo{journal}{Chemical Geology} \bibinfo{volume}{57},
  \bibinfo{pages}{247--267}.
\newblock \DOIprefix\doi{10.1016/0009-2541(86)90053-7}.
\bibitem[{Herd(2019)}]{herd2019reconciling}
\bibinfo{author}{Herd, C.D.K.}, \bibinfo{year}{2019}.
\newblock \bibinfo{title}{{Reconciling redox: making spatial and temporal sense
  of oxygen fugacity variations in martian igneous rocks}}, in:
  \bibinfo{booktitle}{Lunar and Planetary Science Conference}, p.
  \bibinfo{pages}{2746}.
\bibitem[{Herd et~al.(2002)Herd, Borg, Jones and Papike}]{herd2002oxygen}
\bibinfo{author}{Herd, C.D.K.}, \bibinfo{author}{Borg, L.E.},
  \bibinfo{author}{Jones, J.H.}, \bibinfo{author}{Papike, J.J.},
  \bibinfo{year}{2002}.
\newblock \bibinfo{title}{{Oxygen fugacity and geochemical variations in the
  martian basalts: Implications for martian basalt petrogenesis and the
  oxidation state of the upper mantle of Mars}}.
\newblock \bibinfo{journal}{Geochimica et Cosmochimica Acta}
  \bibinfo{volume}{66}, \bibinfo{pages}{2025--2036}.
\newblock \DOIprefix\doi{10.1016/S0016-7037(02)00828-1}.
\bibitem[{Herd et~al.(2001)Herd, Papike and Brearley}]{herd2001oxygen}
\bibinfo{author}{Herd, C.D.K.}, \bibinfo{author}{Papike, J.J.},
  \bibinfo{author}{Brearley, A.J.}, \bibinfo{year}{2001}.
\newblock \bibinfo{title}{{Oxygen fugacity of martian basalts from electron
  microprobe oxygen and TEM-EELS analyses of Fe-Ti oxides}}.
\newblock \bibinfo{journal}{American Mineralogist} \bibinfo{volume}{86},
  \bibinfo{pages}{1015--1024}.
\newblock \DOIprefix\doi{10.2138/am-2001-8-908}.
\bibitem[{Hewins and Herzberg(1996)}]{hewins1996nebular}
\bibinfo{author}{Hewins, R.H.}, \bibinfo{author}{Herzberg, C.T.},
  \bibinfo{year}{1996}.
\newblock \bibinfo{title}{{Nebular turbulence, chondrule formation, and the
  composition of the Earth}}.
\newblock \bibinfo{journal}{Earth and Planetary Science Letters}
  \bibinfo{volume}{144}, \bibinfo{pages}{1--7}.
\newblock \DOIprefix\doi{10.1016/0012-821X(96)00159-8}.
\bibitem[{Hewins and Zanda(2012)}]{hewins2012chondrules}
\bibinfo{author}{Hewins, R.H.}, \bibinfo{author}{Zanda, B.},
  \bibinfo{year}{2012}.
\newblock \bibinfo{title}{{Chondrules: Precursors and interactions with the
  nebular gas}}.
\newblock \bibinfo{journal}{Meteoritics \& Planetary Science}
  \bibinfo{volume}{47}, \bibinfo{pages}{1120--1138}.
\newblock \DOIprefix\doi{10.1111/j.1945-5100.2012.01376.x}.
\bibitem[{Hezel et~al.(2018a)Hezel, Bland, Palme, Jacquet and
  Bigolski}]{hezel2018composition}
\bibinfo{author}{Hezel, D.C.}, \bibinfo{author}{Bland, P.},
  \bibinfo{author}{Palme, H.}, \bibinfo{author}{Jacquet, E.},
  \bibinfo{author}{Bigolski, J.}, \bibinfo{year}{2018}a.
\newblock \bibinfo{title}{{Composition of chondrules and matrix and their
  complementary relationship in chondrites}}, in: \bibinfo{editor}{Russell,
  S.S.}, \bibinfo{editor}{Connolly, Jr., H.C.}, \bibinfo{editor}{Krot, A.N.}
  (Eds.), \bibinfo{booktitle}{Chondrules and the Protoplanetary Disk}.
  \bibinfo{publisher}{Cambridge University Press},
  \bibinfo{address}{Cambridge}. chapter~\bibinfo{chapter}{4}, pp.
  \bibinfo{pages}{91--121}.
\newblock \DOIprefix\doi{10.1017/9781108284073.004}.
\bibitem[{Hezel et~al.(2018b)Hezel, Harak and Libourel}]{hezel2018what}
\bibinfo{author}{Hezel, D.C.}, \bibinfo{author}{Harak, M.},
  \bibinfo{author}{Libourel, G.}, \bibinfo{year}{2018}b.
\newblock \bibinfo{title}{{What we know about elemental bulk chondrule and
  matrix compositions: Presenting the ChondriteDB Database}}.
\newblock \bibinfo{journal}{Chemie der Erde-Geochemistry} \bibinfo{volume}{78},
  \bibinfo{pages}{1--14}.
\newblock \DOIprefix\doi{10.1016/j.chemer.2017.05.003}.
\bibitem[{Hin et~al.(2017)Hin, Coath, Carter, Nimmo, Lai, von Strandmann,
  Willbold, Leinhardt, Walter and Elliott}]{hin2017magnesium}
\bibinfo{author}{Hin, R.C.}, \bibinfo{author}{Coath, C.D.},
  \bibinfo{author}{Carter, P.J.}, \bibinfo{author}{Nimmo, F.},
  \bibinfo{author}{Lai, Y.J.}, \bibinfo{author}{von Strandmann, P.A.E.P.},
  \bibinfo{author}{Willbold, M.}, \bibinfo{author}{Leinhardt, Z.M.},
  \bibinfo{author}{Walter, M.J.}, \bibinfo{author}{Elliott, T.},
  \bibinfo{year}{2017}.
\newblock \bibinfo{title}{{Magnesium isotope evidence that accretional vapour
  loss shapes planetary compositions}}.
\newblock \bibinfo{journal}{Nature} \bibinfo{volume}{549},
  \bibinfo{pages}{511--515}.
\newblock \DOIprefix\doi{10.1038/nature23899}.
\bibitem[{Hosono et~al.(2019)Hosono, Karato, Makino and
  Saitoh}]{hosono2019terrestrial}
\bibinfo{author}{Hosono, N.}, \bibinfo{author}{Karato, S.i.},
  \bibinfo{author}{Makino, J.}, \bibinfo{author}{Saitoh, T.R.},
  \bibinfo{year}{2019}.
\newblock \bibinfo{title}{{Terrestrial magma ocean origin of the Moon}}.
\newblock \bibinfo{journal}{Nature Geoscience} \bibinfo{volume}{12},
  \bibinfo{pages}{418--423}.
\newblock \DOIprefix\doi{10.1038/s41561-019-0354-2}.
\bibitem[{Huang et~al.(2010)Huang, Farka{\v{s}} and
  Jacobsen}]{huang2010calcium}
\bibinfo{author}{Huang, S.}, \bibinfo{author}{Farka{\v{s}}, J.},
  \bibinfo{author}{Jacobsen, S.B.}, \bibinfo{year}{2010}.
\newblock \bibinfo{title}{{Calcium isotopic fractionation between clinopyroxene
  and orthopyroxene from mantle peridotites}}.
\newblock \bibinfo{journal}{Earth and Planetary Science Letters}
  \bibinfo{volume}{292}, \bibinfo{pages}{337--344}.
\newblock \DOIprefix\doi{10.1016/j.epsl.2010.01.042}.
\bibitem[{Humayun and Clayton(1995)}]{humayun1995potassium}
\bibinfo{author}{Humayun, M.}, \bibinfo{author}{Clayton, R.N.},
  \bibinfo{year}{1995}.
\newblock \bibinfo{title}{{Potassium isotope cosmochemistry: Genetic
  implications of volatile element depletion}}.
\newblock \bibinfo{journal}{Geochimica et Cosmochimica Acta}
  \bibinfo{volume}{59}, \bibinfo{pages}{2131--2148}.
\newblock \DOIprefix\doi{10.1016/0016-7037(95)00132-8}.
\bibitem[{Humayun et~al.(2013)Humayun, Nemchin, Zanda, Hewins, Grange, Kennedy,
  Lorand, G{\"{o}}pel, Fieni, Pont and Deldicque}]{humayun2013origin}
\bibinfo{author}{Humayun, M.}, \bibinfo{author}{Nemchin, A.},
  \bibinfo{author}{Zanda, B.}, \bibinfo{author}{Hewins, R.H.},
  \bibinfo{author}{Grange, M.}, \bibinfo{author}{Kennedy, A.},
  \bibinfo{author}{Lorand, J.P.}, \bibinfo{author}{G{\"{o}}pel, C.},
  \bibinfo{author}{Fieni, C.}, \bibinfo{author}{Pont, S.},
  \bibinfo{author}{Deldicque, D.}, \bibinfo{year}{2013}.
\newblock \bibinfo{title}{{Origin and age of the earliest Martian crust from
  meteorite NWA 7533}}.
\newblock \bibinfo{journal}{Nature} \bibinfo{volume}{503},
  \bibinfo{pages}{513--516}.
\newblock \DOIprefix\doi{10.1038/nature12764}.
\bibitem[{Jackson et~al.(2018)Jackson, Gabriel and
  Asphaug}]{jackson2018constraints}
\bibinfo{author}{Jackson, A.P.}, \bibinfo{author}{Gabriel, T.S.J.},
  \bibinfo{author}{Asphaug, E.I.}, \bibinfo{year}{2018}.
\newblock \bibinfo{title}{{Constraints on the pre-impact orbits of Solar system
  giant impactors}}.
\newblock \bibinfo{journal}{Monthly Notices of the Royal Astronomical Society}
  \bibinfo{volume}{474}, \bibinfo{pages}{2924--2936}.
\newblock \DOIprefix\doi{10.1093/mnras/stx2901}.
\bibitem[{Jagoutz et~al.(1979)Jagoutz, Palme, Baddenhausen, Blum, Cendales,
  Dreibus, Spettel, Lorenz and W{\"a}nke}]{jagoutz1979abundances}
\bibinfo{author}{Jagoutz, E.}, \bibinfo{author}{Palme, H.},
  \bibinfo{author}{Baddenhausen, H.}, \bibinfo{author}{Blum, K.},
  \bibinfo{author}{Cendales, M.}, \bibinfo{author}{Dreibus, G.},
  \bibinfo{author}{Spettel, B.}, \bibinfo{author}{Lorenz, V.},
  \bibinfo{author}{W{\"a}nke, H.}, \bibinfo{year}{1979}.
\newblock \bibinfo{title}{{The abundances of major, minor and trace elements in
  the earth's mantle as derived from primitive ultramafic nodules}}, in:
  \bibinfo{booktitle}{Lunar and Planetary Science Conference Proceedings}, pp.
  \bibinfo{pages}{2031--2050}.
\bibitem[{Jaupart et~al.(2015)Jaupart, Labrosse, Lucazeau and
  Mareschal}]{jaupart2015temperatures}
\bibinfo{author}{Jaupart, C.}, \bibinfo{author}{Labrosse, S.},
  \bibinfo{author}{Lucazeau, F.}, \bibinfo{author}{Mareschal, J.C.},
  \bibinfo{year}{2015}.
\newblock \bibinfo{title}{Temperatures, heat, and energy in the mantle of the
  earth}, in: \bibinfo{editor}{Schubert, G.} (Ed.),
  \bibinfo{booktitle}{Treatise on Geophysics (Second Edition)}.
  \bibinfo{publisher}{Elsevier}, \bibinfo{address}{Oxford}.
  volume~\bibinfo{volume}{7}, pp. \bibinfo{pages}{223--270}.
\newblock \DOIprefix\doi{10.1016/B978-0-444-53802-4.00126-3}.
\bibitem[{Javoy(1995)}]{javoy1995integral}
\bibinfo{author}{Javoy, M.}, \bibinfo{year}{1995}.
\newblock \bibinfo{title}{{The integral enstatite chondrite model of the
  Earth}}.
\newblock \bibinfo{journal}{Geophysical Research Letters} \bibinfo{volume}{22},
  \bibinfo{pages}{2219--2222}.
\newblock \DOIprefix\doi{10.1029/95GL02015}.
\bibitem[{Javoy et~al.(2010)Javoy, Kaminski, Guyot, Andrault, Sanloup, Moreira,
  Labrosse, Jambon, Agrinier, Davaille and Jaupart}]{javoy2010chemical}
\bibinfo{author}{Javoy, M.}, \bibinfo{author}{Kaminski, E.},
  \bibinfo{author}{Guyot, F.}, \bibinfo{author}{Andrault, D.},
  \bibinfo{author}{Sanloup, C.}, \bibinfo{author}{Moreira, M.},
  \bibinfo{author}{Labrosse, S.}, \bibinfo{author}{Jambon, A.},
  \bibinfo{author}{Agrinier, P.}, \bibinfo{author}{Davaille, A.},
  \bibinfo{author}{Jaupart, C.}, \bibinfo{year}{2010}.
\newblock \bibinfo{title}{{The chemical composition of the Earth: Enstatite
  chondrite models}}.
\newblock \bibinfo{journal}{Earth and Planetary Science Letters}
  \bibinfo{volume}{293}, \bibinfo{pages}{259--268}.
\newblock \DOIprefix\doi{10.1016/j.epsl.2010.02.033}.
\bibitem[{Johansen et~al.(2015a)Johansen, Jacquet, Cuzzi, Morbidelli and
  Gounelle}]{johansen2015new}
\bibinfo{author}{Johansen, A.}, \bibinfo{author}{Jacquet, E.},
  \bibinfo{author}{Cuzzi, J.N.}, \bibinfo{author}{Morbidelli, A.},
  \bibinfo{author}{Gounelle, M.}, \bibinfo{year}{2015}a.
\newblock \bibinfo{title}{{New paradigms for asteroid formation}}, in:
  \bibinfo{editor}{Michel, P.}, \bibinfo{editor}{DeMeo, F.},
  \bibinfo{editor}{Bottke, W.}, \bibinfo{editor}{Dotson, R.} (Eds.),
  \bibinfo{booktitle}{Asteroids IV}. \bibinfo{publisher}{University of Arizona
  Press}, \bibinfo{address}{Tucson}. volume~\bibinfo{volume}{47}, pp.
  \bibinfo{pages}{471--492}.
\newblock \DOIprefix\doi{10.2458/azu_uapress_9780816532131-ch025}.
\bibitem[{Johansen et~al.(2015b)Johansen, Mac~Low, Lacerda and
  Bizzarro}]{johansen2015growth}
\bibinfo{author}{Johansen, A.}, \bibinfo{author}{Mac~Low, M.M.},
  \bibinfo{author}{Lacerda, P.}, \bibinfo{author}{Bizzarro, M.},
  \bibinfo{year}{2015}b.
\newblock \bibinfo{title}{{Growth of asteroids, planetary embryos, and Kuiper
  belt objects by chondrule accretion}}.
\newblock \bibinfo{journal}{Science Advances} \bibinfo{volume}{1},
  \bibinfo{pages}{e1500109}.
\newblock \DOIprefix\doi{10.1126/sciadv.1500109}.
\bibitem[{Johnson et~al.(2015)Johnson, Minton, Melosh and
  Zuber}]{johnson2015impact}
\bibinfo{author}{Johnson, B.C.}, \bibinfo{author}{Minton, D.A.},
  \bibinfo{author}{Melosh, H.J.}, \bibinfo{author}{Zuber, M.T.},
  \bibinfo{year}{2015}.
\newblock \bibinfo{title}{{Impact jetting as the origin of chondrules}}.
\newblock \bibinfo{journal}{Nature} \bibinfo{volume}{517},
  \bibinfo{pages}{339--341}.
\newblock \DOIprefix\doi{10.1038/nature14105}.
\bibitem[{Jones(1994)}]{jones1994relict}
\bibinfo{author}{Jones, R.H.}, \bibinfo{year}{1994}.
\newblock \bibinfo{title}{Relict grains in chondrules: Evidence for chondrule
  recycling}, in: \bibinfo{editor}{Hewins, R.H.}, \bibinfo{editor}{Jones,
  R.H.}, \bibinfo{editor}{Scott, E.R.D.} (Eds.), \bibinfo{booktitle}{Chondrules
  and the Protoplanetary Disk}, pp. \bibinfo{pages}{163--172}.
\bibitem[{Kato and Moynier(2017)}]{kato2017gallium_Moon}
\bibinfo{author}{Kato, C.}, \bibinfo{author}{Moynier, F.},
  \bibinfo{year}{2017}.
\newblock \bibinfo{title}{{Gallium isotopic evidence for extensive volatile
  loss from the Moon during its formation}}.
\newblock \bibinfo{journal}{Science Advances} \bibinfo{volume}{3},
  \bibinfo{pages}{e1700571}.
\newblock \DOIprefix\doi{10.1126/sciadv.1700571}.
\bibitem[{Katsura et~al.(2010)Katsura, Yoneda, Yamazaki, Yoshino and
  Ito}]{katsura2010adiabatic}
\bibinfo{author}{Katsura, T.}, \bibinfo{author}{Yoneda, A.},
  \bibinfo{author}{Yamazaki, D.}, \bibinfo{author}{Yoshino, T.},
  \bibinfo{author}{Ito, E.}, \bibinfo{year}{2010}.
\newblock \bibinfo{title}{{Adiabatic temperature profile in the mantle}}.
\newblock \bibinfo{journal}{Physics of the Earth and Planetary Interiors}
  \bibinfo{volume}{183}, \bibinfo{pages}{212--218}.
\newblock \DOIprefix\doi{10.1016/j.pepi.2010.07.001}.
\bibitem[{Kerridge(1979)}]{kerridge1979fractionation}
\bibinfo{author}{Kerridge, J.}, \bibinfo{year}{1979}.
\newblock \bibinfo{title}{Fractionation of refractory lithophile elements among
  chondritic meteorites}, in: \bibinfo{booktitle}{Lunar and Planetary Science
  Conference Proceedings}, pp. \bibinfo{pages}{989--996}.
\bibitem[{Kessler‐Silacci et~al.(2006)Kessler‐Silacci, Augereau, Dullemond,
  Geers, Lahuis, {Evans II}, van Dishoeck, Blake, Boogert, Brown, Jorgensen,
  Knez and Pontoppidan}]{kessler2006c2d}
\bibinfo{author}{Kessler‐Silacci, J.}, \bibinfo{author}{Augereau, J.},
  \bibinfo{author}{Dullemond, C.P.}, \bibinfo{author}{Geers, V.},
  \bibinfo{author}{Lahuis, F.}, \bibinfo{author}{{Evans II}, N.J.},
  \bibinfo{author}{van Dishoeck, E.F.}, \bibinfo{author}{Blake, G.A.},
  \bibinfo{author}{Boogert, A.C.A.}, \bibinfo{author}{Brown, J.},
  \bibinfo{author}{Jorgensen, J.K.}, \bibinfo{author}{Knez, C.},
  \bibinfo{author}{Pontoppidan, K.M.}, \bibinfo{year}{2006}.
\newblock \bibinfo{title}{{C2D Spitzer IRS spectra of disks around T Tauri
  stars. I. Silicate emission and grain growth}}.
\newblock \bibinfo{journal}{The Astrophysical Journal} \bibinfo{volume}{639},
  \bibinfo{pages}{275}.
\newblock \DOIprefix\doi{10.1086/499330}.
\bibitem[{Khan et~al.(2018)Khan, Liebske, Rozel, Rivoldini, Nimmo, Connolly,
  Plesa and Giardini}]{khan2018geophysical}
\bibinfo{author}{Khan, A.}, \bibinfo{author}{Liebske, C.},
  \bibinfo{author}{Rozel, A.}, \bibinfo{author}{Rivoldini, A.},
  \bibinfo{author}{Nimmo, F.}, \bibinfo{author}{Connolly, J.A.D.},
  \bibinfo{author}{Plesa, A.C.}, \bibinfo{author}{Giardini, D.},
  \bibinfo{year}{2018}.
\newblock \bibinfo{title}{{A geophysical perspective on the bulk composition of
  Mars}}.
\newblock \bibinfo{journal}{Journal of Geophysical Research: Planets}
  \bibinfo{volume}{123}, \bibinfo{pages}{575--611}.
\newblock \DOIprefix\doi{10.1002/2017JE005371}.
\bibitem[{Kiefer(2003)}]{kiefer2003melting}
\bibinfo{author}{Kiefer, W.S.}, \bibinfo{year}{2003}.
\newblock \bibinfo{title}{{Melting in the Martian mantle: Shergottite formation
  and implications for present-day mantle convection on Mars}}.
\newblock \bibinfo{journal}{Meteoritics \& Planetary Science}
  \bibinfo{volume}{38}, \bibinfo{pages}{1815--1832}.
\newblock \DOIprefix\doi{10.1111/j.1945-5100.2003.tb00017.x}.
\bibitem[{Kimura et~al.(1974)Kimura, Lewis and Anders}]{kimura1974distribution}
\bibinfo{author}{Kimura, K.}, \bibinfo{author}{Lewis, R.S.},
  \bibinfo{author}{Anders, E.}, \bibinfo{year}{1974}.
\newblock \bibinfo{title}{{Distribution of gold and rhenium between nickel-iron
  and silicate melts: implications for the abundance of siderophile elements on
  the Earth and Moon}}.
\newblock \bibinfo{journal}{Geochimica et Cosmochimica Acta}
  \bibinfo{volume}{38}, \bibinfo{pages}{683--701}.
\newblock \DOIprefix\doi{10.1016/0016-7037(74)90144-6}.
\bibitem[{Kita et~al.(2013)Kita, Yin, MacPherson, Ushikubo, Jacobsen,
  Nagashima, Kurahashi, Krot and Jacobsen}]{kita201326al}
\bibinfo{author}{Kita, N.T.}, \bibinfo{author}{Yin, Q.Z.},
  \bibinfo{author}{MacPherson, G.J.}, \bibinfo{author}{Ushikubo, T.},
  \bibinfo{author}{Jacobsen, B.}, \bibinfo{author}{Nagashima, K.},
  \bibinfo{author}{Kurahashi, E.}, \bibinfo{author}{Krot, A.N.},
  \bibinfo{author}{Jacobsen, S.B.}, \bibinfo{year}{2013}.
\newblock \bibinfo{title}{{\ce{^{26}Al}-\ce{^{26}Mg} isotope systematics of the
  first solids in the early solar system}}.
\newblock \bibinfo{journal}{Meteoritics \& Planetary Science}
  \bibinfo{volume}{48}, \bibinfo{pages}{1383--1400}.
\newblock \DOIprefix\doi{10.1111/maps.12141}.
\bibitem[{Kleine et~al.(2009)Kleine, Touboul, Bourdon, Nimmo, Mezger, Palme,
  Jacobsen, Yin and Halliday}]{kleine2009hf}
\bibinfo{author}{Kleine, T.}, \bibinfo{author}{Touboul, M.},
  \bibinfo{author}{Bourdon, B.}, \bibinfo{author}{Nimmo, F.},
  \bibinfo{author}{Mezger, K.}, \bibinfo{author}{Palme, H.},
  \bibinfo{author}{Jacobsen, S.B.}, \bibinfo{author}{Yin, Q.Z.},
  \bibinfo{author}{Halliday, A.N.}, \bibinfo{year}{2009}.
\newblock \bibinfo{title}{{Hf--W chronology of the accretion and early
  evolution of asteroids and terrestrial planets}}.
\newblock \bibinfo{journal}{Geochimica et Cosmochimica Acta}
  \bibinfo{volume}{73}, \bibinfo{pages}{5150--5188}.
\newblock \DOIprefix\doi{10.1016/j.gca.2008.11.047}.
\bibitem[{Konopliv et~al.(2016)Konopliv, Park and
  Folkner}]{konopliv2016improved}
\bibinfo{author}{Konopliv, A.S.}, \bibinfo{author}{Park, R.S.},
  \bibinfo{author}{Folkner, W.M.}, \bibinfo{year}{2016}.
\newblock \bibinfo{title}{{An improved JPL Mars gravity field and orientation
  from Mars orbiter and lander tracking data}}.
\newblock \bibinfo{journal}{Icarus} \bibinfo{volume}{274},
  \bibinfo{pages}{253--260}.
\newblock \DOIprefix\doi{10.1016/j.icarus.2016.02.052}.
\bibitem[{Krot and Wasson(1995)}]{krot1995igneous}
\bibinfo{author}{Krot, A.N.}, \bibinfo{author}{Wasson, J.T.},
  \bibinfo{year}{1995}.
\newblock \bibinfo{title}{{Igneous rims on low-FeO and high-FeO chondrules in
  ordinary chondrites}}.
\newblock \bibinfo{journal}{Geochimica et Cosmochimica Acta}
  \bibinfo{volume}{59}, \bibinfo{pages}{4951--4966}.
\newblock \DOIprefix\doi{10.1016/0016-7037(95)00337-1}.
\bibitem[{Kruijer et~al.(2017a)Kruijer, Burkhardt, Budde and
  Kleine}]{kruijer2017age}
\bibinfo{author}{Kruijer, T.S.}, \bibinfo{author}{Burkhardt, C.},
  \bibinfo{author}{Budde, G.}, \bibinfo{author}{Kleine, T.},
  \bibinfo{year}{2017}a.
\newblock \bibinfo{title}{{Age of Jupiter inferred from the distinct genetics
  and formation times of meteorites}}.
\newblock \bibinfo{journal}{Proceedings of the National Academy of Sciences}
  \bibinfo{volume}{114}, \bibinfo{pages}{6712--6716}.
\newblock \DOIprefix\doi{10.1073/pnas.1704461114}.
\bibitem[{Kruijer et~al.(2020)Kruijer, Kleine and Borg}]{kruijer2019great}
\bibinfo{author}{Kruijer, T.S.}, \bibinfo{author}{Kleine, T.},
  \bibinfo{author}{Borg, L.E.}, \bibinfo{year}{2020}.
\newblock \bibinfo{title}{{The great isotopic dichotomy of the early Solar
  System}}.
\newblock \bibinfo{journal}{Nature Astronomy} \bibinfo{volume}{4},
  \bibinfo{pages}{32--40}.
\newblock \DOIprefix\doi{10.1038/s41550-019-0959-9}.
\bibitem[{Kruijer et~al.(2017b)Kruijer, Kleine, Borg, Brennecka, Irving,
  Bischoff and Agee}]{kruijer2017early}
\bibinfo{author}{Kruijer, T.S.}, \bibinfo{author}{Kleine, T.},
  \bibinfo{author}{Borg, L.E.}, \bibinfo{author}{Brennecka, G.A.},
  \bibinfo{author}{Irving, A.J.}, \bibinfo{author}{Bischoff, A.},
  \bibinfo{author}{Agee, C.B.}, \bibinfo{year}{2017}b.
\newblock \bibinfo{title}{{The early differentiation of Mars inferred from
  Hf--W chronometry}}.
\newblock \bibinfo{journal}{Earth and Planetary Science Letters}
  \bibinfo{volume}{474}, \bibinfo{pages}{345--354}.
\newblock \DOIprefix\doi{10.1016/j.epsl.2017.06.047}.
\bibitem[{Larimer(1967)}]{larimer1967chemical}
\bibinfo{author}{Larimer, J.W.}, \bibinfo{year}{1967}.
\newblock \bibinfo{title}{{Chemical fractionations in meteorites--I.
  Condensation of the elements}}.
\newblock \bibinfo{journal}{Geochimica et Cosmochimica Acta}
  \bibinfo{volume}{31}, \bibinfo{pages}{1215--1238}.
\newblock \DOIprefix\doi{10.1016/S0016-7037(67)80013-9}.
\bibitem[{Larimer(1979)}]{larimer1979condensation}
\bibinfo{author}{Larimer, J.W.}, \bibinfo{year}{1979}.
\newblock \bibinfo{title}{{The condensation and fractionation of refractory
  lithophile elements}}.
\newblock \bibinfo{journal}{Icarus} \bibinfo{volume}{40},
  \bibinfo{pages}{446--454}.
\newblock \DOIprefix\doi{10.1016/0019-1035(79)90038-1}.
\bibitem[{Larimer and Anders(1967)}]{larimer1967chemical2}
\bibinfo{author}{Larimer, J.W.}, \bibinfo{author}{Anders, E.},
  \bibinfo{year}{1967}.
\newblock \bibinfo{title}{{Chemical fractionations in meteorites--II. Abundance
  patterns and their interpretation}}.
\newblock \bibinfo{journal}{Geochimica et Cosmochimica Acta}
  \bibinfo{volume}{31}, \bibinfo{pages}{1239--1270}.
\newblock \DOIprefix\doi{10.1016/S0016-7037(67)80014-0}.
\bibitem[{Laurenz et~al.(2016)Laurenz, Rubie, Frost and
  Vogel}]{laurenz2016importance}
\bibinfo{author}{Laurenz, V.}, \bibinfo{author}{Rubie, D.C.},
  \bibinfo{author}{Frost, D.J.}, \bibinfo{author}{Vogel, A.K.},
  \bibinfo{year}{2016}.
\newblock \bibinfo{title}{{The importance of sulfur for the behavior of
  highly-siderophile elements during Earth's differentiation}}.
\newblock \bibinfo{journal}{Geochimica et Cosmochimica Acta}
  \bibinfo{volume}{194}, \bibinfo{pages}{123--138}.
\newblock \DOIprefix\doi{10.1016/j.gca.2016.08.012}.
\bibitem[{Levison et~al.(2015)Levison, Kretke, Walsh and
  Bottke}]{levison2015growing}
\bibinfo{author}{Levison, H.F.}, \bibinfo{author}{Kretke, K.A.},
  \bibinfo{author}{Walsh, K.J.}, \bibinfo{author}{Bottke, W.F.},
  \bibinfo{year}{2015}.
\newblock \bibinfo{title}{{Growing the terrestrial planets from the gradual
  accumulation of submeter-sized objects}}.
\newblock \bibinfo{journal}{Proceedings of the National Academy of Sciences}
  \bibinfo{volume}{112}, \bibinfo{pages}{14180--14185}.
\newblock \DOIprefix\doi{10.1073/pnas.1513364112}.
\bibitem[{Li et~al.(2016)Li, Dasgupta, Tsuno, Monteleone and
  Shimizu}]{li2016carbon}
\bibinfo{author}{Li, Y.}, \bibinfo{author}{Dasgupta, R.},
  \bibinfo{author}{Tsuno, K.}, \bibinfo{author}{Monteleone, B.},
  \bibinfo{author}{Shimizu, N.}, \bibinfo{year}{2016}.
\newblock \bibinfo{title}{{Carbon and sulfur budget of the silicate Earth
  explained by accretion of differentiated planetary embryos}}.
\newblock \bibinfo{journal}{Nature Geoscience} \bibinfo{volume}{9},
  \bibinfo{pages}{781--785}.
\newblock \DOIprefix\doi{10.1038/ngeo2801}.
\bibitem[{Lillis et~al.(2008)Lillis, Frey and Manga}]{lillis2008rapid}
\bibinfo{author}{Lillis, R.J.}, \bibinfo{author}{Frey, H.V.},
  \bibinfo{author}{Manga, M.}, \bibinfo{year}{2008}.
\newblock \bibinfo{title}{{Rapid decrease in Martian crustal magnetization in
  the Noachian era: Implications for the dynamo and climate of early Mars}}.
\newblock \bibinfo{journal}{Geophysical Research Letters} \bibinfo{volume}{35},
  \bibinfo{pages}{L14203}.
\newblock \DOIprefix\doi{10.1029/2008GL034338}.
\bibitem[{Lock et~al.(2018)Lock, Stewart, Petaev, Leinhardt, Mace, Jacobsen and
  Cuk}]{lock2018origin}
\bibinfo{author}{Lock, S.J.}, \bibinfo{author}{Stewart, S.T.},
  \bibinfo{author}{Petaev, M.I.}, \bibinfo{author}{Leinhardt, Z.},
  \bibinfo{author}{Mace, M.T.}, \bibinfo{author}{Jacobsen, S.B.},
  \bibinfo{author}{Cuk, M.}, \bibinfo{year}{2018}.
\newblock \bibinfo{title}{{The origin of the Moon within a terrestrial
  synestia}}.
\newblock \bibinfo{journal}{Journal of Geophysical Research: Planets}
  \bibinfo{volume}{123}, \bibinfo{pages}{910--951}.
\newblock \DOIprefix\doi{10.1002/2017JE005333}.
\bibitem[{Lodders(2003)}]{lodders2003solar}
\bibinfo{author}{Lodders, K.}, \bibinfo{year}{2003}.
\newblock \bibinfo{title}{{Solar system abundances and condensation
  temperatures of the elements}}.
\newblock \bibinfo{journal}{The Astrophysical Journal} \bibinfo{volume}{591},
  \bibinfo{pages}{1220--1247}.
\newblock \DOIprefix\doi{10.1086/375492}.
\bibitem[{Lodders(2020)}]{lodders2020solar}
\bibinfo{author}{Lodders, K.}, \bibinfo{year}{2020}.
\newblock \bibinfo{title}{{Solar Elemental Abundances}}, in:
  \bibinfo{editor}{Read, P.} (Ed.), \bibinfo{booktitle}{Oxford Research
  Encyclopedia of Planetary Science}. \bibinfo{publisher}{Oxford University
  Press}, \bibinfo{address}{Oxford}, pp. \bibinfo{pages}{1--68}.
\newblock \DOIprefix\doi{10.1093/acrefore/9780190647926.013.145}.
\bibitem[{Longhi et~al.(1992)Longhi, Knittle, Holloway and
  W{\"a}nke}]{longhi1992bulk}
\bibinfo{author}{Longhi, J.}, \bibinfo{author}{Knittle, E.},
  \bibinfo{author}{Holloway, J.R.}, \bibinfo{author}{W{\"a}nke, H.},
  \bibinfo{year}{1992}.
\newblock \bibinfo{title}{{The bulk composition, mineralogy and internal
  structure of Mars}}, in: \bibinfo{editor}{Kieffer, H.H.},
  \bibinfo{editor}{Jakosky, B.M.}, \bibinfo{editor}{Snyder, C.W.},
  \bibinfo{editor}{Matthews, M.S.} (Eds.), \bibinfo{booktitle}{Mars}.
  \bibinfo{publisher}{University of Arizona Press}, \bibinfo{address}{Tuscon},
  pp. \bibinfo{pages}{184--208}.
\bibitem[{Lorand and Conqu{\'e}r{\'e}(1983)}]{lorand1983contribution}
\bibinfo{author}{Lorand, J.P.}, \bibinfo{author}{Conqu{\'e}r{\'e}, F.},
  \bibinfo{year}{1983}.
\newblock \bibinfo{title}{{Contribution {\`a} l'{\'e}tude des sulfures dans les
  enclaves de lherzolite {\`a} spinelle des basaltes alcalins (Massif Central
  et Languedoc, France)}}.
\newblock \bibinfo{journal}{Bulletin de Min{\'e}ralogie} \bibinfo{volume}{106},
  \bibinfo{pages}{585--606}.
\newblock \DOIprefix\doi{10.3406/bulmi.1983.7737}.
\bibitem[{Magna et~al.(2015)Magna, Gussone and Mezger}]{magna2015calcium}
\bibinfo{author}{Magna, T.}, \bibinfo{author}{Gussone, N.},
  \bibinfo{author}{Mezger, K.}, \bibinfo{year}{2015}.
\newblock \bibinfo{title}{{The calcium isotope systematics of Mars}}.
\newblock \bibinfo{journal}{Earth and Planetary Science Letters}
  \bibinfo{volume}{430}, \bibinfo{pages}{86--94}.
\newblock \DOIprefix\doi{10.1016/j.epsl.2015.08.016}.
\bibitem[{Mahan et~al.(2018)Mahan, Moynier, Siebert, Gueguen, Agranier,
  Pringle, Bollard, Connelly and Bizzarro}]{mahan2018volatile}
\bibinfo{author}{Mahan, B.}, \bibinfo{author}{Moynier, F.},
  \bibinfo{author}{Siebert, J.}, \bibinfo{author}{Gueguen, B.},
  \bibinfo{author}{Agranier, A.}, \bibinfo{author}{Pringle, E.A.},
  \bibinfo{author}{Bollard, J.}, \bibinfo{author}{Connelly, J.N.},
  \bibinfo{author}{Bizzarro, M.}, \bibinfo{year}{2018}.
\newblock \bibinfo{title}{{Volatile element evolution of chondrules through
  time}}.
\newblock \bibinfo{journal}{Proceedings of the National Academy of Sciences}
  \bibinfo{volume}{115}, \bibinfo{pages}{8547--8552}.
\newblock \DOIprefix\doi{10.1073/pnas.1807263115}.
\bibitem[{Maltese and Mezger(2020)}]{maltese2020pb}
\bibinfo{author}{Maltese, A.}, \bibinfo{author}{Mezger, K.},
  \bibinfo{year}{2020}.
\newblock \bibinfo{title}{{The Pb isotope evolution of Bulk Silicate Earth:
  Constraints from its accretion and early differentiation history}}.
\newblock \bibinfo{journal}{Geochimica et Cosmochimica Acta}
  \bibinfo{volume}{271}, \bibinfo{pages}{179--193}.
\newblock \DOIprefix\doi{10.1016/j.gca.2019.12.021}.
\bibitem[{Marchi et~al.(2020)Marchi, Walker and
  Canup}]{marchi2020compositionally}
\bibinfo{author}{Marchi, S.}, \bibinfo{author}{Walker, R.J.},
  \bibinfo{author}{Canup, R.M.}, \bibinfo{year}{2020}.
\newblock \bibinfo{title}{{A compositionally heterogeneous martian mantle due
  to late accretion}}.
\newblock \bibinfo{journal}{Science Advances} \bibinfo{volume}{6},
  \bibinfo{pages}{eaay2338}.
\newblock \DOIprefix\doi{10.1126/sciadv.aay2338}.
\bibitem[{Marinova et~al.(2008)Marinova, Aharonson and
  Asphaug}]{marinova2008mega}
\bibinfo{author}{Marinova, M.M.}, \bibinfo{author}{Aharonson, O.},
  \bibinfo{author}{Asphaug, E.}, \bibinfo{year}{2008}.
\newblock \bibinfo{title}{{Mega-impact formation of the Mars hemispheric
  dichotomy}}.
\newblock \bibinfo{journal}{Nature} \bibinfo{volume}{453},
  \bibinfo{pages}{1216--1219}.
\newblock \DOIprefix\doi{10.1038/nature07070}.
\bibitem[{Marty(2012)}]{marty2012origins}
\bibinfo{author}{Marty, B.}, \bibinfo{year}{2012}.
\newblock \bibinfo{title}{{The origins and concentrations of water, carbon,
  nitrogen and noble gases on Earth}}.
\newblock \bibinfo{journal}{Earth and Planetary Science Letters}
  \bibinfo{volume}{313}, \bibinfo{pages}{56--66}.
\newblock \DOIprefix\doi{10.1016/j.epsl.2011.10.040}.
\bibitem[{McCanta et~al.(2009)McCanta, Elkins-Tanton and
  Rutherford}]{mccanta2009expanding}
\bibinfo{author}{McCanta, M.C.}, \bibinfo{author}{Elkins-Tanton, L.},
  \bibinfo{author}{Rutherford, M.J.}, \bibinfo{year}{2009}.
\newblock \bibinfo{title}{{Expanding the application of the Eu-oxybarometer to
  the lherzolitic shergottites and nakhlites: Implications for the oxidation
  state heterogeneity of the Martian interior}}.
\newblock \bibinfo{journal}{Meteoritics \& Planetary Science}
  \bibinfo{volume}{44}, \bibinfo{pages}{725--745}.
\newblock \DOIprefix\doi{10.1111/j.1945-5100.2009.tb00765.x}.
\bibitem[{McDonough(2014)}]{mcdonough2014compositional}
\bibinfo{author}{McDonough, W.F.}, \bibinfo{year}{2014}.
\newblock \bibinfo{title}{{Compositional model for the Earth's core}}, in:
  \bibinfo{editor}{Holland, H.D.}, \bibinfo{editor}{Turekian, K.K.} (Eds.),
  \bibinfo{booktitle}{Treatise on Geochemistry (Second Edition)}.
  \bibinfo{publisher}{Elsevier}, \bibinfo{address}{Oxford}.
  volume~\bibinfo{volume}{3}, pp. \bibinfo{pages}{559--577}.
\newblock \DOIprefix\doi{10.1016/B978-0-08-095975-7.00215-1}.
\bibitem[{McDonough(2016)}]{mcdonough2016composition}
\bibinfo{author}{McDonough, W.F.}, \bibinfo{year}{2016}.
\newblock \bibinfo{title}{{The composition of the lower mantle and core}}, in:
  \bibinfo{editor}{Terasaki, H.}, \bibinfo{editor}{Fischer, R.A.} (Eds.),
  \bibinfo{booktitle}{Deep Earth}. \bibinfo{publisher}{American Geophysical
  Union (AGU)}. chapter~\bibinfo{chapter}{12}, pp. \bibinfo{pages}{145--159}.
\newblock \DOIprefix\doi{10.1002/9781118992487.ch12}.
\bibitem[{McDonough and Sun(1995)}]{mcdonough1995composition}
\bibinfo{author}{McDonough, W.F.}, \bibinfo{author}{Sun, S.s.},
  \bibinfo{year}{1995}.
\newblock \bibinfo{title}{{The composition of the Earth}}.
\newblock \bibinfo{journal}{Chemical Geology} \bibinfo{volume}{120},
  \bibinfo{pages}{223--253}.
\newblock \DOIprefix\doi{10.1016/0009-2541(94)00140-4}.
\bibitem[{McDonough et~al.(1992)McDonough, Sun, Ringwood, Jagoutz and
  Hofmann}]{mcdonough1992potassium}
\bibinfo{author}{McDonough, W.F.}, \bibinfo{author}{Sun, S.s.},
  \bibinfo{author}{Ringwood, A.E.}, \bibinfo{author}{Jagoutz, E.},
  \bibinfo{author}{Hofmann, A.W.}, \bibinfo{year}{1992}.
\newblock \bibinfo{title}{{Potassium, rubidium, and cesium in the Earth and
  Moon and the evolution of the mantle of the Earth}}.
\newblock \bibinfo{journal}{Geochimica et Cosmochimica Acta}
  \bibinfo{volume}{56}, \bibinfo{pages}{1001--1012}.
\newblock \DOIprefix\doi{10.1016/0016-7037(92)90043-I}.
\bibitem[{{McDonough} et~al.(2020){McDonough}, {{\v{S}}r{\'a}mek} and
  {Wipperfurth}}]{mcdonough2020radiogenic}
\bibinfo{author}{{McDonough}, W.F.}, \bibinfo{author}{{{\v{S}}r{\'a}mek}, O.},
  \bibinfo{author}{{Wipperfurth}, S.A.}, \bibinfo{year}{2020}.
\newblock \bibinfo{title}{{Radiogenic power and geoneutrino luminosity of the
  Earth and other terrestrial bodies through time}}.
\newblock \bibinfo{journal}{Geochemistry, Geophysics, Geosystems}
  \bibinfo{volume}{21}, \bibinfo{pages}{e2019GC008865}.
\newblock \DOIprefix\doi{10.1029/2019GC008865}.
\bibitem[{McGovern et~al.(2002)McGovern, Solomon, Smith, Zuber, Simons,
  Wieczorek, Phillips, Neumann, Aharonson and Head}]{mcgovern2002localized}
\bibinfo{author}{McGovern, P.J.}, \bibinfo{author}{Solomon, S.C.},
  \bibinfo{author}{Smith, D.E.}, \bibinfo{author}{Zuber, M.T.},
  \bibinfo{author}{Simons, M.}, \bibinfo{author}{Wieczorek, M.A.},
  \bibinfo{author}{Phillips, R.J.}, \bibinfo{author}{Neumann, G.A.},
  \bibinfo{author}{Aharonson, O.}, \bibinfo{author}{Head, J.W.},
  \bibinfo{year}{2002}.
\newblock \bibinfo{title}{{Localized gravity/topography admittance and
  correlation spectra on Mars: Implications for regional and global
  evolution}}.
\newblock \bibinfo{journal}{Journal of Geophysical Research: Planets}
  \bibinfo{volume}{107}, \bibinfo{pages}{19--1--19--25}.
\newblock \DOIprefix\doi{10.1029/2002JE001854}.
\bibitem[{McKenzie et~al.(2002)McKenzie, Barnett and
  Yuan}]{mckenzie2002relationship}
\bibinfo{author}{McKenzie, D.}, \bibinfo{author}{Barnett, D.N.},
  \bibinfo{author}{Yuan, D.N.}, \bibinfo{year}{2002}.
\newblock \bibinfo{title}{{The relationship between Martian gravity and
  topography}}.
\newblock \bibinfo{journal}{Earth and Planetary Science Letters}
  \bibinfo{volume}{195}, \bibinfo{pages}{1--16}.
\newblock \DOIprefix\doi{10.1016/S0012-821X(01)00555-6}.
\bibitem[{McSween(1977a)}]{mcsween1977carbonaceous}
\bibinfo{author}{McSween, Jr., H.Y.}, \bibinfo{year}{1977}a.
\newblock \bibinfo{title}{{Carbonaceous chondrites of the Ornans type: A
  metamorphic sequence}}.
\newblock \bibinfo{journal}{Geochimica et Cosmochimica Acta}
  \bibinfo{volume}{41}, \bibinfo{pages}{477--491}.
\newblock \DOIprefix\doi{10.1016/0016-7037(77)90286-1}.
\bibitem[{McSween(1977b)}]{mcsween1977petrographic}
\bibinfo{author}{McSween, Jr., H.Y.}, \bibinfo{year}{1977}b.
\newblock \bibinfo{title}{{Petrographic variations among carbonaceous
  chondrites of the Vigarano type}}.
\newblock \bibinfo{journal}{Geochimica et Cosmochimica Acta}
  \bibinfo{volume}{41}, \bibinfo{pages}{1777--1790}.
\newblock \DOIprefix\doi{10.1016/0016-7037(77)90210-1}.
\bibitem[{McSween(1979)}]{mcsween1979alteration}
\bibinfo{author}{McSween, Jr., H.Y.}, \bibinfo{year}{1979}.
\newblock \bibinfo{title}{{Alteration in CM carbonaceous chondrites inferred
  from modal and chemical variations in matrix}}.
\newblock \bibinfo{journal}{Geochimica et Cosmochimica Acta}
  \bibinfo{volume}{43}, \bibinfo{pages}{1761--1770}.
\newblock \DOIprefix\doi{10.1016/0016-7037(79)90024-3}.
\bibitem[{MetBase(1994-2017)}]{metbase}
\bibinfo{author}{MetBase}, \bibinfo{year}{1994-2017}.
\newblock \bibinfo{title}{{Meteorite Information Database}}.
\newblock \URLprefix \url{http://www.metbase.org}. \bibinfo{note}{geoPlatform
  UG, Germany. Accessed: 2019-11-3}.
\bibitem[{Morbidelli et~al.(2020)Morbidelli, Libourel, Palme, Jacobson and
  Rubie}]{morbidelli2020subsolar}
\bibinfo{author}{Morbidelli, A.}, \bibinfo{author}{Libourel, G.},
  \bibinfo{author}{Palme, H.}, \bibinfo{author}{Jacobson, S.A.},
  \bibinfo{author}{Rubie, D.C.}, \bibinfo{year}{2020}.
\newblock \bibinfo{title}{{Subsolar Al/Si and Mg/Si ratios of non-carbonaceous
  chondrites reveal planetesimal formation during early condensation in the
  protoplanetary disk}}.
\newblock \bibinfo{journal}{Earth and Planetary Science Letters}
  \bibinfo{volume}{538}, \bibinfo{pages}{116220}.
\newblock \DOIprefix\doi{10.1016/j.epsl.2020.116220}.
\bibitem[{Morbidelli et~al.(2012)Morbidelli, Lunine, O'Brien, Raymond and
  Walsh}]{morbidelli2012building}
\bibinfo{author}{Morbidelli, A.}, \bibinfo{author}{Lunine, J.I.},
  \bibinfo{author}{O'Brien, D.P.}, \bibinfo{author}{Raymond, S.N.},
  \bibinfo{author}{Walsh, K.J.}, \bibinfo{year}{2012}.
\newblock \bibinfo{title}{{Building terrestrial planets}}.
\newblock \bibinfo{journal}{Annual Review of Earth and Planetary Sciences}
  \bibinfo{volume}{40}, \bibinfo{pages}{251--275}.
\newblock \DOIprefix\doi{10.1146/annurev-earth-042711-105319}.
\bibitem[{Morgan and Anders(1980)}]{morgan1980chemical}
\bibinfo{author}{Morgan, J.W.}, \bibinfo{author}{Anders, E.},
  \bibinfo{year}{1980}.
\newblock \bibinfo{title}{{Chemical composition of Earth, Venus, and Mercury}}.
\newblock \bibinfo{journal}{Proceedings of the National Academy of Sciences}
  \bibinfo{volume}{77}, \bibinfo{pages}{6973--6977}.
\newblock \DOIprefix\doi{10.1073/pnas.77.12.6973}.
\bibitem[{Morris et~al.(2012)Morris, Boley, Desch and
  Athanassiadou}]{morris2012chondrule}
\bibinfo{author}{Morris, M.A.}, \bibinfo{author}{Boley, A.C.},
  \bibinfo{author}{Desch, S.J.}, \bibinfo{author}{Athanassiadou, T.},
  \bibinfo{year}{2012}.
\newblock \bibinfo{title}{{Chondrule formation in bow shocks around eccentric
  planetary embryos}}.
\newblock \bibinfo{journal}{The Astrophysical Journal} \bibinfo{volume}{752},
  \bibinfo{pages}{27}.
\newblock \DOIprefix\doi{10.1088/0004-637X/752/1/27}.
\bibitem[{Moynier et~al.(2012)Moynier, Day, Okui, Yokoyama, Bouvier, Walker and
  Podosek}]{moynier2012planetary}
\bibinfo{author}{Moynier, F.}, \bibinfo{author}{Day, J.M.D.},
  \bibinfo{author}{Okui, W.}, \bibinfo{author}{Yokoyama, T.},
  \bibinfo{author}{Bouvier, A.}, \bibinfo{author}{Walker, R.J.},
  \bibinfo{author}{Podosek, F.A.}, \bibinfo{year}{2012}.
\newblock \bibinfo{title}{{Planetary-scale strontium isotopic heterogeneity and
  the age of volatile depletion of early Solar System materials}}.
\newblock \bibinfo{journal}{The Astrophysical Journal} \bibinfo{volume}{758},
  \bibinfo{pages}{45}.
\newblock \DOIprefix\doi{10.1088/0004-637X/758/1/45}.
\bibitem[{Mukhopadhyay(2012)}]{mukhopadhyay2012early}
\bibinfo{author}{Mukhopadhyay, S.}, \bibinfo{year}{2012}.
\newblock \bibinfo{title}{{Early differentiation and volatile accretion
  recorded in deep-mantle neon and xenon}}.
\newblock \bibinfo{journal}{Nature} \bibinfo{volume}{486},
  \bibinfo{pages}{101--104}.
\newblock \DOIprefix\doi{10.1038/nature11141}.
\bibitem[{Nagashima et~al.(2017)Nagashima, Krot and
  Komatsu}]{nagashima201726al}
\bibinfo{author}{Nagashima, K.}, \bibinfo{author}{Krot, A.N.},
  \bibinfo{author}{Komatsu, M.}, \bibinfo{year}{2017}.
\newblock \bibinfo{title}{{\ce{^{26}Al}-\ce{^{26}Mg} systematics in chondrules
  from Kaba and Yamato 980145 CV3 carbonaceous chondrites}}.
\newblock \bibinfo{journal}{Geochimica et Cosmochimica Acta}
  \bibinfo{volume}{201}, \bibinfo{pages}{303--319}.
\newblock \DOIprefix\doi{10.1016/j.gca.2016.10.030}.
\bibitem[{Nebel et~al.(2011)Nebel, Mezger and van
  Westrenen}]{nebel2011rubidium}
\bibinfo{author}{Nebel, O.}, \bibinfo{author}{Mezger, K.}, \bibinfo{author}{van
  Westrenen, W.}, \bibinfo{year}{2011}.
\newblock \bibinfo{title}{{Rubidium isotopes in primitive chondrites:
  Constraints on Earth's volatile element depletion and lead isotope
  evolution}}.
\newblock \bibinfo{journal}{Earth and Planetary Science Letters}
  \bibinfo{volume}{305}, \bibinfo{pages}{309--316}.
\newblock \DOIprefix\doi{10.1016/j.gca.2010.04.061}.
\bibitem[{Norris and Wood(2017)}]{norris2017earth}
\bibinfo{author}{Norris, C.A.}, \bibinfo{author}{Wood, B.J.},
  \bibinfo{year}{2017}.
\newblock \bibinfo{title}{{Earth's volatile contents established by melting and
  vaporization}}.
\newblock \bibinfo{journal}{Nature} \bibinfo{volume}{549},
  \bibinfo{pages}{507--510}.
\newblock \DOIprefix\doi{10.1038/nature23645}.
\bibitem[{Ohtani and Ringwood(1984)}]{ohtani1984composition}
\bibinfo{author}{Ohtani, E.}, \bibinfo{author}{Ringwood, A.E.},
  \bibinfo{year}{1984}.
\newblock \bibinfo{title}{{Composition of the core, I. Solubility of oxygen in
  molten iron at high temperatures}}.
\newblock \bibinfo{journal}{Earth and Planetary Science Letters}
  \bibinfo{volume}{71}, \bibinfo{pages}{85--93}.
\newblock \DOIprefix\doi{10.1016/0012-821X(84)90054-2}.
\bibitem[{Olsen et~al.(2016)Olsen, Wielandt, Schiller, Van~Kooten and
  Bizzarro}]{olsen2016magnesium}
\bibinfo{author}{Olsen, M.B.}, \bibinfo{author}{Wielandt, D.},
  \bibinfo{author}{Schiller, M.}, \bibinfo{author}{Van~Kooten, E.M.M.E.},
  \bibinfo{author}{Bizzarro, M.}, \bibinfo{year}{2016}.
\newblock \bibinfo{title}{{Magnesium and \ce{^{54}Cr} isotope compositions of
  carbonaceous chondrite chondrules--Insights into early disk processes}}.
\newblock \bibinfo{journal}{Geochimica et Cosmochimica Acta}
  \bibinfo{volume}{191}, \bibinfo{pages}{118--138}.
\newblock \DOIprefix\doi{10.1016/j.gca.2016.07.011}.
\bibitem[{O'Neill(1991a)}]{oneil1991origin}
\bibinfo{author}{O'Neill, H.S.C.}, \bibinfo{year}{1991}a.
\newblock \bibinfo{title}{{The origin of the Moon and the early history of the
  Earth--A chemical model. Part 1: The Moon}}.
\newblock \bibinfo{journal}{Geochimica et Cosmochimica Acta}
  \bibinfo{volume}{55}, \bibinfo{pages}{1135--1157}.
\newblock \DOIprefix\doi{10.1016/0016-7037(91)90168-5}.
\bibitem[{O'Neill(1991b)}]{oneil1991origin_Earth}
\bibinfo{author}{O'Neill, H.S.C.}, \bibinfo{year}{1991}b.
\newblock \bibinfo{title}{{The origin of the Moon and the early history of the
  Earth--A chemical model. Part 2: The Earth}}.
\newblock \bibinfo{journal}{Geochimica et Cosmochimica Acta}
  \bibinfo{volume}{55}, \bibinfo{pages}{1159--1172}.
\newblock \DOIprefix\doi{10.1016/0016-7037(91)90169-6}.
\bibitem[{O'Neill et~al.(1998)O'Neill, Canil and Rubie}]{oneill1998oxide}
\bibinfo{author}{O'Neill, H.S.C.}, \bibinfo{author}{Canil, D.},
  \bibinfo{author}{Rubie, D.C.}, \bibinfo{year}{1998}.
\newblock \bibinfo{title}{{Oxide-metal equilibria to 2500 $ ^{\circ} $C and 25
  GPa: Implications for core formation and the light component in the Earth's
  core}}.
\newblock \bibinfo{journal}{Journal of Geophysical Research: Solid Earth}
  \bibinfo{volume}{103}, \bibinfo{pages}{12239--12260}.
\newblock \DOIprefix\doi{10.1029/97JB02601}.
\bibitem[{O'Neill and Palme(2008)}]{oneill2008collisional}
\bibinfo{author}{O'Neill, H.S.C.}, \bibinfo{author}{Palme, H.},
  \bibinfo{year}{2008}.
\newblock \bibinfo{title}{{Collisional erosion and the non-chondritic
  composition of the terrestrial planets}}.
\newblock \bibinfo{journal}{Philosophical Transactions of the Royal Society of
  London A: Mathematical, Physical and Engineering Sciences}
  \bibinfo{volume}{366}, \bibinfo{pages}{4205--4238}.
\newblock \DOIprefix\doi{10.1098/rsta.2008.0111}.
\bibitem[{O'Rourke and Shim(2019)}]{orourke2019hydrogenation}
\bibinfo{author}{O'Rourke, J.G.}, \bibinfo{author}{Shim, S.H.},
  \bibinfo{year}{2019}.
\newblock \bibinfo{title}{{Hydrogenation of the Martian core by hydrated mantle
  minerals with implications for the early dynamo}}.
\newblock \bibinfo{journal}{Journal of Geophysical Research: Planets}
  \bibinfo{volume}{124}, \bibinfo{pages}{3422--3441}.
\newblock \DOIprefix\doi{10.1029/2019JE005950}.
\bibitem[{Palme and O'Neill(2014)}]{palme2014cosmochemical}
\bibinfo{author}{Palme, H.}, \bibinfo{author}{O'Neill, H.S.C.},
  \bibinfo{year}{2014}.
\newblock \bibinfo{title}{Cosmochemical estimates of mantle composition}, in:
  \bibinfo{editor}{Holland, H.D.}, \bibinfo{editor}{Turekian, K.K.} (Eds.),
  \bibinfo{booktitle}{Treatise on Geochemistry (Second Edition)}.
  \bibinfo{publisher}{Elsevier}, \bibinfo{address}{Oxford}.
  volume~\bibinfo{volume}{3}, pp. \bibinfo{pages}{1--39}.
\newblock \DOIprefix\doi{10.1016/B978-0-08-095975-7.00201-1}.
\bibitem[{Paniello et~al.(2012a)Paniello, Day and Moynier}]{paniello2012zinc}
\bibinfo{author}{Paniello, R.C.}, \bibinfo{author}{Day, J.M.D.},
  \bibinfo{author}{Moynier, F.}, \bibinfo{year}{2012}a.
\newblock \bibinfo{title}{{Zinc isotopic evidence for the origin of the Moon}}.
\newblock \bibinfo{journal}{Nature} \bibinfo{volume}{490},
  \bibinfo{pages}{376--379}.
\newblock \DOIprefix\doi{10.1038/nature11507}.
\bibitem[{Paniello et~al.(2012b)Paniello, Moynier, Beck, Barrat, Podosek and
  Pichat}]{paniello2012zinc_HED}
\bibinfo{author}{Paniello, R.C.}, \bibinfo{author}{Moynier, F.},
  \bibinfo{author}{Beck, P.}, \bibinfo{author}{Barrat, J.A.},
  \bibinfo{author}{Podosek, F.A.}, \bibinfo{author}{Pichat, S.},
  \bibinfo{year}{2012}b.
\newblock \bibinfo{title}{{Zinc isotopes in HEDs: Clues to the formation of
  4-Vesta, and the unique composition of Pecora Escarpment 82502}}.
\newblock \bibinfo{journal}{Geochimica et Cosmochimica Acta}
  \bibinfo{volume}{86}, \bibinfo{pages}{76--87}.
\newblock \DOIprefix\doi{10.1016/j.gca.2012.01.045}.
\bibitem[{Pape et~al.(2019)Pape, Mezger, Bouvier and
  Baumgartner}]{pape2019time}
\bibinfo{author}{Pape, J.}, \bibinfo{author}{Mezger, K.},
  \bibinfo{author}{Bouvier, A.S.}, \bibinfo{author}{Baumgartner, L.P.},
  \bibinfo{year}{2019}.
\newblock \bibinfo{title}{{Time and duration of chondrule formation:
  Constraints from \ce{^{26}Al}-\ce{^{26}Mg} ages of individual chondrules}}.
\newblock \bibinfo{journal}{Geochimica et Cosmochimica Acta}
  \bibinfo{volume}{244}, \bibinfo{pages}{416--436}.
\newblock \DOIprefix\doi{10.1016/j.gca.2018.10.017}.
\bibitem[{Parro et~al.(2017)Parro, Jim{\'e}nez-D{\'\i}az, Mansilla and
  Ruiz}]{parro2017present}
\bibinfo{author}{Parro, L.M.}, \bibinfo{author}{Jim{\'e}nez-D{\'\i}az, A.},
  \bibinfo{author}{Mansilla, F.}, \bibinfo{author}{Ruiz, J.},
  \bibinfo{year}{2017}.
\newblock \bibinfo{title}{{Present-day heat flow model of Mars}}.
\newblock \bibinfo{journal}{Scientific Reports} \bibinfo{volume}{7},
  \bibinfo{pages}{45629}.
\newblock \DOIprefix\doi{10.1038/srep45629}.
\bibitem[{Peplowski et~al.(2011)Peplowski, Evans, Hauck, McCoy, Boynton,
  Gillis-Davis, Ebel, Goldsten, Hamara, Lawrence, McNutt, Nittler, Solomon,
  Rhodes, Sprague, Starr and Stockstill-Cahill}]{peplowski2011radioactive}
\bibinfo{author}{Peplowski, P.N.}, \bibinfo{author}{Evans, L.G.},
  \bibinfo{author}{Hauck, S.A.}, \bibinfo{author}{McCoy, T.J.},
  \bibinfo{author}{Boynton, W.V.}, \bibinfo{author}{Gillis-Davis, J.J.},
  \bibinfo{author}{Ebel, D.S.}, \bibinfo{author}{Goldsten, J.O.},
  \bibinfo{author}{Hamara, D.K.}, \bibinfo{author}{Lawrence, D.J.},
  \bibinfo{author}{McNutt, R.L.}, \bibinfo{author}{Nittler, L.R.},
  \bibinfo{author}{Solomon, S.C.}, \bibinfo{author}{Rhodes, E.A.},
  \bibinfo{author}{Sprague, A.L.}, \bibinfo{author}{Starr, R.D.},
  \bibinfo{author}{Stockstill-Cahill, K.R.}, \bibinfo{year}{2011}.
\newblock \bibinfo{title}{{Radioactive elements on Mercury's surface from
  MESSENGER: Implications for the planet's formation and evolution}}.
\newblock \bibinfo{journal}{Science} \bibinfo{volume}{333},
  \bibinfo{pages}{1850--1852}.
\newblock \DOIprefix\doi{10.1126/science.1211576}.
\bibitem[{Pilipp et~al.(1998)Pilipp, Hartquist, Morfill and
  Levy}]{pilipp1998chondrule}
\bibinfo{author}{Pilipp, W.}, \bibinfo{author}{Hartquist, T.W.},
  \bibinfo{author}{Morfill, G.E.}, \bibinfo{author}{Levy, E.},
  \bibinfo{year}{1998}.
\newblock \bibinfo{title}{{Chondrule formation by lightning in the Protosolar
  Nebula?}}
\newblock \bibinfo{journal}{Astronomy and Astrophysics} \bibinfo{volume}{331},
  \bibinfo{pages}{121--146}.
\bibitem[{Poitrasson et~al.(2004)Poitrasson, Halliday, Lee, Levasseur and
  Teutsch}]{poitrasson2004iron}
\bibinfo{author}{Poitrasson, F.}, \bibinfo{author}{Halliday, A.N.},
  \bibinfo{author}{Lee, D.C.}, \bibinfo{author}{Levasseur, S.},
  \bibinfo{author}{Teutsch, N.}, \bibinfo{year}{2004}.
\newblock \bibinfo{title}{{Iron isotope differences between Earth, Moon, Mars
  and Vesta as possible records of contrasted accretion mechanisms}}.
\newblock \bibinfo{journal}{Earth and Planetary Science Letters}
  \bibinfo{volume}{223}, \bibinfo{pages}{253--266}.
\newblock \DOIprefix\doi{10.1016/j.epsl.2004.04.032}.
\bibitem[{Poole et~al.(2017)Poole, Rehk{\"a}mper, Coles, Goldberg and
  Smith}]{poole2017nucleosynthetic}
\bibinfo{author}{Poole, G.M.}, \bibinfo{author}{Rehk{\"a}mper, M.},
  \bibinfo{author}{Coles, B.J.}, \bibinfo{author}{Goldberg, T.},
  \bibinfo{author}{Smith, C.L.}, \bibinfo{year}{2017}.
\newblock \bibinfo{title}{{Nucleosynthetic molybdenum isotope anomalies in iron
  meteorites--new evidence for thermal processing of solar nebula material}}.
\newblock \bibinfo{journal}{Earth and Planetary Science Letters}
  \bibinfo{volume}{473}, \bibinfo{pages}{215--226}.
\newblock \DOIprefix\doi{10.1016/j.epsl.2017.05.001}.
\bibitem[{Prettyman et~al.(2015)Prettyman, Yamashita, Reedy, McSween,
  Mittlefehldt, Hendricks and Toplis}]{prettyman2015concentrations}
\bibinfo{author}{Prettyman, T.H.}, \bibinfo{author}{Yamashita, N.},
  \bibinfo{author}{Reedy, R.C.}, \bibinfo{author}{McSween, Jr., H.Y.},
  \bibinfo{author}{Mittlefehldt, D.W.}, \bibinfo{author}{Hendricks, J.S.},
  \bibinfo{author}{Toplis, M.J.}, \bibinfo{year}{2015}.
\newblock \bibinfo{title}{{Concentrations of potassium and thorium within
  Vesta's regolith}}.
\newblock \bibinfo{journal}{Icarus} \bibinfo{volume}{259},
  \bibinfo{pages}{39--52}.
\newblock \DOIprefix\doi{10.1016/j.icarus.2015.05.035}.
\bibitem[{Pringle and Moynier(2017)}]{pringle2017rubidium}
\bibinfo{author}{Pringle, E.A.}, \bibinfo{author}{Moynier, F.},
  \bibinfo{year}{2017}.
\newblock \bibinfo{title}{{Rubidium isotopic composition of the Earth,
  meteorites, and the Moon: Evidence for the origin of volatile loss during
  planetary accretion}}.
\newblock \bibinfo{journal}{Earth and Planetary Science Letters}
  \bibinfo{volume}{473}, \bibinfo{pages}{62--70}.
\newblock \DOIprefix\doi{10.1016/j.epsl.2017.05.033}.
\bibitem[{Pringle et~al.(2017)Pringle, Moynier, Beck, Paniello and
  Hezel}]{pringle2017origin}
\bibinfo{author}{Pringle, E.A.}, \bibinfo{author}{Moynier, F.},
  \bibinfo{author}{Beck, P.}, \bibinfo{author}{Paniello, R.},
  \bibinfo{author}{Hezel, D.C.}, \bibinfo{year}{2017}.
\newblock \bibinfo{title}{{The origin of volatile element depletion in early
  solar system material: Clues from Zn isotopes in chondrules}}.
\newblock \bibinfo{journal}{Earth and Planetary Science Letters}
  \bibinfo{volume}{468}, \bibinfo{pages}{62--71}.
\newblock \DOIprefix\doi{10.1016/j.epsl.2017.04.002}.
\bibitem[{Pringle et~al.(2014)Pringle, Moynier, Savage, Badro and
  Barrat}]{pringle2014silicon}
\bibinfo{author}{Pringle, E.A.}, \bibinfo{author}{Moynier, F.},
  \bibinfo{author}{Savage, P.S.}, \bibinfo{author}{Badro, J.},
  \bibinfo{author}{Barrat, J.A.}, \bibinfo{year}{2014}.
\newblock \bibinfo{title}{{Silicon isotopes in angrites and volatile loss in
  planetesimals}}.
\newblock \bibinfo{journal}{Proceedings of the National Academy of Sciences}
  \bibinfo{volume}{111}, \bibinfo{pages}{17029--17032}.
\newblock \DOIprefix\doi{10.1073/pnas.1418889111}.
\bibitem[{Putirka(2016)}]{putirka2016rates}
\bibinfo{author}{Putirka, K.}, \bibinfo{year}{2016}.
\newblock \bibinfo{title}{{Rates and styles of planetary cooling on Earth,
  Moon, Mars, and Vesta, using new models for oxygen fugacity, ferric-ferrous
  ratios, olivine-liquid Fe-Mg exchange, and mantle potential temperature}}.
\newblock \bibinfo{journal}{American Mineralogist} \bibinfo{volume}{101},
  \bibinfo{pages}{819--840}.
\newblock \DOIprefix\doi{10.2138/am-2016-5402}.
\bibitem[{Raymond et~al.(2006)Raymond, Quinn and Lunine}]{raymond2006high}
\bibinfo{author}{Raymond, S.N.}, \bibinfo{author}{Quinn, T.},
  \bibinfo{author}{Lunine, J.I.}, \bibinfo{year}{2006}.
\newblock \bibinfo{title}{{High-resolution simulations of the final assembly of
  Earth-like planets I. Terrestrial accretion and dynamics}}.
\newblock \bibinfo{journal}{Icarus} \bibinfo{volume}{183},
  \bibinfo{pages}{265--282}.
\newblock \DOIprefix\doi{10.1016/j.icarus.2006.03.011}.
\bibitem[{Richter et~al.(2002)Richter, Davis, Ebel and
  Hashimoto}]{richter2002elemental}
\bibinfo{author}{Richter, F.M.}, \bibinfo{author}{Davis, A.M.},
  \bibinfo{author}{Ebel, D.S.}, \bibinfo{author}{Hashimoto, A.},
  \bibinfo{year}{2002}.
\newblock \bibinfo{title}{{Elemental and isotopic fractionation of Type B
  calcium-, aluminum-rich inclusions: Experiments, theoretical considerations,
  and constraints on their thermal evolution}}.
\newblock \bibinfo{journal}{Geochimica et Cosmochimica Acta}
  \bibinfo{volume}{66}, \bibinfo{pages}{521--540}.
\newblock \DOIprefix\doi{10.1016/S0016-7037(01)00782-7}.
\bibitem[{Righter(2019)}]{righter2019volatile}
\bibinfo{author}{Righter, K.}, \bibinfo{year}{2019}.
\newblock \bibinfo{title}{{Volatile element depletion of the Moon--The roles of
  precursors, post-impact disk dynamics, and core formation}}.
\newblock \bibinfo{journal}{Science Advances} \bibinfo{volume}{5},
  \bibinfo{pages}{eaau7658}.
\newblock \DOIprefix\doi{10.1126/sciadv.aau7658}.
\bibitem[{Righter and Chabot(2011)}]{righter2011moderately}
\bibinfo{author}{Righter, K.}, \bibinfo{author}{Chabot, N.L.},
  \bibinfo{year}{2011}.
\newblock \bibinfo{title}{{Moderately and slightly siderophile element
  constraints on the depth and extent of melting in early Mars}}.
\newblock \bibinfo{journal}{Meteoritics \& Planetary Science}
  \bibinfo{volume}{46}, \bibinfo{pages}{157--176}.
\newblock \DOIprefix\doi{10.1111/j.1945-5100.2010.01140.x}.
\bibitem[{Righter et~al.(2015)Righter, Danielson, Pando, Williams, Humayun,
  Hervig and Sharp}]{righter2015highly}
\bibinfo{author}{Righter, K.}, \bibinfo{author}{Danielson, L.R.},
  \bibinfo{author}{Pando, K.M.}, \bibinfo{author}{Williams, J.},
  \bibinfo{author}{Humayun, M.}, \bibinfo{author}{Hervig, R.L.},
  \bibinfo{author}{Sharp, T.G.}, \bibinfo{year}{2015}.
\newblock \bibinfo{title}{{Highly siderophile element (HSE) abundances in the
  mantle of Mars are due to core formation at high pressure and temperature}}.
\newblock \bibinfo{journal}{Meteoritics \& Planetary Science}
  \bibinfo{volume}{50}, \bibinfo{pages}{604--631}.
\newblock \DOIprefix\doi{10.1111/maps.12393}.
\bibitem[{Righter et~al.(2008)Righter, Humayun and
  Danielson}]{righter2008partitioning}
\bibinfo{author}{Righter, K.}, \bibinfo{author}{Humayun, M.},
  \bibinfo{author}{Danielson, L.}, \bibinfo{year}{2008}.
\newblock \bibinfo{title}{{Partitioning of palladium at high pressures and
  temperatures during core formation}}.
\newblock \bibinfo{journal}{Nature Geoscience} \bibinfo{volume}{1},
  \bibinfo{pages}{321--323}.
\newblock \DOIprefix\doi{10.1038/ngeo180}.
\bibitem[{Righter et~al.(2018)Righter, Pando, Humayun, Waeselmann, Yang,
  Boujibar and Danielson}]{righter2018effect}
\bibinfo{author}{Righter, K.}, \bibinfo{author}{Pando, K.},
  \bibinfo{author}{Humayun, M.}, \bibinfo{author}{Waeselmann, N.},
  \bibinfo{author}{Yang, S.}, \bibinfo{author}{Boujibar, A.},
  \bibinfo{author}{Danielson, L.R.}, \bibinfo{year}{2018}.
\newblock \bibinfo{title}{{Effect of silicon on activity coefficients of
  siderophile elements (Au, Pd, Pt, P, Ga, Cu, Zn, and Pb) in liquid Fe: Roles
  of core formation, late sulfide matte, and late veneer in shaping terrestrial
  mantle geochemistry}}.
\newblock \bibinfo{journal}{Geochimica et Cosmochimica Acta}
  \bibinfo{volume}{232}, \bibinfo{pages}{101--123}.
\newblock \DOIprefix\doi{10.1016/j.gca.2018.04.011}.
\bibitem[{Righter et~al.(2019)Righter, Pando, Ross, Righter and
  Lapen}]{righter2019effect}
\bibinfo{author}{Righter, K.}, \bibinfo{author}{Pando, K.},
  \bibinfo{author}{Ross, D.K.}, \bibinfo{author}{Righter, M.},
  \bibinfo{author}{Lapen, T.J.}, \bibinfo{year}{2019}.
\newblock \bibinfo{title}{{Effect of silicon on activity coefficients of Bi,
  Cd, Sn, and Ag in liquid Fe-Si, and implications for differentiation and core
  formation}}.
\newblock \bibinfo{journal}{Meteoritics \& Planetary Science}
  \bibinfo{volume}{54}, \bibinfo{pages}{1379--1394}.
\newblock \DOIprefix\doi{10.1111/maps.13285}.
\bibitem[{Righter et~al.(2016)Righter, Sutton, Danielson, Pando and
  Newville}]{righter2016redox}
\bibinfo{author}{Righter, K.}, \bibinfo{author}{Sutton, S.R.},
  \bibinfo{author}{Danielson, L.}, \bibinfo{author}{Pando, K.},
  \bibinfo{author}{Newville, M.}, \bibinfo{year}{2016}.
\newblock \bibinfo{title}{{Redox variations in the inner solar system with new
  constraints from vanadium XANES in spinels}}.
\newblock \bibinfo{journal}{American Mineralogist} \bibinfo{volume}{101},
  \bibinfo{pages}{1928--1942}.
\newblock \DOIprefix\doi{10.2138/am-2016-5638}.
\bibitem[{Ringwood(1966)}]{ringwood1966chemical}
\bibinfo{author}{Ringwood, A.E.}, \bibinfo{year}{1966}.
\newblock \bibinfo{title}{{Chemical evolution of the terrestrial planets}}.
\newblock \bibinfo{journal}{Geochimica et Cosmochimica Acta}
  \bibinfo{volume}{30}, \bibinfo{pages}{41--104}.
\newblock \DOIprefix\doi{10.1016/0016-7037(66)90090-1}.
\bibitem[{Ringwood(1975)}]{ringwood1975composition}
\bibinfo{author}{Ringwood, A.E.}, \bibinfo{year}{1975}.
\newblock \bibinfo{title}{{Composition and Petrology of the Earth's Mantle}}.
\newblock \bibinfo{publisher}{MacGraw-Hill}, \bibinfo{address}{New York}.
\bibitem[{Rivoldini et~al.(2011)Rivoldini, Van~Hoolst, Verhoeven, Mocquet and
  Dehant}]{rivoldini2011geodesy}
\bibinfo{author}{Rivoldini, A.}, \bibinfo{author}{Van~Hoolst, T.},
  \bibinfo{author}{Verhoeven, O.}, \bibinfo{author}{Mocquet, A.},
  \bibinfo{author}{Dehant, V.}, \bibinfo{year}{2011}.
\newblock \bibinfo{title}{{Geodesy constraints on the interior structure and
  composition of Mars}}.
\newblock \bibinfo{journal}{Icarus} \bibinfo{volume}{213},
  \bibinfo{pages}{451--472}.
\newblock \DOIprefix\doi{10.1016/j.icarus.2011.03.024}.
\bibitem[{Rose and Brenan(2001)}]{rose2001wetting}
\bibinfo{author}{Rose, L.A.}, \bibinfo{author}{Brenan, J.M.},
  \bibinfo{year}{2001}.
\newblock \bibinfo{title}{{Wetting properties of Fe-Ni-Co-Cu-Os melts against
  olivine: Implications for sulfide melt mobility}}.
\newblock \bibinfo{journal}{Economic Geology} \bibinfo{volume}{96},
  \bibinfo{pages}{145--157}.
\newblock \DOIprefix\doi{10.2113/gsecongeo.96.1.145}.
\bibitem[{Rubie et~al.(2011)Rubie, Frost, Mann, Asahara, Nimmo, Tsuno, Kegler,
  Holzheid and Palme}]{rubie2011heterogeneous}
\bibinfo{author}{Rubie, D.C.}, \bibinfo{author}{Frost, D.J.},
  \bibinfo{author}{Mann, U.}, \bibinfo{author}{Asahara, Y.},
  \bibinfo{author}{Nimmo, F.}, \bibinfo{author}{Tsuno, K.},
  \bibinfo{author}{Kegler, P.}, \bibinfo{author}{Holzheid, A.},
  \bibinfo{author}{Palme, H.}, \bibinfo{year}{2011}.
\newblock \bibinfo{title}{{Heterogeneous accretion, composition and
  core--mantle differentiation of the Earth}}.
\newblock \bibinfo{journal}{Earth and Planetary Science Letters}
  \bibinfo{volume}{301}, \bibinfo{pages}{31--42}.
\newblock \DOIprefix\doi{10.1016/j.epsl.2010.11.030}.
\bibitem[{Rubie et~al.(2015)Rubie, Jacobson, Morbidelli, O'Brien, Young,
  de~Vries, Nimmo, Palme and Frost}]{rubie2015accretion}
\bibinfo{author}{Rubie, D.C.}, \bibinfo{author}{Jacobson, S.A.},
  \bibinfo{author}{Morbidelli, A.}, \bibinfo{author}{O'Brien, D.P.},
  \bibinfo{author}{Young, E.D.}, \bibinfo{author}{de~Vries, J.},
  \bibinfo{author}{Nimmo, F.}, \bibinfo{author}{Palme, H.},
  \bibinfo{author}{Frost, D.J.}, \bibinfo{year}{2015}.
\newblock \bibinfo{title}{{Accretion and differentiation of the terrestrial
  planets with implications for the compositions of early-formed Solar System
  bodies and accretion of water}}.
\newblock \bibinfo{journal}{Icarus} \bibinfo{volume}{248},
  \bibinfo{pages}{89--108}.
\newblock \DOIprefix\doi{10.1016/j.icarus.2014.10.015}.
\bibitem[{Rubie et~al.(2016)Rubie, Laurenz, Jacobson, Morbidelli, Palme, Vogel
  and Frost}]{rubie2016highly}
\bibinfo{author}{Rubie, D.C.}, \bibinfo{author}{Laurenz, V.},
  \bibinfo{author}{Jacobson, S.A.}, \bibinfo{author}{Morbidelli, A.},
  \bibinfo{author}{Palme, H.}, \bibinfo{author}{Vogel, A.K.},
  \bibinfo{author}{Frost, D.J.}, \bibinfo{year}{2016}.
\newblock \bibinfo{title}{{Highly siderophile elements were stripped from
  Earth's mantle by iron sulfide segregation}}.
\newblock \bibinfo{journal}{Science} \bibinfo{volume}{353},
  \bibinfo{pages}{1141--1144}.
\newblock \DOIprefix\doi{10.1126/science.aaf6919}.
\bibitem[{Rudnick and Gao(2014)}]{rudnick2014composition}
\bibinfo{author}{Rudnick, R.L.}, \bibinfo{author}{Gao, S.},
  \bibinfo{year}{2014}.
\newblock \bibinfo{title}{Composition of the continental crust}, in:
  \bibinfo{editor}{Holland, H.D.}, \bibinfo{editor}{Turekian, K.K.} (Eds.),
  \bibinfo{booktitle}{Treatise on Geochemistry (Second Edition)}.
  \bibinfo{publisher}{Elsevier}, \bibinfo{address}{Oxford}, pp.
  \bibinfo{pages}{1--51}.
\newblock \DOIprefix\doi{10.1016/B978-0-08-095975-7.00301-6}.
\bibitem[{Russell et~al.(2018)Russell, Connolly and
  Krot}]{russell2018chondrules}
\bibinfo{editor}{Russell, S.S.}, \bibinfo{editor}{Connolly, Jr., H.C.},
  \bibinfo{editor}{Krot, A.N.} (Eds.), \bibinfo{year}{2018}.
\newblock \bibinfo{title}{{Chondrules: Records of Protoplanetary Disk
  Processes}}. volume~\bibinfo{volume}{22}.
\newblock \bibinfo{publisher}{Cambridge University Press},
  \bibinfo{address}{Cambridge}.
\newblock \DOIprefix\doi{10.1017/9781108284073}.
\bibitem[{Sanders and Scott(2018)}]{sanders2018making_splash}
\bibinfo{author}{Sanders, I.S.}, \bibinfo{author}{Scott, E.R.D.},
  \bibinfo{year}{2018}.
\newblock \bibinfo{title}{{Making Chondrules by Splashing Molten
  Planetesimals}}, in: \bibinfo{editor}{Russell, S.S.},
  \bibinfo{editor}{Connolly~Jr., H.C.}, \bibinfo{editor}{Krot, A.N.} (Eds.),
  \bibinfo{booktitle}{Chondrules: Records of Protoplanetary Disk Processes}.
  \bibinfo{publisher}{Cambridge University Press}. Cambridge Planetary Science.
  chapter~\bibinfo{chapter}{14}, p. \bibinfo{pages}{361–374}.
\newblock \DOIprefix\doi{10.1017/9781108284073.014}.
\bibitem[{Sargent et~al.(2008)Sargent, Forrest, Tayrien, McClure, Li, Basu,
  Manoj, Watson, Bohac, Furlan, Kim, Green and Sloan}]{sargent2008silica}
\bibinfo{author}{Sargent, B.A.}, \bibinfo{author}{Forrest, W.J.},
  \bibinfo{author}{Tayrien, C.}, \bibinfo{author}{McClure, M.K.},
  \bibinfo{author}{Li, A.}, \bibinfo{author}{Basu, A.R.},
  \bibinfo{author}{Manoj, P.}, \bibinfo{author}{Watson, D.M.},
  \bibinfo{author}{Bohac, C.J.}, \bibinfo{author}{Furlan, E.},
  \bibinfo{author}{Kim, K.H.}, \bibinfo{author}{Green, J.D.},
  \bibinfo{author}{Sloan, G.C.}, \bibinfo{year}{2008}.
\newblock \bibinfo{title}{{Silica in protoplanetary disks}}.
\newblock \bibinfo{journal}{The Astrophysical Journal} \bibinfo{volume}{690},
  \bibinfo{pages}{1193}.
\newblock \DOIprefix\doi{10.1088/0004-637X/690/2/1193}.
\bibitem[{Schiller et~al.(2018)Schiller, Bizzarro and
  Fernandes}]{schiller2018isotopic}
\bibinfo{author}{Schiller, M.}, \bibinfo{author}{Bizzarro, M.},
  \bibinfo{author}{Fernandes, V.A.}, \bibinfo{year}{2018}.
\newblock \bibinfo{title}{{Isotopic evolution of the protoplanetary disk and
  the building blocks of Earth and the Moon}}.
\newblock \bibinfo{journal}{Nature} \bibinfo{volume}{555},
  \bibinfo{pages}{507--510}.
\newblock \DOIprefix\doi{10.1038/nature25990}.
\bibitem[{Sch{\"o}nb{\"a}chler et~al.(2010)Sch{\"o}nb{\"a}chler, Carlson,
  Horan, Mock and Hauri}]{schonbachler2010heterogeneous}
\bibinfo{author}{Sch{\"o}nb{\"a}chler, M.}, \bibinfo{author}{Carlson, R.W.},
  \bibinfo{author}{Horan, M.F.}, \bibinfo{author}{Mock, T.D.},
  \bibinfo{author}{Hauri, E.H.}, \bibinfo{year}{2010}.
\newblock \bibinfo{title}{{Heterogeneous accretion and the moderately volatile
  element budget of Earth}}.
\newblock \bibinfo{journal}{Science} \bibinfo{volume}{328},
  \bibinfo{pages}{884--887}.
\newblock \DOIprefix\doi{10.1126/science.1186239}.
\bibitem[{Schrader et~al.(2017)Schrader, Nagashima, Krot, Ogliore, Yin, Amelin,
  Stirling and Kaltenbach}]{schrader2017distribution}
\bibinfo{author}{Schrader, D.L.}, \bibinfo{author}{Nagashima, K.},
  \bibinfo{author}{Krot, A.N.}, \bibinfo{author}{Ogliore, R.C.},
  \bibinfo{author}{Yin, Q.Z.}, \bibinfo{author}{Amelin, Y.},
  \bibinfo{author}{Stirling, C.H.}, \bibinfo{author}{Kaltenbach, A.},
  \bibinfo{year}{2017}.
\newblock \bibinfo{title}{{Distribution of \ce{^{26}Al} in the CR chondrite
  chondrule-forming region of the protoplanetary disk}}.
\newblock \bibinfo{journal}{Geochimica et Cosmochimica Acta}
  \bibinfo{volume}{201}, \bibinfo{pages}{275--302}.
\newblock \DOIprefix\doi{10.1016/j.gca.2016.06.023}.
\bibitem[{Scott and Krot(2014)}]{scott2014chondrites}
\bibinfo{author}{Scott, E.R.D.}, \bibinfo{author}{Krot, A.N.},
  \bibinfo{year}{2014}.
\newblock \bibinfo{title}{{Chondrites and their Components}}, in:
  \bibinfo{editor}{Holland, H.D.}, \bibinfo{editor}{Turekian, K.K.} (Eds.),
  \bibinfo{booktitle}{Treatise on Geochemistry (Second Edition)}.
  \bibinfo{publisher}{Elsevier}, \bibinfo{address}{Oxford}.
  volume~\bibinfo{volume}{1}, pp. \bibinfo{pages}{65--137}.
\newblock \DOIprefix\doi{10.1016/B978-0-08-095975-7.00104-2}.
\bibitem[{Shearer et~al.(2011)Shearer, Burger, Sutton, Papike and
  McCubbin}]{shearer2011ree}
\bibinfo{author}{Shearer, C.K.}, \bibinfo{author}{Burger, P.V.},
  \bibinfo{author}{Sutton, S.R.}, \bibinfo{author}{Papike, J.J.},
  \bibinfo{author}{McCubbin, F.}, \bibinfo{year}{2011}.
\newblock \bibinfo{title}{{REE crystal chemistry of phosphates in
  extraterrestrial basalts at different oxygen fugacities: Direct determination
  of europium valence state in merrillite-whitlockite}}, in:
  \bibinfo{booktitle}{Lunar and Planetary Science Conference}, p.
  \bibinfo{pages}{1143}.
\bibitem[{Shibazaki et~al.(2009)Shibazaki, Ohtani, Terasaki, Suzuki and
  Funakoshi}]{shibazaki2009hydrogen}
\bibinfo{author}{Shibazaki, Y.}, \bibinfo{author}{Ohtani, E.},
  \bibinfo{author}{Terasaki, H.}, \bibinfo{author}{Suzuki, A.},
  \bibinfo{author}{Funakoshi, K.i.}, \bibinfo{year}{2009}.
\newblock \bibinfo{title}{{Hydrogen partitioning between iron and ringwoodite:
  Implications for water transport into the Martian core}}.
\newblock \bibinfo{journal}{Earth and Planetary Science Letters}
  \bibinfo{volume}{287}, \bibinfo{pages}{463--470}.
\newblock \DOIprefix\doi{10.1016/j.epsl.2009.08.034}.
\bibitem[{Shukolyukov and Lugmair(2006)}]{shukolyukov2006manganese}
\bibinfo{author}{Shukolyukov, A.}, \bibinfo{author}{Lugmair, G.W.},
  \bibinfo{year}{2006}.
\newblock \bibinfo{title}{{Manganese--chromium isotope systematics of
  carbonaceous chondrites}}.
\newblock \bibinfo{journal}{Earth and Planetary Science Letters}
  \bibinfo{volume}{250}, \bibinfo{pages}{200--213}.
\newblock \DOIprefix\doi{10.1016/j.epsl.2006.07.036}.
\bibitem[{Siebert et~al.(2018)Siebert, Sossi, Blanchard, Mahan, Badro and
  Moynier}]{siebert2018chondritic}
\bibinfo{author}{Siebert, J.}, \bibinfo{author}{Sossi, P.A.},
  \bibinfo{author}{Blanchard, I.}, \bibinfo{author}{Mahan, B.},
  \bibinfo{author}{Badro, J.}, \bibinfo{author}{Moynier, F.},
  \bibinfo{year}{2018}.
\newblock \bibinfo{title}{{Chondritic Mn/Na ratio and limited post-nebular
  volatile loss of the Earth}}.
\newblock \bibinfo{journal}{Earth and Planetary Science Letters}
  \bibinfo{volume}{485}, \bibinfo{pages}{130--139}.
\newblock \DOIprefix\doi{10.1016/j.epsl.2017.12.042}.
\bibitem[{Simon et~al.(2017)Simon, Jordan, Tappa, Schauble, Kohl and
  Young}]{simon2017calcium}
\bibinfo{author}{Simon, J.I.}, \bibinfo{author}{Jordan, M.K.},
  \bibinfo{author}{Tappa, M.J.}, \bibinfo{author}{Schauble, E.A.},
  \bibinfo{author}{Kohl, I.E.}, \bibinfo{author}{Young, E.D.},
  \bibinfo{year}{2017}.
\newblock \bibinfo{title}{{Calcium and titanium isotope fractionation in
  refractory inclusions: tracers of condensation and inheritance in the early
  solar protoplanetary disk}}.
\newblock \bibinfo{journal}{Earth and Planetary Science Letters}
  \bibinfo{volume}{472}, \bibinfo{pages}{277--288}.
\newblock \DOIprefix\doi{10.1016/j.epsl.2017.05.002}.
\bibitem[{Smrekar et~al.(2019)Smrekar, Lognonn{\'e}, Spohn, Banerdt, Breuer,
  Christensen, Dehant, Drilleau, Folkner, Fuji, Garcia, Giardini, Golombek,
  Grott, Gudkova, Johnson, Khan, Langlais, Mittelholz, Mocquet, Myhill,
  Panning, Perrin, Pike, Plesa, Rivoldini, Samuel, St{\"a}hler, van Driel,
  Van~Hoolst, Verhoeven, Weber and Wieczorek}]{smrekar2019pre}
\bibinfo{author}{Smrekar, S.E.}, \bibinfo{author}{Lognonn{\'e}, P.},
  \bibinfo{author}{Spohn, T.}, \bibinfo{author}{Banerdt, W.B.},
  \bibinfo{author}{Breuer, D.}, \bibinfo{author}{Christensen, U.},
  \bibinfo{author}{Dehant, V.}, \bibinfo{author}{Drilleau, M.},
  \bibinfo{author}{Folkner, W.}, \bibinfo{author}{Fuji, N.},
  \bibinfo{author}{Garcia, R.F.}, \bibinfo{author}{Giardini, D.},
  \bibinfo{author}{Golombek, M.}, \bibinfo{author}{Grott, M.},
  \bibinfo{author}{Gudkova, T.}, \bibinfo{author}{Johnson, C.},
  \bibinfo{author}{Khan, A.}, \bibinfo{author}{Langlais, B.},
  \bibinfo{author}{Mittelholz, A.}, \bibinfo{author}{Mocquet, A.},
  \bibinfo{author}{Myhill, R.}, \bibinfo{author}{Panning, M.},
  \bibinfo{author}{Perrin, C.}, \bibinfo{author}{Pike, T.},
  \bibinfo{author}{Plesa, A.C.}, \bibinfo{author}{Rivoldini, A.},
  \bibinfo{author}{Samuel, H.}, \bibinfo{author}{St{\"a}hler, S.C.},
  \bibinfo{author}{van Driel, M.}, \bibinfo{author}{Van~Hoolst, T.},
  \bibinfo{author}{Verhoeven, O.}, \bibinfo{author}{Weber, R.},
  \bibinfo{author}{Wieczorek, M.}, \bibinfo{year}{2019}.
\newblock \bibinfo{title}{{Pre-mission InSights on the interior of Mars}}.
\newblock \bibinfo{journal}{Space Science Reviews} \bibinfo{volume}{215},
  \bibinfo{pages}{3}.
\newblock \DOIprefix\doi{10.1007/s11214-018-0563-9}.
\bibitem[{Sossi et~al.(2019)Sossi, Klemme, O'Neill, Berndt and
  Moynier}]{sossi2019evaporation}
\bibinfo{author}{Sossi, P.A.}, \bibinfo{author}{Klemme, S.},
  \bibinfo{author}{O'Neill, H.S.C.}, \bibinfo{author}{Berndt, J.},
  \bibinfo{author}{Moynier, F.}, \bibinfo{year}{2019}.
\newblock \bibinfo{title}{{Evaporation of moderately volatile elements from
  silicate melts: Experiments and theory}}.
\newblock \bibinfo{journal}{Geochimica et Cosmochimica Acta}
  \bibinfo{volume}{260}, \bibinfo{pages}{204--231}.
\newblock \DOIprefix\doi{10.1016/j.gca.2019.06.021}.
\bibitem[{Sossi et~al.(2016)Sossi, Nebel, Anand and Poitrasson}]{sossi2016on}
\bibinfo{author}{Sossi, P.A.}, \bibinfo{author}{Nebel, O.},
  \bibinfo{author}{Anand, M.}, \bibinfo{author}{Poitrasson, F.},
  \bibinfo{year}{2016}.
\newblock \bibinfo{title}{{On the iron isotope composition of Mars and volatile
  depletion in the terrestrial planets}}.
\newblock \bibinfo{journal}{Earth and Planetary Science Letters}
  \bibinfo{volume}{449}, \bibinfo{pages}{360--371}.
\newblock \DOIprefix\doi{10.1016/j.epsl.2016.05.030}.
\bibitem[{Sossi et~al.(2018)Sossi, Nebel, O'Neill and Moynier}]{sossi2018zinc}
\bibinfo{author}{Sossi, P.A.}, \bibinfo{author}{Nebel, O.},
  \bibinfo{author}{O'Neill, H.S.C.}, \bibinfo{author}{Moynier, F.},
  \bibinfo{year}{2018}.
\newblock \bibinfo{title}{{Zinc isotope composition of the Earth and its
  behaviour during planetary accretion}}.
\newblock \bibinfo{journal}{Chemical Geology} \bibinfo{volume}{477},
  \bibinfo{pages}{73--84}.
\newblock \DOIprefix\doi{10.1016/j.chemgeo.2017.12.006}.
\bibitem[{Steenstra et~al.(2016)Steenstra, Rai, Knibbe, Lin and van
  Westrenen}]{steenstra2016new}
\bibinfo{author}{Steenstra, E.S.}, \bibinfo{author}{Rai, N.},
  \bibinfo{author}{Knibbe, J.S.}, \bibinfo{author}{Lin, Y.H.},
  \bibinfo{author}{van Westrenen, W.}, \bibinfo{year}{2016}.
\newblock \bibinfo{title}{{New geochemical models of core formation in the Moon
  from metal--silicate partitioning of 15 siderophile elements}}.
\newblock \bibinfo{journal}{Earth and Planetary Science Letters}
  \bibinfo{volume}{441}, \bibinfo{pages}{1--9}.
\newblock \DOIprefix\doi{10.1016/j.epsl.2016.02.028}.
\bibitem[{Stolper and Paque(1986)}]{stolper1986crystallization}
\bibinfo{author}{Stolper, E.}, \bibinfo{author}{Paque, J.M.},
  \bibinfo{year}{1986}.
\newblock \bibinfo{title}{{Crystallization sequences of Ca-Al-rich inclusions
  from Allende: The effects of cooling rate and maximum temperature}}.
\newblock \bibinfo{journal}{Geochimica et Cosmochimica Acta}
  \bibinfo{volume}{50}, \bibinfo{pages}{1785--1806}.
\newblock \DOIprefix\doi{10.1016/0016-7037(86)90139-0}.
\bibitem[{Stracke et~al.(2012)Stracke, Palme, Gellissen, M{\"u}nker, Kleine,
  Birbaum, G{\"u}nther, Bourdon and Zipfel}]{stracke2012refractory}
\bibinfo{author}{Stracke, A.}, \bibinfo{author}{Palme, H.},
  \bibinfo{author}{Gellissen, M.}, \bibinfo{author}{M{\"u}nker, C.},
  \bibinfo{author}{Kleine, T.}, \bibinfo{author}{Birbaum, K.},
  \bibinfo{author}{G{\"u}nther, D.}, \bibinfo{author}{Bourdon, B.},
  \bibinfo{author}{Zipfel, J.}, \bibinfo{year}{2012}.
\newblock \bibinfo{title}{{Refractory element fractionation in the Allende
  meteorite: Implications for solar nebula condensation and the chondritic
  composition of planetary bodies}}.
\newblock \bibinfo{journal}{Geochimica et Cosmochimica Acta}
  \bibinfo{volume}{85}, \bibinfo{pages}{114--141}.
\newblock \DOIprefix\doi{10.1016/j.gca.2012.02.006}.
\bibitem[{Suer et~al.(2017)Suer, Siebert, Remusat, Menguy and
  Fiquet}]{suer2017sulfur}
\bibinfo{author}{Suer, T.A.}, \bibinfo{author}{Siebert, J.},
  \bibinfo{author}{Remusat, L.}, \bibinfo{author}{Menguy, N.},
  \bibinfo{author}{Fiquet, G.}, \bibinfo{year}{2017}.
\newblock \bibinfo{title}{{A sulfur-poor terrestrial core inferred from
  metal--silicate partitioning experiments}}.
\newblock \bibinfo{journal}{Earth and Planetary Science Letters}
  \bibinfo{volume}{469}, \bibinfo{pages}{84--97}.
\newblock \DOIprefix\doi{10.1016/j.epsl.2017.04.016}.
\bibitem[{Sugiura and Fujiya(2014)}]{sugiura2014correlated}
\bibinfo{author}{Sugiura, N.}, \bibinfo{author}{Fujiya, W.},
  \bibinfo{year}{2014}.
\newblock \bibinfo{title}{{Correlated accretion ages and
  $\varepsilon$\ce{^{54}Cr} of meteorite parent bodies and the evolution of the
  solar nebula}}.
\newblock \bibinfo{journal}{Meteoritics \& Planetary Science}
  \bibinfo{volume}{49}, \bibinfo{pages}{772--787}.
\newblock \DOIprefix\doi{10.1111/maps.12292}.
\bibitem[{Surkov et~al.(1987)Surkov, Kirnozov, Glazov, Dunchenko, Tatsy and
  Sobornov}]{surkov1987uranium}
\bibinfo{author}{Surkov, Y.A.}, \bibinfo{author}{Kirnozov, F.F.},
  \bibinfo{author}{Glazov, V.N.}, \bibinfo{author}{Dunchenko, A.G.},
  \bibinfo{author}{Tatsy, L.P.}, \bibinfo{author}{Sobornov, O.P.},
  \bibinfo{year}{1987}.
\newblock \bibinfo{title}{{Uranium, thorium, and potassium in the Venusian
  rocks at the landing sites of Vega 1 and 2}}.
\newblock \bibinfo{journal}{Journal of Geophysical Research: Solid Earth}
  \bibinfo{volume}{92}, \bibinfo{pages}{E537--E540}.
\newblock \DOIprefix\doi{10.1029/JB092iB04p0E537}.
\bibitem[{Tait and Day(2018)}]{tait2018chondritic}
\bibinfo{author}{Tait, K.T.}, \bibinfo{author}{Day, J.M.D.},
  \bibinfo{year}{2018}.
\newblock \bibinfo{title}{{Chondritic late accretion to Mars and the nature of
  shergottite reservoirs}}.
\newblock \bibinfo{journal}{Earth and Planetary Science Letters}
  \bibinfo{volume}{494}, \bibinfo{pages}{99--108}.
\newblock \DOIprefix\doi{10.1016/j.epsl.2018.04.040}.
\bibitem[{Taylor(2013)}]{taylor2013bulk}
\bibinfo{author}{Taylor, G.J.}, \bibinfo{year}{2013}.
\newblock \bibinfo{title}{{The bulk composition of Mars}}.
\newblock \bibinfo{journal}{Chemie der Erde-Geochemistry} \bibinfo{volume}{73},
  \bibinfo{pages}{401--420}.
\newblock \DOIprefix\doi{10.1016/j.chemer.2013.09.006}.
\bibitem[{Taylor et~al.(2006)Taylor, Boynton, Br\"{u}ckner, W\"{a}nke, Dreibus,
  Kerry, Keller, Reedy, Evans, Starr, Squyres, Karunatillake, Gasnault,
  Maurice, d'Uston, Englert, Dohm, Baker, Hamara, Janes, Sprague, Kim and
  Drake}]{taylor2006bulk}
\bibinfo{author}{Taylor, G.J.}, \bibinfo{author}{Boynton, W.},
  \bibinfo{author}{Br\"{u}ckner, J.}, \bibinfo{author}{W\"{a}nke, H.},
  \bibinfo{author}{Dreibus, G.}, \bibinfo{author}{Kerry, K.},
  \bibinfo{author}{Keller, J.}, \bibinfo{author}{Reedy, R.},
  \bibinfo{author}{Evans, L.}, \bibinfo{author}{Starr, R.},
  \bibinfo{author}{Squyres, S.}, \bibinfo{author}{Karunatillake, S.},
  \bibinfo{author}{Gasnault, O.}, \bibinfo{author}{Maurice, S.},
  \bibinfo{author}{d'Uston, C.}, \bibinfo{author}{Englert, P.},
  \bibinfo{author}{Dohm, J.}, \bibinfo{author}{Baker, V.},
  \bibinfo{author}{Hamara, D.}, \bibinfo{author}{Janes, D.},
  \bibinfo{author}{Sprague, A.}, \bibinfo{author}{Kim, K.},
  \bibinfo{author}{Drake, D.}, \bibinfo{year}{2006}.
\newblock \bibinfo{title}{{Bulk composition and early differentiation of
  Mars}}.
\newblock \bibinfo{journal}{Journal of Geophysical Research: Planets}
  \bibinfo{volume}{111}, \bibinfo{pages}{E03S10}.
\newblock \DOIprefix\doi{10.1029/2005JE002645}.
\bibitem[{Taylor and McLennan(2009)}]{taylor2009planetary}
\bibinfo{author}{Taylor, S.R.}, \bibinfo{author}{McLennan, S.},
  \bibinfo{year}{2009}.
\newblock \bibinfo{title}{{Planetary Crusts: Their Composition, Origin and
  Evolution}}. volume~\bibinfo{volume}{10}.
\newblock \bibinfo{publisher}{Cambridge University Press},
  \bibinfo{address}{Cambridge}.
\newblock \DOIprefix\doi{10.1017/CBO9780511575358}.
\bibitem[{Tenner et~al.(2018)Tenner, Ushikubo, Nakashima, Schrader, Weisberg,
  Kimura and Kita}]{tenner2018oxygen}
\bibinfo{author}{Tenner, T.J.}, \bibinfo{author}{Ushikubo, T.},
  \bibinfo{author}{Nakashima, D.}, \bibinfo{author}{Schrader, D.L.},
  \bibinfo{author}{Weisberg, M.K.}, \bibinfo{author}{Kimura, M.},
  \bibinfo{author}{Kita, N.T.}, \bibinfo{year}{2018}.
\newblock \bibinfo{title}{Oxygen isotope characteristics of chondrules from
  recent studies by secondary ion mass spectrometry}, in:
  \bibinfo{editor}{Russell, S.S.}, \bibinfo{editor}{Connolly~Jr., H.C.},
  \bibinfo{editor}{Krot, A.N.} (Eds.), \bibinfo{booktitle}{Chondrules: Records
  of Protoplanetary Disk Processes}. \bibinfo{publisher}{Cambridge University
  Press}. Cambridge Planetary Science. chapter~\bibinfo{chapter}{8}, p.
  \bibinfo{pages}{196–246}.
\newblock \DOIprefix\doi{10.1017/9781108284073.008}.
\bibitem[{Tian et~al.(2019)Tian, Chen, Fegley~Jr., Lodders, Barrat, Day and
  Wang}]{tian2019potassium}
\bibinfo{author}{Tian, Z.}, \bibinfo{author}{Chen, H.},
  \bibinfo{author}{Fegley~Jr., B.}, \bibinfo{author}{Lodders, K.},
  \bibinfo{author}{Barrat, J.A.}, \bibinfo{author}{Day, J.M.D.},
  \bibinfo{author}{Wang, K.}, \bibinfo{year}{2019}.
\newblock \bibinfo{title}{{Potassium isotopic compositions of
  howardite-eucrite-diogenite meteorites}}.
\newblock \bibinfo{journal}{Geochimica et Cosmochimica Acta}
  \bibinfo{volume}{266}, \bibinfo{pages}{611--632}.
\newblock \DOIprefix\doi{10.1016/j.gca.2019.08.012}.
\bibitem[{Trinquier et~al.(2007)Trinquier, Birck and
  Allegre}]{trinquier2007widespread}
\bibinfo{author}{Trinquier, A.}, \bibinfo{author}{Birck, J.L.},
  \bibinfo{author}{Allegre, C.J.}, \bibinfo{year}{2007}.
\newblock \bibinfo{title}{{Widespread \ce{^{54}Cr} heterogeneity in the inner
  solar system}}.
\newblock \bibinfo{journal}{The Astrophysical Journal} \bibinfo{volume}{655},
  \bibinfo{pages}{1179}.
\newblock \DOIprefix\doi{10.1086/510360}.
\bibitem[{Trinquier et~al.(2008)Trinquier, Birck, All{\`e}gre, G{\"o}pel and
  Ulfbeck}]{trinquier200853mn}
\bibinfo{author}{Trinquier, A.}, \bibinfo{author}{Birck, J.L.},
  \bibinfo{author}{All{\`e}gre, C.J.}, \bibinfo{author}{G{\"o}pel, C.},
  \bibinfo{author}{Ulfbeck, D.}, \bibinfo{year}{2008}.
\newblock \bibinfo{title}{{\ce{^{53}Mn}--\ce{^{53}Cr} systematics of the early
  Solar System revisited}}.
\newblock \bibinfo{journal}{Geochimica et Cosmochimica Acta}
  \bibinfo{volume}{72}, \bibinfo{pages}{5146--5163}.
\newblock \DOIprefix\doi{10.1016/j.gca.2008.03.023}.
\bibitem[{Trinquier et~al.(2009)Trinquier, Elliott, Ulfbeck, Coath, Krot and
  Bizzarro}]{trinquier2009origin}
\bibinfo{author}{Trinquier, A.}, \bibinfo{author}{Elliott, T.},
  \bibinfo{author}{Ulfbeck, D.}, \bibinfo{author}{Coath, C.},
  \bibinfo{author}{Krot, A.N.}, \bibinfo{author}{Bizzarro, M.},
  \bibinfo{year}{2009}.
\newblock \bibinfo{title}{{Origin of nucleosynthetic isotope heterogeneity in
  the solar protoplanetary disk}}.
\newblock \bibinfo{journal}{Science} \bibinfo{volume}{324},
  \bibinfo{pages}{374--376}.
\newblock \DOIprefix\doi{10.1126/science.1168221}.
\bibitem[{Tsuno et~al.(2018)Tsuno, Grewal and Dasgupta}]{tsuno2018core}
\bibinfo{author}{Tsuno, K.}, \bibinfo{author}{Grewal, D.S.},
  \bibinfo{author}{Dasgupta, R.}, \bibinfo{year}{2018}.
\newblock \bibinfo{title}{{Core-mantle fractionation of carbon in Earth and
  Mars: The effects of sulfur}}.
\newblock \bibinfo{journal}{Geochimica et Cosmochimica Acta}
  \bibinfo{volume}{238}, \bibinfo{pages}{477--495}.
\newblock \DOIprefix\doi{10.1016/j.gca.2018.07.010}.
\bibitem[{Turcotte and Schubert(2014)}]{turcotte2014geodynamics}
\bibinfo{author}{Turcotte, D.}, \bibinfo{author}{Schubert, G.},
  \bibinfo{year}{2014}.
\newblock \bibinfo{title}{{Geodynamics (3rd Edition)}}.
\newblock \bibinfo{publisher}{Cambridge University Press},
  \bibinfo{address}{Cambridge}.
\newblock \DOIprefix\doi{10.1017/cbo9780511843877}.
\bibitem[{Turcotte et~al.(2001)Turcotte, Paul and White}]{turcotte2001thorium}
\bibinfo{author}{Turcotte, D.L.}, \bibinfo{author}{Paul, D.},
  \bibinfo{author}{White, W.M.}, \bibinfo{year}{2001}.
\newblock \bibinfo{title}{{Thorium-uranium systematics require layered mantle
  convection}}.
\newblock \bibinfo{journal}{Journal of Geophysical Research: Solid Earth}
  \bibinfo{volume}{106}, \bibinfo{pages}{4265--4276}.
\newblock \DOIprefix\doi{10.1029/2000JB900409}.
\bibitem[{Urey and Craig(1953)}]{urey1953composition}
\bibinfo{author}{Urey, H.C.}, \bibinfo{author}{Craig, H.},
  \bibinfo{year}{1953}.
\newblock \bibinfo{title}{{The composition of the stone meteorites and the
  origin of the meteorites}}.
\newblock \bibinfo{journal}{Geochimica et Cosmochimica Acta}
  \bibinfo{volume}{4}, \bibinfo{pages}{36--82}.
\newblock \DOIprefix\doi{10.1016/0016-7037(53)90064-7}.
\bibitem[{Villeneuve et~al.(2009)Villeneuve, Chaussidon and
  Libourel}]{villeneuve2009homogeneous}
\bibinfo{author}{Villeneuve, J.}, \bibinfo{author}{Chaussidon, M.},
  \bibinfo{author}{Libourel, G.}, \bibinfo{year}{2009}.
\newblock \bibinfo{title}{{Homogeneous distribution of \ce{^{26}Al} in the
  solar system from the Mg isotopic composition of chondrules}}.
\newblock \bibinfo{journal}{Science} \bibinfo{volume}{325},
  \bibinfo{pages}{985--988}.
\newblock \DOIprefix\doi{10.1126/science.1173907}.
\bibitem[{Wade et~al.(2012)Wade, Wood and Tuff}]{wade2012metal}
\bibinfo{author}{Wade, J.}, \bibinfo{author}{Wood, B.J.},
  \bibinfo{author}{Tuff, J.}, \bibinfo{year}{2012}.
\newblock \bibinfo{title}{{Metal--silicate partitioning of Mo and W at high
  pressures and temperatures: evidence for late accretion of sulphur to the
  Earth}}.
\newblock \bibinfo{journal}{Geochimica et Cosmochimica Acta}
  \bibinfo{volume}{85}, \bibinfo{pages}{58--74}.
\newblock \DOIprefix\doi{10.1016/j.gca.2012.01.010}.
\bibitem[{Wadhwa(2001)}]{wadhwa2001redox}
\bibinfo{author}{Wadhwa, M.}, \bibinfo{year}{2001}.
\newblock \bibinfo{title}{{Redox state of Mars' upper mantle and crust from Eu
  anomalies in shergottite pyroxenes}}.
\newblock \bibinfo{journal}{Science} \bibinfo{volume}{291},
  \bibinfo{pages}{1527--1530}.
\newblock \DOIprefix\doi{10.1126/science.1057594}.
\bibitem[{Wadhwa(2008)}]{wadhwa2008redox}
\bibinfo{author}{Wadhwa, M.}, \bibinfo{year}{2008}.
\newblock \bibinfo{title}{{Redox conditions on small bodies, the Moon and
  Mars}}, in: \bibinfo{editor}{MacPherson, G.J.} (Ed.),
  \bibinfo{booktitle}{Reviews in Mineralogy and Geochemistry}.
  \bibinfo{publisher}{Mineralogical Society of America}.
  volume~\bibinfo{volume}{68}, pp. \bibinfo{pages}{493--510}.
\newblock \DOIprefix\doi{10.2138/rmg.2008.68.17}.
\bibitem[{Walker et~al.(2015)Walker, Bermingham, Liu, Puchtel, Touboul and
  Worsham}]{walker2015search}
\bibinfo{author}{Walker, R.J.}, \bibinfo{author}{Bermingham, K.},
  \bibinfo{author}{Liu, J.}, \bibinfo{author}{Puchtel, I.S.},
  \bibinfo{author}{Touboul, M.}, \bibinfo{author}{Worsham, E.A.},
  \bibinfo{year}{2015}.
\newblock \bibinfo{title}{{In search of late-stage planetary building blocks}}.
\newblock \bibinfo{journal}{Chemical Geology} \bibinfo{volume}{411},
  \bibinfo{pages}{125--142}.
\newblock \DOIprefix\doi{10.1016/j.chemgeo.2015.06.028}.
\bibitem[{Walsh et~al.(2011)Walsh, Morbidelli, Raymond, O'Brien and
  Mandell}]{walsh2011low}
\bibinfo{author}{Walsh, K.J.}, \bibinfo{author}{Morbidelli, A.},
  \bibinfo{author}{Raymond, S.N.}, \bibinfo{author}{O'Brien, D.P.},
  \bibinfo{author}{Mandell, A.M.}, \bibinfo{year}{2011}.
\newblock \bibinfo{title}{{A low mass for Mars from Jupiter's early gas-driven
  migration}}.
\newblock \bibinfo{journal}{Nature} \bibinfo{volume}{475},
  \bibinfo{pages}{206--209}.
\newblock \DOIprefix\doi{10.1038/nature10201}.
\bibitem[{Wang et~al.(2017)Wang, Weiss, Bai, Downey, Wang, Wang, Suavet, Fu and
  Zucolotto}]{wang2017lifetime}
\bibinfo{author}{Wang, H.}, \bibinfo{author}{Weiss, B.P.},
  \bibinfo{author}{Bai, X.N.}, \bibinfo{author}{Downey, B.G.},
  \bibinfo{author}{Wang, J.}, \bibinfo{author}{Wang, J.},
  \bibinfo{author}{Suavet, C.}, \bibinfo{author}{Fu, R.R.},
  \bibinfo{author}{Zucolotto, M.E.}, \bibinfo{year}{2017}.
\newblock \bibinfo{title}{{Lifetime of the solar nebula constrained by
  meteorite paleomagnetism}}.
\newblock \bibinfo{journal}{Science} \bibinfo{volume}{355},
  \bibinfo{pages}{623--627}.
\newblock \DOIprefix\doi{10.1126/science.aaf5043}.
\bibitem[{Wang and Becker(2017)}]{wang2017chalcophile}
\bibinfo{author}{Wang, Z.}, \bibinfo{author}{Becker, H.}, \bibinfo{year}{2017}.
\newblock \bibinfo{title}{{Chalcophile elements in Martian meteorites indicate
  low sulfur content in the Martian interior and a volatile element-depleted
  late veneer}}.
\newblock \bibinfo{journal}{Earth and Planetary Science Letters}
  \bibinfo{volume}{463}, \bibinfo{pages}{56--68}.
\newblock \DOIprefix\doi{10.1016/j.epsl.2017.01.023}.
\bibitem[{W{\"a}nke(1981)}]{wanke1981constitution}
\bibinfo{author}{W{\"a}nke, H.}, \bibinfo{year}{1981}.
\newblock \bibinfo{title}{{Constitution of terrestrial planets}}.
\newblock \bibinfo{journal}{Philosophical Transactions of the Royal Society of
  London. Series A: Mathematical and Physical Sciences} \bibinfo{volume}{303},
  \bibinfo{pages}{287--302}.
\newblock \DOIprefix\doi{10.1098/rsta.1981.0203}.
\bibitem[{W{\"a}nke and Dreibus(1988)}]{wanke1988chemical}
\bibinfo{author}{W{\"a}nke, H.}, \bibinfo{author}{Dreibus, G.},
  \bibinfo{year}{1988}.
\newblock \bibinfo{title}{{Chemical composition and accretion history of
  terrestrial planets}}.
\newblock \bibinfo{journal}{Philosophical Transactions of the Royal Society of
  London. Series A, Mathematical and Physical Sciences} \bibinfo{volume}{325},
  \bibinfo{pages}{545--557}.
\newblock \DOIprefix\doi{10.1098/rsta.1988.0067}.
\bibitem[{W{\"a}nke and Dreibus(1994)}]{wanke1994chemistry}
\bibinfo{author}{W{\"a}nke, H.}, \bibinfo{author}{Dreibus, G.},
  \bibinfo{year}{1994}.
\newblock \bibinfo{title}{{Chemistry and accretion history of Mars}}.
\newblock \bibinfo{journal}{Philosophical Transactions of the Royal Society of
  London. Series A, Mathematical and Physical Sciences} \bibinfo{volume}{349},
  \bibinfo{pages}{285--293}.
\newblock \DOIprefix\doi{10.1098/rsta.1994.0132}.
\bibitem[{W{\"a}nke et~al.(1984)W{\"a}nke, Dreibus and
  Jagoutz}]{wanke1984mantle}
\bibinfo{author}{W{\"a}nke, H.}, \bibinfo{author}{Dreibus, G.},
  \bibinfo{author}{Jagoutz, E.}, \bibinfo{year}{1984}.
\newblock \bibinfo{title}{{Mantle chemistry and accretion history of the
  Earth}}, in: \bibinfo{editor}{Kr{\"o}ner, A.}, \bibinfo{editor}{Hanson,
  G.N.}, \bibinfo{editor}{Goodwin, A.M.} (Eds.), \bibinfo{booktitle}{Archaean
  Geochemistry}. \bibinfo{publisher}{Springer}, pp. \bibinfo{pages}{1--24}.
\newblock \DOIprefix\doi{10.1007/978-3-642-70001-9_1}.
\bibitem[{Warren(2005)}]{warren2005new}
\bibinfo{author}{Warren, P.H.}, \bibinfo{year}{2005}.
\newblock \bibinfo{title}{{"New" lunar meteorites: Implications for composition
  of the global lunar surface, lunar crust, and the bulk Moon}}.
\newblock \bibinfo{journal}{Meteoritics \& Planetary Science}
  \bibinfo{volume}{40}, \bibinfo{pages}{477--506}.
\newblock \DOIprefix\doi{10.1111/j.1945-5100.2005.tb00395.x}.
\bibitem[{Warren(2008)}]{warren2008depleted}
\bibinfo{author}{Warren, P.H.}, \bibinfo{year}{2008}.
\newblock \bibinfo{title}{{A depleted, not ideally chondritic bulk Earth: The
  explosive-volcanic basalt loss hypothesis}}.
\newblock \bibinfo{journal}{Geochimica et Cosmochimica Acta}
  \bibinfo{volume}{72}, \bibinfo{pages}{2217--2235}.
\newblock \DOIprefix\doi{10.1016/j.gca.2007.11.038}.
\bibitem[{Warren(2011)}]{warren2011stable}
\bibinfo{author}{Warren, P.H.}, \bibinfo{year}{2011}.
\newblock \bibinfo{title}{{Stable-isotopic anomalies and the accretionary
  assemblage of the Earth and Mars: A subordinate role for carbonaceous
  chondrites}}.
\newblock \bibinfo{journal}{Earth and Planetary Science Letters}
  \bibinfo{volume}{311}, \bibinfo{pages}{93--100}.
\newblock \DOIprefix\doi{10.1016/j.epsl.2011.08.047}.
\bibitem[{Wasserburg et~al.(1964)Wasserburg, MacDonald, Hoyle and
  Fowler}]{wasserburg1964relative}
\bibinfo{author}{Wasserburg, G.J.}, \bibinfo{author}{MacDonald, G.J.F.},
  \bibinfo{author}{Hoyle, F.}, \bibinfo{author}{Fowler, W.A.},
  \bibinfo{year}{1964}.
\newblock \bibinfo{title}{{Relative contributions of uranium, thorium, and
  potassium to heat production in the Earth}}.
\newblock \bibinfo{journal}{Science} \bibinfo{volume}{143},
  \bibinfo{pages}{465--467}.
\newblock \DOIprefix\doi{10.1126/science.143.3605.465}.
\bibitem[{Wasserburg et~al.(1963)Wasserburg, Mazor and
  Zartman}]{wasserburg1963isotopic}
\bibinfo{author}{Wasserburg, G.J.}, \bibinfo{author}{Mazor, E.},
  \bibinfo{author}{Zartman, R.E.}, \bibinfo{year}{1963}.
\newblock \bibinfo{title}{{Isotopic and chemical composition of some
  terrestrial natural gases}}, in: \bibinfo{editor}{Geiss, J.},
  \bibinfo{editor}{Goldberg, E.D.} (Eds.), \bibinfo{booktitle}{Earth Science
  and Meteorites}. \bibinfo{publisher}{North-Holland Pub. Co.}, pp.
  \bibinfo{pages}{219--240}.
\bibitem[{Wasson and Kallemeyn(1988)}]{wasson1988compositions}
\bibinfo{author}{Wasson, J.T.}, \bibinfo{author}{Kallemeyn, G.W.},
  \bibinfo{year}{1988}.
\newblock \bibinfo{title}{{Compositions of chondrites}}.
\newblock \bibinfo{journal}{Philosophical Transactions of the Royal Society of
  London A: Mathematical, Physical and Engineering Sciences}
  \bibinfo{volume}{325}, \bibinfo{pages}{535--544}.
\newblock \DOIprefix\doi{10.1098/rsta.1988.0066}.
\bibitem[{Wasson et~al.(1995)Wasson, Krot, Lee and Rubin}]{wasson1995compound}
\bibinfo{author}{Wasson, J.T.}, \bibinfo{author}{Krot, A.N.},
  \bibinfo{author}{Lee, M.S.}, \bibinfo{author}{Rubin, A.E.},
  \bibinfo{year}{1995}.
\newblock \bibinfo{title}{{Compound chondrules}}.
\newblock \bibinfo{journal}{Geochimica et Cosmochimica Acta}
  \bibinfo{volume}{59}, \bibinfo{pages}{1847--1869}.
\newblock \DOIprefix\doi{10.1016/0016-7037(95)00087-G}.
\bibitem[{Wasson and Richardson(2001)}]{wasson2001fractionation}
\bibinfo{author}{Wasson, J.T.}, \bibinfo{author}{Richardson, J.W.},
  \bibinfo{year}{2001}.
\newblock \bibinfo{title}{{Fractionation trends among IVA iron meteorites:
  Contrasts with IIIAB trends}}.
\newblock \bibinfo{journal}{Geochimica et Cosmochimica Acta}
  \bibinfo{volume}{65}, \bibinfo{pages}{951--970}.
\newblock \DOIprefix\doi{10.1016/S0016-7037(00)00597-4}.
\bibitem[{Weisberg et~al.(2011)Weisberg, Ebel, Connolly, Kita and
  Ushikubo}]{weisberg2011petrology}
\bibinfo{author}{Weisberg, M.K.}, \bibinfo{author}{Ebel, D.S.},
  \bibinfo{author}{Connolly, H.C.}, \bibinfo{author}{Kita, N.T.},
  \bibinfo{author}{Ushikubo, T.}, \bibinfo{year}{2011}.
\newblock \bibinfo{title}{{Petrology and oxygen isotope compositions of
  chondrules in E3 chondrites}}.
\newblock \bibinfo{journal}{Geochimica et Cosmochimica Acta}
  \bibinfo{volume}{75}, \bibinfo{pages}{6556--6569}.
\newblock \DOIprefix\doi{10.1016/j.gca.2011.08.040}.
\bibitem[{Wheeler et~al.(2011)Wheeler, Walker and McDonough}]{wheeler2011pd}
\bibinfo{author}{Wheeler, K.T.}, \bibinfo{author}{Walker, D.},
  \bibinfo{author}{McDonough, W.F.}, \bibinfo{year}{2011}.
\newblock \bibinfo{title}{{Pd and Ag metal-silicate partitioning applied to
  Earth differentiation and core-mantle exchange}}.
\newblock \bibinfo{journal}{Meteoritics \& Planetary Science}
  \bibinfo{volume}{46}, \bibinfo{pages}{199--217}.
\newblock \DOIprefix\doi{10.1111/j.1945-5100.2010.01145.x}.
\bibitem[{Wiechert and Halliday(2007)}]{wiechert2007non}
\bibinfo{author}{Wiechert, U.}, \bibinfo{author}{Halliday, A.N.},
  \bibinfo{year}{2007}.
\newblock \bibinfo{title}{{Non-chondritic magnesium and the origins of the
  inner terrestrial planets}}.
\newblock \bibinfo{journal}{Earth and Planetary Science Letters}
  \bibinfo{volume}{256}, \bibinfo{pages}{360--371}.
\newblock \DOIprefix\doi{10.1016/j.epsl.2007.01.007}.
\bibitem[{Wiechert et~al.(2001)Wiechert, Halliday, Lee, Snyder, Taylor and
  Rumble}]{wiechert2001oxygen}
\bibinfo{author}{Wiechert, U.}, \bibinfo{author}{Halliday, A.N.},
  \bibinfo{author}{Lee, D.C.}, \bibinfo{author}{Snyder, G.A.},
  \bibinfo{author}{Taylor, L.A.}, \bibinfo{author}{Rumble, D.},
  \bibinfo{year}{2001}.
\newblock \bibinfo{title}{{Oxygen isotopes and the Moon-forming giant impact}}.
\newblock \bibinfo{journal}{Science} \bibinfo{volume}{294},
  \bibinfo{pages}{345--348}.
\newblock \DOIprefix\doi{10.1126/science.1063037}.
\bibitem[{Wieczorek and Zuber(2004)}]{wieczorek2004thickness}
\bibinfo{author}{Wieczorek, M.A.}, \bibinfo{author}{Zuber, M.T.},
  \bibinfo{year}{2004}.
\newblock \bibinfo{title}{{Thickness of the Martian crust: Improved constraints
  from geoid-to-topography ratios}}.
\newblock \bibinfo{journal}{Journal of Geophysical Research: Planets}
  \bibinfo{volume}{109}.
\newblock \DOIprefix\doi{10.1029/2003JE002153}.
\bibitem[{Willig et~al.(2020)Willig, Stracke, Beier and
  Salters}]{willig2020constraints}
\bibinfo{author}{Willig, M.}, \bibinfo{author}{Stracke, A.},
  \bibinfo{author}{Beier, C.}, \bibinfo{author}{Salters, V.J.M.},
  \bibinfo{year}{2020}.
\newblock \bibinfo{title}{{Constraints on mantle evolution from Ce-Nd-Hf
  isotope systematics}}.
\newblock \bibinfo{journal}{Geochimica et Cosmochimica Acta}
  \bibinfo{volume}{272}, \bibinfo{pages}{36--53}.
\newblock \DOIprefix\doi{10.1016/j.gca.2019.12.029}.
\bibitem[{Wipperfurth et~al.(2018)Wipperfurth, Guo, {\v{S}}r{\'a}mek and
  McDonough}]{wipperfurth2018earth}
\bibinfo{author}{Wipperfurth, S.A.}, \bibinfo{author}{Guo, M.},
  \bibinfo{author}{{\v{S}}r{\'a}mek, O.}, \bibinfo{author}{McDonough, W.F.},
  \bibinfo{year}{2018}.
\newblock \bibinfo{title}{{Earth's chondritic Th/U: Negligible fractionation
  during accretion, core formation, and crust-mantle differentiation}}.
\newblock \bibinfo{journal}{Earth and Planetary Science Letters}
  \bibinfo{volume}{498}, \bibinfo{pages}{196--202}.
\newblock \DOIprefix\doi{10.1016/j.epsl.2018.06.029}.
\bibitem[{Wipperfurth et~al.(2020)Wipperfurth, {\v{S}}r{\'a}mek and
  McDonough}]{wipperfurth2019reference}
\bibinfo{author}{Wipperfurth, S.A.}, \bibinfo{author}{{\v{S}}r{\'a}mek, O.},
  \bibinfo{author}{McDonough, W.F.}, \bibinfo{year}{2020}.
\newblock \bibinfo{title}{Reference models for lithospheric geoneutrino
  signal}.
\newblock \bibinfo{journal}{Journal of Geophysical Research: Solid Earth}
  \bibinfo{volume}{125}, \bibinfo{pages}{e2019JB018433}.
\newblock \DOIprefix\doi{10.1029/2019JB018433}.
\bibitem[{Wombacher et~al.(2008)Wombacher, Rehk{\"a}mper, Mezger, Bischoff and
  M{\"u}nker}]{wombacher2008cadmium}
\bibinfo{author}{Wombacher, F.}, \bibinfo{author}{Rehk{\"a}mper, M.},
  \bibinfo{author}{Mezger, K.}, \bibinfo{author}{Bischoff, A.},
  \bibinfo{author}{M{\"u}nker, C.}, \bibinfo{year}{2008}.
\newblock \bibinfo{title}{{Cadmium stable isotope cosmochemistry}}.
\newblock \bibinfo{journal}{Geochimica et Cosmochimica Acta}
  \bibinfo{volume}{72}, \bibinfo{pages}{646--667}.
\newblock \DOIprefix\doi{10.1016/j.gca.2007.10.024}.
\bibitem[{Wood et~al.(2019)Wood, Smythe and Harrison}]{wood2019condensation}
\bibinfo{author}{Wood, B.J.}, \bibinfo{author}{Smythe, D.J.},
  \bibinfo{author}{Harrison, T.}, \bibinfo{year}{2019}.
\newblock \bibinfo{title}{{The condensation temperatures of the elements: A
  reappraisal}}.
\newblock \bibinfo{journal}{American Mineralogist} \bibinfo{volume}{104},
  \bibinfo{pages}{844--856}.
\newblock \DOIprefix\doi{10.2138/am-2019-6852CCBY}.
\bibitem[{Wood et~al.(2009)Wood, Wade and Kilburn}]{wood2008core}
\bibinfo{author}{Wood, B.J.}, \bibinfo{author}{Wade, J.},
  \bibinfo{author}{Kilburn, M.R.}, \bibinfo{year}{2009}.
\newblock \bibinfo{title}{{Core formation and the oxidation state of the Earth:
  Additional constraints from Nb, V and Cr partitioning}}.
\newblock \bibinfo{journal}{Geochimica et Cosmochimica Acta}
  \bibinfo{volume}{72}, \bibinfo{pages}{1415--1426}.
\newblock \DOIprefix\doi{10.1016/j.gca.2007.11.036}.
\bibitem[{Yang et~al.(2015)Yang, Humayun, Righter, Jefferson, Fields and
  Irving}]{yang2015siderophile}
\bibinfo{author}{Yang, S.}, \bibinfo{author}{Humayun, M.},
  \bibinfo{author}{Righter, K.}, \bibinfo{author}{Jefferson, G.},
  \bibinfo{author}{Fields, D.}, \bibinfo{author}{Irving, A.J.},
  \bibinfo{year}{2015}.
\newblock \bibinfo{title}{{Siderophile and chalcophile element abundances in
  shergottites: Implications for Martian core formation}}.
\newblock \bibinfo{journal}{Meteoritics \& Planetary Science}
  \bibinfo{volume}{50}, \bibinfo{pages}{691--714}.
\newblock \DOIprefix\doi{10.1111/maps.12384}.
\bibitem[{Yoder et~al.(2003)Yoder, Konopliv, Yuan, Standish and
  Folkner}]{yoder2003fluid}
\bibinfo{author}{Yoder, C.F.}, \bibinfo{author}{Konopliv, A.S.},
  \bibinfo{author}{Yuan, D.N.}, \bibinfo{author}{Standish, E.M.},
  \bibinfo{author}{Folkner, W.M.}, \bibinfo{year}{2003}.
\newblock \bibinfo{title}{{Fluid core size of Mars from detection of the solar
  tide}}.
\newblock \bibinfo{journal}{Science} \bibinfo{volume}{300},
  \bibinfo{pages}{299--303}.
\newblock \DOIprefix\doi{10.1126/science.1079645}.
\bibitem[{Yoshizaki and McDonough(2020)}]{yoshizaki2019mars_long}
\bibinfo{author}{Yoshizaki, T.}, \bibinfo{author}{McDonough, W.F.},
  \bibinfo{year}{2020}.
\newblock \bibinfo{title}{{The composition of Mars}}.
\newblock \bibinfo{journal}{Geochimica et Cosmochimica Acta}
  \bibinfo{volume}{273}, \bibinfo{pages}{137--162}.
\newblock \DOIprefix\doi{10.1016/j.gca.2020.01.011}.
\bibitem[{Yoshizaki et~al.(2019)Yoshizaki, Nakashima, Nakamura, Park, Sakamoto,
  Ishida and Itoh}]{yoshizaki2019nebular}
\bibinfo{author}{Yoshizaki, T.}, \bibinfo{author}{Nakashima, D.},
  \bibinfo{author}{Nakamura, T.}, \bibinfo{author}{Park, C.},
  \bibinfo{author}{Sakamoto, N.}, \bibinfo{author}{Ishida, H.},
  \bibinfo{author}{Itoh, S.}, \bibinfo{year}{2019}.
\newblock \bibinfo{title}{{Nebular history of an ultrarefractory phase bearing
  CAI from a reduced type CV chondrite}}.
\newblock \bibinfo{journal}{Geochimica et Cosmochimica Acta}
  \bibinfo{volume}{252}, \bibinfo{pages}{39--60}.
\newblock \DOIprefix\doi{10.1016/j.gca.2019.02.034}.
\bibitem[{Young and Galy(2004)}]{young2004isotope}
\bibinfo{author}{Young, E.D.}, \bibinfo{author}{Galy, A.},
  \bibinfo{year}{2004}.
\newblock \bibinfo{title}{{The isotope geochemistry and cosmochemistry of
  magnesium}}.
\newblock \bibinfo{journal}{Reviews in Mineralogy and Geochemistry}
  \bibinfo{volume}{55}, \bibinfo{pages}{197--230}.
\newblock \DOIprefix\doi{10.2138/gsrmg.55.1.197}.
\bibitem[{Young et~al.(2016)Young, Kohl, Warren, Rubie, Jacobson and
  Morbidelli}]{young2016oxygen}
\bibinfo{author}{Young, E.D.}, \bibinfo{author}{Kohl, I.E.},
  \bibinfo{author}{Warren, P.H.}, \bibinfo{author}{Rubie, D.C.},
  \bibinfo{author}{Jacobson, S.A.}, \bibinfo{author}{Morbidelli, A.},
  \bibinfo{year}{2016}.
\newblock \bibinfo{title}{{Oxygen isotopic evidence for vigorous mixing during
  the Moon-forming giant impact}}.
\newblock \bibinfo{journal}{Science} \bibinfo{volume}{351},
  \bibinfo{pages}{493--496}.
\newblock \DOIprefix\doi{10.1126/science.aad0525}.
\bibitem[{Young et~al.(2019)Young, Shahar, Nimmo, Schlichting, Schauble, Tang
  and Labidi}]{young2019near}
\bibinfo{author}{Young, E.D.}, \bibinfo{author}{Shahar, A.},
  \bibinfo{author}{Nimmo, F.}, \bibinfo{author}{Schlichting, H.E.},
  \bibinfo{author}{Schauble, E.A.}, \bibinfo{author}{Tang, H.},
  \bibinfo{author}{Labidi, J.}, \bibinfo{year}{2019}.
\newblock \bibinfo{title}{{Near-equilibrium isotope fractionation during
  planetesimal evaporation}}.
\newblock \bibinfo{journal}{Icarus} \bibinfo{volume}{323},
  \bibinfo{pages}{1--15}.
\newblock \DOIprefix\doi{10.1016/j.icarus.2019.01.012}.
\bibitem[{Zanda et~al.(2018)Zanda, Lewin and Humayun}]{zanda2018chondritic}
\bibinfo{author}{Zanda, B.}, \bibinfo{author}{Lewin, {\'E}.},
  \bibinfo{author}{Humayun, M.}, \bibinfo{year}{2018}.
\newblock \bibinfo{title}{{The chondritic assemblage}}, in:
  \bibinfo{editor}{Russell, S.S.}, \bibinfo{editor}{Connolly, Jr., H.C.},
  \bibinfo{editor}{Krot, A.N.} (Eds.), \bibinfo{booktitle}{{Chondrules: Records
  of Protoplanetary Disk Processes}}. \bibinfo{publisher}{Cambridge University
  Press}, \bibinfo{address}{Cambridge}. volume~\bibinfo{volume}{22}.
  chapter~\bibinfo{chapter}{5}, pp. \bibinfo{pages}{122--150}.
\newblock \DOIprefix\doi{10.1017/9781108284073.005}.
\bibitem[{Zhang et~al.(2012)Zhang, Dauphas, Davis, Leya and
  Fedkin}]{zhang2012proto}
\bibinfo{author}{Zhang, J.}, \bibinfo{author}{Dauphas, N.},
  \bibinfo{author}{Davis, A.M.}, \bibinfo{author}{Leya, I.},
  \bibinfo{author}{Fedkin, A.}, \bibinfo{year}{2012}.
\newblock \bibinfo{title}{{The proto-Earth as a significant source of lunar
  material}}.
\newblock \bibinfo{journal}{Nature Geoscience} \bibinfo{volume}{5},
  \bibinfo{pages}{251--255}.
\newblock \DOIprefix\doi{10.1038/ngeo1429}.
\bibitem[{Zhu et~al.(2020)Zhu, Moynier, Schiller and Bizzarro}]{zhu2020dating}
\bibinfo{author}{Zhu, K.}, \bibinfo{author}{Moynier, F.},
  \bibinfo{author}{Schiller, M.}, \bibinfo{author}{Bizzarro, M.},
  \bibinfo{year}{2020}.
\newblock \bibinfo{title}{{Dating and tracing the origin of enstatite chondrite
  chondrules with Cr isotopes}}.
\newblock \bibinfo{journal}{The Astrophysical Journal Letters}
  \bibinfo{volume}{894}, \bibinfo{pages}{L26}.
\newblock \DOIprefix\doi{10.3847/2041-8213/ab8dca}.
\bibitem[{Zuber et~al.(2000)Zuber, Solomon, Phillips, Smith, Tyler, Aharonson,
  Balmino, Banerdt, Head, Johnson, Lemoine, McGovern, Neumann, Rowlands and
  Zhong}]{zuber2000internal}
\bibinfo{author}{Zuber, M.T.}, \bibinfo{author}{Solomon, S.C.},
  \bibinfo{author}{Phillips, R.J.}, \bibinfo{author}{Smith, D.E.},
  \bibinfo{author}{Tyler, G.L.}, \bibinfo{author}{Aharonson, O.},
  \bibinfo{author}{Balmino, G.}, \bibinfo{author}{Banerdt, W.B.},
  \bibinfo{author}{Head, J.W.}, \bibinfo{author}{Johnson, C.L.},
  \bibinfo{author}{Lemoine, F.G.}, \bibinfo{author}{McGovern, P.J.},
  \bibinfo{author}{Neumann, G.A.}, \bibinfo{author}{Rowlands, D.D.},
  \bibinfo{author}{Zhong, S.}, \bibinfo{year}{2000}.
\newblock \bibinfo{title}{{Internal structure and early thermal evolution of
  Mars from Mars Global Surveyor topography and gravity}}.
\newblock \bibinfo{journal}{Science} \bibinfo{volume}{287},
  \bibinfo{pages}{1788--1793}.
\newblock \DOIprefix\doi{10.1126/science.287.5459.1788}.

\end{thebibliography}


\begin{thebibliography}{124}
\expandafter\ifx\csname natexlab\endcsname\relax\def\natexlab#1{#1}\fi
\providecommand{\url}[1]{\texttt{#1}}
\providecommand{\href}[2]{#2}
\providecommand{\path}[1]{#1}
\providecommand{\DOIprefix}{doi:}
\providecommand{\ArXivprefix}{arXiv:}
\providecommand{\URLprefix}{URL: }
\providecommand{\Pubmedprefix}{pmid:}
\providecommand{\doi}[1]{\href{http://dx.doi.org/#1}{\path{#1}}}
\providecommand{\Pubmed}[1]{\href{pmid:#1}{\path{#1}}}
\providecommand{\bibinfo}[2]{#2}
\ifx\xfnm\relax \def\xfnm[#1]{\unskip,\space#1}\fi
\bibitem[{Albarede(2009)}]{albarede2009volatile}
\bibinfo{author}{Albarede, F.}, \bibinfo{year}{2009}.
\newblock \bibinfo{title}{{Volatile accretion history of the terrestrial
  planets and dynamic implications}}.
\newblock \bibinfo{journal}{Nature} \bibinfo{volume}{461},
  \bibinfo{pages}{1227--1233}.
\newblock \DOIprefix\doi{10.1038/nature08477}.
\bibitem[{Alexander(2019a)}]{alexander2019quantitative_CC}
\bibinfo{author}{Alexander, C.M.O'D.}, \bibinfo{year}{2019}a.
\newblock \bibinfo{title}{{Quantitative models for the elemental and isotopic
  fractionations in chondrites: The carbonaceous chondrites}}.
\newblock \bibinfo{journal}{Geochimica et Cosmochimica Acta}
  \bibinfo{volume}{254}, \bibinfo{pages}{277--309}.
\newblock \DOIprefix\doi{10.1016/j.gca.2019.02.008}.
\bibitem[{Alexander(2019b)}]{alexander2019quantitative_NC}
\bibinfo{author}{Alexander, C.M.O'D.}, \bibinfo{year}{2019}b.
\newblock \bibinfo{title}{{Quantitative models for the elemental and isotopic
  fractionations in the chondrites: The non-carbonaceous chondrites}}.
\newblock \bibinfo{journal}{Geochimica et Cosmochimica Acta}
  \bibinfo{volume}{254}, \bibinfo{pages}{246--276}.
\newblock \DOIprefix\doi{10.1016/j.gca.2019.01.026}.
\bibitem[{Alexander et~al.(2012)Alexander, Bowden, Fogel, Howard, Herd and
  Nittler}]{alexander2012provenances}
\bibinfo{author}{Alexander, C.M.O'D.}, \bibinfo{author}{Bowden, R.},
  \bibinfo{author}{Fogel, M.L.}, \bibinfo{author}{Howard, K.T.},
  \bibinfo{author}{Herd, C.D.K.}, \bibinfo{author}{Nittler, L.R.},
  \bibinfo{year}{2012}.
\newblock \bibinfo{title}{{The provenances of asteroids, and their
  contributions to the volatile inventories of the terrestrial planets}}.
\newblock \bibinfo{journal}{Science} \bibinfo{volume}{337},
  \bibinfo{pages}{721--723}.
\newblock \DOIprefix\doi{10.1126/science.1223474}.
\bibitem[{Armytage et~al.(2013)Armytage, Jephcoat, Bouhifd and
  Porcelli}]{armytage2013metal}
\bibinfo{author}{Armytage, R.M.}, \bibinfo{author}{Jephcoat, A.P.},
  \bibinfo{author}{Bouhifd, M.A.}, \bibinfo{author}{Porcelli, D.},
  \bibinfo{year}{2013}.
\newblock \bibinfo{title}{{Metal--silicate partitioning of iodine at high
  pressures and temperatures: Implications for the Earth's core and $
  \mathrm{^{129*}Xe} $ budgets}}.
\newblock \bibinfo{journal}{Earth and Planetary Science Letters}
  \bibinfo{volume}{373}, \bibinfo{pages}{140--149}.
\newblock \DOIprefix\doi{10.1016/j.epsl.2013.04.031}.
\bibitem[{Ballhaus et~al.(2017)Ballhaus, Fonseca, M{\"u}nker, Rohrbach, Nagel,
  Speelmanns, Helmy, Zirner, Vogel and Heuser}]{ballhaus2017great}
\bibinfo{author}{Ballhaus, C.}, \bibinfo{author}{Fonseca, R.O.C.},
  \bibinfo{author}{M{\"u}nker, C.}, \bibinfo{author}{Rohrbach, A.},
  \bibinfo{author}{Nagel, T.}, \bibinfo{author}{Speelmanns, I.M.},
  \bibinfo{author}{Helmy, H.M.}, \bibinfo{author}{Zirner, A.},
  \bibinfo{author}{Vogel, A.K.}, \bibinfo{author}{Heuser, A.},
  \bibinfo{year}{2017}.
\newblock \bibinfo{title}{{The great sulfur depletion of Earth's mantle is not
  a signature of mantle--core equilibration}}.
\newblock \bibinfo{journal}{Contributions to Mineralogy and Petrology}
  \bibinfo{volume}{172}, \bibinfo{pages}{68}.
\newblock \DOIprefix\doi{10.1007/s00410-017-1388-3}.
\bibitem[{Barrat et~al.(2016)Barrat, Dauphas, Gillet, Bollinger, Etoubleau,
  Bischoff and Yamaguchi}]{barrat2016evidence}
\bibinfo{author}{Barrat, J.A.}, \bibinfo{author}{Dauphas, N.},
  \bibinfo{author}{Gillet, P.}, \bibinfo{author}{Bollinger, C.},
  \bibinfo{author}{Etoubleau, J.}, \bibinfo{author}{Bischoff, A.},
  \bibinfo{author}{Yamaguchi, A.}, \bibinfo{year}{2016}.
\newblock \bibinfo{title}{{Evidence from Tm anomalies for non-CI refractory
  lithophile element proportions in terrestrial planets and achondrites}}.
\newblock \bibinfo{journal}{Geochimica et Cosmochimica Acta}
  \bibinfo{volume}{176}, \bibinfo{pages}{1--17}.
\newblock \DOIprefix\doi{10.1016/j.gca.2015.12.004}.
\bibitem[{Boujibar et~al.(2014)Boujibar, Andrault, Bouhifd, Bolfan-Casanova,
  Devidal and Trcera}]{boujibar2014metal}
\bibinfo{author}{Boujibar, A.}, \bibinfo{author}{Andrault, D.},
  \bibinfo{author}{Bouhifd, M.A.}, \bibinfo{author}{Bolfan-Casanova, N.},
  \bibinfo{author}{Devidal, J.L.}, \bibinfo{author}{Trcera, N.},
  \bibinfo{year}{2014}.
\newblock \bibinfo{title}{{Metal--silicate partitioning of sulphur, new
  experimental and thermodynamic constraints on planetary accretion}}.
\newblock \bibinfo{journal}{Earth and Planetary Science Letters}
  \bibinfo{volume}{391}, \bibinfo{pages}{42--54}.
\newblock \DOIprefix\doi{10.1016/j.epsl.2014.01.021}.
\bibitem[{Braukm{\"u}ller et~al.(2019)Braukm{\"u}ller, Wombacher, Funk and
  M{\"u}nker}]{braukmuller2019earth}
\bibinfo{author}{Braukm{\"u}ller, N.}, \bibinfo{author}{Wombacher, F.},
  \bibinfo{author}{Funk, C.}, \bibinfo{author}{M{\"u}nker, C.},
  \bibinfo{year}{2019}.
\newblock \bibinfo{title}{{Earth's volatile element depletion pattern inherited
  from a carbonaceous chondrite-like source}}.
\newblock \bibinfo{journal}{Nature Geoscience} \bibinfo{volume}{12},
  \bibinfo{pages}{564--568}.
\newblock \DOIprefix\doi{10.1038/s41561-019-0375-x}.
\bibitem[{Braukm{\"u}ller et~al.(2018)Braukm{\"u}ller, Wombacher, Hezel,
  Escoube and M{\"u}nker}]{braukmuller2018chemical}
\bibinfo{author}{Braukm{\"u}ller, N.}, \bibinfo{author}{Wombacher, F.},
  \bibinfo{author}{Hezel, D.C.}, \bibinfo{author}{Escoube, R.},
  \bibinfo{author}{M{\"u}nker, C.}, \bibinfo{year}{2018}.
\newblock \bibinfo{title}{{The chemical composition of carbonaceous chondrites:
  Implications for volatile element depletion, complementarity and
  alteration}}.
\newblock \bibinfo{journal}{Geochimica et Cosmochimica Acta}
  \bibinfo{volume}{239}, \bibinfo{pages}{17--48}.
\newblock \DOIprefix\doi{10.1016/j.gca.2018.07.023}.
\bibitem[{Brenan(2015)}]{brenan2015se}
\bibinfo{author}{Brenan, J.M.}, \bibinfo{year}{2015}.
\newblock \bibinfo{title}{{Se--Te fractionation by sulfide--silicate melt
  partitioning: implications for the composition of mantle-derived magmas and
  their melting residues}}.
\newblock \bibinfo{journal}{Earth and Planetary Science Letters}
  \bibinfo{volume}{422}, \bibinfo{pages}{45--57}.
\newblock \DOIprefix\doi{10.1016/j.epsl.2015.04.011}.
\bibitem[{Budde et~al.(2019)Budde, Burkhardt and Kleine}]{budde2019molybdenum}
\bibinfo{author}{Budde, G.}, \bibinfo{author}{Burkhardt, C.},
  \bibinfo{author}{Kleine, T.}, \bibinfo{year}{2019}.
\newblock \bibinfo{title}{{Molybdenum isotopic evidence for the late accretion
  of outer Solar System material to Earth}}.
\newblock \bibinfo{journal}{Nature Astronomy} \bibinfo{volume}{3},
  \bibinfo{pages}{736--741}.
\newblock \DOIprefix\doi{10.1038/s41550-019-0779-y}.
\bibitem[{Cassen(1996)}]{cassen1996models}
\bibinfo{author}{Cassen, P.}, \bibinfo{year}{1996}.
\newblock \bibinfo{title}{{Models for the fractionation of moderately volatile
  elements in the solar nebula}}.
\newblock \bibinfo{journal}{Meteoritics \& Planetary Science}
  \bibinfo{volume}{31}, \bibinfo{pages}{793--806}.
\newblock \DOIprefix\doi{10.1111/j.1945-5100.1996.tb02114.x}.
\bibitem[{Cassen(2001)}]{cassen2001nebular}
\bibinfo{author}{Cassen, P.}, \bibinfo{year}{2001}.
\newblock \bibinfo{title}{{Nebular thermal evolution and the properties of
  primitive planetary materials}}.
\newblock \bibinfo{journal}{Meteoritics \& Planetary Science}
  \bibinfo{volume}{36}, \bibinfo{pages}{671--700}.
\newblock \DOIprefix\doi{10.1111/j.1945-5100.2001.tb01908.x}.
\bibitem[{Chase(1998)}]{chase1998nist}
\bibinfo{author}{Chase, Jr., M.W.}, \bibinfo{year}{1998}.
\newblock \bibinfo{title}{{NIST-JANAF Thermochemical Tables, 4th Edition}}.
\newblock \DOIprefix\doi{10.18434/T42S31}.
\bibitem[{Chou(1978)}]{chou1978fractionation}
\bibinfo{author}{Chou, C.L.}, \bibinfo{year}{1978}.
\newblock \bibinfo{title}{{Fractionation of siderophile elements in the Earth's
  upper mantle}}, in: \bibinfo{booktitle}{Lunar and Planetary Science
  Conference Proceedings}, pp. \bibinfo{pages}{219--230}.
\bibitem[{Clay et~al.(2017)Clay, Burgess, Busemann, Ruzi{\'e}-Hamilton,
  Joachim, Day and Ballentine}]{clay2017halogens}
\bibinfo{author}{Clay, P.L.}, \bibinfo{author}{Burgess, R.},
  \bibinfo{author}{Busemann, H.}, \bibinfo{author}{Ruzi{\'e}-Hamilton, L.},
  \bibinfo{author}{Joachim, B.}, \bibinfo{author}{Day, J.M.D.},
  \bibinfo{author}{Ballentine, C.J.}, \bibinfo{year}{2017}.
\newblock \bibinfo{title}{{Halogens in chondritic meteorites and terrestrial
  accretion}}.
\newblock \bibinfo{journal}{Nature} \bibinfo{volume}{551},
  \bibinfo{pages}{614--618}.
\newblock \DOIprefix\doi{10.1038/nature24625}.
\bibitem[{Dahl and Stevenson(2010)}]{dahl2010turbulent}
\bibinfo{author}{Dahl, T.W.}, \bibinfo{author}{Stevenson, D.J.},
  \bibinfo{year}{2010}.
\newblock \bibinfo{title}{{Turbulent mixing of metal and silicate during planet
  accretion—And interpretation of the Hf--W chronometer}}.
\newblock \bibinfo{journal}{Earth and Planetary Science Letters}
  \bibinfo{volume}{295}, \bibinfo{pages}{177--186}.
\newblock \DOIprefix\doi{10.1016/j.epsl.2010.03.038}.
\bibitem[{Dauphas and Pourmand(2011)}]{dauphas2011hf}
\bibinfo{author}{Dauphas, N.}, \bibinfo{author}{Pourmand, A.},
  \bibinfo{year}{2011}.
\newblock \bibinfo{title}{{Hf--W--Th evidence for rapid growth of Mars and its
  status as a planetary embryo}}.
\newblock \bibinfo{journal}{Nature} \bibinfo{volume}{473},
  \bibinfo{pages}{489--492}.
\newblock \DOIprefix\doi{10.1038/nature10077}.
\bibitem[{Dauphas and Pourmand(2015)}]{dauphas2015thulium}
\bibinfo{author}{Dauphas, N.}, \bibinfo{author}{Pourmand, A.},
  \bibinfo{year}{2015}.
\newblock \bibinfo{title}{{Thulium anomalies and rare earth element patterns in
  meteorites and Earth: Nebular fractionation and the nugget effect}}.
\newblock \bibinfo{journal}{Geochimica et Cosmochimica Acta}
  \bibinfo{volume}{163}, \bibinfo{pages}{234--261}.
\newblock \DOIprefix\doi{10.1016/j.gca.2015.03.037}.
\bibitem[{Deguen et~al.(2014)Deguen, Landeau and Olson}]{deguen2014turbulent}
\bibinfo{author}{Deguen, R.}, \bibinfo{author}{Landeau, M.},
  \bibinfo{author}{Olson, P.}, \bibinfo{year}{2014}.
\newblock \bibinfo{title}{{Turbulent metal--silicate mixing, fragmentation, and
  equilibration in magma oceans}}.
\newblock \bibinfo{journal}{Earth and Planetary Science Letters}
  \bibinfo{volume}{391}, \bibinfo{pages}{274--287}.
\newblock \DOIprefix\doi{10.1016/j.epsl.2014.02.007}.
\bibitem[{Desch et~al.(2018)Desch, Kalyaan and Alexander}]{desch2017effect}
\bibinfo{author}{Desch, S.J.}, \bibinfo{author}{Kalyaan, A.},
  \bibinfo{author}{Alexander, C.M.O'D.}, \bibinfo{year}{2018}.
\newblock \bibinfo{title}{{The effect of Jupiter's formation on the
  distribution of refractory elements and inclusions in meteorites}}.
\newblock \bibinfo{journal}{The Astrophysical Journal Supplement Series}
  \bibinfo{volume}{238}, \bibinfo{pages}{11}.
\newblock \DOIprefix\doi{10.3847/1538-4365/aad95f}.
\bibitem[{Dreibus and Palme(1996)}]{dreibus1996cosmochemical}
\bibinfo{author}{Dreibus, G.}, \bibinfo{author}{Palme, H.},
  \bibinfo{year}{1996}.
\newblock \bibinfo{title}{{Cosmochemical constraints on the sulfur content in
  the Earth's core}}.
\newblock \bibinfo{journal}{Geochimica et Cosmochimica Acta}
  \bibinfo{volume}{60}, \bibinfo{pages}{1125--1130}.
\newblock \DOIprefix\doi{10.1016/0016-7037(96)00028-2}.
\bibitem[{Dreibus et~al.(1979)Dreibus, Spettel and
  W{\"a}nke}]{dreibus1979halogens}
\bibinfo{author}{Dreibus, G.}, \bibinfo{author}{Spettel, B.},
  \bibinfo{author}{W{\"a}nke, H.}, \bibinfo{year}{1979}.
\newblock \bibinfo{title}{{Halogens in meteorites and their primordial
  abundances}}.
\newblock \bibinfo{journal}{Physics and Chemistry of the Earth}
  \bibinfo{volume}{11}, \bibinfo{pages}{33--38}.
\newblock \DOIprefix\doi{10.1016/0079-1946(79)90005-3}.
\bibitem[{Fegley and Lodders(2018)}]{fegley2018volatile}
\bibinfo{author}{Fegley, Jr., B.}, \bibinfo{author}{Lodders, K.},
  \bibinfo{year}{2018}.
\newblock \bibinfo{title}{{Volatile element chemistry in the solar nebula -
  revisited 40 years later}}, in: \bibinfo{booktitle}{Nuclei in the Cosmos XV},
  p. \bibinfo{pages}{107}.
\bibitem[{Fegley et~al.(2020)Fegley, Lodders and Jacobsen}]{fegley2020volatile}
\bibinfo{author}{Fegley, Jr., B.}, \bibinfo{author}{Lodders, K.},
  \bibinfo{author}{Jacobsen, N.S.}, \bibinfo{year}{2020}.
\newblock \bibinfo{title}{{Volatile element chemistry during accretion of the
  Earth}}.
\newblock \bibinfo{journal}{Geochemistry} \bibinfo{volume}{80},
  \bibinfo{pages}{125594}.
\newblock \DOIprefix\doi{10.1016/j.chemer.2019.125594}.
\bibitem[{Fischer-G{\"o}dde et~al.(2020)Fischer-G{\"o}dde, Elfers, M{\"u}nker,
  Szilas, Maier, Messling, Morishita, Van~Kranendonk and
  Smithies}]{fischer2020ruthenium}
\bibinfo{author}{Fischer-G{\"o}dde, M.}, \bibinfo{author}{Elfers, B.M.},
  \bibinfo{author}{M{\"u}nker, C.}, \bibinfo{author}{Szilas, K.},
  \bibinfo{author}{Maier, W.D.}, \bibinfo{author}{Messling, N.},
  \bibinfo{author}{Morishita, T.}, \bibinfo{author}{Van~Kranendonk, M.},
  \bibinfo{author}{Smithies, H.}, \bibinfo{year}{2020}.
\newblock \bibinfo{title}{{Ruthenium isotope vestige of Earth's pre-late-veneer
  mantle preserved in Archaean rocks}}.
\newblock \bibinfo{journal}{Nature} \bibinfo{volume}{579},
  \bibinfo{pages}{240--244}.
\newblock \DOIprefix\doi{10.1038/s41586-020-2069-3}.
\bibitem[{Floss et~al.(1996)Floss, El~Goresy, Zinner, Kransel, Rammensee and
  Palme}]{floss1996elemental}
\bibinfo{author}{Floss, C.}, \bibinfo{author}{El~Goresy, A.},
  \bibinfo{author}{Zinner, E.}, \bibinfo{author}{Kransel, G.},
  \bibinfo{author}{Rammensee, W.}, \bibinfo{author}{Palme, H.},
  \bibinfo{year}{1996}.
\newblock \bibinfo{title}{{Elemental and isotopic fractionations produced
  through evaporation of the Allende CV chondrilte: Implications for the origin
  of HAL-type hibonite inclusions}}.
\newblock \bibinfo{journal}{Geochimica et Cosmochimica Acta}
  \bibinfo{volume}{60}, \bibinfo{pages}{1975--1997}.
\newblock \DOIprefix\doi{10.1016/0016-7037(96)00068-3}.
\bibitem[{Gaetani and Grove(1997)}]{gaetani1997partitioning}
\bibinfo{author}{Gaetani, G.A.}, \bibinfo{author}{Grove, T.L.},
  \bibinfo{year}{1997}.
\newblock \bibinfo{title}{{Partitioning of moderately siderophile elements
  among olivine, silicate melt, and sulfide melt: Constraints on core formation
  in the Earth and Mars}}.
\newblock \bibinfo{journal}{Geochimica et Cosmochimica Acta}
  \bibinfo{volume}{61}, \bibinfo{pages}{1829--1846}.
\newblock \DOIprefix\doi{10.1016/S0016-7037(97)00033-1}.
\bibitem[{Gaetani and Grove(1999)}]{gaetani1999wetting}
\bibinfo{author}{Gaetani, G.A.}, \bibinfo{author}{Grove, T.L.},
  \bibinfo{year}{1999}.
\newblock \bibinfo{title}{{Wetting of mantle olivine by sulfide melt:
  implications for Re/Os ratios in mantle peridotite and late-stage core
  formation}}.
\newblock \bibinfo{journal}{Earth and Planetary Science Letters}
  \bibinfo{volume}{169}, \bibinfo{pages}{147--163}.
\newblock \DOIprefix\doi{10.1016/S0012-821X(99)00062-X}.
\bibitem[{Gellissen et~al.(2019)Gellissen, Holzheid, Kegler and
  Palme}]{gellissen2019heating}
\bibinfo{author}{Gellissen, M.}, \bibinfo{author}{Holzheid, A.},
  \bibinfo{author}{Kegler, P.}, \bibinfo{author}{Palme, H.},
  \bibinfo{year}{2019}.
\newblock \bibinfo{title}{{Heating experiments relevant to the depletion of Na,
  K and Mn in the Earth and other planetary bodies}}.
\newblock \bibinfo{journal}{Geochemistry} \bibinfo{volume}{79},
  \bibinfo{pages}{125540}.
\newblock \DOIprefix\doi{10.1016/j.chemer.2019.125540}.
\bibitem[{Herzog et~al.(2009)Herzog, Moynier, Albar{\`e}de and
  Berezhnoy}]{herzog2009isotopic}
\bibinfo{author}{Herzog, G.F.}, \bibinfo{author}{Moynier, F.},
  \bibinfo{author}{Albar{\`e}de, F.}, \bibinfo{author}{Berezhnoy, A.A.},
  \bibinfo{year}{2009}.
\newblock \bibinfo{title}{{Isotopic and elemental abundances of copper and zinc
  in lunar samples, Zagami, Pele's hairs, and a terrestrial basalt}}.
\newblock \bibinfo{journal}{Geochimica et Cosmochimica Acta}
  \bibinfo{volume}{73}, \bibinfo{pages}{5884--5904}.
\newblock \DOIprefix\doi{10.1016/j.gca.2009.05.067}.
\bibitem[{Hezel et~al.(2018)Hezel, Harak and Libourel}]{hezel2018what}
\bibinfo{author}{Hezel, D.C.}, \bibinfo{author}{Harak, M.},
  \bibinfo{author}{Libourel, G.}, \bibinfo{year}{2018}.
\newblock \bibinfo{title}{{What we know about elemental bulk chondrule and
  matrix compositions: Presenting the ChondriteDB Database}}.
\newblock \bibinfo{journal}{Chemie der Erde-Geochemistry} \bibinfo{volume}{78},
  \bibinfo{pages}{1--14}.
\newblock \DOIprefix\doi{10.1016/j.chemer.2017.05.003}.
\bibitem[{Hezel et~al.(2008)Hezel, Russell, Ross and Kearsley}]{hezel2008modal}
\bibinfo{author}{Hezel, D.C.}, \bibinfo{author}{Russell, S.S.},
  \bibinfo{author}{Ross, A.J.}, \bibinfo{author}{Kearsley, A.T.},
  \bibinfo{year}{2008}.
\newblock \bibinfo{title}{{Modal abundances of CAIs: Implications for bulk
  chondrite element abundances and fractionations}}.
\newblock \bibinfo{journal}{Meteoritics \& Planetary Science}
  \bibinfo{volume}{43}, \bibinfo{pages}{1879--1894}.
\newblock \DOIprefix\doi{10.1111/j.1945-5100.2008.tb00649.x}.
\bibitem[{Humayun and Clayton(1995)}]{humayun1995potassium}
\bibinfo{author}{Humayun, M.}, \bibinfo{author}{Clayton, R.N.},
  \bibinfo{year}{1995}.
\newblock \bibinfo{title}{{Potassium isotope cosmochemistry: Genetic
  implications of volatile element depletion}}.
\newblock \bibinfo{journal}{Geochimica et Cosmochimica Acta}
  \bibinfo{volume}{59}, \bibinfo{pages}{2131--2148}.
\newblock \DOIprefix\doi{10.1016/0016-7037(95)00132-8}.
\bibitem[{Jackson et~al.(2018)Jackson, Bennett, Du, Cottrell and
  Fei}]{jackson2018early}
\bibinfo{author}{Jackson, C.R.}, \bibinfo{author}{Bennett, N.R.},
  \bibinfo{author}{Du, Z.}, \bibinfo{author}{Cottrell, E.},
  \bibinfo{author}{Fei, Y.}, \bibinfo{year}{2018}.
\newblock \bibinfo{title}{{Early episodes of high-pressure core formation
  preserved in plume mantle}}.
\newblock \bibinfo{journal}{Nature} \bibinfo{volume}{553},
  \bibinfo{pages}{491--495}.
\newblock \DOIprefix\doi{10.1038/nature25446}.
\bibitem[{Jacobsen(2005)}]{jacobsen2005hf}
\bibinfo{author}{Jacobsen, S.B.}, \bibinfo{year}{2005}.
\newblock \bibinfo{title}{{The Hf-W isotopic system and the origin of the Earth
  and Moon}}.
\newblock \bibinfo{journal}{Annual Review of Earth and Planetary Sciences}
  \bibinfo{volume}{33}, \bibinfo{pages}{531--570}.
\newblock \DOIprefix\doi{10.1146/annurev.earth.33.092203.122614}.
\bibitem[{Jaupart et~al.(2015)Jaupart, Labrosse, Lucazeau and
  Mareschal}]{jaupart2015temperatures}
\bibinfo{author}{Jaupart, C.}, \bibinfo{author}{Labrosse, S.},
  \bibinfo{author}{Lucazeau, F.}, \bibinfo{author}{Mareschal, J.C.},
  \bibinfo{year}{2015}.
\newblock \bibinfo{title}{Temperatures, heat, and energy in the mantle of the
  earth}, in: \bibinfo{editor}{Schubert, G.} (Ed.),
  \bibinfo{booktitle}{Treatise on Geophysics (Second Edition)}.
  \bibinfo{publisher}{Elsevier}, \bibinfo{address}{Oxford}.
  volume~\bibinfo{volume}{7}, pp. \bibinfo{pages}{223--270}.
\newblock \DOIprefix\doi{10.1016/B978-0-444-53802-4.00126-3}.
\bibitem[{Jones and Drake(1986)}]{jones1986geochemical}
\bibinfo{author}{Jones, J.H.}, \bibinfo{author}{Drake, M.J.},
  \bibinfo{year}{1986}.
\newblock \bibinfo{title}{{Geochemical constraints on core formation in the
  Earth}}.
\newblock \bibinfo{journal}{Nature} \bibinfo{volume}{322},
  \bibinfo{pages}{221--228}.
\newblock \DOIprefix\doi{10.1038/322221a0}.
\bibitem[{Khan and Connolly(2008)}]{khan2008constraining}
\bibinfo{author}{Khan, A.}, \bibinfo{author}{Connolly, J.A.D.},
  \bibinfo{year}{2008}.
\newblock \bibinfo{title}{{Constraining the composition and thermal state of
  Mars from inversion of geophysical data}}.
\newblock \bibinfo{journal}{Journal of Geophysical Research: Planets}
  \bibinfo{volume}{113}, \bibinfo{pages}{E07003}.
\newblock \DOIprefix\doi{10.1029/2007JE002996}.
\bibitem[{Khan et~al.(2018)Khan, Liebske, Rozel, Rivoldini, Nimmo, Connolly,
  Plesa and Giardini}]{khan2018geophysical}
\bibinfo{author}{Khan, A.}, \bibinfo{author}{Liebske, C.},
  \bibinfo{author}{Rozel, A.}, \bibinfo{author}{Rivoldini, A.},
  \bibinfo{author}{Nimmo, F.}, \bibinfo{author}{Connolly, J.A.D.},
  \bibinfo{author}{Plesa, A.C.}, \bibinfo{author}{Giardini, D.},
  \bibinfo{year}{2018}.
\newblock \bibinfo{title}{{A geophysical perspective on the bulk composition of
  Mars}}.
\newblock \bibinfo{journal}{Journal of Geophysical Research: Planets}
  \bibinfo{volume}{123}, \bibinfo{pages}{575--611}.
\newblock \DOIprefix\doi{10.1002/2017JE005371}.
\bibitem[{Kimura et~al.(1974)Kimura, Lewis and Anders}]{kimura1974distribution}
\bibinfo{author}{Kimura, K.}, \bibinfo{author}{Lewis, R.S.},
  \bibinfo{author}{Anders, E.}, \bibinfo{year}{1974}.
\newblock \bibinfo{title}{{Distribution of gold and rhenium between nickel-iron
  and silicate melts: implications for the abundance of siderophile elements on
  the Earth and Moon}}.
\newblock \bibinfo{journal}{Geochimica et Cosmochimica Acta}
  \bibinfo{volume}{38}, \bibinfo{pages}{683--701}.
\newblock \DOIprefix\doi{10.1016/0016-7037(74)90144-6}.
\bibitem[{Kiseeva and Wood(2013)}]{kiseeva2013simple}
\bibinfo{author}{Kiseeva, E.S.}, \bibinfo{author}{Wood, B.J.},
  \bibinfo{year}{2013}.
\newblock \bibinfo{title}{{A simple model for chalcophile element partitioning
  between sulphide and silicate liquids with geochemical applications}}.
\newblock \bibinfo{journal}{Earth and Planetary Science Letters}
  \bibinfo{volume}{383}, \bibinfo{pages}{68--81}.
\newblock \DOIprefix\doi{10.1016/j.epsl.2013.09.034}.
\bibitem[{Kleine et~al.(2020)Kleine, Budde, Burkhardt, Kruijer, Worsham,
  Morbidelli and Nimmo}]{kleine2020non}
\bibinfo{author}{Kleine, T.}, \bibinfo{author}{Budde, G.},
  \bibinfo{author}{Burkhardt, C.}, \bibinfo{author}{Kruijer, T.S.},
  \bibinfo{author}{Worsham, E.A.}, \bibinfo{author}{Morbidelli, A.},
  \bibinfo{author}{Nimmo, F.}, \bibinfo{year}{2020}.
\newblock \bibinfo{title}{{The Non-carbonaceous--Carbonaceous Meteorite
  Dichotomy}}.
\newblock \bibinfo{journal}{Space Science Review} \bibinfo{volume}{216},
  \bibinfo{pages}{55}.
\newblock \DOIprefix\doi{10.1007/s11214-020-00675-w}.
\bibitem[{Kleine et~al.(2009)Kleine, Touboul, Bourdon, Nimmo, Mezger, Palme,
  Jacobsen, Yin and Halliday}]{kleine2009hf}
\bibinfo{author}{Kleine, T.}, \bibinfo{author}{Touboul, M.},
  \bibinfo{author}{Bourdon, B.}, \bibinfo{author}{Nimmo, F.},
  \bibinfo{author}{Mezger, K.}, \bibinfo{author}{Palme, H.},
  \bibinfo{author}{Jacobsen, S.B.}, \bibinfo{author}{Yin, Q.Z.},
  \bibinfo{author}{Halliday, A.N.}, \bibinfo{year}{2009}.
\newblock \bibinfo{title}{{Hf--W chronology of the accretion and early
  evolution of asteroids and terrestrial planets}}.
\newblock \bibinfo{journal}{Geochimica et Cosmochimica Acta}
  \bibinfo{volume}{73}, \bibinfo{pages}{5150--5188}.
\newblock \DOIprefix\doi{10.1016/j.gca.2008.11.047}.
\bibitem[{Klock and Palme(1988)}]{klock1988partitioning}
\bibinfo{author}{Klock, W.}, \bibinfo{author}{Palme, H.}, \bibinfo{year}{1988}.
\newblock \bibinfo{title}{{Partitioning of siderophile and chalcophile elements
  between sulfide, olivine, and glass in a naturally reduced basalt from Disko
  Island, Greenland}}, in: \bibinfo{booktitle}{Proceedings of Lunar and
  Planetary Science Conference}, pp. \bibinfo{pages}{471--483}.
\bibitem[{Kramers(2003)}]{kramers2003volatile}
\bibinfo{author}{Kramers, J.D.}, \bibinfo{year}{2003}.
\newblock \bibinfo{title}{{Volatile element abundance patterns and an early
  liquid water ocean on Earth}}.
\newblock \bibinfo{journal}{Precambrian Research} \bibinfo{volume}{126},
  \bibinfo{pages}{379--394}.
\newblock \DOIprefix\doi{10.1016/S0301-9268(03)00106-2}.
\bibitem[{Li and Aud{\'e}tat(2015)}]{li2015effects}
\bibinfo{author}{Li, Y.}, \bibinfo{author}{Aud{\'e}tat, A.},
  \bibinfo{year}{2015}.
\newblock \bibinfo{title}{{Effects of temperature, silicate melt composition,
  and oxygen fugacity on the partitioning of V, Mn, Co, Ni, Cu, Zn, As, Mo, Ag,
  Sn, Sb, W, Au, Pb, and Bi between sulfide phases and silicate melt}}.
\newblock \bibinfo{journal}{Geochimica et Cosmochimica Acta}
  \bibinfo{volume}{162}, \bibinfo{pages}{25--45}.
\newblock \DOIprefix\doi{10.1016/j.gca.2015.04.036}.
\bibitem[{Lodders(2003)}]{lodders2003solar}
\bibinfo{author}{Lodders, K.}, \bibinfo{year}{2003}.
\newblock \bibinfo{title}{{Solar system abundances and condensation
  temperatures of the elements}}.
\newblock \bibinfo{journal}{The Astrophysical Journal} \bibinfo{volume}{591},
  \bibinfo{pages}{1220--1247}.
\newblock \DOIprefix\doi{10.1086/375492}.
\bibitem[{Lodders(2020)}]{lodders2020solar}
\bibinfo{author}{Lodders, K.}, \bibinfo{year}{2020}.
\newblock \bibinfo{title}{{Solar Elemental Abundances}}, in:
  \bibinfo{editor}{Read, P.} (Ed.), \bibinfo{booktitle}{Oxford Research
  Encyclopedia of Planetary Science}. \bibinfo{publisher}{Oxford University
  Press}, \bibinfo{address}{Oxford}, pp. \bibinfo{pages}{1--68}.
\newblock \DOIprefix\doi{10.1093/acrefore/9780190647926.013.145}.
\bibitem[{Lodders and Fegley(1997)}]{lodders1997oxygen}
\bibinfo{author}{Lodders, K.}, \bibinfo{author}{Fegley, Jr., B.},
  \bibinfo{year}{1997}.
\newblock \bibinfo{title}{{An oxygen isotope model for the composition of
  Mars}}.
\newblock \bibinfo{journal}{Icarus} \bibinfo{volume}{126},
  \bibinfo{pages}{373--394}.
\newblock \DOIprefix\doi{10.1006/icar.1996.5653}.
\bibitem[{Lodders and Palme(1991)}]{lodders1991chalcophile}
\bibinfo{author}{Lodders, K.}, \bibinfo{author}{Palme, H.},
  \bibinfo{year}{1991}.
\newblock \bibinfo{title}{{On the chalcophile character of molybdenum:
  determination of sulfide/silicate partition coefficients of Mo and W}}.
\newblock \bibinfo{journal}{Earth and Planetary Science Letters}
  \bibinfo{volume}{103}, \bibinfo{pages}{311--324}.
\newblock \DOIprefix\doi{10.1016/0012-821X(91)90169-I}.
\bibitem[{Lorand and Conqu{\'e}r{\'e}(1983)}]{lorand1983contribution}
\bibinfo{author}{Lorand, J.P.}, \bibinfo{author}{Conqu{\'e}r{\'e}, F.},
  \bibinfo{year}{1983}.
\newblock \bibinfo{title}{{Contribution {\`a} l'{\'e}tude des sulfures dans les
  enclaves de lherzolite {\`a} spinelle des basaltes alcalins (Massif Central
  et Languedoc, France)}}.
\newblock \bibinfo{journal}{Bulletin de Min{\'e}ralogie} \bibinfo{volume}{106},
  \bibinfo{pages}{585--606}.
\newblock \DOIprefix\doi{10.3406/bulmi.1983.7737}.
\bibitem[{Mason and Martin(1977)}]{mason1977geochemical}
\bibinfo{author}{Mason, B.}, \bibinfo{author}{Martin, P.M.},
  \bibinfo{year}{1977}.
\newblock \bibinfo{title}{{Geochemical differences among components of the
  Allende meteorite}}.
\newblock \bibinfo{journal}{Smithsonian Contributions to the Earth Sciences}
  \bibinfo{volume}{19}, \bibinfo{pages}{84--95}.
\newblock \DOIprefix\doi{10.5479/si.00810274.22.1}.
\bibitem[{Masuda and Tanaka(1979)}]{masuda1979experimental}
\bibinfo{author}{Masuda, A.}, \bibinfo{author}{Tanaka, T.},
  \bibinfo{year}{1979}.
\newblock \bibinfo{title}{{Experimental studies on behaviors of major and minor
  lithophile elements in vaporization under evacuated condition}}.
\newblock \bibinfo{journal}{Meteoritics} \bibinfo{volume}{14},
  \bibinfo{pages}{13--28}.
\newblock \DOIprefix\doi{10.1111/j.1945-5100.1979.tb00475.x}.
\bibitem[{McDonough(2014)}]{mcdonough2014compositional}
\bibinfo{author}{McDonough, W.F.}, \bibinfo{year}{2014}.
\newblock \bibinfo{title}{{Compositional model for the Earth's core}}, in:
  \bibinfo{editor}{Holland, H.D.}, \bibinfo{editor}{Turekian, K.K.} (Eds.),
  \bibinfo{booktitle}{Treatise on Geochemistry (Second Edition)}.
  \bibinfo{publisher}{Elsevier}, \bibinfo{address}{Oxford}.
  volume~\bibinfo{volume}{3}, pp. \bibinfo{pages}{559--577}.
\newblock \DOIprefix\doi{10.1016/B978-0-08-095975-7.00215-1}.
\bibitem[{McDonough and Sun(1995)}]{mcdonough1995composition}
\bibinfo{author}{McDonough, W.F.}, \bibinfo{author}{Sun, S.s.},
  \bibinfo{year}{1995}.
\newblock \bibinfo{title}{{The composition of the Earth}}.
\newblock \bibinfo{journal}{Chemical Geology} \bibinfo{volume}{120},
  \bibinfo{pages}{223--253}.
\newblock \DOIprefix\doi{10.1016/0009-2541(94)00140-4}.
\bibitem[{{McDonough} et~al.(2020){McDonough}, {{\v{S}}r{\'a}mek} and
  {Wipperfurth}}]{mcdonough2020radiogenic}
\bibinfo{author}{{McDonough}, W.F.}, \bibinfo{author}{{{\v{S}}r{\'a}mek}, O.},
  \bibinfo{author}{{Wipperfurth}, S.A.}, \bibinfo{year}{2020}.
\newblock \bibinfo{title}{{Radiogenic power and geoneutrino luminosity of the
  Earth and other terrestrial bodies through time}}.
\newblock \bibinfo{journal}{Geochemistry, Geophysics, Geosystems}
  \bibinfo{volume}{21}, \bibinfo{pages}{e2019GC008865}.
\newblock \DOIprefix\doi{10.1029/2019GC008865}.
\bibitem[{MetBase(1994-2017)}]{metbase}
\bibinfo{author}{MetBase}, \bibinfo{year}{1994-2017}.
\newblock \bibinfo{title}{{Meteorite Information Database}}.
\newblock \URLprefix \url{http://www.metbase.org}. \bibinfo{note}{geoPlatform
  UG, Germany. Accessed: 2019-11-3}.
\bibitem[{Morgan and Anders(1979)}]{morgan1979chemical}
\bibinfo{author}{Morgan, J.W.}, \bibinfo{author}{Anders, E.},
  \bibinfo{year}{1979}.
\newblock \bibinfo{title}{{Chemical composition of Mars}}.
\newblock \bibinfo{journal}{Geochimica et Cosmochimica Acta}
  \bibinfo{volume}{43}, \bibinfo{pages}{1601--1610}.
\newblock \DOIprefix\doi{10.1016/0016-7037(79)90180-7}.
\bibitem[{Nimmo and Agnor(2006)}]{nimmo2006isotopic}
\bibinfo{author}{Nimmo, F.}, \bibinfo{author}{Agnor, C.B.},
  \bibinfo{year}{2006}.
\newblock \bibinfo{title}{{Isotopic outcomes of N-body accretion simulations:
  constraints on equilibration processes during large impacts from Hf/W
  observations}}.
\newblock \bibinfo{journal}{Earth and Planetary Science Letters}
  \bibinfo{volume}{243}, \bibinfo{pages}{26--43}.
\newblock \DOIprefix\doi{10.1016/j.epsl.2005.12.009}.
\bibitem[{Nimmo and Faul(2013)}]{nimmo2013dissipation}
\bibinfo{author}{Nimmo, F.}, \bibinfo{author}{Faul, U.H.},
  \bibinfo{year}{2013}.
\newblock \bibinfo{title}{{Dissipation at tidal and seismic frequencies in a
  melt-free, anhydrous Mars}}.
\newblock \bibinfo{journal}{Journal of Geophysical Research: Planets}
  \bibinfo{volume}{118}, \bibinfo{pages}{2558--2569}.
\newblock \DOIprefix\doi{10.1002/2013JE004499}.
\bibitem[{Norman et~al.(2006)Norman, Yaxley, Bennett and
  Brandon}]{norman2006magnesium}
\bibinfo{author}{Norman, M.D.}, \bibinfo{author}{Yaxley, G.M.},
  \bibinfo{author}{Bennett, V.C.}, \bibinfo{author}{Brandon, A.D.},
  \bibinfo{year}{2006}.
\newblock \bibinfo{title}{{Magnesium isotopic composition of olivine from the
  Earth, Mars, Moon, and pallasite parent body}}.
\newblock \bibinfo{journal}{Geophysical Research Letters} \bibinfo{volume}{33},
  \bibinfo{pages}{L15202}.
\newblock \DOIprefix\doi{10.1029/2006GL026446}.
\bibitem[{Notsu et~al.(1978)Notsu, Onuma, Nishida and Nagasawa}]{notsu1978high}
\bibinfo{author}{Notsu, K.}, \bibinfo{author}{Onuma, N.},
  \bibinfo{author}{Nishida, N.}, \bibinfo{author}{Nagasawa, H.},
  \bibinfo{year}{1978}.
\newblock \bibinfo{title}{{High temperature heating of the Allende meteorite}}.
\newblock \bibinfo{journal}{Geochimica et Cosmochimica Acta}
  \bibinfo{volume}{42}, \bibinfo{pages}{903--907}.
\newblock \DOIprefix\doi{10.1016/0016-7037(78)90101-1}.
\bibitem[{O'Neill(1991a)}]{oneil1991origin}
\bibinfo{author}{O'Neill, H.S.C.}, \bibinfo{year}{1991}a.
\newblock \bibinfo{title}{{The origin of the Moon and the early history of the
  Earth--A chemical model. Part 1: The Moon}}.
\newblock \bibinfo{journal}{Geochimica et Cosmochimica Acta}
  \bibinfo{volume}{55}, \bibinfo{pages}{1135--1157}.
\newblock \DOIprefix\doi{10.1016/0016-7037(91)90168-5}.
\bibitem[{O'Neill(1991b)}]{oneil1991origin_Earth}
\bibinfo{author}{O'Neill, H.S.C.}, \bibinfo{year}{1991}b.
\newblock \bibinfo{title}{{The origin of the Moon and the early history of the
  Earth--A chemical model. Part 2: The Earth}}.
\newblock \bibinfo{journal}{Geochimica et Cosmochimica Acta}
  \bibinfo{volume}{55}, \bibinfo{pages}{1159--1172}.
\newblock \DOIprefix\doi{10.1016/0016-7037(91)90169-6}.
\bibitem[{O'Neill and Palme(2008)}]{oneill2008collisional}
\bibinfo{author}{O'Neill, H.S.C.}, \bibinfo{author}{Palme, H.},
  \bibinfo{year}{2008}.
\newblock \bibinfo{title}{{Collisional erosion and the non-chondritic
  composition of the terrestrial planets}}.
\newblock \bibinfo{journal}{Philosophical Transactions of the Royal Society of
  London A: Mathematical, Physical and Engineering Sciences}
  \bibinfo{volume}{366}, \bibinfo{pages}{4205--4238}.
\newblock \DOIprefix\doi{10.1098/rsta.2008.0111}.
\bibitem[{Palme and O'Neill(2014)}]{palme2014cosmochemical}
\bibinfo{author}{Palme, H.}, \bibinfo{author}{O'Neill, H.S.C.},
  \bibinfo{year}{2014}.
\newblock \bibinfo{title}{Cosmochemical estimates of mantle composition}, in:
  \bibinfo{editor}{Holland, H.D.}, \bibinfo{editor}{Turekian, K.K.} (Eds.),
  \bibinfo{booktitle}{Treatise on Geochemistry (Second Edition)}.
  \bibinfo{publisher}{Elsevier}, \bibinfo{address}{Oxford}.
  volume~\bibinfo{volume}{3}, pp. \bibinfo{pages}{1--39}.
\newblock \DOIprefix\doi{10.1016/B978-0-08-095975-7.00201-1}.
\bibitem[{Paniello et~al.(2012)Paniello, Day and Moynier}]{paniello2012zinc}
\bibinfo{author}{Paniello, R.C.}, \bibinfo{author}{Day, J.M.D.},
  \bibinfo{author}{Moynier, F.}, \bibinfo{year}{2012}.
\newblock \bibinfo{title}{{Zinc isotopic evidence for the origin of the Moon}}.
\newblock \bibinfo{journal}{Nature} \bibinfo{volume}{490},
  \bibinfo{pages}{376--379}.
\newblock \DOIprefix\doi{10.1038/nature11507}.
\bibitem[{Parro et~al.(2017)Parro, Jim{\'e}nez-D{\'\i}az, Mansilla and
  Ruiz}]{parro2017present}
\bibinfo{author}{Parro, L.M.}, \bibinfo{author}{Jim{\'e}nez-D{\'\i}az, A.},
  \bibinfo{author}{Mansilla, F.}, \bibinfo{author}{Ruiz, J.},
  \bibinfo{year}{2017}.
\newblock \bibinfo{title}{{Present-day heat flow model of Mars}}.
\newblock \bibinfo{journal}{Scientific Reports} \bibinfo{volume}{7},
  \bibinfo{pages}{45629}.
\newblock \DOIprefix\doi{10.1038/srep45629}.
\bibitem[{Patten et~al.(2013)Patten, Barnes, Mathez and
  Jenner}]{patten2013partition}
\bibinfo{author}{Patten, C.}, \bibinfo{author}{Barnes, S.J.},
  \bibinfo{author}{Mathez, E.A.}, \bibinfo{author}{Jenner, F.E.},
  \bibinfo{year}{2013}.
\newblock \bibinfo{title}{{Partition coefficients of chalcophile elements
  between sulfide and silicate melts and the early crystallization history of
  sulfide liquid: LA-ICP-MS analysis of MORB sulfide droplets}}.
\newblock \bibinfo{journal}{Chemical Geology} \bibinfo{volume}{358},
  \bibinfo{pages}{170--188}.
\newblock \DOIprefix\doi{10.1016/j.chemgeo.2013.08.040}.
\bibitem[{Poitrasson et~al.(2004)Poitrasson, Halliday, Lee, Levasseur and
  Teutsch}]{poitrasson2004iron}
\bibinfo{author}{Poitrasson, F.}, \bibinfo{author}{Halliday, A.N.},
  \bibinfo{author}{Lee, D.C.}, \bibinfo{author}{Levasseur, S.},
  \bibinfo{author}{Teutsch, N.}, \bibinfo{year}{2004}.
\newblock \bibinfo{title}{{Iron isotope differences between Earth, Moon, Mars
  and Vesta as possible records of contrasted accretion mechanisms}}.
\newblock \bibinfo{journal}{Earth and Planetary Science Letters}
  \bibinfo{volume}{223}, \bibinfo{pages}{253--266}.
\newblock \DOIprefix\doi{10.1016/j.epsl.2004.04.032}.
\bibitem[{Righter(2011)}]{righter2011prediction}
\bibinfo{author}{Righter, K.}, \bibinfo{year}{2011}.
\newblock \bibinfo{title}{{Prediction of metal--silicate partition coefficients
  for siderophile elements: an update and assessment of PT conditions for
  metal--silicate equilibrium during accretion of the Earth}}.
\newblock \bibinfo{journal}{Earth and Planetary Science Letters}
  \bibinfo{volume}{304}, \bibinfo{pages}{158--167}.
\newblock \DOIprefix\doi{10.1016/j.epsl.2011.01.028}.
\bibitem[{Righter and Chabot(2011)}]{righter2011moderately}
\bibinfo{author}{Righter, K.}, \bibinfo{author}{Chabot, N.L.},
  \bibinfo{year}{2011}.
\newblock \bibinfo{title}{{Moderately and slightly siderophile element
  constraints on the depth and extent of melting in early Mars}}.
\newblock \bibinfo{journal}{Meteoritics \& Planetary Science}
  \bibinfo{volume}{46}, \bibinfo{pages}{157--176}.
\newblock \DOIprefix\doi{10.1111/j.1945-5100.2010.01140.x}.
\bibitem[{Righter and Drake(1997)}]{righter1997metal}
\bibinfo{author}{Righter, K.}, \bibinfo{author}{Drake, M.J.},
  \bibinfo{year}{1997}.
\newblock \bibinfo{title}{{Metal-silicate equilibrium in a homogeneously
  accreting earth: new results for Re}}.
\newblock \bibinfo{journal}{Earth and Planetary Science Letters}
  \bibinfo{volume}{146}, \bibinfo{pages}{541--553}.
\newblock \DOIprefix\doi{10.1016/S0012-821X(96)00243-9}.
\bibitem[{Righter et~al.(2017)Righter, Nickodem, Pando, Danielson, Boujibar,
  Righter and Lapen}]{righter2017distribution}
\bibinfo{author}{Righter, K.}, \bibinfo{author}{Nickodem, K.},
  \bibinfo{author}{Pando, K.}, \bibinfo{author}{Danielson, L.},
  \bibinfo{author}{Boujibar, A.}, \bibinfo{author}{Righter, M.},
  \bibinfo{author}{Lapen, T.J.}, \bibinfo{year}{2017}.
\newblock \bibinfo{title}{{Distribution of Sb, As, Ge, and In between metal and
  silicate during accretion and core formation in the Earth}}.
\newblock \bibinfo{journal}{Geochimica et Cosmochimica Acta}
  \bibinfo{volume}{198}, \bibinfo{pages}{1--16}.
\newblock \DOIprefix\doi{10.1016/j.gca.2016.10.045}.
\bibitem[{Righter et~al.(2018a)Righter, Pando, Humayun, Waeselmann, Yang,
  Boujibar and Danielson}]{righter2018effect}
\bibinfo{author}{Righter, K.}, \bibinfo{author}{Pando, K.},
  \bibinfo{author}{Humayun, M.}, \bibinfo{author}{Waeselmann, N.},
  \bibinfo{author}{Yang, S.}, \bibinfo{author}{Boujibar, A.},
  \bibinfo{author}{Danielson, L.R.}, \bibinfo{year}{2018}a.
\newblock \bibinfo{title}{{Effect of silicon on activity coefficients of
  siderophile elements (Au, Pd, Pt, P, Ga, Cu, Zn, and Pb) in liquid Fe: Roles
  of core formation, late sulfide matte, and late veneer in shaping terrestrial
  mantle geochemistry}}.
\newblock \bibinfo{journal}{Geochimica et Cosmochimica Acta}
  \bibinfo{volume}{232}, \bibinfo{pages}{101--123}.
\newblock \DOIprefix\doi{10.1016/j.gca.2018.04.011}.
\bibitem[{Righter et~al.(2018b)Righter, Pando, Marin, Ross, Righter, Danielson,
  Lapen and Lee}]{righter2018volatile}
\bibinfo{author}{Righter, K.}, \bibinfo{author}{Pando, K.},
  \bibinfo{author}{Marin, N.}, \bibinfo{author}{Ross, D.K.},
  \bibinfo{author}{Righter, M.}, \bibinfo{author}{Danielson, L.},
  \bibinfo{author}{Lapen, T.J.}, \bibinfo{author}{Lee, C.},
  \bibinfo{year}{2018}b.
\newblock \bibinfo{title}{{Volatile element signatures in the mantles of Earth,
  Moon, and Mars: Core formation fingerprints from Bi, Cd, In, and Sn}}.
\newblock \bibinfo{journal}{Meteoritics \& Planetary Science}
  \bibinfo{volume}{53}, \bibinfo{pages}{284--305}.
\newblock \DOIprefix\doi{10.1016/j.gca.2017.05.024}.
\bibitem[{Righter et~al.(2010)Righter, Pando, Danielson and
  Lee}]{righter2010partitioning}
\bibinfo{author}{Righter, K.}, \bibinfo{author}{Pando, K.M.},
  \bibinfo{author}{Danielson, L.}, \bibinfo{author}{Lee, C.T.},
  \bibinfo{year}{2010}.
\newblock \bibinfo{title}{{Partitioning of Mo, P and other siderophile elements
  (Cu, Ga, Sn, Ni, Co, Cr, Mn, V, and W) between metal and silicate melt as a
  function of temperature and silicate melt composition}}.
\newblock \bibinfo{journal}{Earth and Planetary Science Letters}
  \bibinfo{volume}{291}, \bibinfo{pages}{1--9}.
\newblock \DOIprefix\doi{10.1016/j.epsl.2009.12.018}.
\bibitem[{Rivoldini et~al.(2011)Rivoldini, Van~Hoolst, Verhoeven, Mocquet and
  Dehant}]{rivoldini2011geodesy}
\bibinfo{author}{Rivoldini, A.}, \bibinfo{author}{Van~Hoolst, T.},
  \bibinfo{author}{Verhoeven, O.}, \bibinfo{author}{Mocquet, A.},
  \bibinfo{author}{Dehant, V.}, \bibinfo{year}{2011}.
\newblock \bibinfo{title}{{Geodesy constraints on the interior structure and
  composition of Mars}}.
\newblock \bibinfo{journal}{Icarus} \bibinfo{volume}{213},
  \bibinfo{pages}{451--472}.
\newblock \DOIprefix\doi{10.1016/j.icarus.2011.03.024}.
\bibitem[{Rose and Brenan(2001)}]{rose2001wetting}
\bibinfo{author}{Rose, L.A.}, \bibinfo{author}{Brenan, J.M.},
  \bibinfo{year}{2001}.
\newblock \bibinfo{title}{{Wetting properties of Fe-Ni-Co-Cu-Os melts against
  olivine: Implications for sulfide melt mobility}}.
\newblock \bibinfo{journal}{Economic Geology} \bibinfo{volume}{96},
  \bibinfo{pages}{145--157}.
\newblock \DOIprefix\doi{10.2113/gsecongeo.96.1.145}.
\bibitem[{Rose-Weston et~al.(2009)Rose-Weston, Brenan, Fei, Secco and
  Frost}]{rose2009effect}
\bibinfo{author}{Rose-Weston, L.}, \bibinfo{author}{Brenan, J.M.},
  \bibinfo{author}{Fei, Y.}, \bibinfo{author}{Secco, R.A.},
  \bibinfo{author}{Frost, D.J.}, \bibinfo{year}{2009}.
\newblock \bibinfo{title}{{Effect of pressure, temperature, and oxygen fugacity
  on the metal-silicate partitioning of Te, Se, and S: Implications for earth
  differentiation}}.
\newblock \bibinfo{journal}{Geochimica et Cosmochimica Acta}
  \bibinfo{volume}{73}, \bibinfo{pages}{4598--4615}.
\newblock \DOIprefix\doi{10.1016/j.gca.2009.04.028}.
\bibitem[{Rubie et~al.(2015)Rubie, Jacobson, Morbidelli, O'Brien, Young,
  de~Vries, Nimmo, Palme and Frost}]{rubie2015accretion}
\bibinfo{author}{Rubie, D.C.}, \bibinfo{author}{Jacobson, S.A.},
  \bibinfo{author}{Morbidelli, A.}, \bibinfo{author}{O'Brien, D.P.},
  \bibinfo{author}{Young, E.D.}, \bibinfo{author}{de~Vries, J.},
  \bibinfo{author}{Nimmo, F.}, \bibinfo{author}{Palme, H.},
  \bibinfo{author}{Frost, D.J.}, \bibinfo{year}{2015}.
\newblock \bibinfo{title}{{Accretion and differentiation of the terrestrial
  planets with implications for the compositions of early-formed Solar System
  bodies and accretion of water}}.
\newblock \bibinfo{journal}{Icarus} \bibinfo{volume}{248},
  \bibinfo{pages}{89--108}.
\newblock \DOIprefix\doi{10.1016/j.icarus.2014.10.015}.
\bibitem[{Rubie et~al.(2016)Rubie, Laurenz, Jacobson, Morbidelli, Palme, Vogel
  and Frost}]{rubie2016highly}
\bibinfo{author}{Rubie, D.C.}, \bibinfo{author}{Laurenz, V.},
  \bibinfo{author}{Jacobson, S.A.}, \bibinfo{author}{Morbidelli, A.},
  \bibinfo{author}{Palme, H.}, \bibinfo{author}{Vogel, A.K.},
  \bibinfo{author}{Frost, D.J.}, \bibinfo{year}{2016}.
\newblock \bibinfo{title}{{Highly siderophile elements were stripped from
  Earth's mantle by iron sulfide segregation}}.
\newblock \bibinfo{journal}{Science} \bibinfo{volume}{353},
  \bibinfo{pages}{1141--1144}.
\newblock \DOIprefix\doi{10.1126/science.aaf6919}.
\bibitem[{Rubie et~al.(2003)Rubie, Melosh, Reid, Liebske and
  Righter}]{rubie2003mechanisms}
\bibinfo{author}{Rubie, D.C.}, \bibinfo{author}{Melosh, H.J.},
  \bibinfo{author}{Reid, J.E.}, \bibinfo{author}{Liebske, C.},
  \bibinfo{author}{Righter, K.}, \bibinfo{year}{2003}.
\newblock \bibinfo{title}{{Mechanisms of metal--silicate equilibration in the
  terrestrial magma ocean}}.
\newblock \bibinfo{journal}{Earth and Planetary Science Letters}
  \bibinfo{volume}{205}, \bibinfo{pages}{239--255}.
\newblock \DOIprefix\doi{10.1016/S0012-821X(02)01044-0}.
\bibitem[{Rubin(2011)}]{rubin2011origin}
\bibinfo{author}{Rubin, A.E.}, \bibinfo{year}{2011}.
\newblock \bibinfo{title}{{Origin of the differences in
  refractory-lithophile-element abundances among chondrite groups}}.
\newblock \bibinfo{journal}{Icarus} \bibinfo{volume}{213},
  \bibinfo{pages}{547--558}.
\newblock \DOIprefix\doi{10.1016/j.icarus.2011.04.003}.
\bibitem[{Rudnick and Gao(2014)}]{rudnick2014composition}
\bibinfo{author}{Rudnick, R.L.}, \bibinfo{author}{Gao, S.},
  \bibinfo{year}{2014}.
\newblock \bibinfo{title}{Composition of the continental crust}, in:
  \bibinfo{editor}{Holland, H.D.}, \bibinfo{editor}{Turekian, K.K.} (Eds.),
  \bibinfo{booktitle}{Treatise on Geochemistry (Second Edition)}.
  \bibinfo{publisher}{Elsevier}, \bibinfo{address}{Oxford}, pp.
  \bibinfo{pages}{1--51}.
\newblock \DOIprefix\doi{10.1016/B978-0-08-095975-7.00301-6}.
\bibitem[{Samuel(2012)}]{samuel2012reevaluation}
\bibinfo{author}{Samuel, H.}, \bibinfo{year}{2012}.
\newblock \bibinfo{title}{{A re-evaluation of metal diapir breakup and
  equilibration in terrestrial magma oceans}}.
\newblock \bibinfo{journal}{Earth and Planetary Science Letters}
  \bibinfo{volume}{313}, \bibinfo{pages}{105--114}.
\newblock \DOIprefix\doi{10.1016/j.epsl.2011.11.001}.
\bibitem[{Samuel et~al.(2019)Samuel, Lognonn{\'e}, Panning and
  Lainey}]{samuel2019rheology}
\bibinfo{author}{Samuel, H.}, \bibinfo{author}{Lognonn{\'e}, P.},
  \bibinfo{author}{Panning, M.}, \bibinfo{author}{Lainey, V.},
  \bibinfo{year}{2019}.
\newblock \bibinfo{title}{{The rheology and thermal history of Mars revealed by
  the orbital evolution of Phobos}}.
\newblock \bibinfo{journal}{Nature} \bibinfo{volume}{569},
  \bibinfo{pages}{523--527}.
\newblock \DOIprefix\doi{10.1038/s41586-019-1202-7}.
\bibitem[{Scott and Krot(2014)}]{scott2014chondrites}
\bibinfo{author}{Scott, E.R.D.}, \bibinfo{author}{Krot, A.N.},
  \bibinfo{year}{2014}.
\newblock \bibinfo{title}{{Chondrites and their Components}}, in:
  \bibinfo{editor}{Holland, H.D.}, \bibinfo{editor}{Turekian, K.K.} (Eds.),
  \bibinfo{booktitle}{Treatise on Geochemistry (Second Edition)}.
  \bibinfo{publisher}{Elsevier}, \bibinfo{address}{Oxford}.
  volume~\bibinfo{volume}{1}, pp. \bibinfo{pages}{65--137}.
\newblock \DOIprefix\doi{10.1016/B978-0-08-095975-7.00104-2}.
\bibitem[{Sharp and Draper(2013)}]{sharp2013chlorine}
\bibinfo{author}{Sharp, Z.D.}, \bibinfo{author}{Draper, D.S.},
  \bibinfo{year}{2013}.
\newblock \bibinfo{title}{{The chlorine abundance of Earth: implications for a
  habitable planet}}.
\newblock \bibinfo{journal}{Earth and Planetary Science Letters}
  \bibinfo{volume}{369}, \bibinfo{pages}{71--77}.
\newblock \DOIprefix\doi{10.1016/j.epsl.2013.03.005}.
\bibitem[{Siebert et~al.(2011)Siebert, Corgne and
  Ryerson}]{siebert2011systematics}
\bibinfo{author}{Siebert, J.}, \bibinfo{author}{Corgne, A.},
  \bibinfo{author}{Ryerson, F.J.}, \bibinfo{year}{2011}.
\newblock \bibinfo{title}{{Systematics of metal--silicate partitioning for many
  siderophile elements applied to Earth's core formation}}.
\newblock \bibinfo{journal}{Geochimica et Cosmochimica Acta}
  \bibinfo{volume}{75}, \bibinfo{pages}{1451--1489}.
\newblock \DOIprefix\doi{10.1016/j.gca.2010.12.013}.
\bibitem[{Siebert et~al.(2018)Siebert, Sossi, Blanchard, Mahan, Badro and
  Moynier}]{siebert2018chondritic}
\bibinfo{author}{Siebert, J.}, \bibinfo{author}{Sossi, P.A.},
  \bibinfo{author}{Blanchard, I.}, \bibinfo{author}{Mahan, B.},
  \bibinfo{author}{Badro, J.}, \bibinfo{author}{Moynier, F.},
  \bibinfo{year}{2018}.
\newblock \bibinfo{title}{{Chondritic Mn/Na ratio and limited post-nebular
  volatile loss of the Earth}}.
\newblock \bibinfo{journal}{Earth and Planetary Science Letters}
  \bibinfo{volume}{485}, \bibinfo{pages}{130--139}.
\newblock \DOIprefix\doi{10.1016/j.epsl.2017.12.042}.
\bibitem[{Sossi and Fegley(2018)}]{sossi2018thermodynamics}
\bibinfo{author}{Sossi, P.A.}, \bibinfo{author}{Fegley, Jr., B.},
  \bibinfo{year}{2018}.
\newblock \bibinfo{title}{{Thermodynamics of element volatility and its
  application to planetary processes}}.
\newblock \bibinfo{journal}{Reviews in Mineralogy and Geochemistry}
  \bibinfo{volume}{84}, \bibinfo{pages}{393--459}.
\newblock \DOIprefix\doi{10.2138/rmg.2018.84.11}.
\bibitem[{Sossi et~al.(2019)Sossi, Klemme, O'Neill, Berndt and
  Moynier}]{sossi2019evaporation}
\bibinfo{author}{Sossi, P.A.}, \bibinfo{author}{Klemme, S.},
  \bibinfo{author}{O'Neill, H.S.C.}, \bibinfo{author}{Berndt, J.},
  \bibinfo{author}{Moynier, F.}, \bibinfo{year}{2019}.
\newblock \bibinfo{title}{{Evaporation of moderately volatile elements from
  silicate melts: Experiments and theory}}.
\newblock \bibinfo{journal}{Geochimica et Cosmochimica Acta}
  \bibinfo{volume}{260}, \bibinfo{pages}{204--231}.
\newblock \DOIprefix\doi{10.1016/j.gca.2019.06.021}.
\bibitem[{Sossi et~al.(2016)Sossi, Nebel, Anand and Poitrasson}]{sossi2016on}
\bibinfo{author}{Sossi, P.A.}, \bibinfo{author}{Nebel, O.},
  \bibinfo{author}{Anand, M.}, \bibinfo{author}{Poitrasson, F.},
  \bibinfo{year}{2016}.
\newblock \bibinfo{title}{{On the iron isotope composition of Mars and volatile
  depletion in the terrestrial planets}}.
\newblock \bibinfo{journal}{Earth and Planetary Science Letters}
  \bibinfo{volume}{449}, \bibinfo{pages}{360--371}.
\newblock \DOIprefix\doi{10.1016/j.epsl.2016.05.030}.
\bibitem[{Sossi et~al.(2018)Sossi, Nebel, O'Neill and Moynier}]{sossi2018zinc}
\bibinfo{author}{Sossi, P.A.}, \bibinfo{author}{Nebel, O.},
  \bibinfo{author}{O'Neill, H.S.C.}, \bibinfo{author}{Moynier, F.},
  \bibinfo{year}{2018}.
\newblock \bibinfo{title}{{Zinc isotope composition of the Earth and its
  behaviour during planetary accretion}}.
\newblock \bibinfo{journal}{Chemical Geology} \bibinfo{volume}{477},
  \bibinfo{pages}{73--84}.
\newblock \DOIprefix\doi{10.1016/j.chemgeo.2017.12.006}.
\bibitem[{Steenstra et~al.(2020)Steenstra, van Haaster, van Mulligen,
  Flemetakis, Berndt, Klemme and van Westrenen}]{steenstra2020experimental}
\bibinfo{author}{Steenstra, E.S.}, \bibinfo{author}{van Haaster, F.},
  \bibinfo{author}{van Mulligen, R.}, \bibinfo{author}{Flemetakis, S.},
  \bibinfo{author}{Berndt, J.}, \bibinfo{author}{Klemme, S.},
  \bibinfo{author}{van Westrenen, W.}, \bibinfo{year}{2020}.
\newblock \bibinfo{title}{{An experimental assessment of the chalcophile
  behavior of F, Cl, Br and I: implications for the fate of halogens during
  planetary accretion and the formation of magmatic ore deposits}}.
\newblock \bibinfo{journal}{Geochimica et Cosmochimica Acta}
  \bibinfo{volume}{273}, \bibinfo{pages}{275--290}.
\newblock \DOIprefix\doi{10.1016/j.gca.2020.01.006}.
\bibitem[{Stracke et~al.(2012)Stracke, Palme, Gellissen, M{\"u}nker, Kleine,
  Birbaum, G{\"u}nther, Bourdon and Zipfel}]{stracke2012refractory}
\bibinfo{author}{Stracke, A.}, \bibinfo{author}{Palme, H.},
  \bibinfo{author}{Gellissen, M.}, \bibinfo{author}{M{\"u}nker, C.},
  \bibinfo{author}{Kleine, T.}, \bibinfo{author}{Birbaum, K.},
  \bibinfo{author}{G{\"u}nther, D.}, \bibinfo{author}{Bourdon, B.},
  \bibinfo{author}{Zipfel, J.}, \bibinfo{year}{2012}.
\newblock \bibinfo{title}{{Refractory element fractionation in the Allende
  meteorite: Implications for solar nebula condensation and the chondritic
  composition of planetary bodies}}.
\newblock \bibinfo{journal}{Geochimica et Cosmochimica Acta}
  \bibinfo{volume}{85}, \bibinfo{pages}{114--141}.
\newblock \DOIprefix\doi{10.1016/j.gca.2012.02.006}.
\bibitem[{Suer et~al.(2017)Suer, Siebert, Remusat, Menguy and
  Fiquet}]{suer2017sulfur}
\bibinfo{author}{Suer, T.A.}, \bibinfo{author}{Siebert, J.},
  \bibinfo{author}{Remusat, L.}, \bibinfo{author}{Menguy, N.},
  \bibinfo{author}{Fiquet, G.}, \bibinfo{year}{2017}.
\newblock \bibinfo{title}{{A sulfur-poor terrestrial core inferred from
  metal--silicate partitioning experiments}}.
\newblock \bibinfo{journal}{Earth and Planetary Science Letters}
  \bibinfo{volume}{469}, \bibinfo{pages}{84--97}.
\newblock \DOIprefix\doi{10.1016/j.epsl.2017.04.016}.
\bibitem[{Taylor(2013)}]{taylor2013bulk}
\bibinfo{author}{Taylor, G.J.}, \bibinfo{year}{2013}.
\newblock \bibinfo{title}{{The bulk composition of Mars}}.
\newblock \bibinfo{journal}{Chemie der Erde-Geochemistry} \bibinfo{volume}{73},
  \bibinfo{pages}{401--420}.
\newblock \DOIprefix\doi{10.1016/j.chemer.2013.09.006}.
\bibitem[{Taylor and McLennan(2009)}]{taylor2009planetary}
\bibinfo{author}{Taylor, S.R.}, \bibinfo{author}{McLennan, S.},
  \bibinfo{year}{2009}.
\newblock \bibinfo{title}{{Planetary Crusts: Their Composition, Origin and
  Evolution}}. volume~\bibinfo{volume}{10}.
\newblock \bibinfo{publisher}{Cambridge University Press},
  \bibinfo{address}{Cambridge}.
\newblock \DOIprefix\doi{10.1017/CBO9780511575358}.
\bibitem[{Turcotte and Schubert(2014)}]{turcotte2014geodynamics}
\bibinfo{author}{Turcotte, D.}, \bibinfo{author}{Schubert, G.},
  \bibinfo{year}{2014}.
\newblock \bibinfo{title}{{Geodynamics (3rd Edition)}}.
\newblock \bibinfo{publisher}{Cambridge University Press},
  \bibinfo{address}{Cambridge}.
\newblock \DOIprefix\doi{10.1017/cbo9780511843877}.
\bibitem[{Varas-Reus et~al.(2019)Varas-Reus, K{\"o}nig, Yierpan, Lorand and
  Schoenberg}]{varas2019selenium}
\bibinfo{author}{Varas-Reus, M.I.}, \bibinfo{author}{K{\"o}nig, S.},
  \bibinfo{author}{Yierpan, A.}, \bibinfo{author}{Lorand, J.P.},
  \bibinfo{author}{Schoenberg, R.}, \bibinfo{year}{2019}.
\newblock \bibinfo{title}{{Selenium isotopes as tracers of a late volatile
  contribution to Earth from the outer Solar System}}.
\newblock \bibinfo{journal}{Nature Geoscience} \bibinfo{volume}{12},
  \bibinfo{pages}{779--782}.
\newblock \DOIprefix\doi{10.1038/s41561-019-0414-7}.
\bibitem[{Wade et~al.(2012)Wade, Wood and Tuff}]{wade2012metal}
\bibinfo{author}{Wade, J.}, \bibinfo{author}{Wood, B.J.},
  \bibinfo{author}{Tuff, J.}, \bibinfo{year}{2012}.
\newblock \bibinfo{title}{{Metal--silicate partitioning of Mo and W at high
  pressures and temperatures: evidence for late accretion of sulphur to the
  Earth}}.
\newblock \bibinfo{journal}{Geochimica et Cosmochimica Acta}
  \bibinfo{volume}{85}, \bibinfo{pages}{58--74}.
\newblock \DOIprefix\doi{10.1016/j.gca.2012.01.010}.
\bibitem[{Walker et~al.(2015)Walker, Bermingham, Liu, Puchtel, Touboul and
  Worsham}]{walker2015search}
\bibinfo{author}{Walker, R.J.}, \bibinfo{author}{Bermingham, K.},
  \bibinfo{author}{Liu, J.}, \bibinfo{author}{Puchtel, I.S.},
  \bibinfo{author}{Touboul, M.}, \bibinfo{author}{Worsham, E.A.},
  \bibinfo{year}{2015}.
\newblock \bibinfo{title}{{In search of late-stage planetary building blocks}}.
\newblock \bibinfo{journal}{Chemical Geology} \bibinfo{volume}{411},
  \bibinfo{pages}{125--142}.
\newblock \DOIprefix\doi{10.1016/j.chemgeo.2015.06.028}.
\bibitem[{Walker et~al.(2002)Walker, Horan, Morgan, Becker, Grossman and
  Rubin}]{walker2002comparative}
\bibinfo{author}{Walker, R.J.}, \bibinfo{author}{Horan, M.F.},
  \bibinfo{author}{Morgan, J.W.}, \bibinfo{author}{Becker, H.},
  \bibinfo{author}{Grossman, J.N.}, \bibinfo{author}{Rubin, A.E.},
  \bibinfo{year}{2002}.
\newblock \bibinfo{title}{{Comparative \ce{^{187}Re}-\ce{^{187}Os} systematics
  of chondrites: Implications regarding early solar system processes}}.
\newblock \bibinfo{journal}{Geochimica et Cosmochimica Acta}
  \bibinfo{volume}{66}, \bibinfo{pages}{4187--4201}.
\newblock \DOIprefix\doi{10.1016/S0016-7037(02)01003-7}.
\bibitem[{Wang and Becker(2013)}]{wang2013ratios}
\bibinfo{author}{Wang, Z.}, \bibinfo{author}{Becker, H.}, \bibinfo{year}{2013}.
\newblock \bibinfo{title}{{Ratios of S, Se and Te in the silicate Earth require
  a volatile-rich late veneer}}.
\newblock \bibinfo{journal}{Nature} \bibinfo{volume}{499},
  \bibinfo{pages}{328--331}.
\newblock \DOIprefix\doi{10.1038/nature12285}.
\bibitem[{W{\"a}nke and Dreibus(1994)}]{wanke1994chemistry}
\bibinfo{author}{W{\"a}nke, H.}, \bibinfo{author}{Dreibus, G.},
  \bibinfo{year}{1994}.
\newblock \bibinfo{title}{{Chemistry and accretion history of Mars}}.
\newblock \bibinfo{journal}{Philosophical Transactions of the Royal Society of
  London. Series A, Mathematical and Physical Sciences} \bibinfo{volume}{349},
  \bibinfo{pages}{285--293}.
\newblock \DOIprefix\doi{10.1098/rsta.1994.0132}.
\bibitem[{W{\"a}nke et~al.(1984)W{\"a}nke, Dreibus and
  Jagoutz}]{wanke1984mantle}
\bibinfo{author}{W{\"a}nke, H.}, \bibinfo{author}{Dreibus, G.},
  \bibinfo{author}{Jagoutz, E.}, \bibinfo{year}{1984}.
\newblock \bibinfo{title}{{Mantle chemistry and accretion history of the
  Earth}}, in: \bibinfo{editor}{Kr{\"o}ner, A.}, \bibinfo{editor}{Hanson,
  G.N.}, \bibinfo{editor}{Goodwin, A.M.} (Eds.), \bibinfo{booktitle}{Archaean
  Geochemistry}. \bibinfo{publisher}{Springer}, pp. \bibinfo{pages}{1--24}.
\newblock \DOIprefix\doi{10.1007/978-3-642-70001-9_1}.
\bibitem[{Wasson(1985)}]{wasson1985meteorites}
\bibinfo{author}{Wasson, J.T.}, \bibinfo{year}{1985}.
\newblock \bibinfo{title}{{Meteorites: Their Record of Early Solar-System
  History}}.
\newblock \bibinfo{publisher}{W.H. Freeman and Company}, \bibinfo{address}{New
  York}.
\bibitem[{Wasson and Kallemeyn(1988)}]{wasson1988compositions}
\bibinfo{author}{Wasson, J.T.}, \bibinfo{author}{Kallemeyn, G.W.},
  \bibinfo{year}{1988}.
\newblock \bibinfo{title}{{Compositions of chondrites}}.
\newblock \bibinfo{journal}{Philosophical Transactions of the Royal Society of
  London A: Mathematical, Physical and Engineering Sciences}
  \bibinfo{volume}{325}, \bibinfo{pages}{535--544}.
\newblock \DOIprefix\doi{10.1098/rsta.1988.0066}.
\bibitem[{Wipperfurth et~al.(2020)Wipperfurth, {\v{S}}r{\'a}mek and
  McDonough}]{wipperfurth2019reference}
\bibinfo{author}{Wipperfurth, S.A.}, \bibinfo{author}{{\v{S}}r{\'a}mek, O.},
  \bibinfo{author}{McDonough, W.F.}, \bibinfo{year}{2020}.
\newblock \bibinfo{title}{Reference models for lithospheric geoneutrino
  signal}.
\newblock \bibinfo{journal}{Journal of Geophysical Research: Solid Earth}
  \bibinfo{volume}{125}, \bibinfo{pages}{e2019JB018433}.
\newblock \DOIprefix\doi{10.1029/2019JB018433}.
\bibitem[{Wohlers and Wood(2017)}]{wohlers2017uranium}
\bibinfo{author}{Wohlers, A.}, \bibinfo{author}{Wood, B.J.},
  \bibinfo{year}{2017}.
\newblock \bibinfo{title}{{Uranium, thorium and REE partitioning into sulfide
  liquids: Implications for reduced S-rich bodies}}.
\newblock \bibinfo{journal}{Geochimica et Cosmochimica Acta}
  \bibinfo{volume}{205}, \bibinfo{pages}{226--244}.
\newblock \DOIprefix\doi{10.1016/j.gca.2017.01.050}.
\bibitem[{Wombacher et~al.(2008)Wombacher, Rehk{\"a}mper, Mezger, Bischoff and
  M{\"u}nker}]{wombacher2008cadmium}
\bibinfo{author}{Wombacher, F.}, \bibinfo{author}{Rehk{\"a}mper, M.},
  \bibinfo{author}{Mezger, K.}, \bibinfo{author}{Bischoff, A.},
  \bibinfo{author}{M{\"u}nker, C.}, \bibinfo{year}{2008}.
\newblock \bibinfo{title}{{Cadmium stable isotope cosmochemistry}}.
\newblock \bibinfo{journal}{Geochimica et Cosmochimica Acta}
  \bibinfo{volume}{72}, \bibinfo{pages}{646--667}.
\newblock \DOIprefix\doi{10.1016/j.gca.2007.10.024}.
\bibitem[{Wood and Halliday(2010)}]{wood2010lead}
\bibinfo{author}{Wood, B.J.}, \bibinfo{author}{Halliday, A.N.},
  \bibinfo{year}{2010}.
\newblock \bibinfo{title}{{The lead isotopic age of the Earth can be explained
  by core formation alone}}.
\newblock \bibinfo{journal}{Nature} \bibinfo{volume}{465},
  \bibinfo{pages}{767--770}.
\newblock \DOIprefix\doi{10.1038/nature09072}.
\bibitem[{Wood et~al.(2014)Wood, Kiseeva and Mirolo}]{wood2014accretion}
\bibinfo{author}{Wood, B.J.}, \bibinfo{author}{Kiseeva, E.S.},
  \bibinfo{author}{Mirolo, F.J.}, \bibinfo{year}{2014}.
\newblock \bibinfo{title}{{Accretion and core formation: The effects of sulfur
  on metal--silicate partition coefficients}}.
\newblock \bibinfo{journal}{Geochimica et Cosmochimica Acta}
  \bibinfo{volume}{145}, \bibinfo{pages}{248--267}.
\newblock \DOIprefix\doi{10.1016/j.gca.2014.09.002}.
\bibitem[{Wood et~al.(2019)Wood, Smythe and Harrison}]{wood2019condensation}
\bibinfo{author}{Wood, B.J.}, \bibinfo{author}{Smythe, D.J.},
  \bibinfo{author}{Harrison, T.}, \bibinfo{year}{2019}.
\newblock \bibinfo{title}{{The condensation temperatures of the elements: A
  reappraisal}}.
\newblock \bibinfo{journal}{American Mineralogist} \bibinfo{volume}{104},
  \bibinfo{pages}{844--856}.
\newblock \DOIprefix\doi{10.2138/am-2019-6852CCBY}.
\bibitem[{Yang et~al.(2015)Yang, Humayun, Righter, Jefferson, Fields and
  Irving}]{yang2015siderophile}
\bibinfo{author}{Yang, S.}, \bibinfo{author}{Humayun, M.},
  \bibinfo{author}{Righter, K.}, \bibinfo{author}{Jefferson, G.},
  \bibinfo{author}{Fields, D.}, \bibinfo{author}{Irving, A.J.},
  \bibinfo{year}{2015}.
\newblock \bibinfo{title}{{Siderophile and chalcophile element abundances in
  shergottites: Implications for Martian core formation}}.
\newblock \bibinfo{journal}{Meteoritics \& Planetary Science}
  \bibinfo{volume}{50}, \bibinfo{pages}{691--714}.
\newblock \DOIprefix\doi{10.1111/maps.12384}.
\bibitem[{Yi et~al.(2000)Yi, Halliday, Alt, Lee, Rehk{\"a}mper, Garcia,
  Langmuir and Su}]{yi2000cadmium}
\bibinfo{author}{Yi, W.}, \bibinfo{author}{Halliday, A.N.},
  \bibinfo{author}{Alt, J.C.}, \bibinfo{author}{Lee, D.C.},
  \bibinfo{author}{Rehk{\"a}mper, M.}, \bibinfo{author}{Garcia, M.O.},
  \bibinfo{author}{Langmuir, C.H.}, \bibinfo{author}{Su, Y.},
  \bibinfo{year}{2000}.
\newblock \bibinfo{title}{{Cadmium, indium, tin, tellurium, and sulfur in
  oceanic basalts: Implications for chalcophile element fractionation in the
  Earth}}.
\newblock \bibinfo{journal}{Journal of Geophysical Research: Solid Earth}
  \bibinfo{volume}{105}, \bibinfo{pages}{18927--18948}.
\newblock \DOIprefix\doi{10.1029/2000JB900152}.
\bibitem[{Yoshizaki et~al.(2018)Yoshizaki, Ash, Yokoyama, Lipella and
  McDonough}]{yoshizaki2018chemically}
\bibinfo{author}{Yoshizaki, T.}, \bibinfo{author}{Ash, R.D.},
  \bibinfo{author}{Yokoyama, T.}, \bibinfo{author}{Lipella, M.D.},
  \bibinfo{author}{McDonough, W.F.}, \bibinfo{year}{2018}.
\newblock \bibinfo{title}{{Chemically defining the building blocks of the
  Earth}}.
\newblock \bibinfo{journal}{arXiv preprint arXiv:1812.11717} .
\bibitem[{Yoshizaki and McDonough(2020)}]{yoshizaki2019mars_long}
\bibinfo{author}{Yoshizaki, T.}, \bibinfo{author}{McDonough, W.F.},
  \bibinfo{year}{2020}.
\newblock \bibinfo{title}{{The composition of Mars}}.
\newblock \bibinfo{journal}{Geochimica et Cosmochimica Acta}
  \bibinfo{volume}{273}, \bibinfo{pages}{137--162}.
\newblock \DOIprefix\doi{10.1016/j.gca.2020.01.011}.
\bibitem[{Zambardi et~al.(2013)Zambardi, Poitrasson, Corgne, M{\'e}heut,
  Quitt{\'e} and Anand}]{zambardi2013silicon}
\bibinfo{author}{Zambardi, T.}, \bibinfo{author}{Poitrasson, F.},
  \bibinfo{author}{Corgne, A.}, \bibinfo{author}{M{\'e}heut, M.},
  \bibinfo{author}{Quitt{\'e}, G.}, \bibinfo{author}{Anand, M.},
  \bibinfo{year}{2013}.
\newblock \bibinfo{title}{{Silicon isotope variations in the inner solar
  system: Implications for planetary formation, differentiation and
  composition}}.
\newblock \bibinfo{journal}{Geochimica et Cosmochimica Acta}
  \bibinfo{volume}{121}, \bibinfo{pages}{67--83}.
\newblock \DOIprefix\doi{10.1016/j.gca.2013.06.040}.
\bibitem[{Zolotov and Mironenko(2007)}]{zolotov2007hydrogen}
\bibinfo{author}{Zolotov, M.Y.}, \bibinfo{author}{Mironenko, M.V.},
  \bibinfo{year}{2007}.
\newblock \bibinfo{title}{{Hydrogen chloride as a source of acid fluids in
  parent bodies of chondrites}}, in: \bibinfo{booktitle}{Lunar and Planetary
  Science Conference}, p. \bibinfo{pages}{2340}.

\end{thebibliography}

\renewcommand{\refname}{Supplementary references}
\putbib[myrefs]
\end{bibunit}

\end{appendices}

\end{document}